\documentclass[twocolumn]{aastex701}

\usepackage{amsmath}
\usepackage{graphicx}
\usepackage{subcaption}
\usepackage{booktabs}
\usepackage{multirow}
\usepackage{threeparttable}
\usepackage{float}
\usepackage{hyperref}
\usepackage[noabbrev]{cleveref}

\begin{document}

\title{Constraining gamma-ray burst viewing angles with Swift/XRT afterglow light curves}

\correspondingauthor{Shuang-Xi Yi, Yuan-Chuan Zou}

\author{Cheng-Jie Sun}
\affiliation{School of Physics and Physical Engineering, Qufu Normal University, Qufu 273165, China}
\email{}

\author[orcid=0000-0003-0672-5646]{Shuang-Xi Yi$^{\dag}$}
\affiliation{School of Physics and Physical Engineering, Qufu Normal University, Qufu 273165, China}
\email[show]{yisx2015@qfnu.edu.cn}

\author{Lin Zhou}
\affiliation{Department of Astronomy, School of Physics, Huazhong University of Science and Technology, Wuhan 430074, China}
\email{}

\author{Yuan-Chuan Zou$^{\dag}$}
\affiliation{Department of Astronomy, School of Physics, Huazhong University of Science and Technology, Wuhan 430074, China}
\email[show]{zouyc@hust.edu.cn}

\author{Yu-Peng Yang}
\affiliation{School of Physics and Physical Engineering, Qufu Normal University, Qufu 273165, China}
\email{ypyang@qfnu.edu.cn}

\author{Si-Ji Xin}
\affiliation{School of Physics and Physical Engineering, Qufu Normal University, Qufu 273165, China}
\email{}

\author{Yan-Kun Qu}
\affiliation{School of Physics and Physical Engineering, Qufu Normal University, Qufu 273165, China}
\email{}

\author{Wen-Long Zhang}
\affiliation{Purple Mountain Observatory, Chinese Academy of Sciences, Nanjing 210023, China}
\affiliation{School of Astronomy and Space Sciences, University of Science and Technology of China, Hefei 230026, China}
\email{}

\author{Fa-Yin Wang}
\affiliation{School of Astronomy and Space Science, Nanjing University, Nanjing 210023, China}
\email{fayinwang@nju.edu.cn}

\begin{abstract}
  Gamma-ray bursts (GRBs) are among the most energetic phenomena in the universe, and their afterglow light curves provide insights into jet geometry and viewing angle. The interaction between relativistic ejecta and the circumburst medium produces afterglow emission, and jet break features offer crucial information about viewing geometry.
  To constrain GRB viewing angles by analyzing jet break features in the Swift X-Ray Telescope afterglow light curves, we constructed and compared two models: a simplified geometric model without high-latitude emission (model~1) and a comprehensive model with high-latitude emission (model~2). Employing these two models, we examined viewing angles and off-axis ratios ($q = \theta_{\text{obs}} / \theta_{\text{jet}}$) under different density profiles, and evaluated the impact of high-latitude emission on our modeling approach.
  Both models were applied to a selected sample of 20 GRBs in the interstellar medium (ISM) and 20 GRBs in the wind medium, with criteria ensuring that the jet breaks were attributed to the edge effect and that data coverage was sufficient. Markov Chain Monte Carlo methods were employed for parameter estimation.
  Model comparison based on the reduced chi-squared ($\chi^2_{\rm red}$) and the Bayesian information criterion (BIC) indicates that model~1 provides a better fit to the observed light curves for all GRBs in the sample. We find that most GRBs in the sample have small off-axis ratios ($\bar{q} = 0.1851$ for model~1), suggesting that the viewing angles are generally close to the jet axis; the distribution of viewing angles in logarithmic space follows an approximately Gaussian pattern. A Kolmogorov-Smirnov test reveals no significant difference in the distribution of off-axis ratios between ISM and wind medium. 
  We find no significant difference in the off-axis ratio between bursts with and without an X-ray plateau. While the viewing angle decreases significantly with redshift, the off-axis ratio shows no significant evolution, consistent with the degree of off-axis alignment being independent of cosmic epoch.
\end{abstract}


\section{Introduction}

Gamma-ray bursts (GRBs) can be classified into long-duration gamma-ray bursts (LGRBs; $T_{90} > 2$ s) and short-duration gamma-ray bursts (SGRBs; $T_{90} \leq 2$ s) based on their duration. Astronomers generally believe LGRBs originate from the core collapse of massive stars, while SGRBs arise from compact binary mergers \citep{1993ApJ...413L.101K, 1993ApJ...405..273W}. Progenitor types determine the properties of the circumburst medium: progenitors of LGRBs usually lose mass via stellar winds in their late evolutionary stages, forming a wind medium where density decreases with radius ($n \propto R^{-k}$, with a theoretical expectation of $k = 2$); on the other hand, SGRBs often occur in a uniform interstellar medium (ISM, $k = 0$). However, actual observations show that the density profile of LGRBs often deviates from standard theoretical predictions. For example, \citet{2013ApJ...776..120Y} found that the density parameter $k$ ranges between 0.4 and 1.4, with a typical value of $k \sim 1$, indicating a complex medium that may reflect new mass loss processes experienced by the progenitor in its final stages. 

The interaction between relativistic ejecta and the circumburst medium drives afterglow radiation \citep{1999ApJ...525..737R}. When a relativistic outflow propagates through a power-law medium ($n \propto R^{-k}$), its deceleration is dominated by the forward shock (FS). This shock accelerates ambient electrons to relativistic energies and amplifies magnetic fields, generating multi-wavelength afterglow emission via synchrotron radiation \citep{1997ApJ...485L...5W, 1997ApJ...476..232M}. The X-ray afterglow light curve typically exhibits a multi-segment structure, primarily consisting of five phases: (i) an initial steep decay phase, considered the tail of the prompt emission and explained by the ``curvature effect'' (which describes the geometric time delay of radiation from different parts of the relativistically expanding shell); (ii) a plateau or shallow decay phase, often related to energy injection or refreshed shocks; (iii) a normal decay phase, representing the standard evolution of the external shock in the ambient medium; (iv) a late-time steep decay phase that begins following a transition known as the jet break, which is identified by a steepening in the light curve decay; and (v) X-ray flares, likely resulting from central engine reactivation \citep{2006ApJ...642..389N, 2006ApJ...642..354Z, 2015ApJ...807...92Y, 2021MNRAS.507.1047Y}. 

Extensive statistical studies have revealed tight correlations among afterglow parameters, such as the luminosity--time correlation of the X-ray plateau phase. The intrinsic nature of this correlation has been confirmed after correcting for observational biases \citep{2008MNRAS.391L..79D, 2010ApJ...722L.215D, 2013ApJ...774..157D}. Similar plateau features have been identified across broad energy ranges, including high-energy Fermi Large Area Telescope light curves. These correlations have further been applied as cosmological probes \citep{2020ApJ...905L..26D, 2021ApJS..255...13D}, within the broader effort to address the Hubble constant tension \citep{2021ApJ...912..150D}. Complementarily, the jet break phase provides a direct geometric probe of jet structure. \citet{1997ApJ...487L...1R} first established that when the bulk Lorentz factor $\Gamma$ decreases to approximately $1/\theta_{\text{jet}}$, the finite angular extent of the jet becomes visible, producing a characteristic jet break in the afterglow light curve. This ``edge effect'' provides a direct link between light curve morphology and jet geometry, with the magnitude of the break $\Delta\alpha = (3-k)/(4-k)$ determined by the density profile of the circumburst medium \citep{2007RMxAC..27..140G, 2006RPPh...69.2259M}; the associated closure relations have recently been tested across multiple wavelengths \citep{2025ApJ...978...51D}. Off-axis viewing geometries produce smoother, more gradual transitions \citep{2010arXiv1012.5101G, 2022MNRAS.515..555B}. Off-axis afterglow modeling has been extensively explored across X-ray and high-energy bands, suggesting that afterglow features such as breaks and flares may be interpreted within a unified off-axis framework \citep{2020ApJ...905..112F, 2022ApJ...940..189F}.

The Large High Altitude Air Shower Observatory (LHAASO) recorded the TeV afterglow of the most luminous gamma-ray burst GRB 221009A, which displayed a pronounced jet break at roughly $670\,$s. \citet{2023Sci...380.1390L} reported a peak time $t_{\rm peak}\approx18\,\mathrm{s}$ and a jet half-opening angle $\theta_{\rm jet}\approx0.014\,\mathrm{rad}$. Building on these findings, \citet{2024MNRAS.532.2189Z} carried out a detailed analysis of the viewing angle, assuming a top-hat jet propagating through a uniform ISM. They formulated and compared two models to constrain the viewing angle: the first was a simplified geometric model that relates the time-dependent observable jet area to the growth of the relativistic beaming cone ($1/\Gamma$), adopting ISM scaling $\Gamma\propto t^{-3/8}$; and the second was a more comprehensive model that integrates radiation from different jet elements under the relativistic beaming effect. By fitting the LHAASO light curve with Markov Chain Monte Carlo (MCMC) methods, both models converged on very small viewing angle estimates, $\theta_{\rm obs}=9.4\times10^{-4}\,\mathrm{rad}$ and $\theta_{\rm obs}=5.96\times10^{-3}\,\mathrm{rad}$, both substantially smaller than the jet half-opening angle, thereby supporting an on-axis viewing geometry for this event.

Building upon the work of \citet{2024MNRAS.532.2189Z}, we enhanced the model's universality to achieve two core research objectives. First, unlike that work, which analyzed a single TeV source, we applied the improved two models to multiple X-ray afterglow light curve samples from the Swift X-Ray Telescope (XRT) archive to analyze the impact of high-latitude emission on viewing angle inference. Second, we generalized the model from a uniform ISM scenario to incorporate a wind medium, enabling the investigation of viewing angle distributions under different density profiles. 

The structure of this paper is as follows. 
Section~\ref{sec:Two top-hat jet models} presents the two top-hat jet models and the MCMC methodology, including the underlying physical assumptions and parameter estimation framework. 
Section~\ref{sec:Data sample} describes the sample selection criteria. 
Section~\ref{sec:Results} presents the fitting results and statistical analyses, covering model comparison, parameter distributions across environments, and potential correlations with other afterglow properties. 
Section~\ref{sec:Conclusion and discussion} summarizes the key findings, discusses selection biases and other limitations, and outlines directions for future work.

\section{Two top-hat jet models}\label{sec:Two top-hat jet models}

The top-hat jet model assumes a conical jet structure with uniform energy per solid angle within the jet opening angle and a sharp cutoff at the boundary \citep{1997ApJ...487L...1R, 2024MNRAS.530.2877L}. While real GRB jets are expected to possess angular structure rather than an ideal sharp edge \citep{2010arXiv1012.5101G}, this approximation remains widely adopted because it captures the essential jet break geometry with minimal free parameters. When the line of sight lies close to the jet axis, the light curve evolution around the jet break is comparable between top-hat and structured jets, as emission is dominated by the bright jet core where the edge effect is the primary cause of the break \citep{2022MNRAS.515..555B}. We therefore adopt the top-hat jet as a reasonable baseline for comparing the two models, acknowledging that systematic biases may arise for events with larger off-axis ratios or complex jet structures.

We assume the top-hat jet to undergo adiabatic evolution during the period encompassing the jet break, neglecting lateral expansion. The detected radiation originates from a narrow cone ($\theta(t) \approx 1/\Gamma(t)$) aligned with the line of sight; when the spectral index remains constant across a jet break, this indicates that the break is predominantly caused by the edge effect \citep{2007RMxAC..27..140G, 2010arXiv1012.5101G}. We verify in Section~\ref{sec:Data sample} that the spectral index remains constant across the jet break and that significant lateral expansion has not yet set in at the break for all GRBs in our sample. The bulk Lorentz factor $\Gamma(t)$ evolves as proposed by \citet{1998ApJ...497L..17S, 2019pgrb.book.....Z} for the ISM:
\begin{equation}
\begin{split}
\Gamma(t) &= \left( \frac{17 E (1+z)^3}{1024\pi n m_p c^5 t^3} \right)^{1/8} \\ 
&= 436 E_{52}^{1/8} n_0^{-1/8} \left(\frac{t}{1+z}\right)^{-3/8},
\label{eq:bm_ism}
\end{split}
\end{equation}
where the blastwave energy $E$ relates to the isotropic gamma-ray energy via $E = E_{\gamma,\text{iso}}/\eta_\gamma$, with $\eta_\gamma$ representing the gamma-ray radiation efficiency \citep{2010ApJ...725.2209L}. Here $E_{52} \equiv E/10^{52}$ erg is the normalized blastwave energy. For a wind medium with a density profile $n \propto R^{-2}$ ($k=2$), $\Gamma(t)$ follows \citep{2000ApJ...536..195C}:
\begin{equation}
\Gamma(t) = 5.9 \left( \frac{1+z}{2} \right)^{1/4} E_{52}^{1/4} A_*^{-1/4} t_{\text{day}}^{-1/4},
\label{eq:bm_wind}
\end{equation}
where $A_*$ is the wind density parameter.

\subsection{Top-hat jet model without high-latitude emission}

\begin{figure}[H]
\centering
\includegraphics[width=0.45\textwidth]{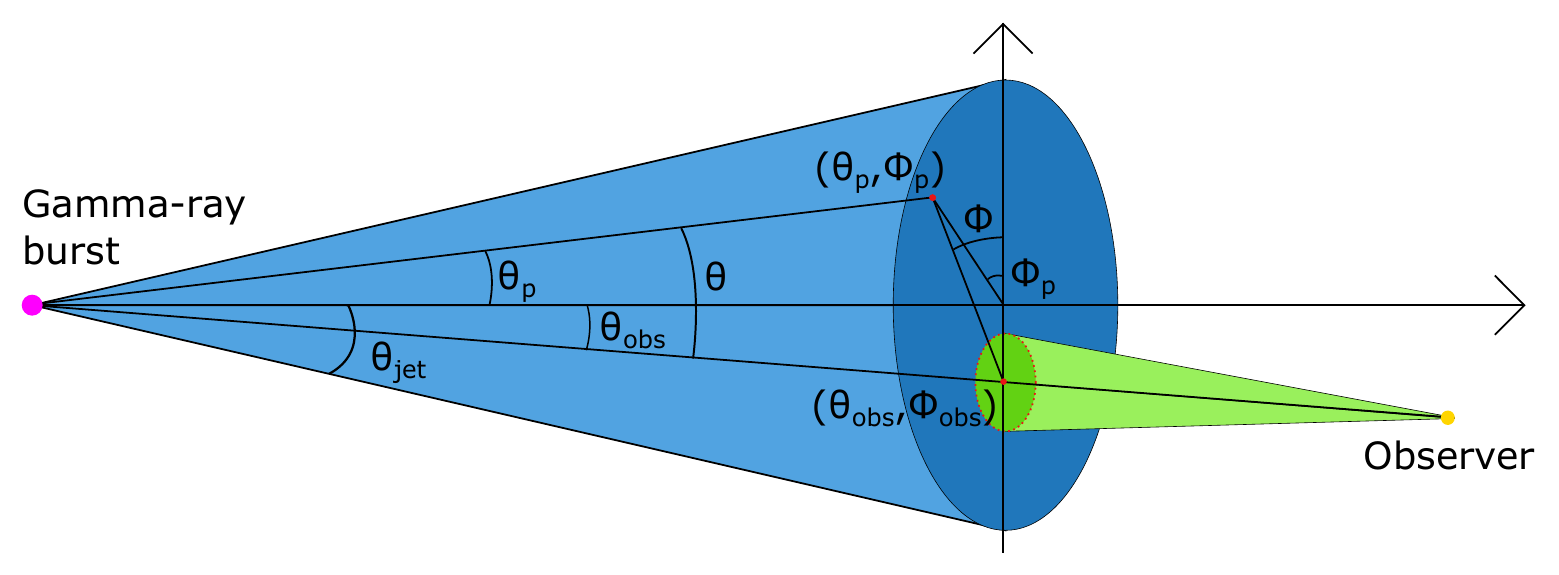}
\caption{Top-hat jet geometry. The jet (blue cone) has half-opening angle $\theta_{\mathrm{jet}}$; the observer's line of sight intersects the forward shock at $(\theta_{\mathrm{obs}}, \phi_{\mathrm{obs}})$. Red dots mark the emitting unit $(\theta_p, \phi_p)$.}
\label{fig:3D_schematic}
\end{figure}

Given the jet's narrowness, this simplified geometric model approximates the surface perpendicular to the jet axis at a certain radius as an equal-arrival-time surface, neglecting the delayed arrival of radiation from high-latitude regions due to extended photon paths. For an off-axis viewing geometry, this model provides the following physical picture: as the forward shock sweeps up the ambient medium and decelerates, its Lorentz factor $\Gamma$ decreases, causing the effective radiating region to expand. At the characteristic time $t_{\text{in}}$, the theoretical visible region first becomes tangent to the actual jet boundary; at $t_{\text{out}}$, it becomes tangent again and fully covers the jet surface \citep{2024MNRAS.532.2189Z}. Thereafter, no new effective radiating regions appear within the jet. This geometric relationship is illustrated schematically in Fig.~\ref{fig:3D_schematic}.

\begin{figure}[H]
\centering
\includegraphics[width=0.45\textwidth]{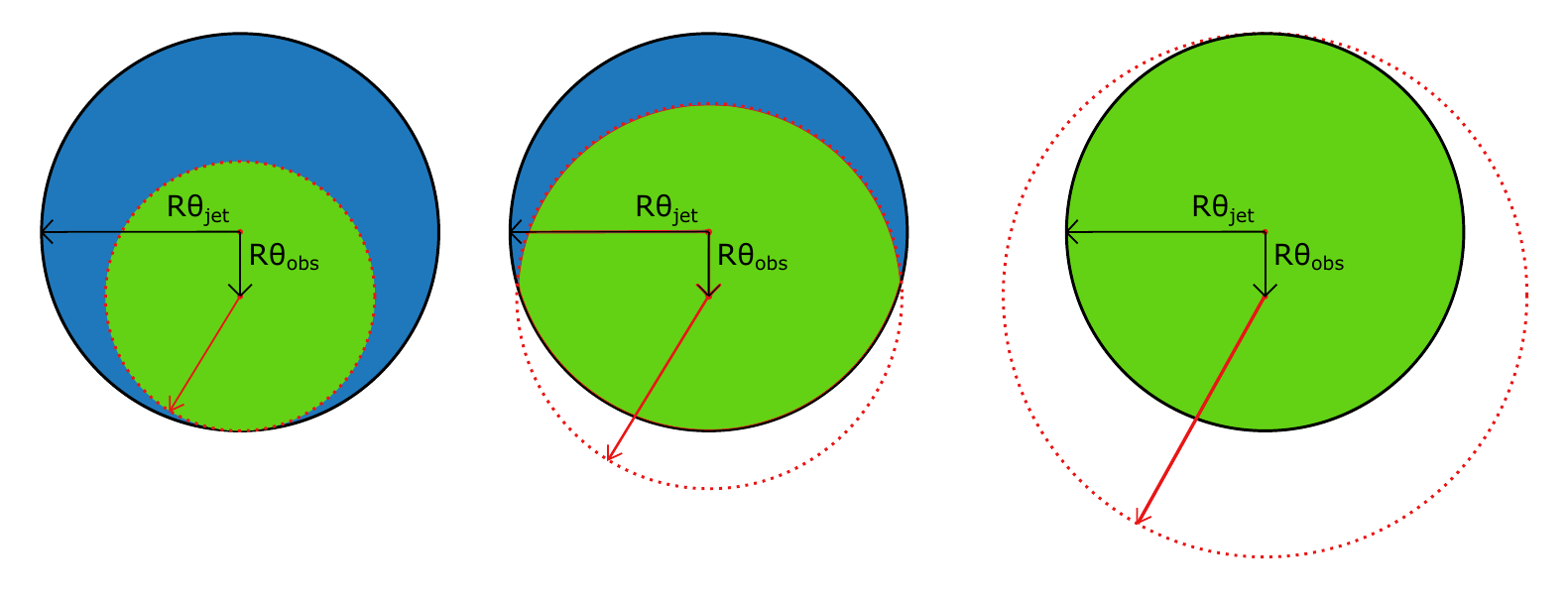}
\caption{Schematic of the jet's visible region evolution. The diagram shows the FS plane (blue circle), along with the theoretical (red dashed) and actual observable areas (green shaded).}
\label{fig:2D_evolution}
\end{figure}

These characteristic times satisfy the geometric conditions \citep{2024MNRAS.532.2189Z}:
\begin{equation}
\Gamma^{-1}(t_{\text{in}}) = \theta_{\text{jet}} - \theta_{\text{obs}}, \quad
\Gamma^{-1}(t_{\text{out}}) = \theta_{\text{jet}} + \theta_{\text{obs}}.
\end{equation}
The flux density is expressed as:
\begin{equation}
F_{\nu}(t) = S_{\text{ratio}} \cdot F_{c} t^{-\alpha_{1}},
\end{equation}
where $t^{-\alpha_1}$ describes the intrinsic power-law decay, $F_c$ is a normalization factor, and $S_{\text{ratio}}$ represents the ratio of the jet's visible region to the theoretically visible region \citep{2024MNRAS.532.2189Z}:
\begin{equation}
S_{\text{ratio}} = \left\{
\begin{array}{ll}
1 & t < t_{\text{in}}, \\
S_{\text{overlap}}/(\pi R^{2} \theta^{2}) & t_{\text{in}} < t < t_{\text{out}}, \\
(\theta_{\text{jet}}/\theta)^{2} & t > t_{\text{out}}.
\end{array}
\right.
\end{equation}
The theoretical evolution of the visible region and jet region for an off-axis observer in the top-hat jet model is shown schematically in Fig.~\ref{fig:2D_evolution}, where the overlapping area is calculated \citep{2024MNRAS.532.2189Z}: \footnote{Notice that the original formula in eq. (6) of \cite{2024MNRAS.532.2189Z} has two typos, which have been corrected here.}

\begin{equation}
\begin{split}
S_{\text{overlap}} = R^{2} \bigg[ \theta_{\text{jet}}^{2} \arccos\left( \frac{\theta_{\text{obs}}^{2} + \theta_{\text{jet}}^{2} - \theta^{2}}{2\theta_{\text{obs}}\theta_{\text{jet}}} \right) \\
+ \theta^{2} \arccos\left( \frac{\theta_{\text{obs}}^{2} + \theta^{2} - \theta_{\text{jet}}^{2}}{2\theta_{\text{obs}}\theta} \right) \\
- 2 \sqrt{p(p-\theta)(p-\theta_{\text{obs}})(p-\theta_{\text{jet}})} \bigg],
\end{split}
\label{eq:s_overlap}
\end{equation}
where $p$ is the semi-perimeter parameter defined as $p = (\theta + \theta_{\text{obs}} + \theta_{\text{jet}})/2$. This model has four free parameters: $\{\theta_{\text{jet}}, \theta_{\text{obs}}, \alpha_1, F_c\}$. It is expressed in logarithmic form:
\begin{equation}
\log F_{\nu}(t) = -\alpha_{1}\log t + C_1 + \log S_{\text{ratio}},
\label{eq:model1_log}
\end{equation}
with $C_1 = \log F_c$.

\subsection{Top-hat jet model with high-latitude emission}

To evaluate the influence of high-latitude emission on viewing angle inference, we developed a comprehensive model that incorporated this effect. The total flux was computed by integrating over all radiating elements on the equal-arrival-time surface. The transformation for a single radiating element followed \citet{2002ApJ...570L..61G}:
\begin{equation}
F_{\nu}(\theta_{\text{obs}}, t) = a^{3} F_{\nu/a}(0, at),
\end{equation}
where the relativistic transformation factor $a$ is defined as $a = (1-\beta)/(1-\beta\cos\theta)$, with $\beta = v/c$. This transformation accounts for three physical effects: the relativistic Doppler shift in frequency from $\nu$ to $\nu/a$, the time compression or dilation from $t$ to $at$, and the intensity amplification represented by the $a^3$ factor. The angle $\theta$ between the transmitting unit (with coordinates $\theta_p$, $\phi_p$) and the line of sight is determined by the spherical coordinate transformation:
\begin{equation}
\cos\theta = \cos\theta_{\text{obs}}\cos\theta_p + \sin\theta_{\text{obs}}\sin\theta_p\cos\phi,
\end{equation}
where $\phi$ is the azimuthal angle difference between the transmitting unit and the line of sight. This geometric relationship is illustrated schematically in Fig.~\ref{fig:3D_schematic}.

Assuming the flux density follows a power-law form \citep{2002ApJ...568..820G}:
\begin{equation}
f_{\nu}(0, t) \propto t^{-\alpha_{2}} \nu^{-\beta_{2}}.
\end{equation}
Integrating over the equal-arrival-time surface and accounting for the XRT band response yields \citep{2024MNRAS.532.2189Z}:
\begin{equation}
F_{\text{band}} = t^{-\alpha_2} C_2 \iint a^{3+\beta_2-\alpha_2} \mathrm{d}S,
\label{eq:fband}
\end{equation}
where the surface element is $\mathrm{d}S = R^2\sin\theta_p \, \mathrm{d}\theta_p \, \mathrm{d}\phi$, with $R$ the shock radius as in Equation~(\ref{eq:s_overlap}), which approximates the actual paraboloidal equal-arrival-time surface using spherical surface elements through fine sampling of polar and azimuthal angles. The normalization factor includes the energy band integration:
\begin{equation}
C_2 = \int_{0.3\,\text{keV}}^{10\,\text{keV}} F_{\nu,\mathrm{const}} \nu^{-\beta_{2}} \mathrm{d}\nu.
\end{equation}
The integration limit $\theta_p \in [0, \theta_{\text{jet}}]$ ensures that only emissions within the jet boundary are included. The model parameter set is $\{C_2, \alpha_2, \beta_2, \theta_{\text{obs}}, \theta_{\text{jet}}\}$. In practice we sample $\log C_2$ rather than $C_2$, in analogy with $C_1 = \log F_c$ in model~1; the values labelled $C_2$ in the corner plots are therefore logarithmic.

We emphasize that $\alpha_1$ and $\alpha_2$ are not the same quantity, and their best-fit values should not be compared directly. In model~1, $S_{\rm ratio} = 1$ before $t_{\rm in}$, so $\alpha_1$ is the observed pre-break decay index. In model~2, the angular integral in Equation~(\ref{eq:fband}) is itself time dependent: before the break the emission is dominated by the region within $\theta_p \lesssim 1/\Gamma$, whose solid angle scales as $\Gamma^{-2}$. The observed pre-break decay index is therefore $\alpha_2 - 3/4$ for the ISM and $\alpha_2 - 1/2$ for the wind medium, whereas $\alpha_2$ is the local (per unit area) decay index of the emitting surface. This accounts for the systematic offset of roughly $0.75$ (ISM) and $0.5$ (wind) between the fitted $\alpha_2$ and $\alpha_1$ values.

\subsection{Model fitting and parameter estimation}

We employed the MCMC method for parameter estimation. The MCMC method constructs Markov chains within the parameter space whose stationary distributions converge to the target posterior distribution. The implementation was based on the emcee library, and visualization of the posterior distributions was performed using the corner package \citep{2013PASP..125..306F, 2016JOSS....1...24F}. 
While $\theta_{\text{obs}}$ describes the absolute viewing angle, its physical significance depends on the jet opening angle. We therefore introduced the dimensionless off-axis ratio $q = \theta_{\text{obs}} / \theta_{\text{jet}}$ to characterize the degree of off-axis deviation in a way that is directly comparable across bursts with different jet sizes. In the following, we use $\theta_{\text{obs}}$ when discussing absolute angular distributions and $q$ for statistical comparisons.
To reduce model complexity, we created time-sliced spectra covering the before and after the jet break epochs for each GRB event through the Swift-XRT spectra repository; the energy spectral index $\beta_2$ was then set to the photon index minus 1 \citep{2009MNRAS.397.1177E, 2009ApJ...698...43R}. The jet opening angle $\theta_{\text{jet}}$ was approximated using the following empirical formulas \citep{2020ApJ...900..112Z}:
\begin{equation}
\begin{split}
\theta_{\text{jet,ISM}} = 0.076 \left( \frac{T_{\text{jet}}}{1 \text{day}} \right)^{3/8} \left( \frac{1+z}{2} \right)^{-3/8} E_{\gamma,\text{iso},53}^{-1/8} \\
\times \left( \frac{\eta_\gamma}{0.2} \right)^{1/8} n_0^{1/8}
\end{split}
\label{eq:theta_ism}
\end{equation}
for ISM, and for wind medium:
\begin{equation}
\begin{split}
\theta_{\text{jet,wind}} = 0.12 \left( \frac{T_{\text{jet}}}{1 \text{day}} \right)^{1/4} \left( \frac{1+z}{2} \right)^{-1/4} E_{\gamma,\text{iso},52}^{-1/4} \\
\times \left( \frac{\eta_\gamma}{0.2} \right)^{1/4} A_*^{1/4}.
\end{split}
\label{eq:theta_wind}
\end{equation}
Here, $T_{\text{jet}}$ denotes the jet break time. Following \citet{2020ApJ...900..112Z}, we adopted $\eta_\gamma = 0.2$, ISM number density $n_0 = 1 \text{ cm}^{-3}$, and wind parameter $A_* = 1$. The isotropic gamma-ray energy $E_{\gamma,\text{iso}}$ for each GRB was obtained from the literature as listed in Table~\ref{tab:table_1}. With the above settings, the free parameters of model~1 were reduced to $\{q_1, \alpha_1, C_1\}$, and those of model~2 to $\{q_2, \alpha_2, C_2\}$.

Within the Bayesian inference framework, we assigned prior probability distributions to the model parameters to incorporate physical knowledge and constraints prior to obtaining observational data. For the core parameter $q = \theta_{\text{obs}} / \theta_{\text{jet}}$, its prior was based on the fundamental physical assumption that the jet axes of GRBs are randomly oriented in the universe. This assumption implied that, prior to observing any specific burst, the orientation of its jet axis was isotropic in space. In spherical coordinates, the probability that a jet axis points toward any infinitesimal solid angle element was proportional to the size of that element. All directions forming an angle $\theta_{\text{obs}}$ with the observer's line of sight constituted a ``latitude circle,'' with a corresponding solid angle element of $d\Omega = 2\pi \sin(\theta_{\text{obs}}) d\theta_{\text{obs}}$. Consequently, the prior probability density of the viewing angle $\theta_{\text{obs}}$ satisfied $p(\theta_{\text{obs}}) \propto \sin(\theta_{\text{obs}})$. Through the variable transformation $q = \theta_{\text{obs}} / \theta_{\text{jet}}$, we obtained the prior distribution for the off-axis ratio $q$ as $p(q) \propto \sin(q \cdot \theta_{\text{jet}})$. This prior reflected the relative probabilities of observing different off-axis angles under the assumption of random jet orientation. For other parameters (such as the normalization factors $C_1$, $C_2$ and decay indices $\alpha_1$, $\alpha_2$), we adopted uniform priors within their physically plausible ranges. For each GRB, the MCMC fitting was performed using all available data points within the fitting interval. We used 90 walkers with 4000 burn-in and 6000 production steps. Convergence was confirmed by inspecting the trace plots. Crucially, the two models were fitted to exactly the same data points over exactly the same time interval, using the same Gaussian likelihood $\ln\mathcal{L} \propto -\chi^2/2$, so that the reduced chi-squared ($\chi^2_{\rm red}$) and Bayesian information criterion ($\mathrm{BIC} = \chi^2 + k \ln N$, where $k$ is the number of free parameters and $N$ is the number of data points, which ranges from 24 to 494 across the sample) reported in Table~\ref{tab:table_2} are directly comparable between them.

\section{Data sample}\label{sec:Data sample}

\begin{table*}[htbp]
    \centering
    \caption{GRB sample parameters.}
    \label{tab:table_1}
    \begin{tabular}{cccccccccccc}
        \toprule
        GRB & $z$ & $\Delta\alpha$ & $\Gamma_1$ & $\Gamma_2$ & $E_{\mathrm{iso},52}$ & $\theta_{\mathrm{jet}}$ & $\Gamma_{\rm break}$ & $\Gamma_{\rm trans}$ & $t_{\rm break}$(s) & $t_{\rm trans}$(s) & Ref. \\
        \midrule
        050318  & 1.440  & $0.84^{+0.24}_{-0.20}$  & $1.98^{+0.14}_{-0.13}$  & $1.91^{+0.14}_{-0.10}$  & $2.3^{+0.23}_{-0.23}$  & $0.041^{+0.004}_{-0.004}$  & $24.3$ & $9.9$ & $12100$ & $132618$  & 1 \\
        050505*  & 4.270  & $0.65^{+0.17}_{-0.12}$  & $1.97^{+0.08}_{-0.08}$  & $2.05^{+0.14}_{-0.13}$  & $16^{+1.1}_{-1.1}$  & $0.039^{+0.005}_{-0.002}$  & $25.5$ & $10.4$ & $44000$ & $482838$  & 1 \\
        060313$^{\dag}$  & 0.700  & $0.94^{+0.14}_{-0.13}$  & $1.93^{+0.21}_{-0.13}$  & $2.17^{+0.24}_{-0.23}$  & $2.9$  & $0.038^{+0.004}_{-0.004}$  & $26.1$ & $10.6$ & $7586$ & $83289$  & 2 \\
        061121*  & 1.314  & $0.68^{+0.05}_{-0.05}$  & $1.82^{+0.08}_{-0.08}$  & $1.74^{+0.10}_{-0.10}$  & $23.5^{+2.7}_{-2.7}$  & $0.039^{+0.003}_{-0.003}$  & $25.1$ & $10.2$ & $23000$ & $252283$  & 1 \\
        061222A*  & 2.088  & $0.85^{+0.08}_{-0.08}$  & $1.82^{+0.08}_{-0.08}$  & $2.02^{+0.12}_{-0.12}$  & $30^{+6.4}_{-6.4}$  & $0.048^{+0.004}_{-0.003}$  & $20.8$ & $8.5$ & $55000$ & $599776$  & 1 \\
        070318  & 0.840  & $0.85^{+0.50}_{-0.30}$  & $1.97^{+0.13}_{-0.12}$  & $1.90^{+0.40}_{-0.40}$  & $3.64^{+0.17}_{-0.17}$  & $0.135^{+0.010}_{-0.007}$  & $7.3$ & $3.1$ & $260000$ & $2504378$  & 1 \\        
        070420$^{\dag}$  & 2.500  & $0.75^{+0.20}_{-0.18}$  & $1.84^{+0.34}_{-0.25}$  & $1.90^{+0.33}_{-0.28}$  & $20.56$  & $0.044^{+0.007}_{-0.006}$  & $22.2$ & $9.1$ & $46000$ & $502796$  & 3 \\
        080710  & 0.850  & $0.89^{+0.26}_{-0.20}$  & $1.75^{+0.16}_{-0.06}$  & $2.08^{+0.22}_{-0.21}$  & $1.68^{+0.22}_{-0.22}$  & $0.055^{+0.007}_{-0.005}$  & $17.8$ & $7.3$ & $19000$ & $205818$  & 5 \\
        090516*  & 4.109  & $0.93^{+0.15}_{-0.11}$  & $2.02^{+0.11}_{-0.11}$  & $2.07^{+0.12}_{-0.11}$  & $72^{+14}_{-14}$  & $0.022^{+0.002}_{-0.002}$  & $43.8$ & $17.8$ & $16600$ & $183662$  & 1 \\
        100425A*  & 1.755  & $0.66^{+0.28}_{-0.32}$  & $2.01^{+0.22}_{-0.16}$  & $1.80^{+0.50}_{-0.30}$  & $0.37^{+0.07}_{-0.07}$  & $0.082^{+0.013}_{-0.031}$  & $12.1$ & $5.0$ & $48000$ & $505095$  & 9 \\
        110213A*  & 1.460  & $0.88^{+0.11}_{-0.09}$  & $1.80^{+0.09}_{-0.05}$  & $1.98^{+0.09}_{-0.09}$  & $5.78^{+0.81}_{-0.81}$  & $0.032^{+0.002}_{-0.002}$  & $30.8$ & $12.5$ & $8872$ & $97735$  & 1 \\
        110709B$^{\dag}$  & 0.750  & $0.61^{+0.06}_{-0.06}$  & $2.05^{+0.07}_{-0.07}$  & $2.00^{+0.11}_{-0.11}$  & $17$  & $0.064^{+0.004}_{-0.004}$  & $15.5$ & $6.4$ & $56000$ & $601857$  & 7 \\
        110818A  & 3.360  & $0.99^{+0.72}_{-0.51}$  & $1.97^{+0.17}_{-0.17}$  & $1.97^{+0.38}_{-0.27}$  & $19.2^{+1.3}_{-1.3}$  & $0.037^{+0.005}_{-0.009}$  & $26.5$ & $10.8$ & $35000$ & $384419$  & 8 \\
        121024A*  & 2.298  & $0.81^{+0.28}_{-0.26}$  & $1.92^{+0.13}_{-0.13}$  & $1.76^{+0.37}_{-0.28}$  & $1.38^{+0.13}_{-0.13}$  & $0.058^{+0.009}_{-0.012}$  & $17.0$ & $7.0$ & $36000$ & $388999$  & 9 \\
        121211A*  & 1.023  & $0.60^{+0.29}_{-0.20}$  & $1.91^{+0.12}_{-0.12}$  & $2.30^{+0.40}_{-0.40}$  & $0.49^{+0.1}_{-0.1}$  & $0.074^{+0.013}_{-0.011}$  & $13.4$ & $5.5$ & $29600$ & $314511$  & 1 \\
        130606A*  & 5.910  & $0.70^{+0.20}_{-0.15}$  & $1.65^{+0.12}_{-0.10}$  & $1.81^{+0.25}_{-0.12}$  & $28.3^{+5.1}_{-5.1}$  & $0.025^{+0.003}_{-0.003}$  & $39.3$ & $16.0$ & $22000$ & $243158$  & 1 \\
        160131A  & 0.970  & $0.77^{+0.22}_{-0.18}$  & $1.97^{+0.06}_{-0.06}$  & $1.91^{+0.25}_{-0.24}$  & $87^{+6.6}_{-6.6}$  & $0.059^{+0.004}_{-0.004}$  & $16.6$ & $6.8$ & $91000$ & $982027$  & 5 \\
        161017A  & 2.013  & $0.72^{+0.13}_{-0.11}$  & $2.00^{+0.13}_{-0.12}$  & $1.95^{+0.17}_{-0.15}$  & $6.87^{+0.72}_{-0.72}$  & $0.045^{+0.009}_{-0.003}$  & $22.0$ & $9.0$ & $28100$ & $307058$  & 5 \\
        210610B  & 1.130  & $0.93^{+0.34}_{-0.25}$  & $1.84^{+0.12}_{-0.11}$  & $1.94^{+0.29}_{-0.27}$  & $46.2^{+3.6}_{-3.5}$  & $0.060^{+0.006}_{-0.007}$  & $16.3$ & $6.7$ & $83000$ & $894892$  & 8 \\
        210722A  & 1.145  & $0.77^{+0.13}_{-0.13}$  & $1.80^{+0.09}_{-0.08}$  & $1.90^{+0.28}_{-0.23}$  & $2.08^{+0.22}_{-0.22}$  & $0.041^{+0.003}_{-0.004}$  & $24.3$ & $9.9$ & $10300$ & $112889$  & 4 \\
        050915A  & 2.527  & $0.44^{+0.50}_{-0.15}$  & $2.10^{+0.50}_{-0.40}$  & $1.73^{+0.33}_{-0.21}$  & $1.8^{+1.3}_{-1.3}$  & $0.052^{+0.029}_{-0.025}$  & $20.3$ & $12.0$ & $9727$ & $78986$  & 6 \\
        060210*  & 3.910  & $0.58^{+0.05}_{-0.05}$  & $2.00^{+0.07}_{-0.07}$  & $2.00^{+0.08}_{-0.08}$  & $32.2^{+3.2}_{-3.2}$  & $0.030^{+0.002}_{-0.002}$  & $35.2$ & $20.8$ & $27000$ & $221277$  & 1 \\
        070508*  & 0.820  & $0.47^{+0.16}_{-0.16}$  & $1.70^{+0.13}_{-0.13}$  & $1.98^{+0.30}_{-0.27}$  & $7.74^{+0.29}_{-0.29}$  & $0.061^{+0.006}_{-0.014}$  & $17.3$ & $10.3$ & $41000$ & $331204$  & 1 \\
        080413B  & 1.100  & $0.50^{+0.12}_{-0.10}$  & $1.82^{+0.08}_{-0.08}$  & $1.98^{+0.17}_{-0.17}$  & $1.61^{+0.27}_{-0.27}$  & $0.095^{+0.011}_{-0.008}$  & $11.1$ & $6.7$ & $57400$ & $451492$  & 1 \\
        090424  & 0.544  & $0.35^{+0.09}_{-0.09}$  & $1.86^{+0.09}_{-0.09}$  & $1.93^{+0.19}_{-0.18}$  & $4.07^{+0.41}_{-0.41}$  & $0.121^{+0.015}_{-0.015}$  & $8.8$ & $5.3$ & $280000$ & $2141984$  & 1 \\
        090618*  & 0.540  & $0.40^{+0.15}_{-0.13}$  & $1.79^{+0.05}_{-0.05}$  & $1.68^{+0.19}_{-0.18}$  & $28.6^{+2.9}_{-2.9}$  & $0.076^{+0.008}_{-0.008}$  & $13.9$ & $8.3$ & $310000$ & $2478080$  & 1 \\
        091020  & 1.710  & $0.42^{+0.05}_{-0.05}$  & $1.99^{+0.13}_{-0.13}$  & $2.05^{+0.10}_{-0.10}$  & $8.4^{+1.1}_{-1.1}$  & $0.036^{+0.004}_{-0.003}$  & $29.2$ & $17.3$ & $8185$ & $66940$  & 1 \\
        091029*  & 2.752  & $0.53^{+0.17}_{-0.12}$  & $2.05^{+0.13}_{-0.12}$  & $1.99^{+0.13}_{-0.13}$  & $7.97^{+0.82}_{-0.82}$  & $0.044^{+0.004}_{-0.004}$  & $24.2$ & $14.3$ & $23000$ & $187520$  & 1 \\
        110422A  & 1.770  & $0.51^{+0.06}_{-0.06}$  & $1.74^{+0.09}_{-0.09}$  & $1.91^{+0.11}_{-0.10}$  & $79.8^{+8.2}_{-8.2}$  & $0.020^{+0.002}_{-0.001}$  & $52.8$ & $31.2$ & $7430$ & $61049$  & 1 \\
        111008A*  & 4.990  & $0.35^{+0.25}_{-0.11}$  & $1.83^{+0.10}_{-0.10}$  & $1.76^{+0.13}_{-0.12}$  & $24.7^{+1.2}_{-1.2}$  & $0.034^{+0.003}_{-0.008}$  & $30.8$ & $18.2$ & $43000$ & $351908$  & 1 \\
        120712A*  & 4.175  & $0.53^{+0.13}_{-0.12}$  & $1.97^{+0.16}_{-0.13}$  & $2.32^{+0.20}_{-0.19}$  & $21.2^{+2.1}_{-2.1}$  & $0.020^{+0.003}_{-0.003}$  & $53.3$ & $31.5$ & $3556$ & $29219$  & 1 \\
        120909A  & 3.930  & $0.40^{+1.02}_{-0.16}$  & $1.90^{+0.13}_{-0.13}$  & $2.10^{+0.16}_{-0.15}$  & $87^{+10}_{-10}$  & $0.021^{+0.004}_{-0.006}$  & $50.0$ & $29.5$ & $18000$ & $147862$  & 1 \\
        130505A  & 2.270  & $0.58^{+0.05}_{-0.04}$  & $1.72^{+0.06}_{-0.06}$  & $1.79^{+0.07}_{-0.07}$  & $347^{+35}_{-35}$  & $0.017^{+0.001}_{-0.001}$  & $60.8$ & $35.9$ & $21800$ & $179211$  & 1 \\
        140206A*  & 2.730  & $0.37^{+0.17}_{-0.09}$  & $1.80^{+0.06}_{-0.06}$  & $1.77^{+0.11}_{-0.11}$  & $29.69^{+3.05}_{-3.05}$  & $0.041^{+0.008}_{-0.003}$  & $25.9$ & $15.3$ & $64000$ & $522477$  & 4 \\
        140419A  & 3.956  & $0.55^{+0.05}_{-0.05}$  & $1.81^{+0.11}_{-0.11}$  & $1.95^{+0.08}_{-0.08}$  & $228^{+18}_{-18}$  & $0.014^{+0.001}_{-0.001}$  & $73.8$ & $43.6$ & $9977$ & $82059$  & 8 \\
        140423A  & 3.260  & $0.50^{+0.18}_{-0.14}$  & $2.05^{+0.10}_{-0.10}$  & $1.94^{+0.23}_{-0.22}$  & $50.3^{+2}_{-2}$  & $0.028^{+0.002}_{-0.002}$  & $38.3$ & $22.6$ & $26000$ & $213236$  & 8 \\
        140512A*  & 0.725  & $0.55^{+0.11}_{-0.09}$  & $1.77^{+0.07}_{-0.07}$  & $1.89^{+0.11}_{-0.10}$  & $7.25^{+0.61}_{-0.61}$  & $0.057^{+0.011}_{-0.003}$  & $18.5$ & $11.0$ & $28000$ & $226716$  & 5 \\
        150403A*  & 2.060  & $0.38^{+0.20}_{-0.12}$  & $1.71^{+0.04}_{-0.04}$  & $1.70^{+0.18}_{-0.17}$  & $17.78^{+0.41}_{-0.41}$  & $0.071^{+0.012}_{-0.006}$  & $15.0$ & $8.9$ & $282000$ & $2263665$  & 9 \\
        200829A$^{\dag}$  & 1.290  & $0.45^{+0.06}_{-0.05}$  & $1.83^{+0.12}_{-0.11}$  & $1.92^{+0.10}_{-0.09}$  & $126^{+2.9}_{-3.0}$  & $0.022^{+0.001}_{-0.001}$  & $48.4$ & $28.6$ & $13800$ & $113344$  & 8 \\
        220101A  & 4.610  & $0.57^{+0.06}_{-0.05}$  & $1.61^{+0.05}_{-0.05}$  & $1.80^{+0.09}_{-0.09}$  & $324^{+21}_{-20}$  & $0.021^{+0.001}_{-0.001}$  & $51.4$ & $30.3$ & $68400$ & $561946$  & 8 \\
        \bottomrule
    \end{tabular}
\begin{tablenotes}
    \item Ref. (1) \citet{2018JHEAp..18...21W}; (2) \citet{2015ApJ...815..102F}; (3) \citet{2014ApJ...785...29S}; (4) \citet{2021ApJ...908..242D}; (5) \citet{2016MNRAS.458.1921S}; (6) \citet{2022ApJ...924...69Y}; (7) \citet{2013A&A...552L...5P}; (8) \citet{2025ApJ...980..241A}; (9) \citet{2023ApJ...949L...4L}.
    \item Notes. $\Gamma_1$ and $\Gamma_2$ represent the photon indices before and after the jet break, respectively; $\Delta\alpha$ represents the slope difference across the jet break. The temporal and spectral parameters are derived from the Swift-XRT catalogue and spectra repository \citep{2007A&A...469..379E, 2009MNRAS.397.1177E}; the last column gives the reference from which the adopted $E_{\mathrm{iso},52}$ value is taken. An asterisk marks the bursts classified as having an X-ray plateau in the catalogue of \citet{2022ApJ...924...69Y}. A dagger marks the four bursts without a spectroscopic redshift: for GRB 060313 and GRB 200829A the quoted value is photometric, while for GRB 070420 and GRB 110709B it is estimated from empirical relations in the corresponding reference. $\Gamma_{\rm break}$ is Equation~(\ref{eq:bm_ism}) or (\ref{eq:bm_wind}) evaluated at $t_{\rm break}$; $\Gamma_{\rm trans}$ and $t_{\rm trans}$ mark the onset of significant lateral spreading (Section~\ref{sec:Data sample}).
\end{tablenotes}
\end{table*}

To ensure the reliability of model fitting, we established a rigorous set of selection criteria for GRB jet break samples based on theoretical expectations and data quality. The core objective is to confirm that the light curve break is dominated by the edge effect and to ensure sufficient observational data for accurately constraining the model parameters. The specific procedure and criteria are as follows.

\begin{figure}[htbp]
\centering
\parbox{\columnwidth}{\centering
  \begin{subfigure}[b]{0.48\columnwidth}
    \includegraphics[width=\linewidth]{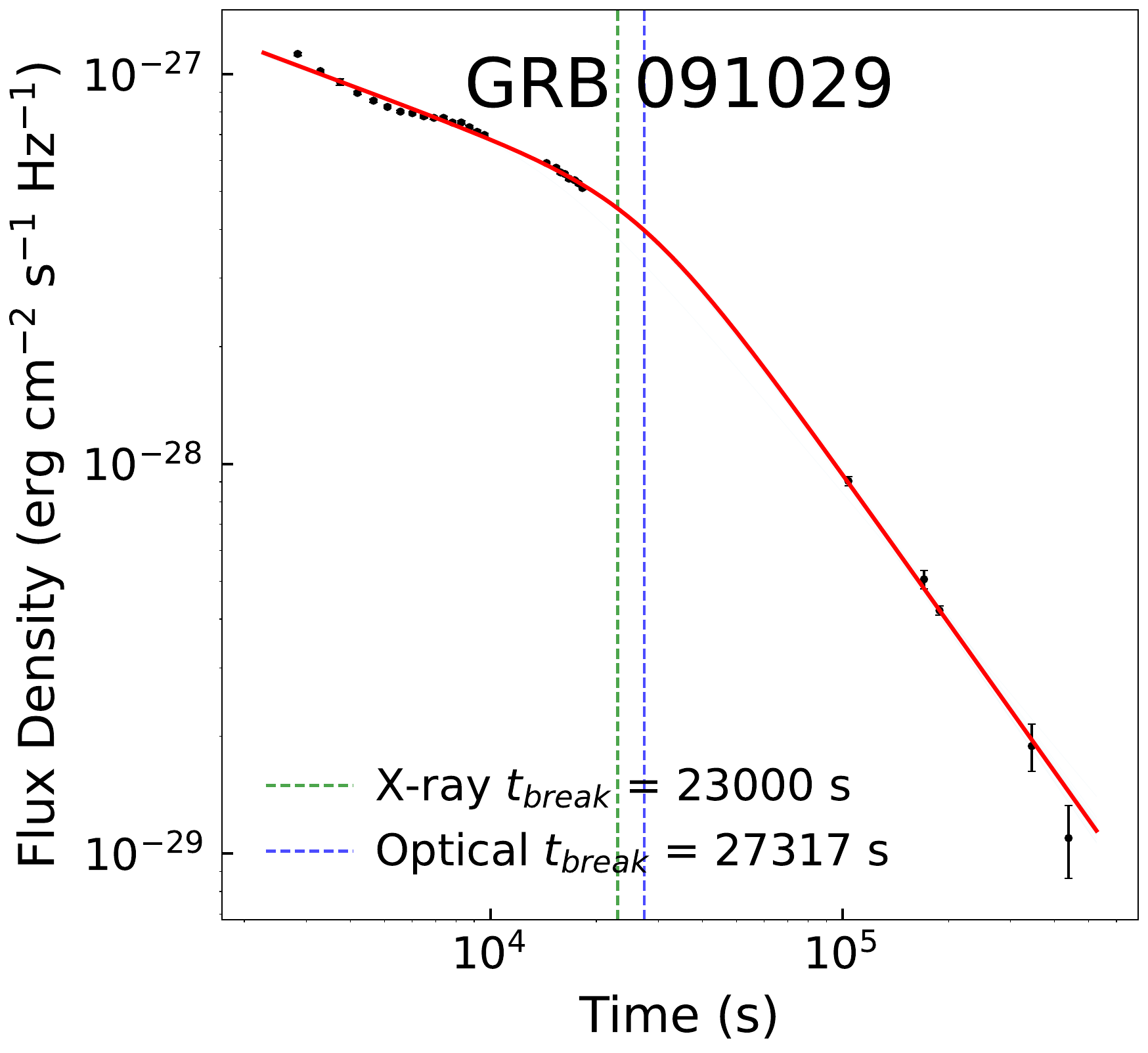}
  \end{subfigure}%
  \hfill
  \begin{subfigure}[b]{0.48\columnwidth}
    \includegraphics[width=\linewidth]{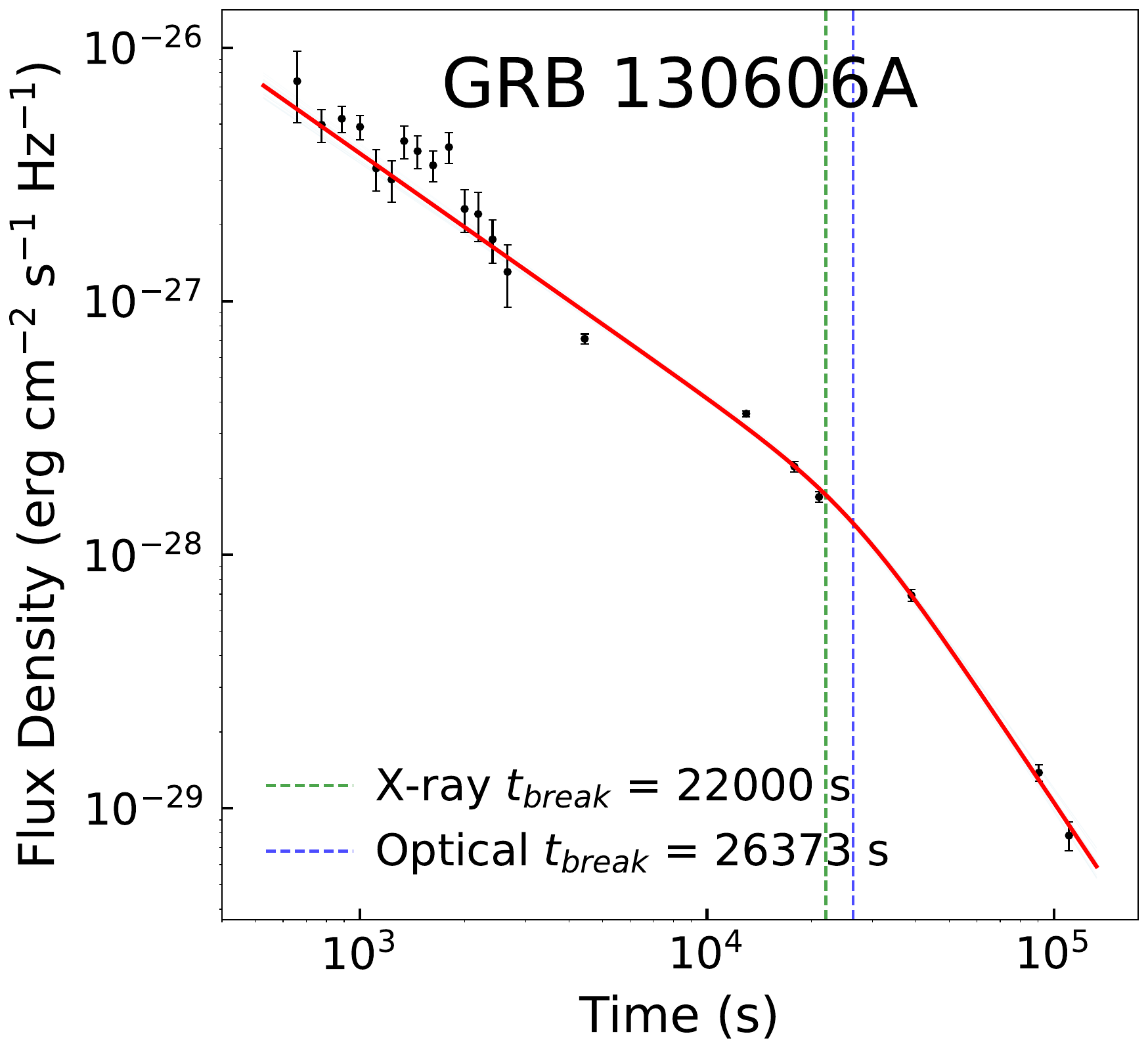}
  \end{subfigure}%
  \\
  \begin{subfigure}[b]{0.48\columnwidth}
    \includegraphics[width=\linewidth]{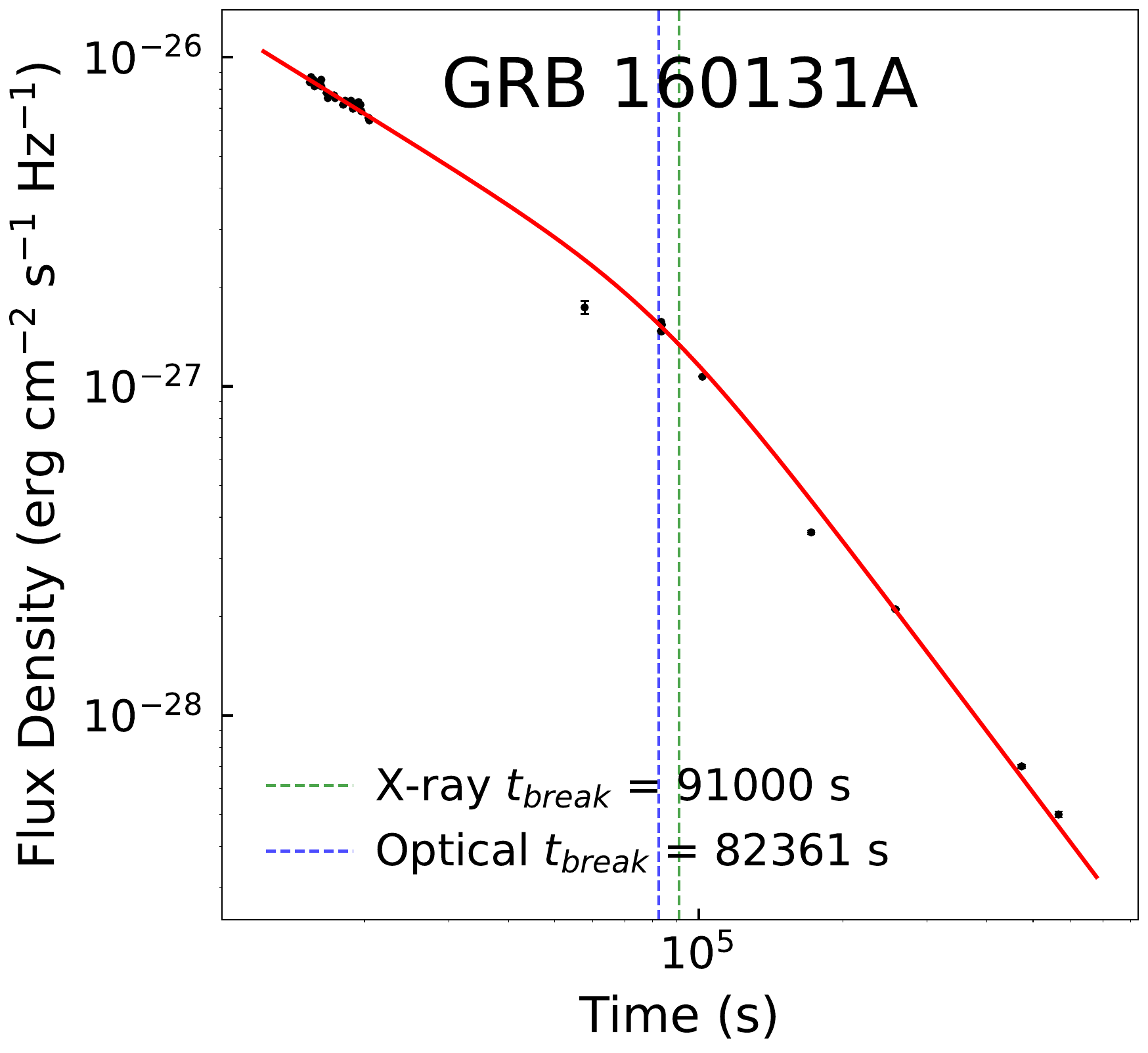}
  \end{subfigure}%
  \hfill
  \begin{subfigure}[b]{0.48\columnwidth}
    \includegraphics[width=\linewidth]{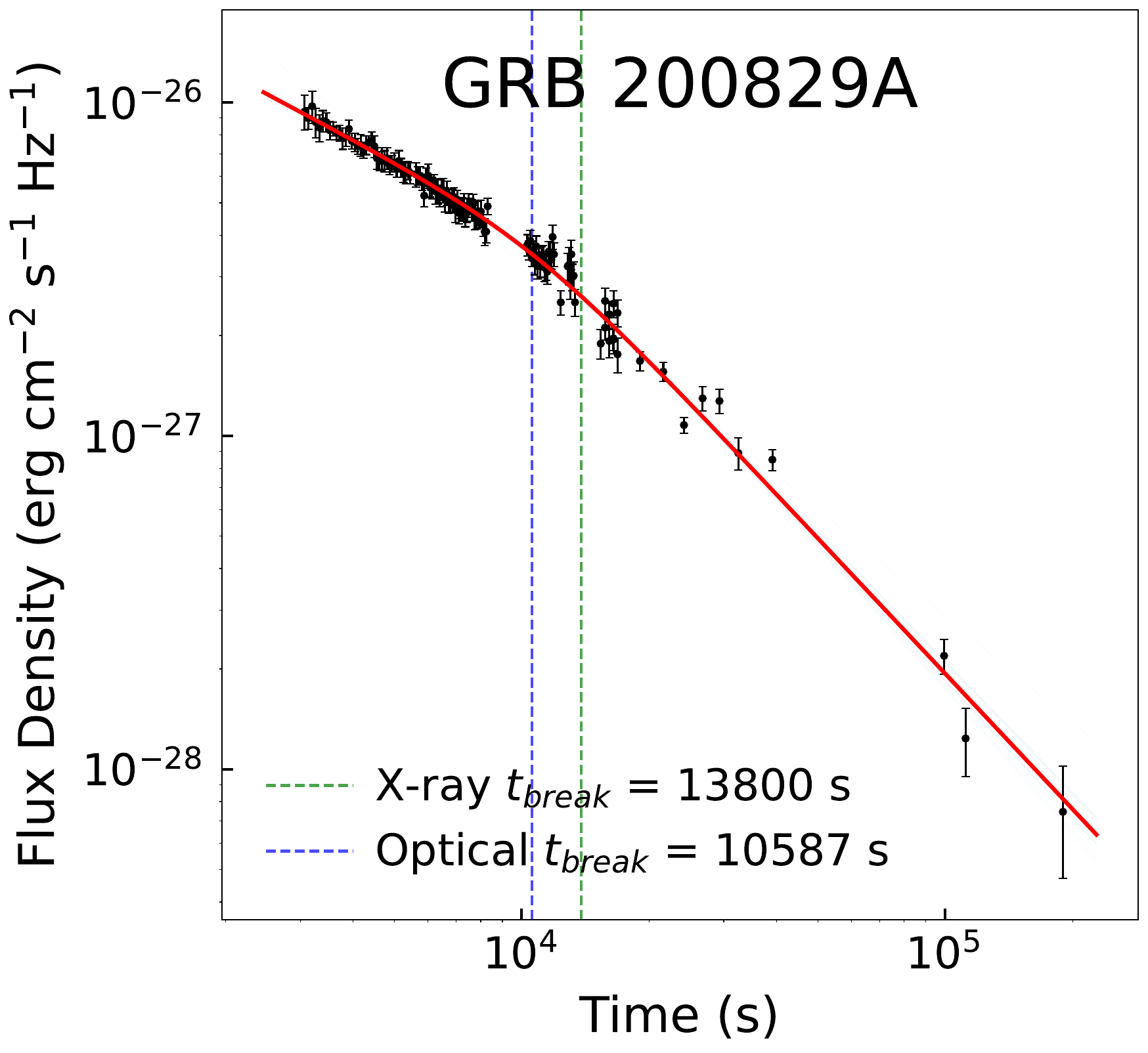}
  \end{subfigure}%
}
  \caption{Optical broken power-law fits for four GRBs with sufficient data coverage across the jet break: GRB~091029 (r$'$ band), GRB~130606A ($i$ band), GRB~160131A ($i'$ band), and GRB~200829A ($R$ band).}
  \label{fig:optical_bpl}
\end{figure}

First, in the stage of jet break identification and confirmation, we searched for decay break features in the XRT catalog \citep{2007A&A...469..379E, 2009MNRAS.397.1177E}. An event was considered to harbor a potential jet break if it satisfied one of the following morphological conditions: the light curve exhibited a complete i-ii-iii-iv segment structure, a ii-iii-iv segment structure, or a clear two-segment structure with a significant steepening, requiring either $\alpha_2 > 1.5$, or $\alpha_2 > 1.2$ together with $t_{\text{break}} > 10^4\,\mathrm{s}$. For events satisfying the above morphological conditions, we further analyzed the X-ray spectral evolution before and after the break. Specifically, we created time-sliced spectra covering the epochs before and after the jet break for each GRB event through the Swift-XRT spectra repository \citep{2009MNRAS.397.1177E}. We attributed the corresponding temporal break to an edge effect and retained the event in our sample only when the photon index variation before and after the jet break was consistent within uncertainties or its change was less than 0.2 \citep{2006ApJ...642..354Z, 2009ApJ...698...43R}. The photon indices before and after the break were listed in Table~\ref{tab:table_1}. A geometric jet break should be achromatic across wavelengths. To test this property, we cross-matched the sample with the optical GRB catalog of \citet{2024MNRAS.533.4023D}. Four bursts (GRB 091029, 130606A, 160131A, and 200829A) possess sufficiently dense optical coverage spanning the break epoch. Broken power-law fits to their optical light curves yield break times consistent with the X-ray values within uncertainties (Fig.~\ref{fig:optical_bpl}, Table~\ref{tab:optical_bpl}), confirming achromatic behavior for these events. Given that only 4 of 40 GRBs have sufficient optical data, we did not impose multi-wavelength achromaticity as a formal selection criterion.

Second, we considered the theoretical models and the requirements for initial parameters. Different theoretical models yield different predictions for the change in decay index at the jet break \citep{2003ApJ...591.1086G}. Within model~1, the ISM requires $\Delta\alpha \approx 0.75$, whereas the wind medium requires $\Delta\alpha \approx 0.5$. Accordingly, during preliminary screening, for ISM events we required $\Delta\alpha$ to lie within the range 0.6--1.0, and for wind medium events within 0.34--0.6. We note that this step assigns the circumburst medium from the observed break sharpness itself, and that a few bursts sit close to the dividing value $\Delta\alpha = 0.6$ with uncertainties that straddle it (GRB 110709B, 060210 and 130505A). Re-assigning these borderline events to the other subsample changes the mean off-axis ratio of each group by less than 0.03 and leaves the K-S result of Section~\ref{sec:Results} unchanged. We further note that a fixed $k = 0$ or $k = 2$ is an idealization: \citet{2013ApJ...776..120Y} found a typical $k \sim 1$, and intermediate density profiles would propagate into the inferred $q$ as an additional systematic. Furthermore, our model required redshift and isotropic gamma-ray energy $E_{\gamma,\text{iso}}$ as input parameters; therefore, we only selected events with a known redshift and a previously derived $E_{\gamma,\text{iso}}$ value. Four bursts (GRB 060313, 070420, 110709B and 200829A, marked with a dagger in Table~\ref{tab:table_1}) do not have a spectroscopic redshift and are retained with the photometric or empirically estimated values quoted in the corresponding references; removing them changes the sample-mean off-axis ratio by less than 0.005. While these criteria are well-motivated theoretically, they significantly limit the sample size. Both models assume negligible lateral expansion. To verify this assumption for our sample, we adopt the numerical calibration of \citet{2018ApJ...865...94D}, where significant lateral spreading begins when the four-velocity reaches
\begin{equation}
u_{\rm trans} = \frac{1}{Q_k \, \theta_{\rm jet}},
\label{eq:u_spread}
\end{equation}
with $Q_k = 2.5$ for the ISM and $Q_k = 1.6$ for the wind medium. Converting $\Gamma_{\rm trans} = \sqrt{1 + u_{\rm trans}^2}$ to an observer-frame time $t_{\rm trans}$ via Equations~(\ref{eq:bm_ism})--(\ref{eq:bm_wind}), we find a median $t_{\rm trans}/t_{\rm break} \simeq 11$ for the ISM (range 9.6--11.1) and $\simeq 8$ for the wind medium (range 7.7--8.2) across the sample, so that lateral spreading has not yet set in at the jet break epoch. Note that $\Gamma_{\rm break}$, obtained from Equations~(\ref{eq:bm_ism})--(\ref{eq:bm_wind}) at $t_{\rm break}$, agrees with $1/\theta_{\rm jet}$ to within $\lesssim 10$ per cent, as expected from the jet-break condition; the ratio $t_{\rm trans}/t_{\rm break}$ is consequently set almost entirely by $Q_k$ and by the deceleration index, and varies little from burst to burst. The corresponding Lorentz factors $\Gamma_{\rm break}$ and $\Gamma_{\rm trans}$ at these two epochs, together with $t_{\rm break}$ and $t_{\rm trans}$, are listed for each GRB in Table~\ref{tab:table_1}. For the bursts whose fitting interval extends beyond $t_{\rm trans}$, the latest data points may already be mildly affected by spreading; we verified, however, that truncating the fits at $t_{\rm trans}$ changes the inferred $q$ by a median of only $\sim 5$ per cent and keeps it within the $1\sigma$ interval of the full-interval fit for 39 of the 40 bursts. The only exception is GRB~060210, which extends to $\sim 7\,t_{\rm trans}$ and barely exceeds the $95\%$ interval, without affecting the aggregate statistics.

Finally, in the assessment of data quality and fitting robustness, light curve data points are often discontinuous due to observational gaps, such as satellite passages through the South Atlantic Anomaly (SAA) or Earth occultation. To ensure sufficient data points for robustly characterizing the break feature, we imposed quantitative constraints from both data coverage and fitting result quality. On one hand, we strictly confine the model fitting interval to the ``normal decay'' and ``late steeper decay'' phases, minimizing the influence of early-time emission or potential late-time rebrightening on the fitting results. Meanwhile, based on the flare identification list provided by the XRT catalog, we excluded all time intervals potentially contaminated by flare activity to avoid misjudgment of jet breaks caused by central engine reactivation \citep{2007A&A...469..379E, 2009MNRAS.397.1177E, 2016ApJS..224...20Y, 2021MNRAS.507.1047Y}. To evaluate data coverage near the break, we defined an assessment window centered on the break time $t_{\text{break}}$, with a width equal to one-fifth of the logarithmic time span of the entire fitting interval. This window was divided into 5 bins, and we required a data coverage fraction of at least 0.6 within it, thereby excluding samples whose break shape could not be reliably characterized due to severe data gaps. On the other hand, regarding the reliability of parameter constraints, we employed an MCMC method to obtain the posterior distribution of the off-axis ratio $q$ for each burst and calculated its coefficient of variation (CV). This coefficient is defined as the ratio of the standard deviation to the mean of the posterior distribution. We systematically excluded bursts for which either model returned CV $> 0.7$. This threshold is empirical rather than derived, and is intended only to remove events whose $q$ posterior is essentially unconstrained by the data; adopting 0.6 instead changes the sample by 5 bursts and leaves the aggregate statistics unchanged.

Following the multi-step selection criteria described above, we ultimately obtained a refined sample of 40 GRBs, comprising 20 bursts occurring in the ISM and 20 in a wind medium; the equal split between the two media is a coincidence of the selection rather than an imposed balance. In both Table~\ref{tab:table_1} and Table~\ref{tab:table_2}, the bursts are ordered by circumburst medium: the rows above GRB~050915A belong to the ISM subsample, and GRB~050915A itself together with the rows below it belong to the wind subsample.

\begin{table}[htbp]
\centering
\caption{Optical and X-ray jet break times.}
\label{tab:optical_bpl}
\begin{tabular}{cccc}
\toprule
GRB & Band & $t_{\rm break, opt}$ ($10^{4}$ s) & $t_{\rm break, X}$ ($10^{4}$ s)\\ \midrule
091029   & r$'$ & $2.7^{+0.2}_{-0.2}$   & $2.3^{+0.6}_{-0.6}$ \\
130606A  & $i$   & $2.6^{+0.3}_{-0.3}$  & $2.2^{+0.7}_{-0.5}$ \\
160131A  & $i'$  & $8.2^{+0.1}_{-0.1}$    & $9.1^{+1.3}_{-1.5}$ \\
200829A  & $R$   & $1.1^{+0.1}_{-0.1}$    & $1.4^{+0.3}_{-0.3}$ \\
\bottomrule
\end{tabular}
\end{table}

\section{Results}\label{sec:Results}

\begin{figure}[htbp]
\centering
  \begin{subfigure}[b]{0.235\textwidth}
    \includegraphics[width=\linewidth]{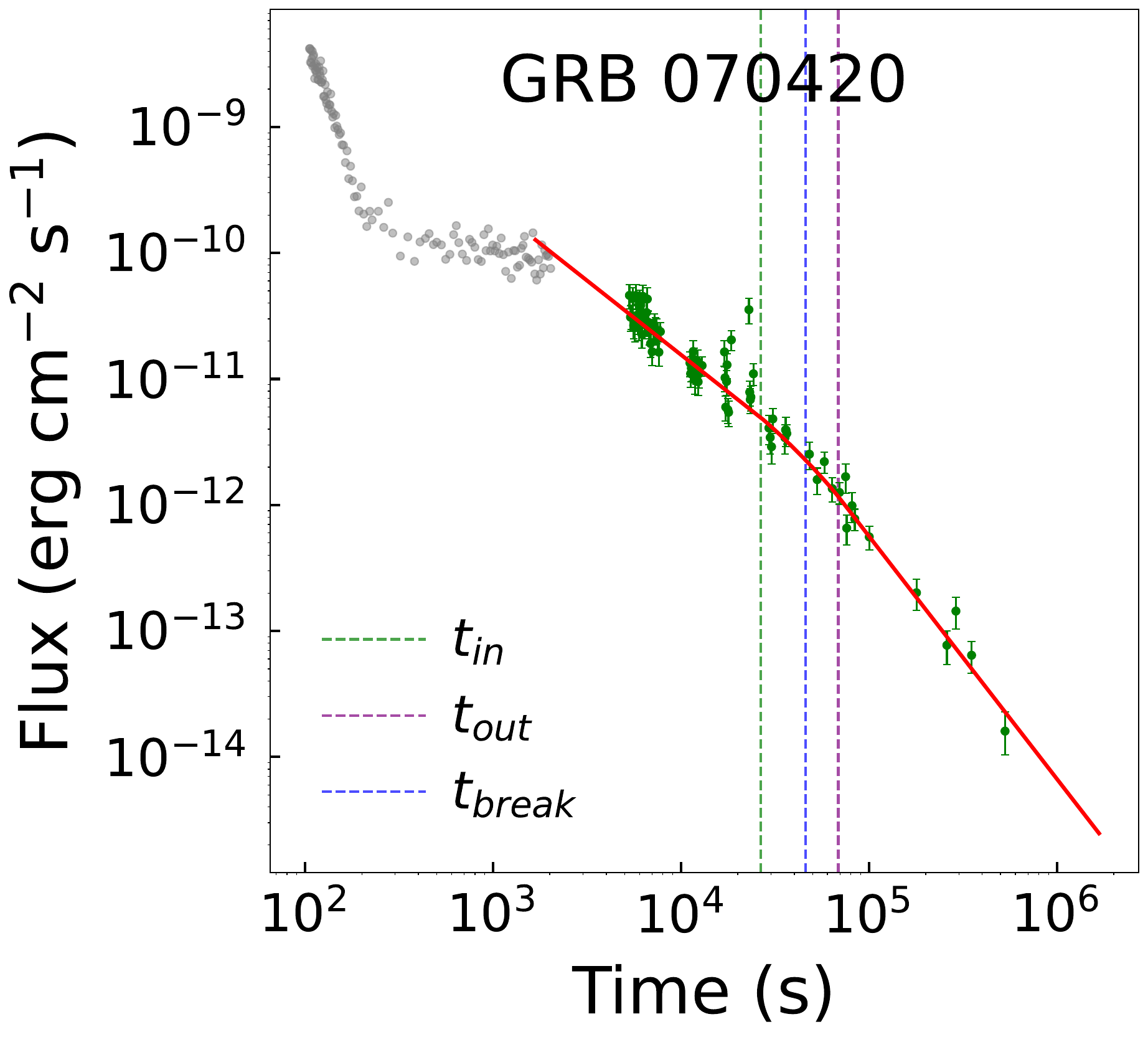}
  \end{subfigure}%
  \begin{subfigure}[b]{0.255\textwidth}
    \includegraphics[width=\linewidth]{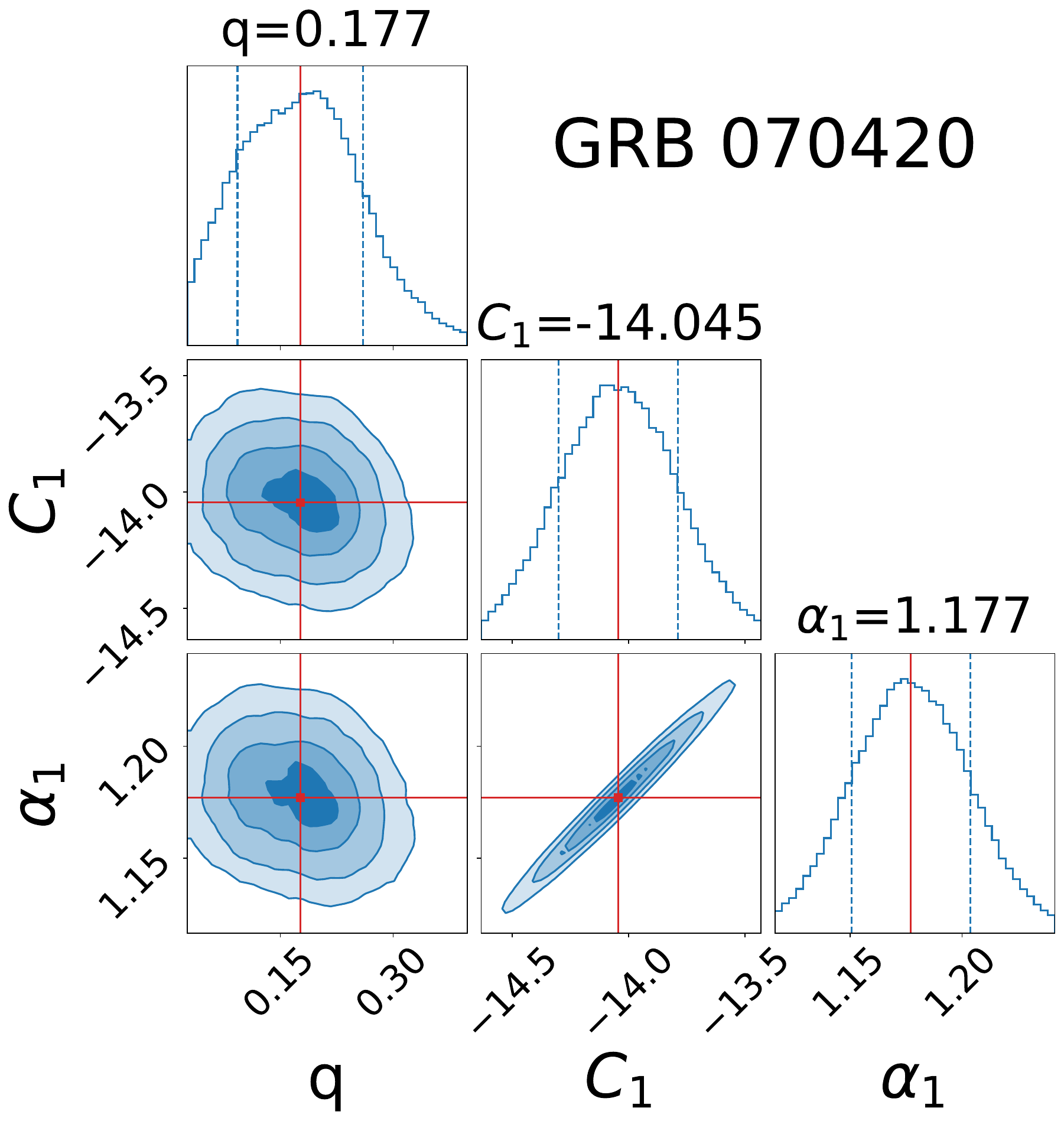}
  \end{subfigure}%
  \\
  \begin{subfigure}[b]{0.235\textwidth}
    \includegraphics[width=\linewidth]{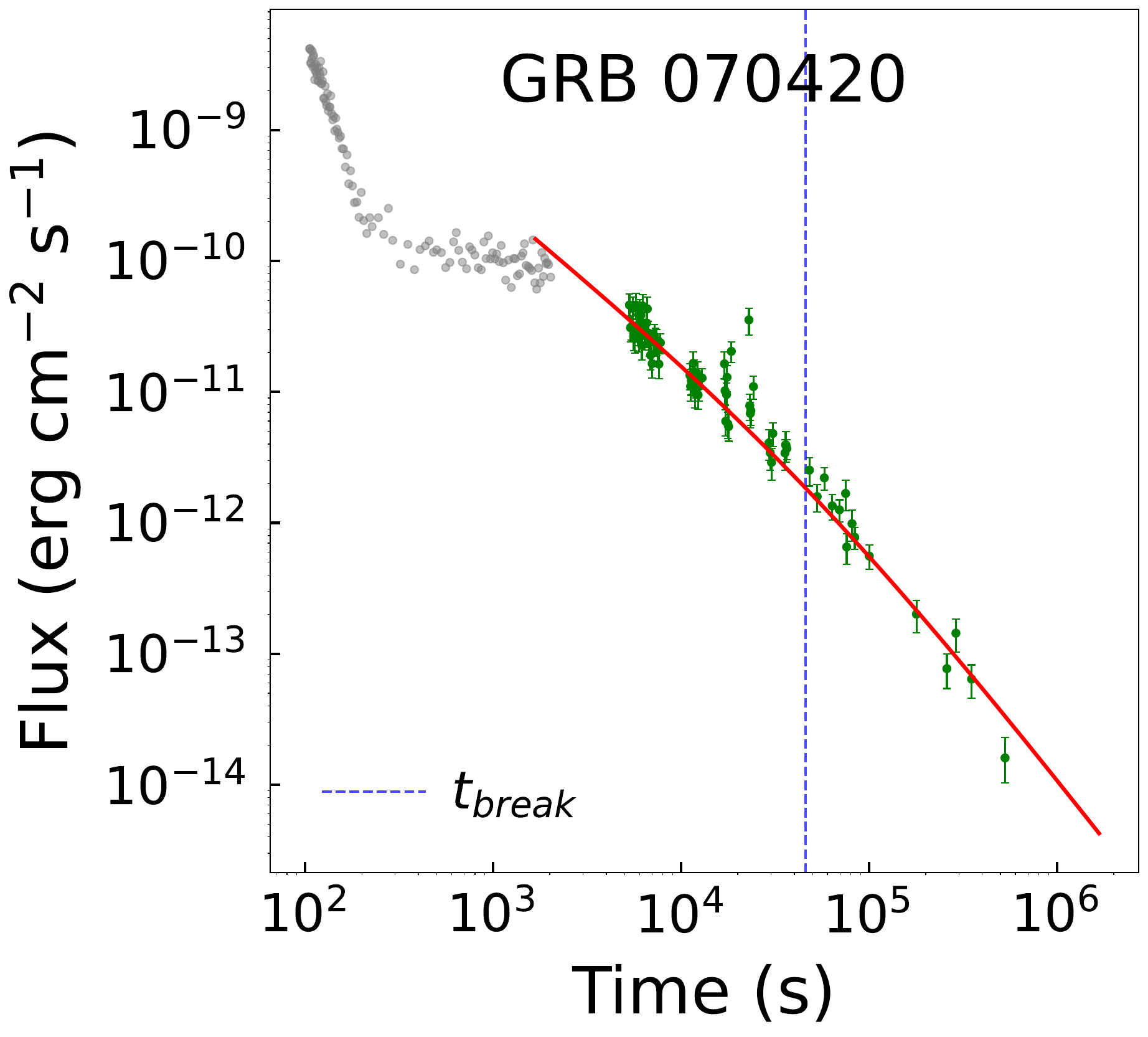}
  \end{subfigure}%
  \begin{subfigure}[b]{0.255\textwidth}
    \includegraphics[width=\linewidth]{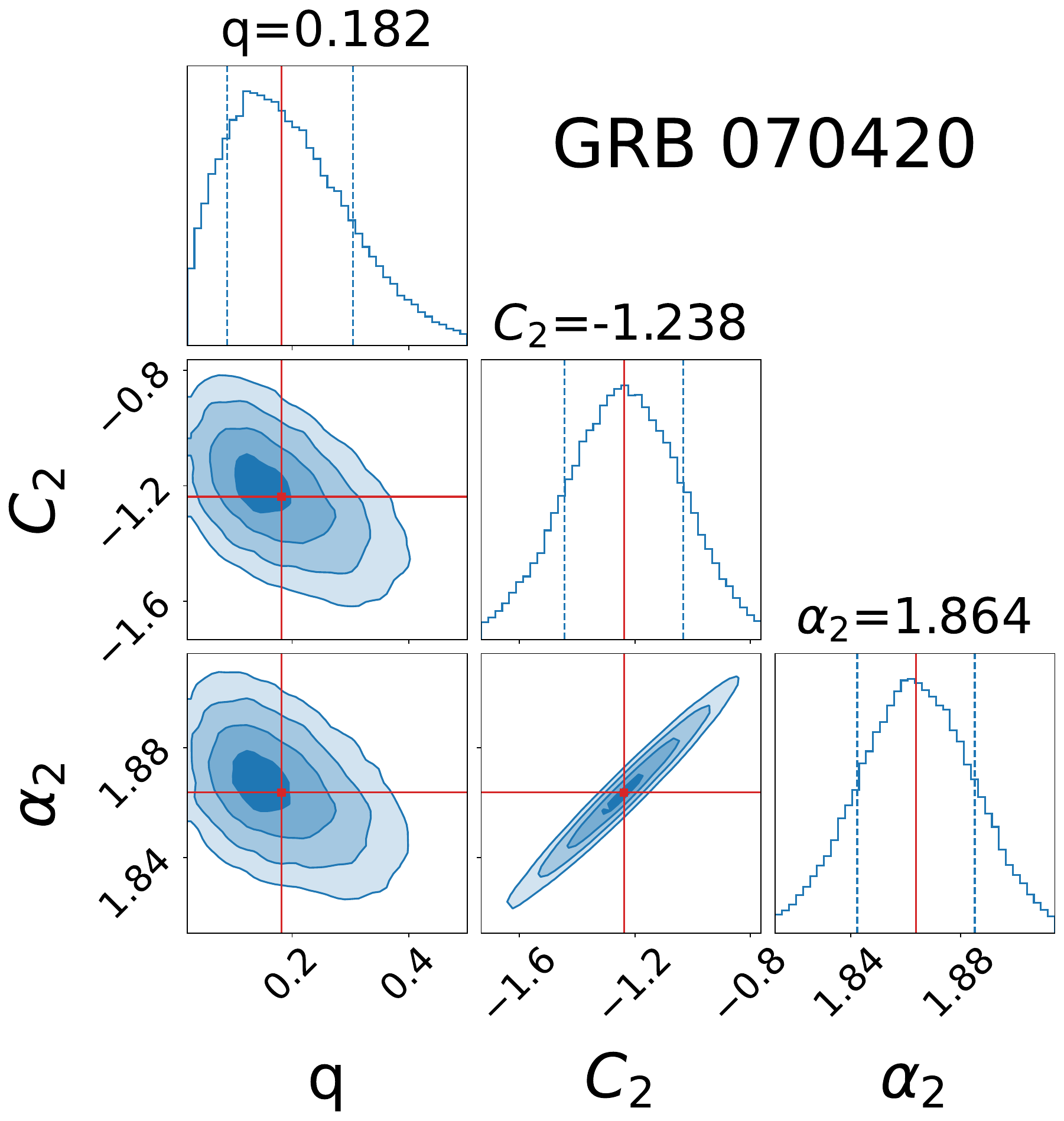}
  \end{subfigure}%
  \\
  \begin{subfigure}[b]{0.235\textwidth}
    \includegraphics[width=\linewidth]{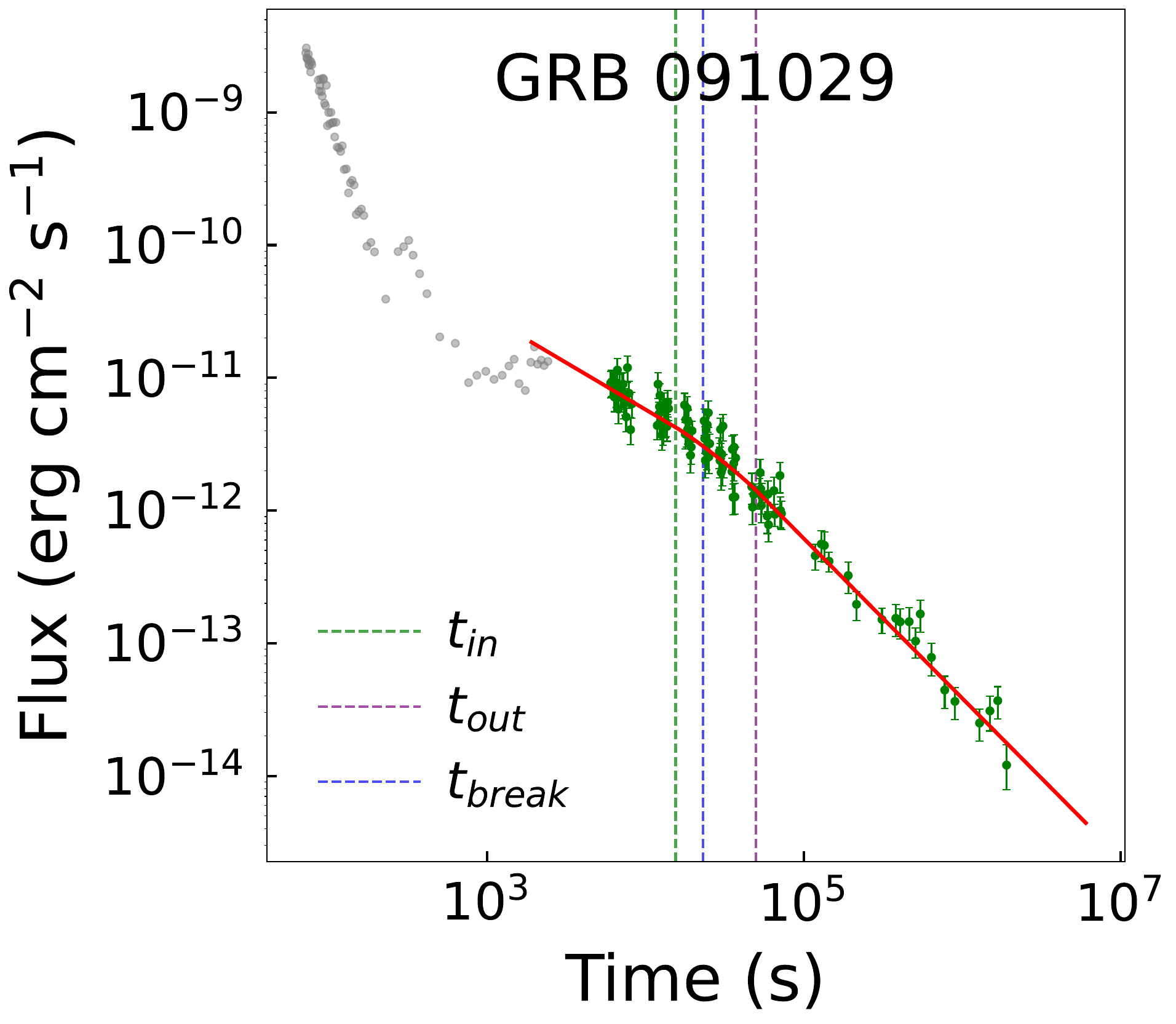}
  \end{subfigure}%
    \begin{subfigure}[b]{0.255\textwidth}
    \includegraphics[width=\linewidth]{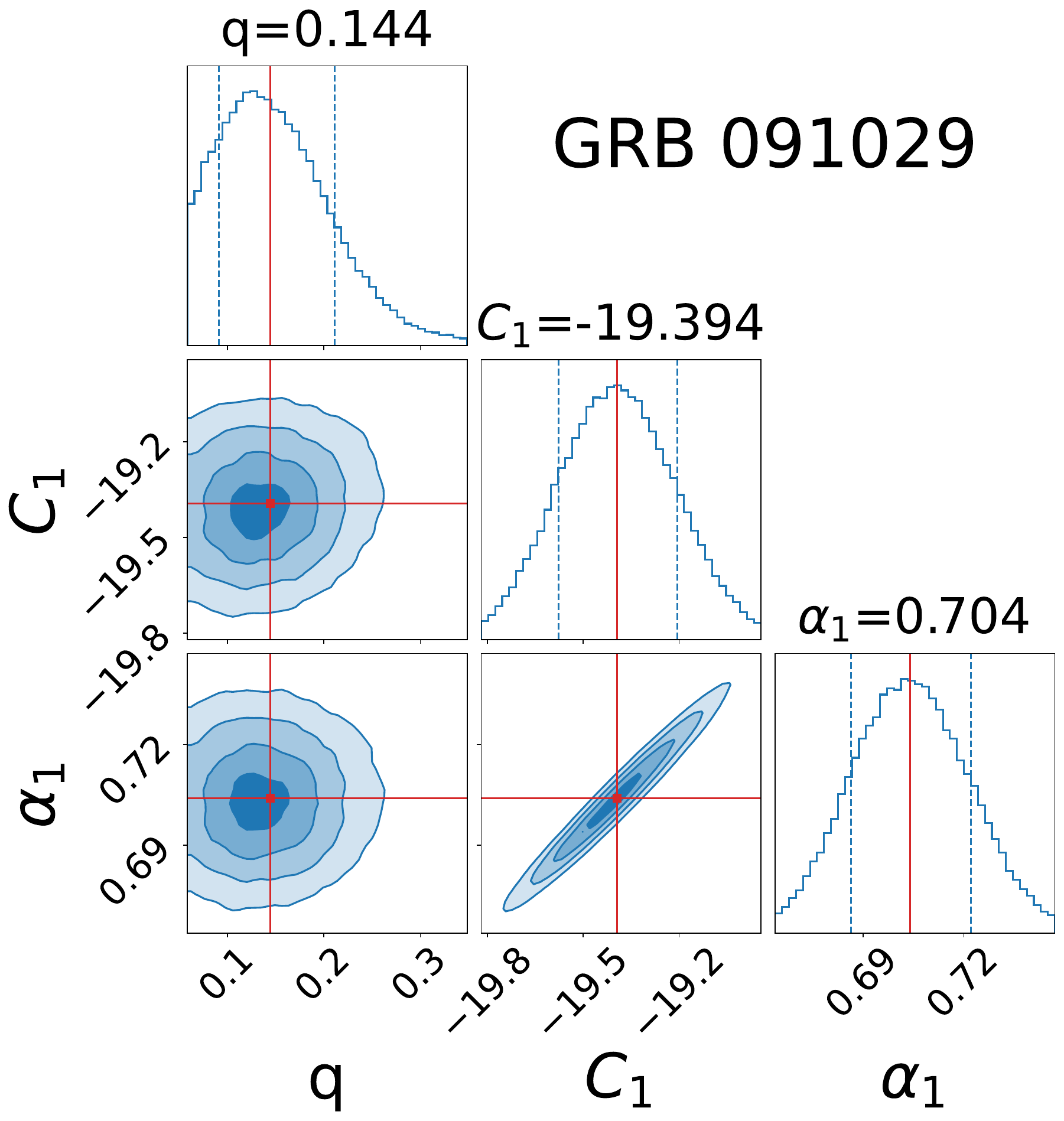}
  \end{subfigure}%
  \\
  \begin{subfigure}[b]{0.235\textwidth}
    \includegraphics[width=\linewidth]{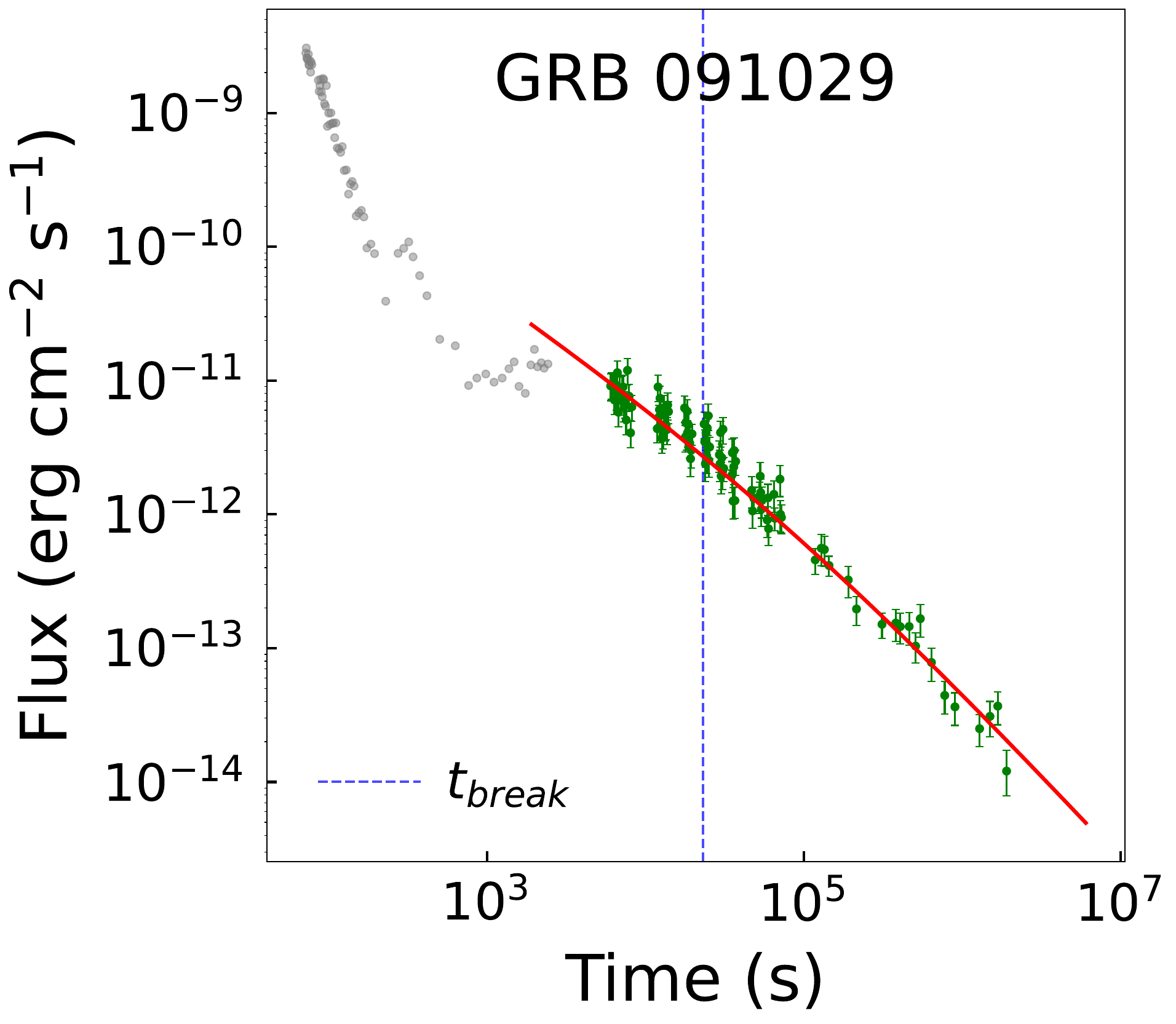}
  \end{subfigure}%
    \begin{subfigure}[b]{0.255\textwidth}
    \includegraphics[width=\linewidth]{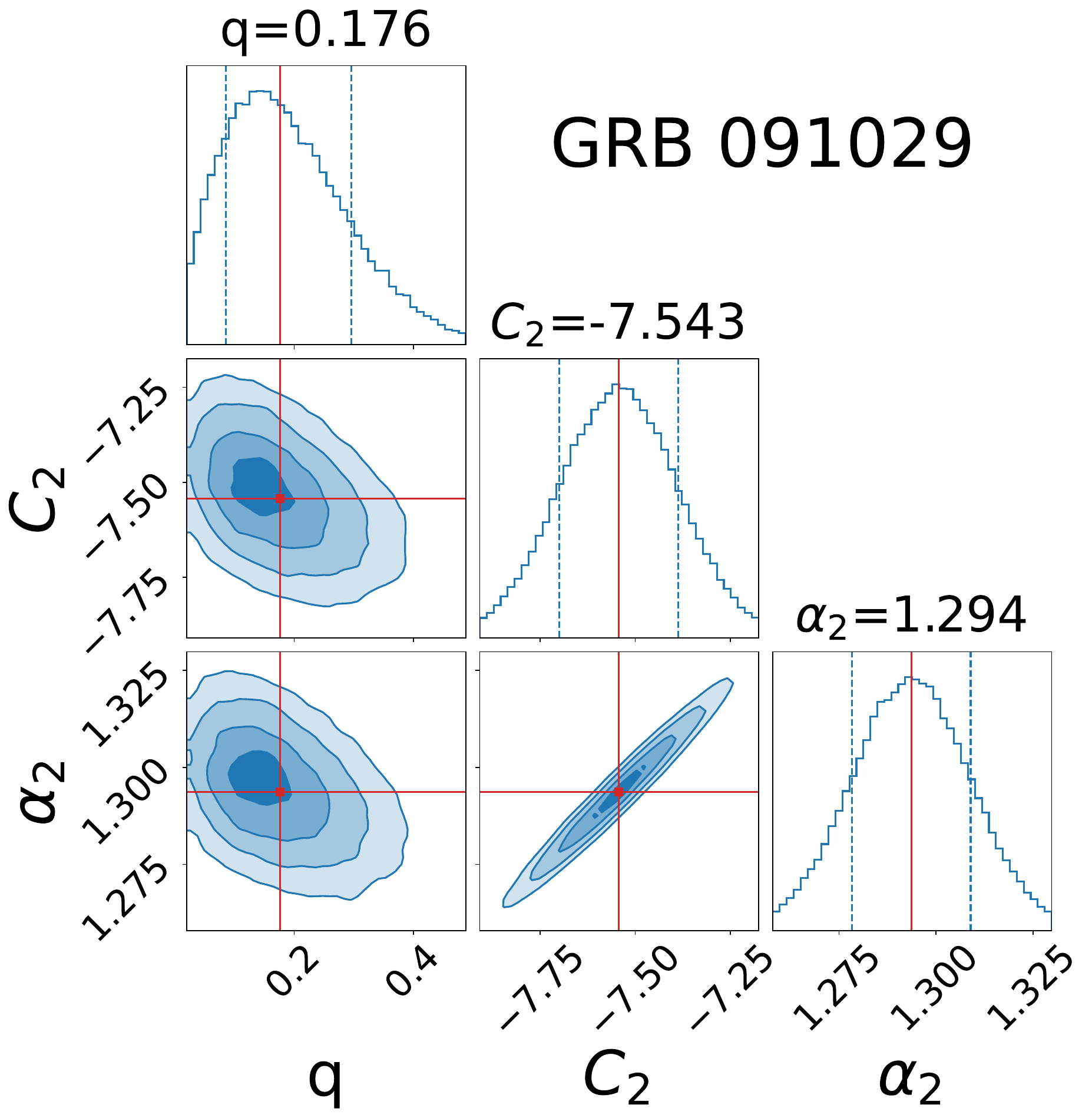}
  \end{subfigure}%
  \caption{Comparison of model fitting for representative GRBs in ISM (top: GRB 070420) and wind (bottom: GRB 091029) media. For each GRB, the upper and lower rows show results for model~1 and model~2, respectively.}
  \label{fig:pdf_figures_simple}
\end{figure}

We applied both models to fit the afterglow light curves of 20 GRBs in the ISM and 20 GRBs in a wind medium. In Fig.~\ref{fig:pdf_figures_simple}, we present representative cases from each medium to illustrate the two fitting approaches: GRB 070420 (ISM) and GRB 091029 (wind). The model fits for both media align closely with the observed data points. Moreover, the MCMC sampling yields broadly consistent estimates for the key parameters between the two models, with a small systematic offset that we quantify below.

The detailed fitting results for each GRB are presented in Table~\ref{tab:table_2}, which summarizes the key parameters, including the off-axis ratio $q$, and viewing angle $\theta_{\text{obs}}$ for both models, along with their uncertainties derived from the MCMC posterior distributions. Through case-by-case analysis of individual samples and statistical analysis of the entire sample set, we obtained a deeper understanding of the behavioral differences between the two models. For individual events in both interstellar medium and stellar wind environments, either model may return the larger $q$: on a burst-by-burst basis the inter-model difference is comparable to the random uncertainty introduced by data quality, so single-object results are primarily driven by case-specific characteristics. In the aggregate, however, a systematic offset is present. Model~1 gives the larger $q$ in 26 of the 40 bursts, the median ratio is $q_1/q_2 = 1.24$, and the sample means are $\bar{q}_1 = 0.1851$ against $\bar{q}_2 = 0.1545$ (paired Wilcoxon signed-rank test, $p = 0.048$). Model~1 tends to overestimate $q$ by roughly 20 per cent relative to model~2. The physical origin of this offset is discussed below.

Table~\ref{tab:table_2} also reports the $\chi^2_{\rm red}$ and BIC values for each burst. Both models share the same number of free parameters ($k = 3$) and are fitted to the same data, so the difference $\Delta\mathrm{BIC} = \mathrm{BIC}_2 - \mathrm{BIC}_1$, listed in the last column, directly reflects the $\chi^2$ difference. For all 40 bursts, model~1 yields lower $\chi^2_{\rm red}$ and BIC values than model~2 ($\Delta\mathrm{BIC} > 0$ without exception), indicating a systematically better description of the observed light curves. This preference reflects how the two models treat the jet break in relation to our sample selection. Model~2 performs a full two-dimensional angular integration weighted by a continuous Doppler factor $a=(1-\beta)/(1-\beta\cos\theta)$, which decays smoothly with angle but never vanishes. This replaces the discrete geometric crossing in model~1 with a gradual flux redistribution as the beaming cone widens, inherently producing a smoother jet break. Our sample, however, is selected via $\Delta\alpha$ constraints to preferentially retain bursts with well-defined, sharp breaks, penalizing model~2 in the goodness-of-fit metrics despite its more complete physical treatment. Given its systematically lower BIC values and better match to the sample, model~1 is adopted as the preferred framework for the statistical analyses that follow. We note, however, that model~1 lacks the Doppler-weighted angular integration and therefore attributes all observed break smoothness solely to geometric off-axis effects, systematically overestimating $q$.

A related caveat concerns model~2 itself. The coefficient of variation $q_{\rm cv,2}$ in Table~\ref{tab:table_2} lies close to $0.55$ for the large majority of bursts, essentially independently of data quality, whereas $q_{\rm cv,1}$ varies from $0.17$ to $0.60$ and tracks the quality of the individual light curve. This indicates that the $q$ posterior of model~2 retains much of the shape of the prior, and that the $q_2$ values are better read as upper limits than as detections. The offset between $q_1$ and $q_2$ quantified above should therefore be taken as an indication of the size of the model-dependent systematic on $q$, rather than as a calibration of one model against the other.

\begin{table*}[htbp]
\centering
\caption{GRB fitting results.}
\label{tab:table_2}
\begin{tabular}{cccccccccccc}
\toprule
GRB & $q_{1}$ & $\theta_{\mathrm{obs,1}}$ & $q_{\mathrm{cv,1}}$ & $\chi^{2}_{\mathrm{red,1}}$ & $\mathrm{BIC}_{1}$ & $q_{2}$ & $\theta_{\mathrm{obs,2}}$ & $q_{\mathrm{cv,2}}$ & $\chi^{2}_{\mathrm{red,2}}$ & $\mathrm{BIC}_{2}$ & $\Delta\mathrm{BIC}$ \\
\midrule
050318 & $0.135^{+0.075}_{-0.066}$ & $0.006^{+0.003}_{-0.003}$ & $0.489$ & $0.9$ & $76.0$ & $0.233^{+0.170}_{-0.120}$ & $0.010^{+0.007}_{-0.005}$ & $0.624$ & $1.3$ & $103.5$ & $27.5$ \\
050505 & $0.230^{+0.063}_{-0.078}$ & $0.009^{+0.002}_{-0.003}$ & $0.321$ & $0.9$ & $145.5$ & $0.158^{+0.106}_{-0.081}$ & $0.006^{+0.004}_{-0.003}$ & $0.549$ & $1.0$ & $169.8$ & $24.3$ \\
060313 & $0.142^{+0.061}_{-0.063}$ & $0.005^{+0.002}_{-0.002}$ & $0.424$ & $0.8$ & $47.7$ & $0.080^{+0.058}_{-0.042}$ & $0.003^{+0.002}_{-0.002}$ & $0.585$ & $2.0$ & $103.4$ & $55.7$ \\
061121 & $0.318^{+0.045}_{-0.052}$ & $0.013^{+0.002}_{-0.002}$ & $0.165$ & $0.8$ & $180.3$ & $0.047^{+0.035}_{-0.025}$ & $0.002^{+0.001}_{-0.001}$ & $0.590$ & $1.2$ & $257.1$ & $76.8$ \\
061222A & $0.161^{+0.040}_{-0.055}$ & $0.008^{+0.002}_{-0.003}$ & $0.312$ & $1.1$ & $309.8$ & $0.027^{+0.021}_{-0.015}$ & $0.001^{+0.001}_{-0.001}$ & $0.612$ & $1.7$ & $488.8$ & $179.0$ \\
070318 & $0.141^{+0.099}_{-0.072}$ & $0.019^{+0.013}_{-0.010}$ & $0.598$ & $1.6$ & $102.6$ & $0.217^{+0.144}_{-0.114}$ & $0.029^{+0.019}_{-0.015}$ & $0.536$ & $1.8$ & $116.5$ & $13.9$ \\
070420 & $0.177^{+0.083}_{-0.084}$ & $0.008^{+0.004}_{-0.004}$ & $0.461$ & $1.4$ & $121.4$ & $0.182^{+0.123}_{-0.093}$ & $0.008^{+0.006}_{-0.004}$ & $0.549$ & $1.7$ & $143.2$ & $21.8$ \\
080710 & $0.156^{+0.068}_{-0.070}$ & $0.009^{+0.004}_{-0.004}$ & $0.427$ & $1.2$ & $79.4$ & $0.183^{+0.123}_{-0.094}$ & $0.010^{+0.007}_{-0.005}$ & $0.550$ & $1.7$ & $111.0$ & $31.6$ \\
090516 & $0.077^{+0.047}_{-0.038}$ & $0.002^{+0.001}_{-0.001}$ & $0.492$ & $1.1$ & $157.2$ & $0.120^{+0.083}_{-0.062}$ & $0.003^{+0.002}_{-0.001}$ & $0.555$ & $1.7$ & $228.1$ & $70.9$ \\
100425A & $0.385^{+0.128}_{-0.168}$ & $0.031^{+0.010}_{-0.014}$ & $0.389$ & $0.8$ & $27.3$ & $0.194^{+0.126}_{-0.101}$ & $0.016^{+0.010}_{-0.008}$ & $0.530$ & $1.0$ & $30.8$ & $3.5$ \\
110213A & $0.075^{+0.044}_{-0.034}$ & $0.002^{+0.001}_{-0.001}$ & $0.464$ & $1.0$ & $217.2$ & $0.076^{+0.054}_{-0.040}$ & $0.002^{+0.002}_{-0.001}$ & $0.572$ & $1.6$ & $352.6$ & $135.4$ \\
110709B & $0.105^{+0.055}_{-0.044}$ & $0.007^{+0.004}_{-0.003}$ & $0.476$ & $1.0$ & $369.4$ & $0.048^{+0.035}_{-0.026}$ & $0.003^{+0.002}_{-0.002}$ & $0.584$ & $1.1$ & $418.9$ & $49.5$ \\
110818A & $0.272^{+0.146}_{-0.128}$ & $0.010^{+0.006}_{-0.005}$ & $0.451$ & $1.2$ & $40.1$ & $0.251^{+0.163}_{-0.130}$ & $0.009^{+0.006}_{-0.005}$ & $0.531$ & $1.5$ & $49.0$ & $8.9$ \\
121024A & $0.254^{+0.097}_{-0.108}$ & $0.015^{+0.006}_{-0.006}$ & $0.391$ & $1.0$ & $55.6$ & $0.144^{+0.099}_{-0.075}$ & $0.008^{+0.006}_{-0.004}$ & $0.551$ & $1.3$ & $70.9$ & $15.3$ \\
121211A & $0.255^{+0.135}_{-0.128}$ & $0.019^{+0.010}_{-0.009}$ & $0.455$ & $0.9$ & $70.7$ & $0.218^{+0.139}_{-0.111}$ & $0.016^{+0.010}_{-0.008}$ & $0.528$ & $1.1$ & $80.1$ & $9.4$ \\
130606A & $0.163^{+0.084}_{-0.079}$ & $0.004^{+0.002}_{-0.002}$ & $0.489$ & $1.1$ & $69.5$ & $0.251^{+0.139}_{-0.121}$ & $0.006^{+0.004}_{-0.003}$ & $0.480$ & $1.4$ & $84.3$ & $14.8$ \\
160131A & $0.085^{+0.053}_{-0.040}$ & $0.005^{+0.003}_{-0.002}$ & $0.535$ & $1.0$ & $438.4$ & $0.189^{+0.041}_{-0.040}$ & $0.011^{+0.002}_{-0.002}$ & $0.215$ & $1.0$ & $456.2$ & $17.8$ \\
161017A & $0.249^{+0.063}_{-0.081}$ & $0.011^{+0.003}_{-0.004}$ & $0.312$ & $0.9$ & $71.1$ & $0.154^{+0.104}_{-0.080}$ & $0.007^{+0.005}_{-0.004}$ & $0.551$ & $1.3$ & $98.9$ & $27.8$ \\
210610B & $0.168^{+0.067}_{-0.078}$ & $0.010^{+0.004}_{-0.005}$ & $0.419$ & $0.9$ & $113.6$ & $0.156^{+0.094}_{-0.077}$ & $0.009^{+0.006}_{-0.005}$ & $0.511$ & $1.2$ & $144.1$ & $30.5$ \\
210722A & $0.166^{+0.136}_{-0.090}$ & $0.007^{+0.006}_{-0.004}$ & $0.567$ & $0.9$ & $57.7$ & $0.128^{+0.090}_{-0.067}$ & $0.005^{+0.004}_{-0.003}$ & $0.566$ & $1.3$ & $80.4$ & $22.7$ \\
050915A & $0.314^{+0.126}_{-0.123}$ & $0.016^{+0.007}_{-0.006}$ & $0.379$ & $1.0$ & $31.1$ & $0.265^{+0.164}_{-0.135}$ & $0.014^{+0.009}_{-0.007}$ & $0.521$ & $1.3$ & $38.7$ & $7.6$ \\
060210 & $0.107^{+0.044}_{-0.032}$ & $0.003^{+0.001}_{-0.001}$ & $0.339$ & $1.0$ & $263.6$ & $0.075^{+0.054}_{-0.039}$ & $0.002^{+0.002}_{-0.001}$ & $0.566$ & $1.4$ & $356.9$ & $93.3$\\
070508 & $0.265^{+0.102}_{-0.102}$ & $0.016^{+0.006}_{-0.006}$ & $0.361$ & $0.9$ & $124.7$ & $0.172^{+0.117}_{-0.089}$ & $0.011^{+0.007}_{-0.005}$ & $0.548$ & $1.0$ & $139.3$ & $14.6$ \\
080413B & $0.181^{+0.065}_{-0.068}$ & $0.017^{+0.006}_{-0.007}$ & $0.349$ & $1.1$ & $172.8$ & $0.116^{+0.087}_{-0.061}$ & $0.011^{+0.008}_{-0.006}$ & $0.583$ & $1.2$ & $200.7$ & $27.9$ \\
090424 & $0.258^{+0.105}_{-0.110}$ & $0.031^{+0.013}_{-0.013}$ & $0.417$ & $1.2$ & $166.8$ & $0.171^{+0.116}_{-0.089}$ & $0.021^{+0.014}_{-0.011}$ & $0.542$ & $1.2$ & $171.3$ & $4.5$ \\
090618 & $0.240^{+0.137}_{-0.104}$ & $0.018^{+0.010}_{-0.008}$ & $0.421$ & $1.1$ & $364.4$ & $0.213^{+0.135}_{-0.108}$ & $0.016^{+0.010}_{-0.008}$ & $0.521$ & $1.1$ & $374.6$ & $10.2$ \\
091020 & $0.242^{+0.067}_{-0.065}$ & $0.009^{+0.002}_{-0.002}$ & $0.265$ & $1.1$ & $235.9$ & $0.079^{+0.055}_{-0.041}$ & $0.003^{+0.002}_{-0.002}$ & $0.565$ & $1.3$ & $277.7$ & $41.8$ \\
091029 & $0.145^{+0.067}_{-0.053}$ & $0.006^{+0.003}_{-0.002}$ & $0.406$ & $1.1$ & $130.0$ & $0.177^{+0.119}_{-0.091}$ & $0.008^{+0.005}_{-0.004}$ & $0.552$ & $1.3$ & $150.5$ & $20.5$ \\
110422A & $0.122^{+0.055}_{-0.044}$ & $0.002^{+0.001}_{-0.001}$ & $0.370$ & $1.0$ & $265.1$ & $0.092^{+0.065}_{-0.048}$ & $0.002^{+0.001}_{-0.001}$ & $0.565$ & $1.3$ & $333.1$ & $68.0$ \\
111008A & $0.266^{+0.069}_{-0.077}$ & $0.009^{+0.002}_{-0.003}$ & $0.280$ & $0.9$ & $109.0$ & $0.230^{+0.149}_{-0.117}$ & $0.008^{+0.005}_{-0.004}$ & $0.541$ & $1.0$ & $119.4$ & $10.4$ \\
120712A & $0.144^{+0.079}_{-0.058}$ & $0.003^{+0.002}_{-0.001}$ & $0.432$ & $1.2$ & $96.1$ & $0.193^{+0.134}_{-0.100}$ & $0.004^{+0.003}_{-0.002}$ & $0.549$ & $1.6$ & $119.6$ & $23.5$ \\
120909A & $0.208^{+0.067}_{-0.067}$ & $0.004^{+0.001}_{-0.001}$ & $0.319$ & $0.9$ & $104.9$ & $0.273^{+0.185}_{-0.139}$ & $0.006^{+0.004}_{-0.003}$ & $0.574$ & $1.0$ & $118.8$ & $13.9$ \\
130505A & $0.083^{+0.027}_{-0.019}$ & $0.001^{+0.001}_{-0.001}$ & $0.253$ & $1.0$ & $412.0$ & $0.022^{+0.018}_{-0.012}$ & $0.001^{+0.001}_{-0.001}$ & $0.624$ & $2.0$ & $802.6$ & $390.6$ \\
140206A & $0.261^{+0.048}_{-0.062}$ & $0.011^{+0.002}_{-0.003}$ & $0.226$ & $1.0$ & $257.0$ & $0.148^{+0.100}_{-0.076}$ & $0.006^{+0.004}_{-0.003}$ & $0.546$ & $1.1$ & $282.3$ & $25.3$ \\
140419A & $0.152^{+0.037}_{-0.045}$ & $0.002^{+0.001}_{-0.001}$ & $0.266$ & $0.9$ & $281.4$ & $0.049^{+0.036}_{-0.026}$ & $0.001^{+0.001}_{-0.001}$ & $0.587$ & $1.4$ & $434.3$ & $152.9$ \\
140423A & $0.135^{+0.082}_{-0.051}$ & $0.004^{+0.002}_{-0.001}$ & $0.456$ & $1.1$ & $107.2$ & $0.292^{+0.192}_{-0.150}$ & $0.008^{+0.005}_{-0.004}$ & $0.564$ & $1.3$ & $122.9$ & $15.7$ \\
140512A & $0.128^{+0.046}_{-0.040}$ & $0.007^{+0.003}_{-0.002}$ & $0.321$ & $1.1$ & $245.5$ & $0.184^{+0.129}_{-0.095}$ & $0.011^{+0.007}_{-0.005}$ & $0.696$ & $1.3$ & $284.5$ & $39.0$ \\
150403A & $0.145^{+0.054}_{-0.046}$ & $0.010^{+0.004}_{-0.003}$ & $0.374$ & $1.1$ & $429.1$ & $0.261^{+0.146}_{-0.128}$ & $0.018^{+0.010}_{-0.009}$ & $0.482$ & $1.2$ & $441.7$ & $12.6$ \\
200829A & $0.171^{+0.048}_{-0.049}$ & $0.004^{+0.001}_{-0.001}$ & $0.282$ & $0.9$ & $441.6$ & $0.084^{+0.059}_{-0.044}$ & $0.002^{+0.001}_{-0.001}$ & $0.566$ & $1.0$ & $512.6$ & $71.0$ \\
220101A & $0.123^{+0.041}_{-0.042}$ & $0.003^{+0.001}_{-0.001}$ & $0.310$ & $1.2$ & $513.2$ & $0.031^{+0.024}_{-0.017}$ & $0.001^{+0.001}_{-0.001}$ & $0.614$ & $1.5$ & $629.1$ & $115.9$ \\

\bottomrule
\end{tabular}
\begin{tablenotes}
    \item Notes. Subscripts 1 and 2 refer to model~1 and model~2, respectively. $q_{\mathrm{cv}}$ is the coefficient of variation of the $q$ posterior. $\Delta\mathrm{BIC} = \mathrm{BIC}_2 - \mathrm{BIC}_1$; a positive value favours model~1. BIC values depend on the number of data points and are therefore comparable between the two models for a given burst, but not between different bursts.
\end{tablenotes}
\end{table*}

\begin{figure}[htbp]
\centering
  \begin{subfigure}[b]{0.35\textwidth}
    \includegraphics[width=\linewidth]{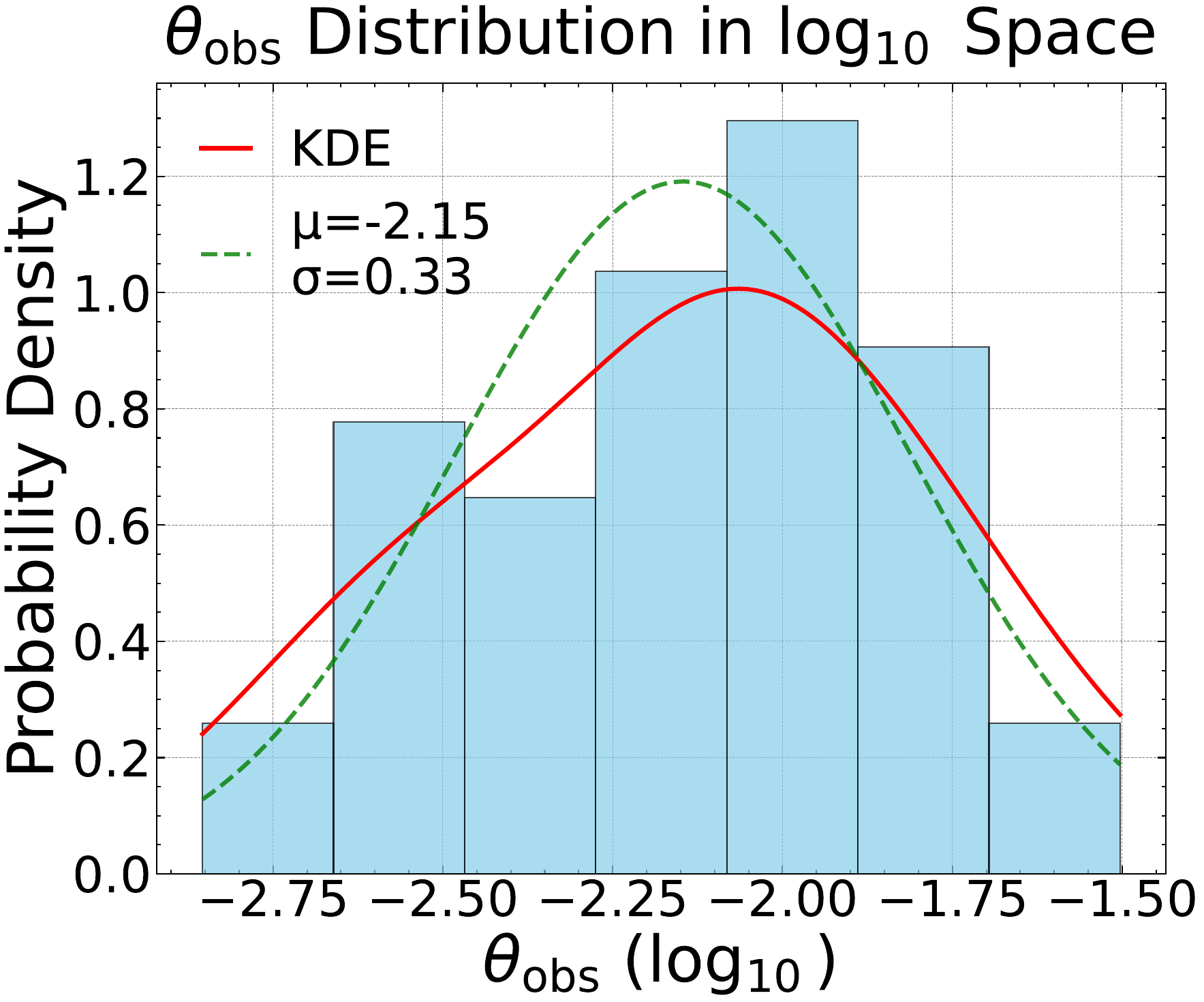}
  \end{subfigure}%
  \caption{Distribution of the viewing angle $\theta_{\rm obs}$ inferred from model~1, plotted in $\log_{10}$ space. The red solid line is the kernel density estimate and the green dashed line is the Gaussian fit.}
  \label{fig:Parameter distributions}
\end{figure}

Fig.~\ref{fig:Parameter distributions} shows the distribution of viewing angles $\theta_{\text{obs}}$ for model~1, plotted in logarithmic space for all samples. The kernel density estimation (KDE) curve exhibits a shape that closely approximates a Gaussian distribution and peaks at small angles, indicating that the GRB samples generally have small off-axis angles. The fitting results of model~1 yield mean off-axis ratios of $\bar{q} = 0.1858$ in ISM and $\bar{q} = 0.1845$ in the wind medium. We performed a Kolmogorov-Smirnov (K-S) test for the off-axis ratio $q$ and obtained a statistic of $D = 0.1500$ with a probability $p = 0.9831$, which indicates no significant difference between the ISM and wind medium distributions. This indicates comparable viewing angle characteristics for both GRB populations despite distinct circumburst media.

We examined whether the presence of an X-ray plateau in the afterglow is related to the inferred off-axis ratio. Of the 40 GRBs in our sample, 36 (those observed before 2019) have reliable plateau classifications from the catalog of \citet{2022ApJ...924...69Y}; among these, 18 exhibit an X-ray plateau (marked with asterisks in Table~\ref{tab:table_1}). A two-sample K-S test on the $q$ distributions of the two groups (18 with and 18 without a plateau) yields $D = 0.2222$, $p = 0.7810$, indicating no significant difference; we caution that with this sample size the test has limited power against a modest offset. Within the top-hat approximation, we therefore find no evidence that the off-axis ratio depends on the presence of an X-ray plateau, consistent with the plateau being primarily a central-engine phenomenon \citep{2006ApJ...642..389N}. However, this does not rule out correlations in off-axis structured jets, where plateau-like features can arise from delayed core emission \citep{2020MNRAS.492.2847B}.

\begin{figure}[htbp]
\centering
  \begin{subfigure}[b]{0.39\textwidth}
    \includegraphics[width=\linewidth]{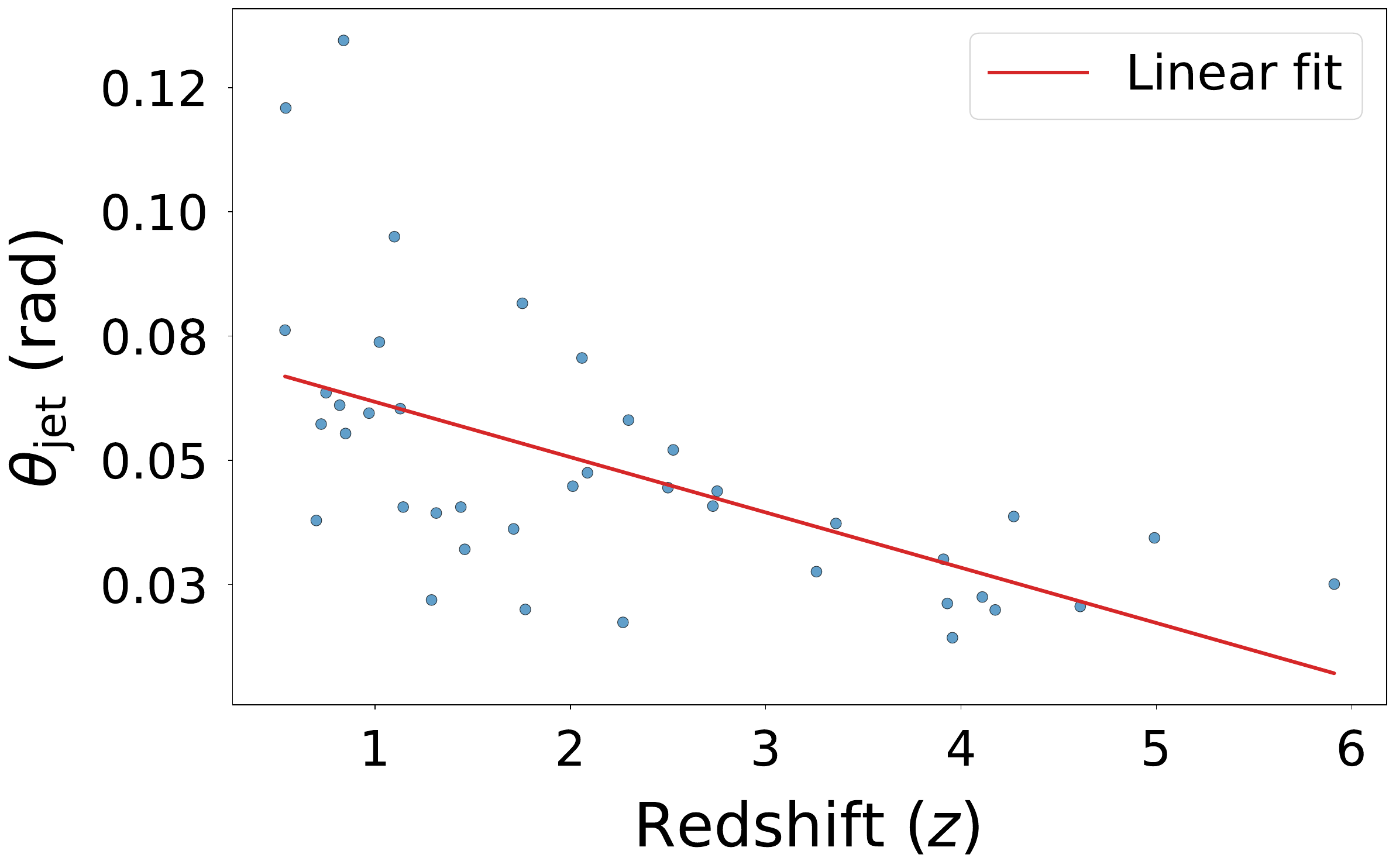}
    \caption{}
  \end{subfigure}
  \begin{subfigure}[b]{0.39\textwidth}
    \includegraphics[width=\linewidth]{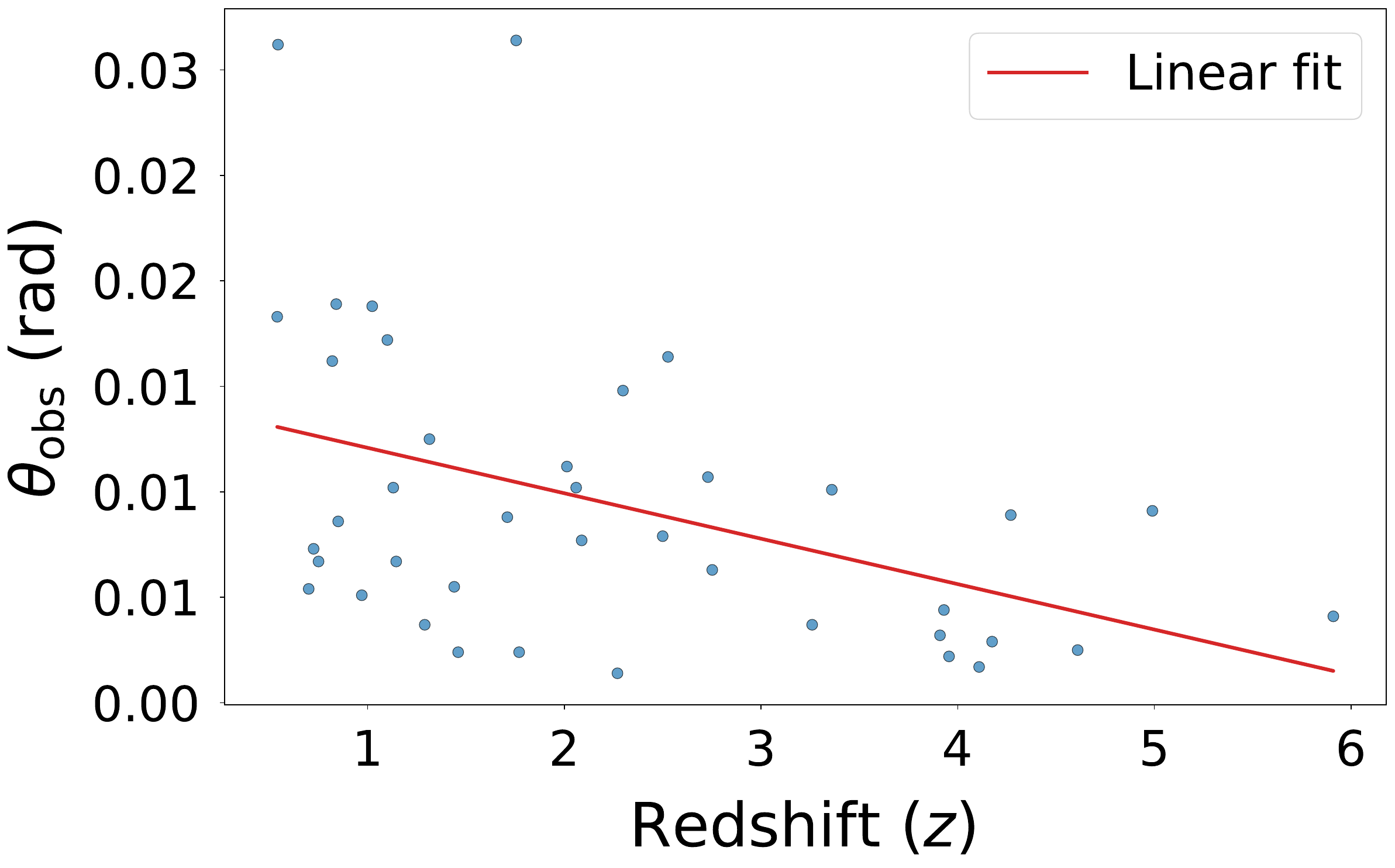}
    \caption{}
  \end{subfigure}
  \begin{subfigure}[b]{0.39\textwidth}
    \includegraphics[width=\linewidth]{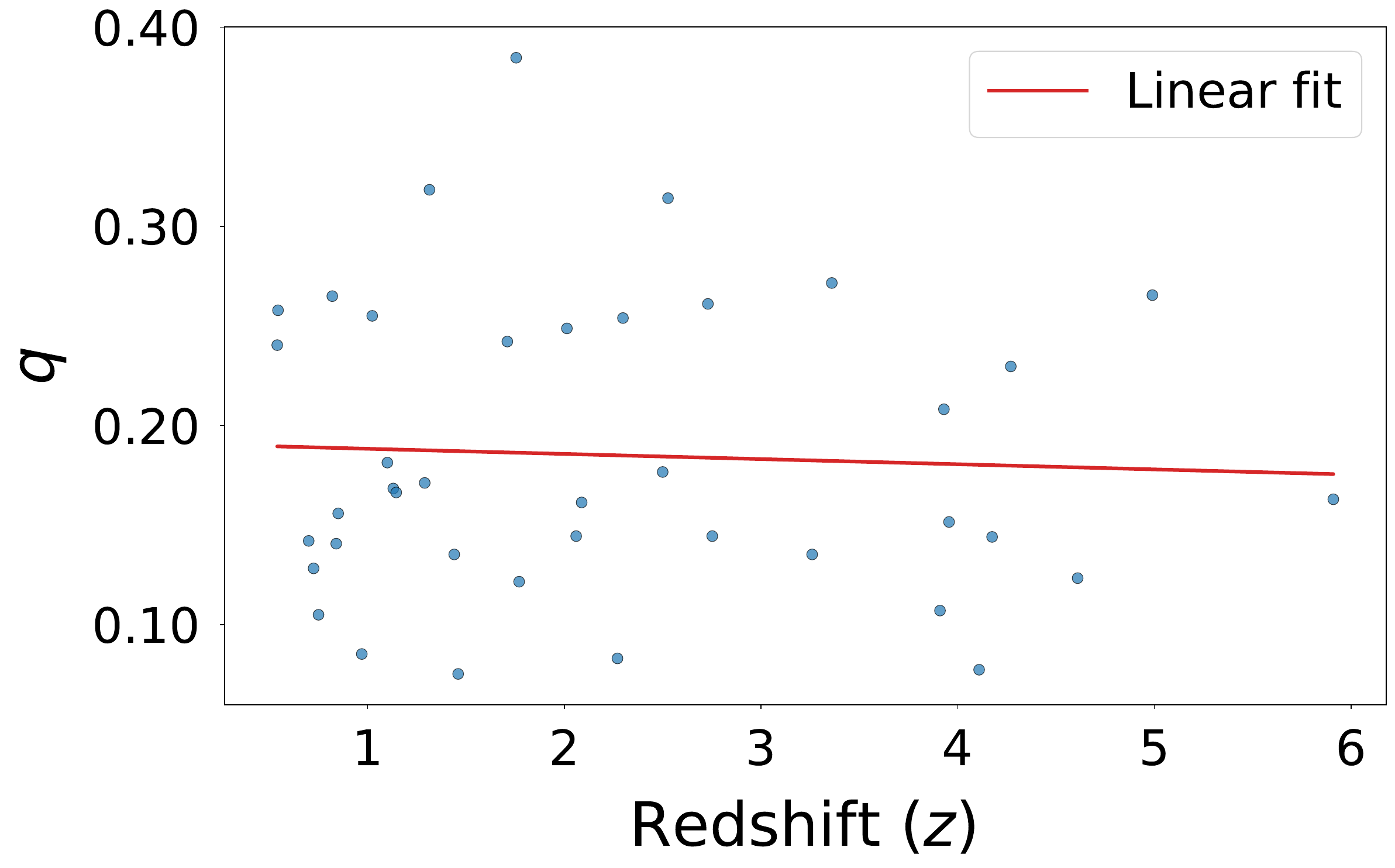}
    \caption{}
  \end{subfigure}
  \caption{Redshift evolution of jet properties from model~1: (a)~$\theta_{\rm jet}$, (b)~$\theta_{\rm obs}$ and (c)~off-axis ratio $q$.}
  \label{fig:evolution}
\end{figure}

We further investigated the redshift evolution of jet properties using model~1, which provides a statistically preferred description of the light curves (Fig.~\ref{fig:evolution}). A significant Spearman rank anti-correlation is found between $\theta_{\rm jet}$ and redshift ($\rho = -0.6664$, $p < 0.0001$), consistent with previous studies showing that higher-redshift GRBs possess narrower jets \citep{2019ApJ...871..118L, 2019MNRAS.488.5823L, 2020MNRAS.494.4371L}. This trend must, however, be interpreted with care. Our $\theta_{\rm jet}$ is not measured independently but derived from Equations~(\ref{eq:theta_ism})--(\ref{eq:theta_wind}), which contain explicit factors of $(1+z)^{-3/8}$ (ISM) or $(1+z)^{-1/4}$ (wind) and of $E_{\gamma,\rm iso}^{-1/8}$ or $E_{\gamma,\rm iso}^{-1/4}$, while $E_{\gamma,\rm iso}$ itself increases with redshift in a flux-limited sample. Part of the anti-correlation is therefore induced by construction and by Malmquist bias. Controlling for $E_{\gamma,\rm iso}$, the partial Spearman coefficient is $\rho_{\rm partial} = -0.5681$ ($p = 0.0002$), indicating that the correlation persists even after removing the effect of $E_{\gamma,\rm iso}$. Since $\theta_{\rm obs} = q\,\theta_{\rm jet}$ and $q$ is uncorrelated with redshift (see below), the viewing angle simply inherits this trend ($\rho = -0.4289$, $p = 0.0058$) rather than constituting an independent result. In contrast, the off-axis ratio $q$ shows no significant evolution ($\rho = -0.0035$, $p = 0.9830$), indicating that the degree of off-axis alignment is independent of redshift. This independence supports the robustness of our aggregate $q$ statistics against redshift-dependent selection effects, while the observed $\theta_{\rm jet}$ evolution underscores the importance of accounting for beaming-angle evolution in GRB population studies \citep{2020MNRAS.498.5041L}.

\section{Conclusion and discussion}\label{sec:Conclusion and discussion}

Based on the Swift/XRT sample, we performed MCMC fitting and statistical analysis of the X-ray afterglow light curves for 40 GRBs using two top-hat jet models: model~1 (without high-latitude emission) and model~2 (with high-latitude emission). Our analysis yields the following conclusions:

\begin{itemize}
    \item Model comparison based on $\chi^2_{\rm red}$ and BIC prefers model~1 for all 40 bursts; we adopt it for the statistical analysis, noting that it overestimates $q$ by roughly 20 per cent relative to model~2.
    \item The inferred viewing angles are close to the jet axis for most events, with a mean off-axis ratio $\bar{q} = 0.1851$. The logarithmic distribution of viewing angles is approximately Gaussian.
    \item A K-S test finds no significant difference between the off-axis ratio distributions in ISM and wind media ($D = 0.1500$, $p = 0.9831$).
    \item No significant difference in the off-axis ratio is found between bursts with and without an X-ray plateau ($D = 0.2222$, $p = 0.7810$).
    \item While both $\theta_{\rm jet}$ and $\theta_{\rm obs}$ exhibit significant anti-correlations with redshift ($\rho = -0.6664$, $p < 0.0001$; $\rho = -0.4289$, $p = 0.0058$), the off-axis ratio $q$ shows no significant evolution ($\rho = -0.0035$, $p = 0.9830$), consistent with the degree of off-axis alignment being independent of cosmic epoch. We stress that the $\theta_{\rm jet}$ trend is partly induced by the explicit redshift and energy dependence of Equations~(\ref{eq:theta_ism})--(\ref{eq:theta_wind}).
\end{itemize}

Our study has several limitations. The selection criteria may introduce biases: requiring a reliable jet break identification preferentially retains GRBs with clean afterglow morphology, while the exclusion of flare-contaminated intervals and data-coverage requirements favor smoother light curves. The $\Delta\alpha$ constraints further select events compatible with the top-hat jet framework, so the resulting sample likely spans a narrower range of viewing angle behavior than an unbiased population. Additionally, only 4 of 40 GRBs have sufficient optical coverage to verify multi-wavelength achromaticity at the jet break; this criterion could not be imposed as a formal selection requirement. The reported $q$ values are subject to model-dependent systematics. Model~1 lacks the physical smoothing from an angular-dependent $\Gamma$ and therefore tends to overestimate $q$; model~2, conversely, inherently smooths the jet break through its Doppler-weighted angular integration, which is penalized by a sample selected for sharp breaks; its $q$ posteriors remain prior-dominated, so its $q$ values are effectively upper limits. The circumburst medium is assigned from the measured $\Delta\alpha$ rather than from independent evidence, and is restricted to $k = 0$ or $k = 2$, whereas intermediate profiles are known to occur. Four bursts lack a spectroscopic redshift. The absence of a plateau--$q$ correlation and the redshift independence of $q$ are established within the top-hat framework and may differ in structured jet geometries. Environmental simplifications (fixed $n_0$, $A_*$) and the neglect of lateral jet expansion introduce additional systematic uncertainties.

Future work could extend this methodology in several directions. First, relaxing the top-hat approximation to incorporate structured jet profiles with angular-dependent energy and Lorentz factor distributions would enable more realistic modeling of off-axis emission. Second, given the achromatic nature of jet breaks, multi-wavelength joint fitting could provide tighter constraints on the viewing angle by simultaneously modeling the break across different energy bands, improving both the precision of the inferred parameters and the robustness against single-band systematics.

\begin{acknowledgments}
We thank the anonymous referee for constructive comments that improved the quality of this manuscript. This work made use of data supplied by the UK Swift Science Data Centre at the University of Leicester. This work is supported by Shandong Provincial Natural Science Foundation (ZR2025MS47, ZR2025MS16), the Natural Science Foundation of Jiangxi Province of China (grant No. 20242BAB26012) and the China Manned Space Project (CMS-CSST-2021-A12).
\end{acknowledgments}

\clearpage
\onecolumngrid
%\section{Appendices}
\appendix
\label{sec:Appendices}
The following figures show the model fitting results and posterior distributions for all GRBs in our sample.
\begin{figure}[htbp]
\centering
  \begin{subfigure}[b]{0.23\textwidth}
    \includegraphics[width=\linewidth]{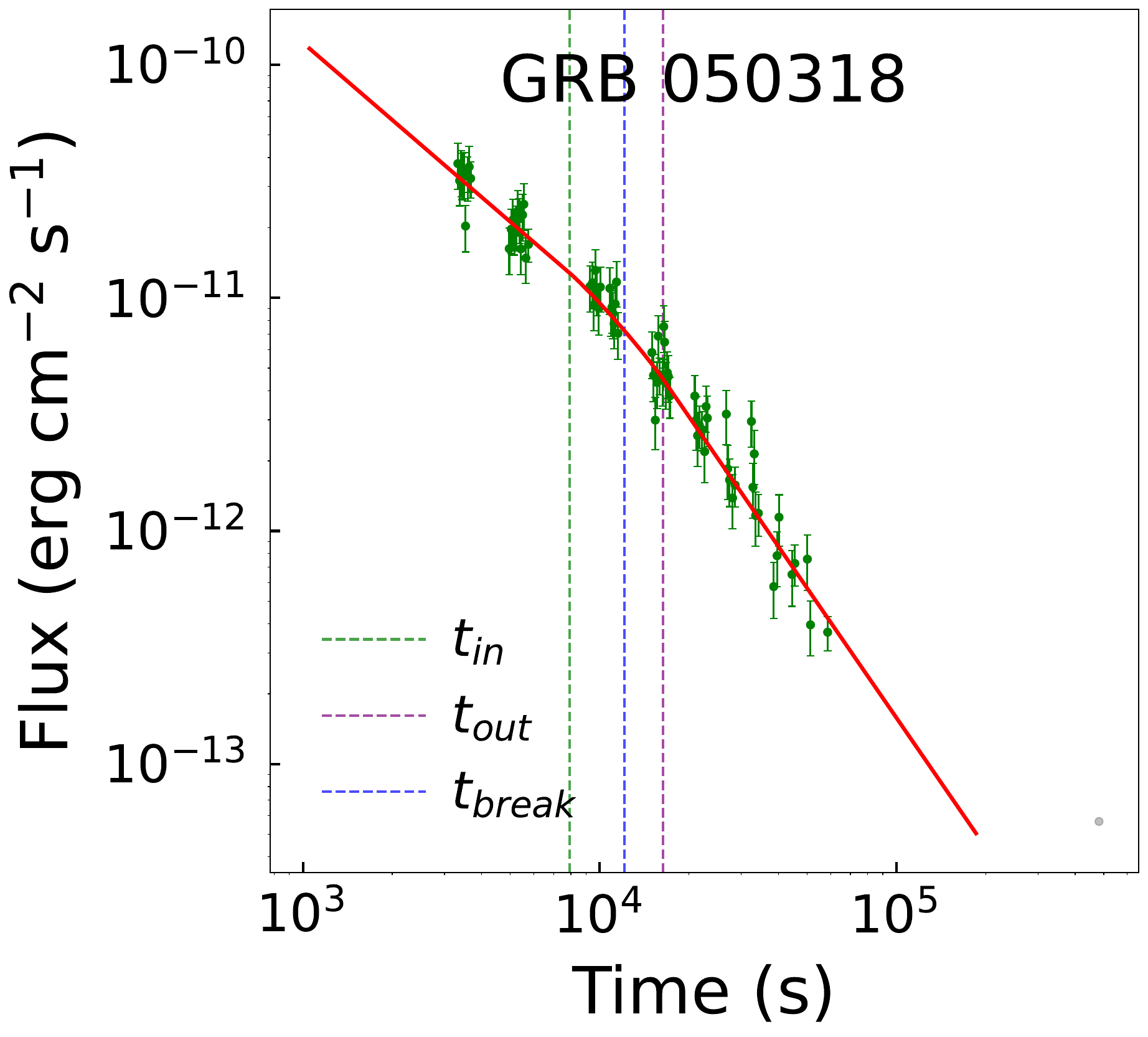}
  \end{subfigure}%
  \hspace{0.01\textwidth}
  \begin{subfigure}[b]{0.23\textwidth}
    \includegraphics[width=\linewidth]{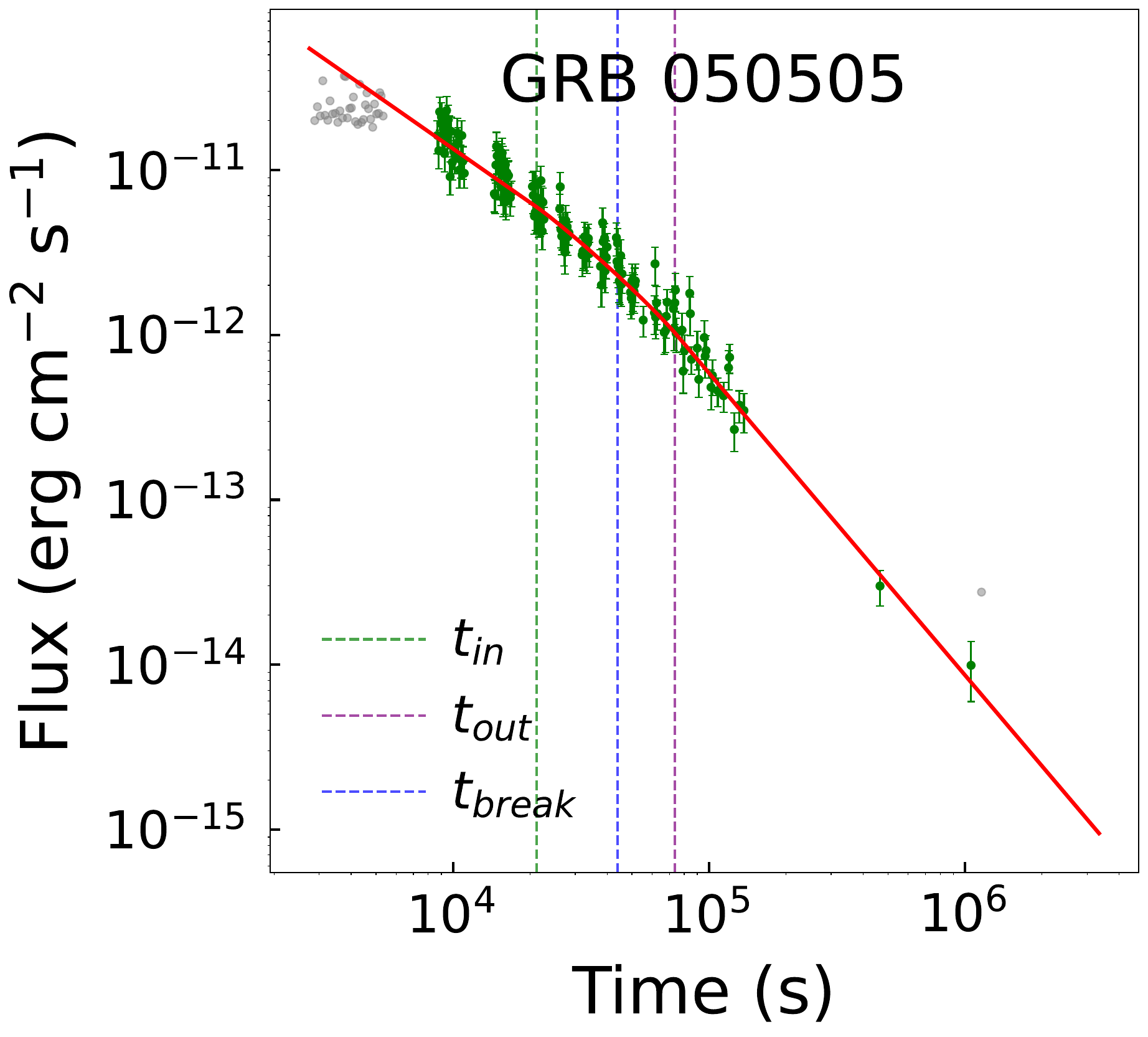}
  \end{subfigure}%
  \hspace{0.01\textwidth}
  \begin{subfigure}[b]{0.23\textwidth}
    \includegraphics[width=\linewidth]{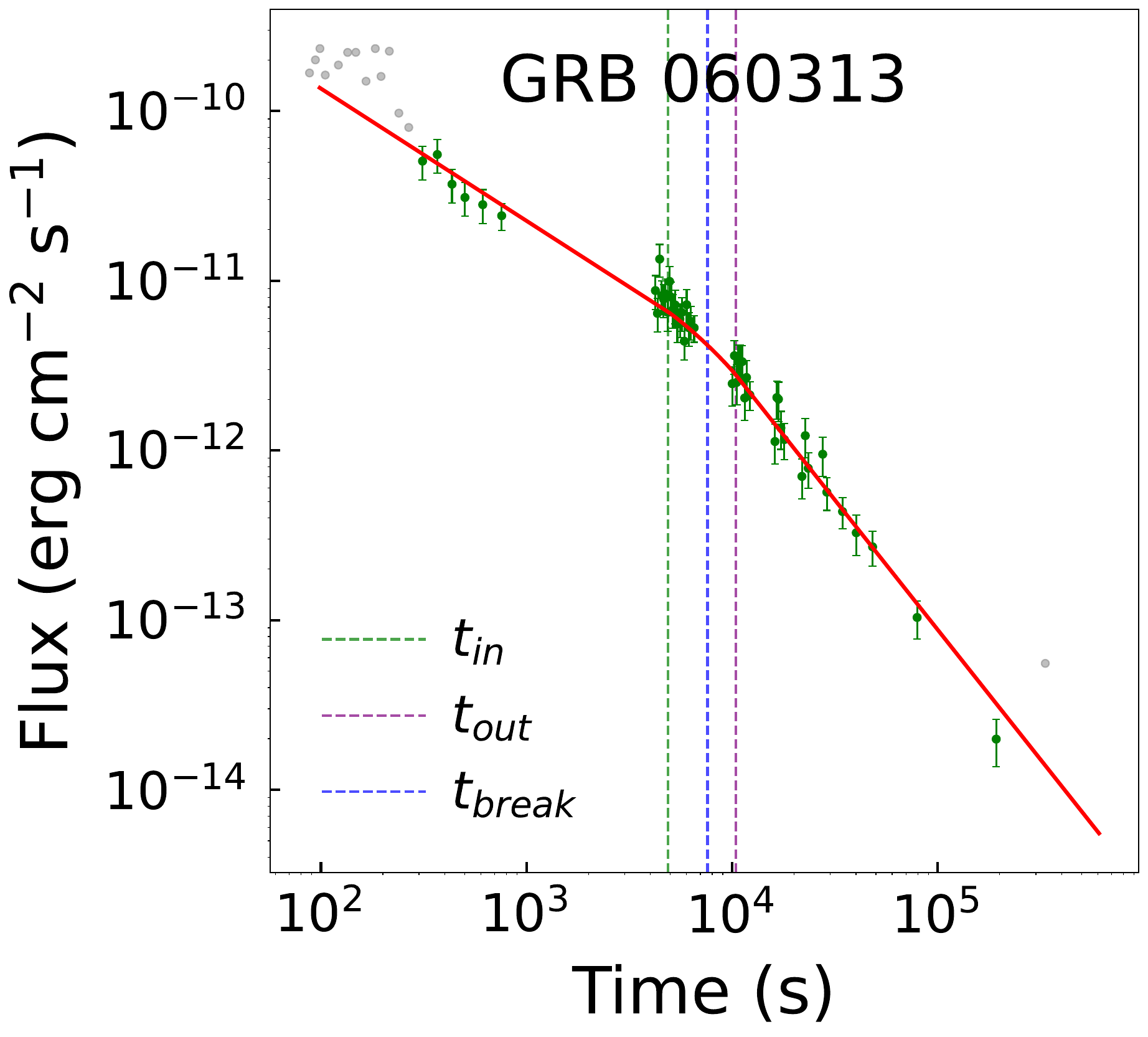}
  \end{subfigure}%
  \hspace{0.01\textwidth}
  \begin{subfigure}[b]{0.23\textwidth}
    \includegraphics[width=\linewidth]{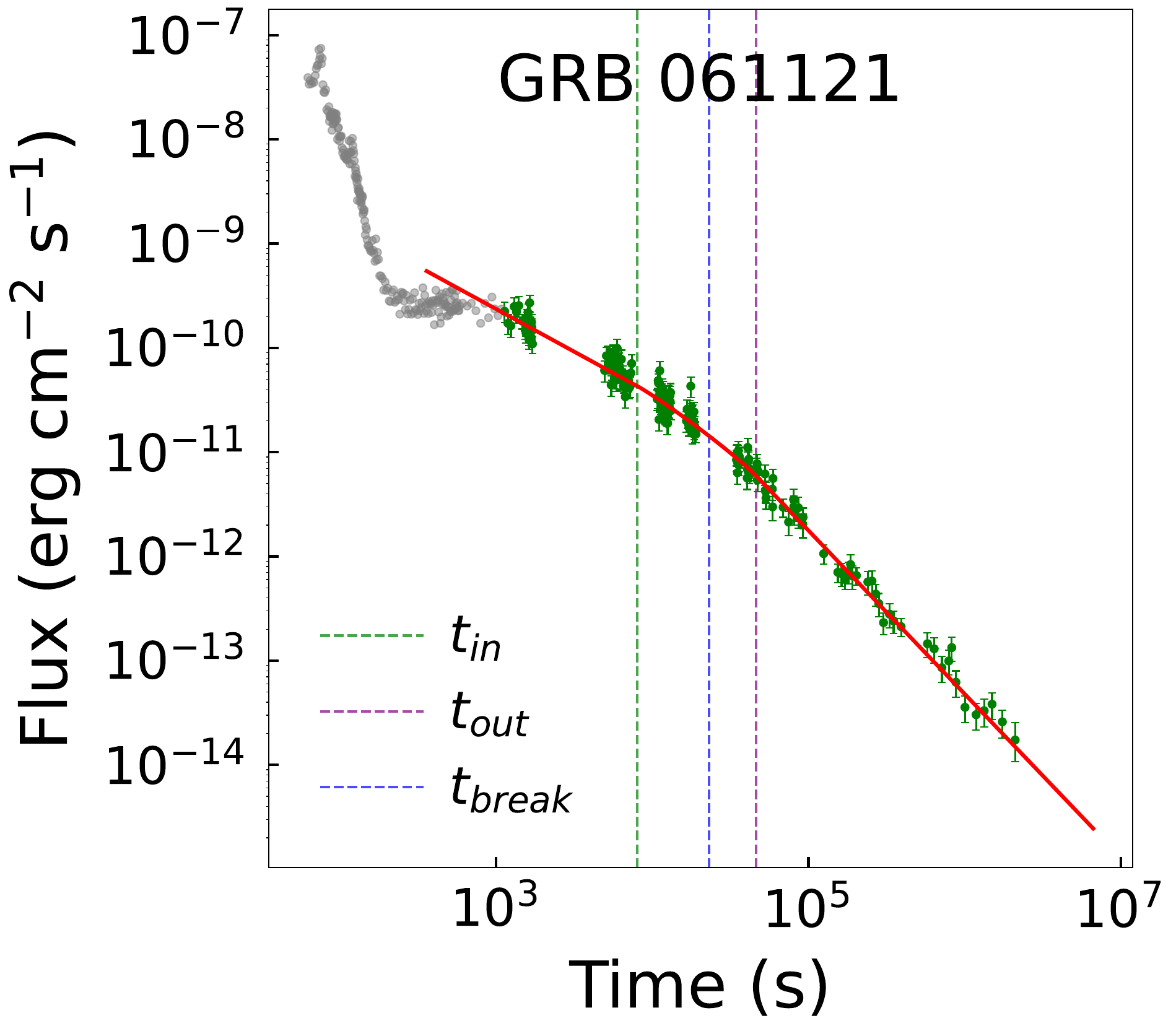}
  \end{subfigure}%
  \\%
  \begin{subfigure}[b]{0.23\textwidth}
    \includegraphics[width=\linewidth]{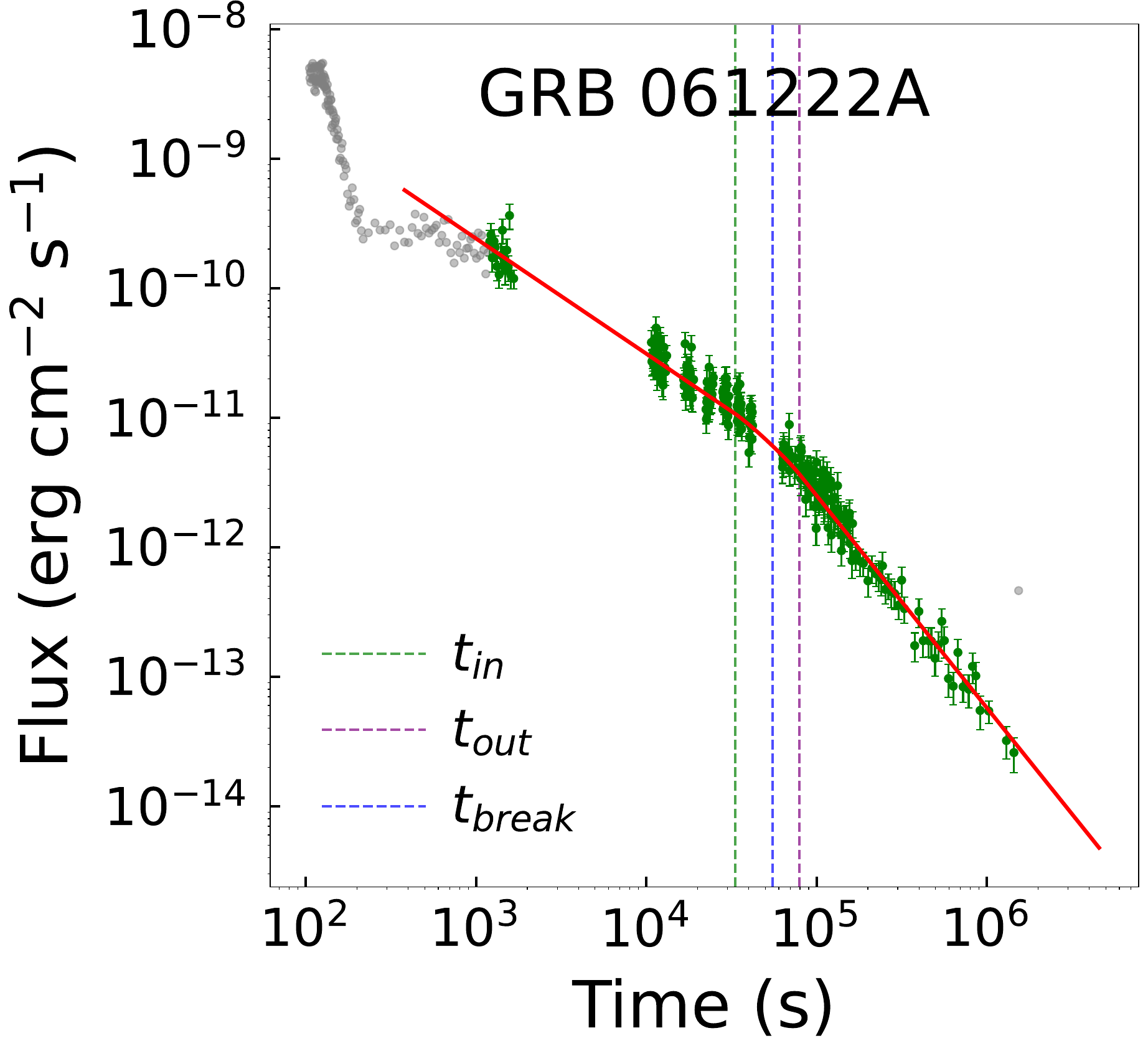}
  \end{subfigure}%
  \hspace{0.01\textwidth}
  \begin{subfigure}[b]{0.23\textwidth}
    \includegraphics[width=\linewidth]{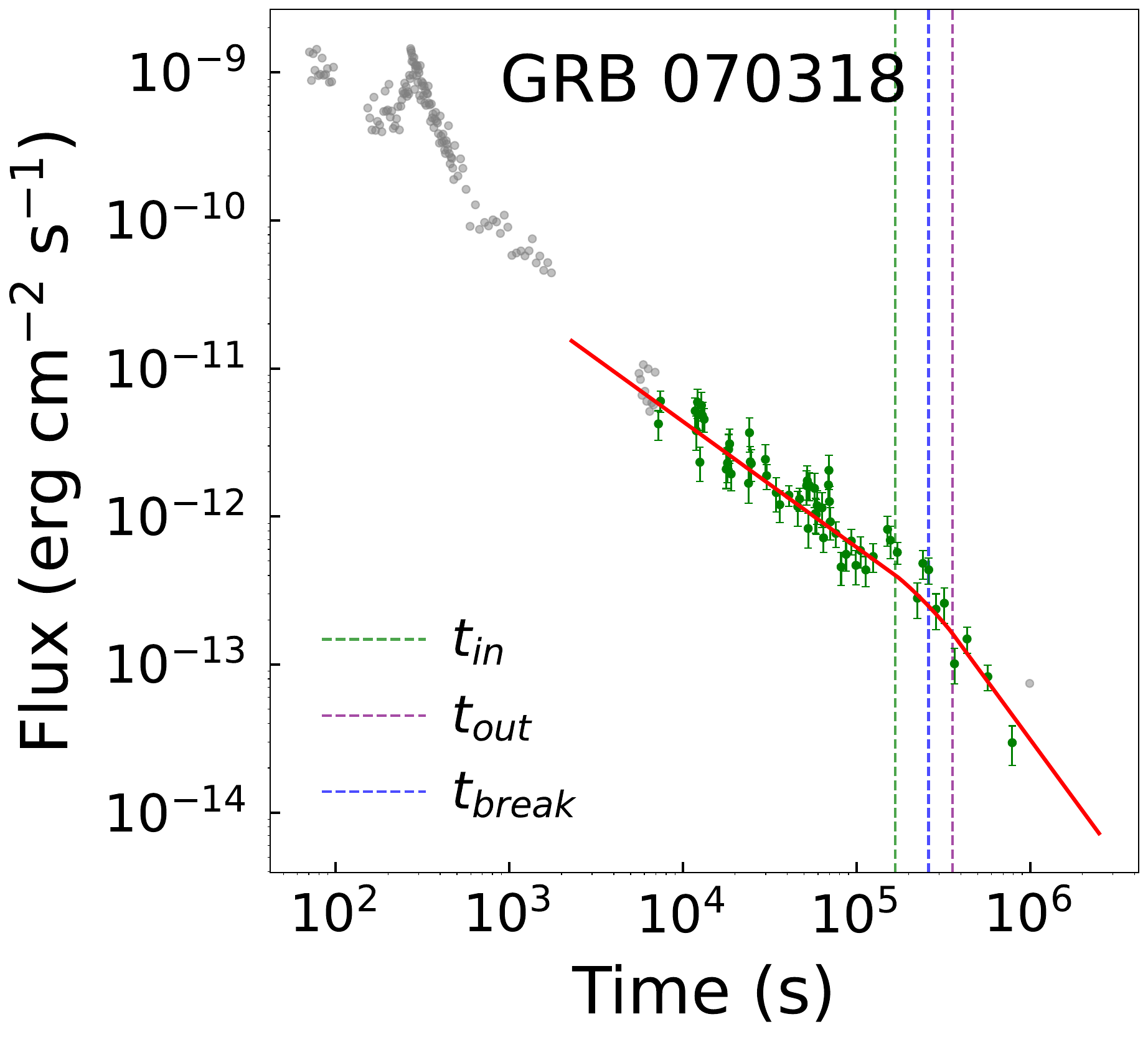}
  \end{subfigure}%
  \hspace{0.01\textwidth}
  \begin{subfigure}[b]{0.23\textwidth}
    \includegraphics[width=\linewidth]{result_070420_10_flux_vs_time.pdf}
  \end{subfigure}%
  \hspace{0.01\textwidth}
  \begin{subfigure}[b]{0.23\textwidth}
    \includegraphics[width=\linewidth]{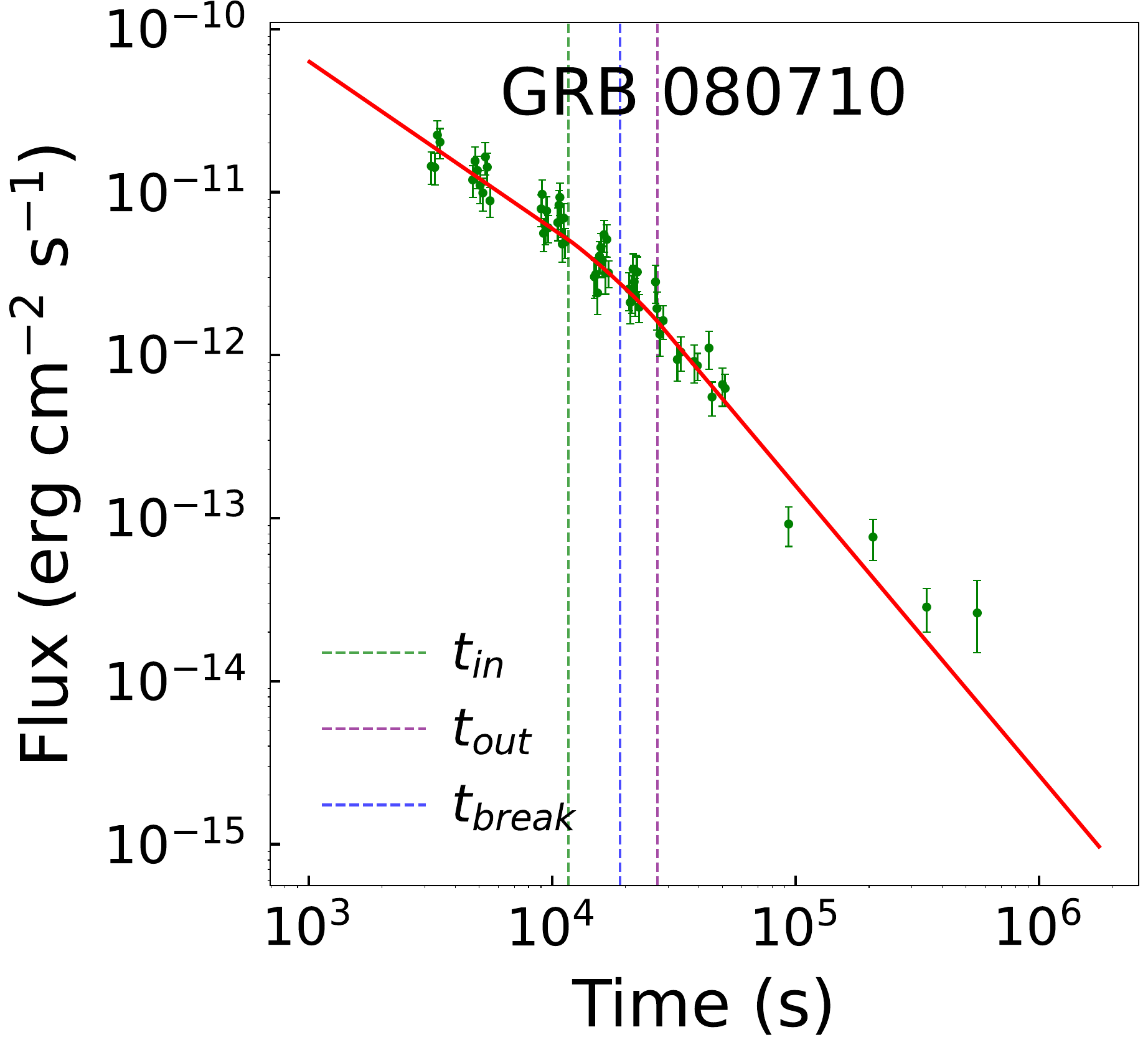}
  \end{subfigure}%
  \\%
  \begin{subfigure}[b]{0.23\textwidth}
    \includegraphics[width=\linewidth]{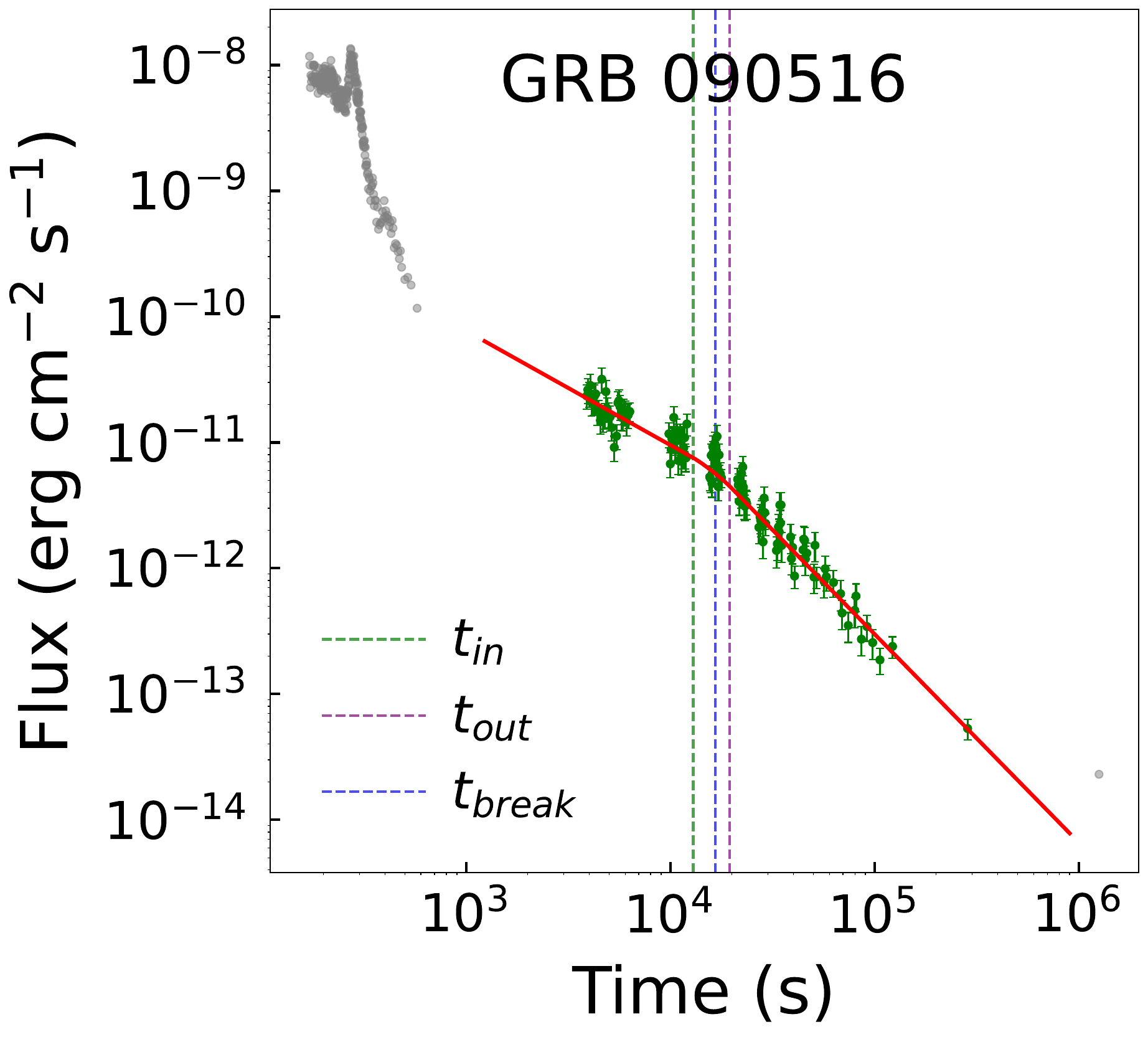}
  \end{subfigure}%
  \hspace{0.01\textwidth}
  \begin{subfigure}[b]{0.23\textwidth}
    \includegraphics[width=\linewidth]{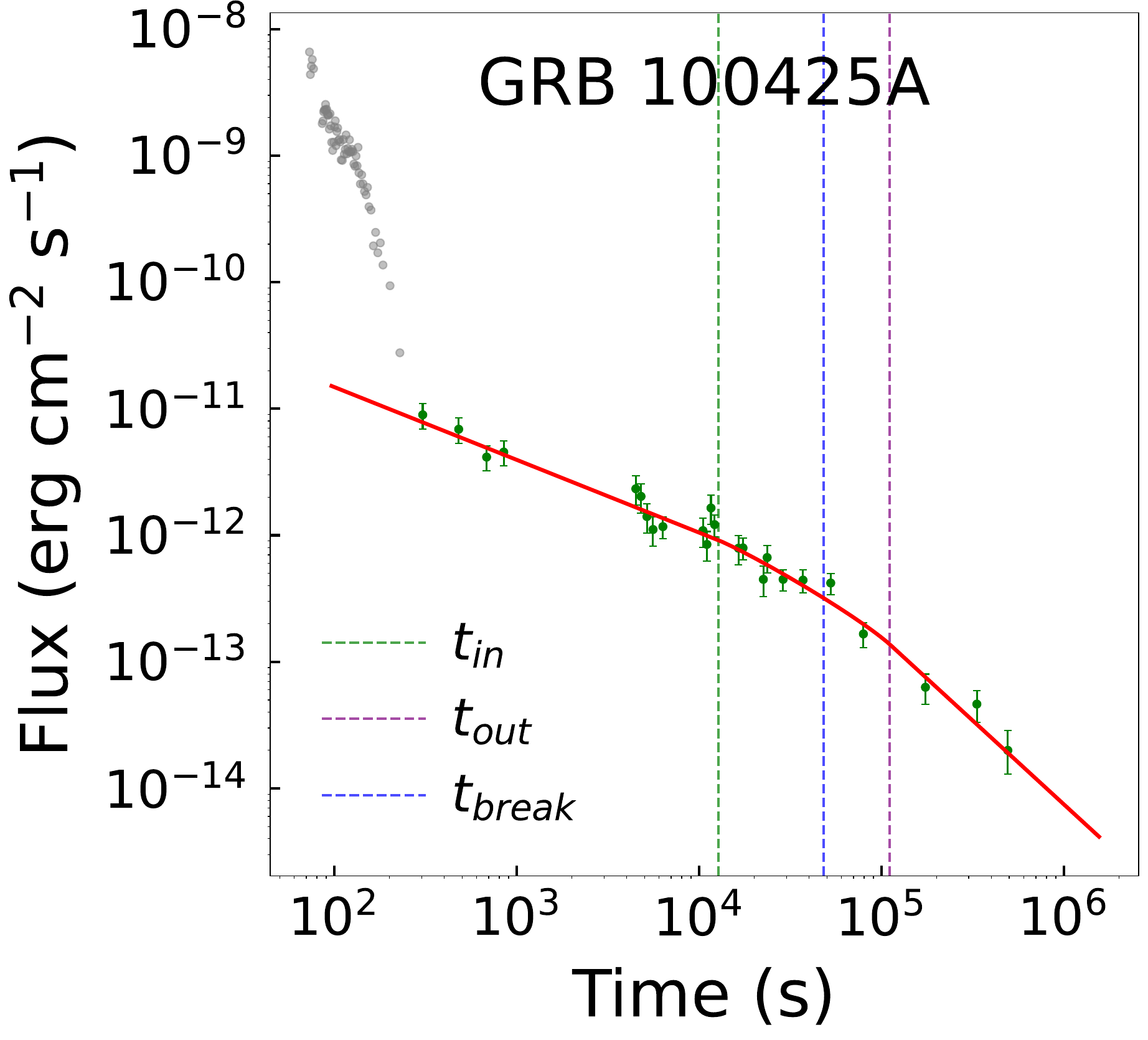}
  \end{subfigure}%
  \hspace{0.01\textwidth}
  \begin{subfigure}[b]{0.23\textwidth}
    \includegraphics[width=\linewidth]{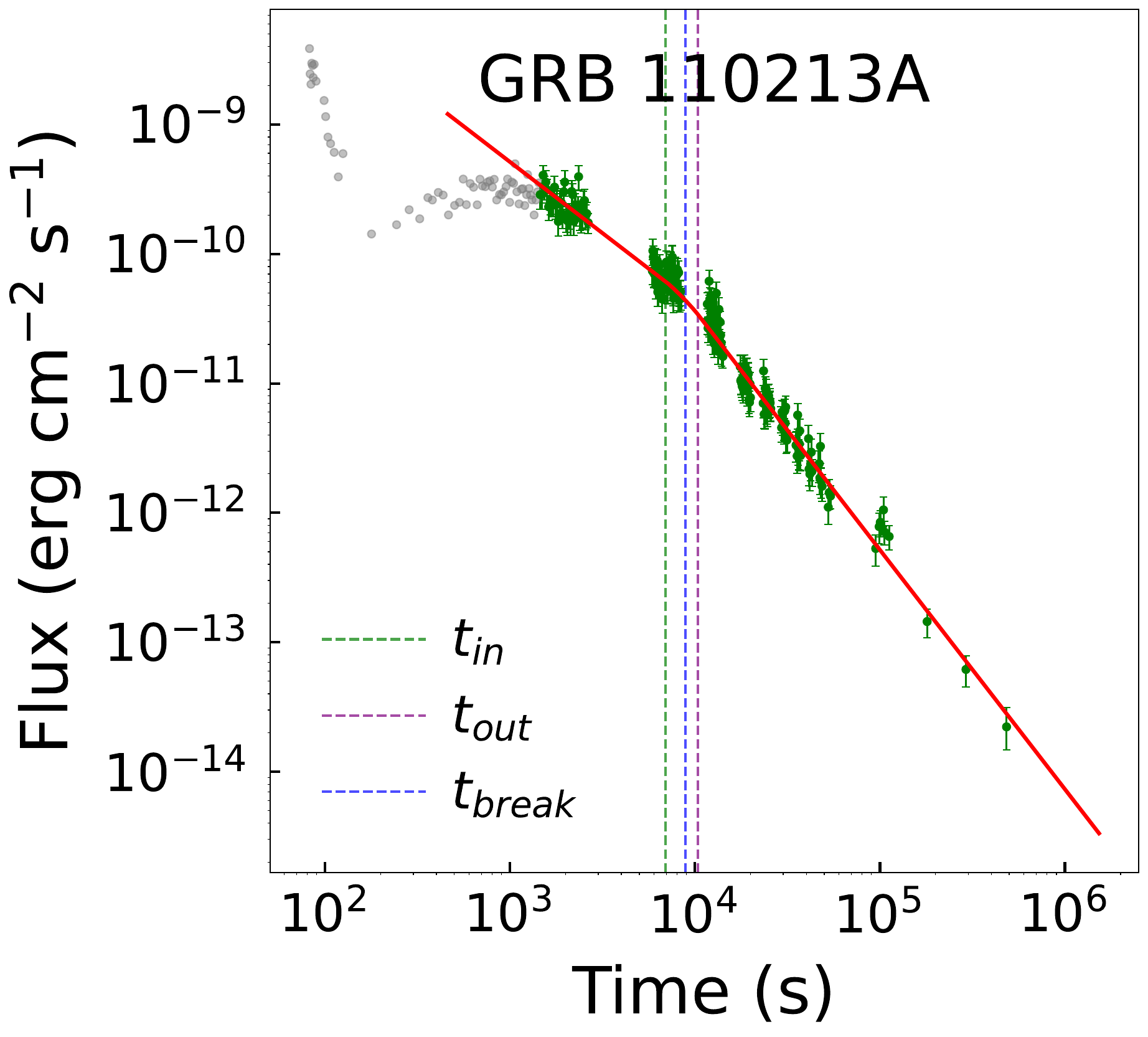}
  \end{subfigure}%
  \hspace{0.01\textwidth}
  \begin{subfigure}[b]{0.23\textwidth}
    \includegraphics[width=\linewidth]{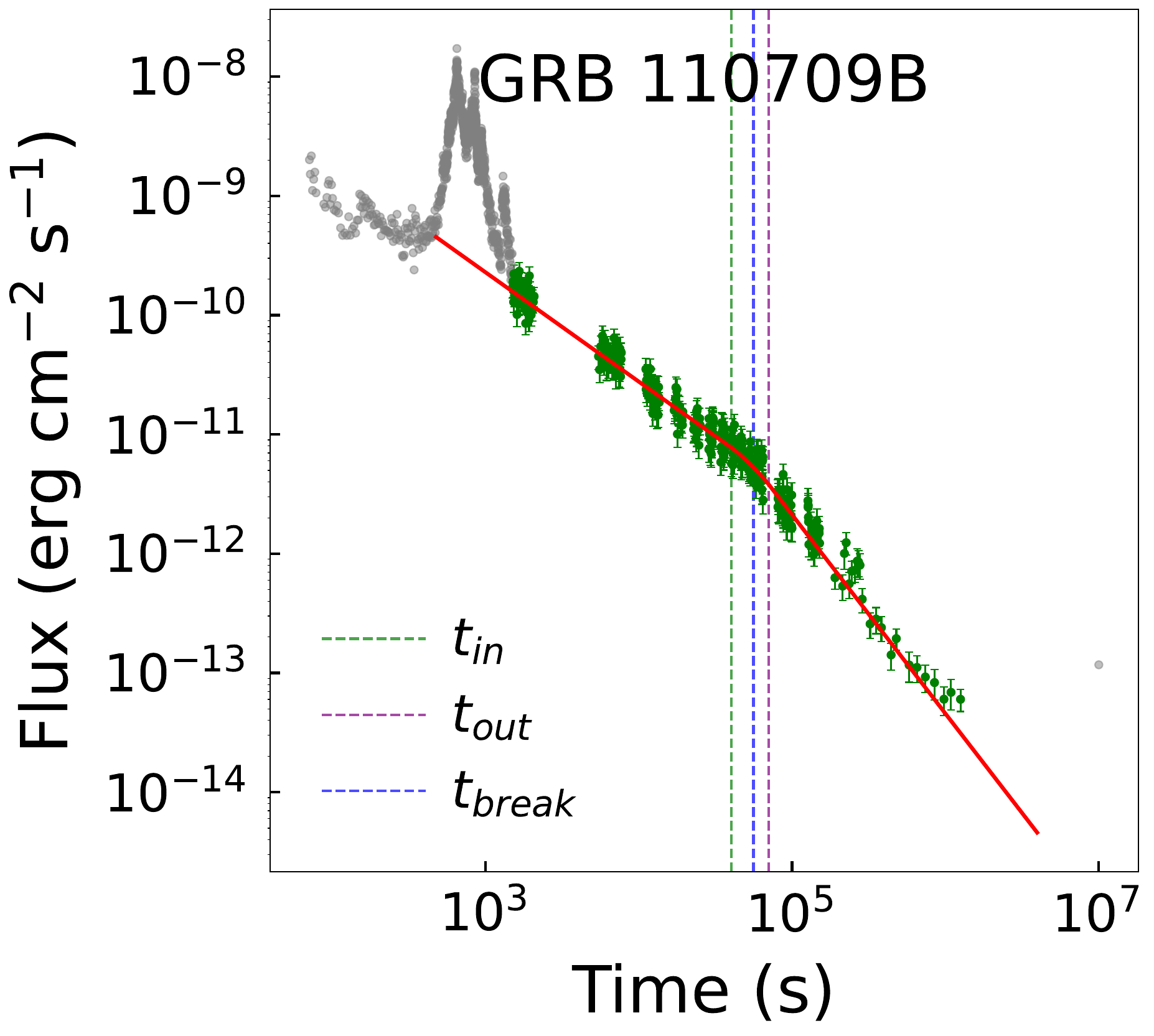}
  \end{subfigure}%
  \\%
  \begin{subfigure}[b]{0.23\textwidth}
    \includegraphics[width=\linewidth]{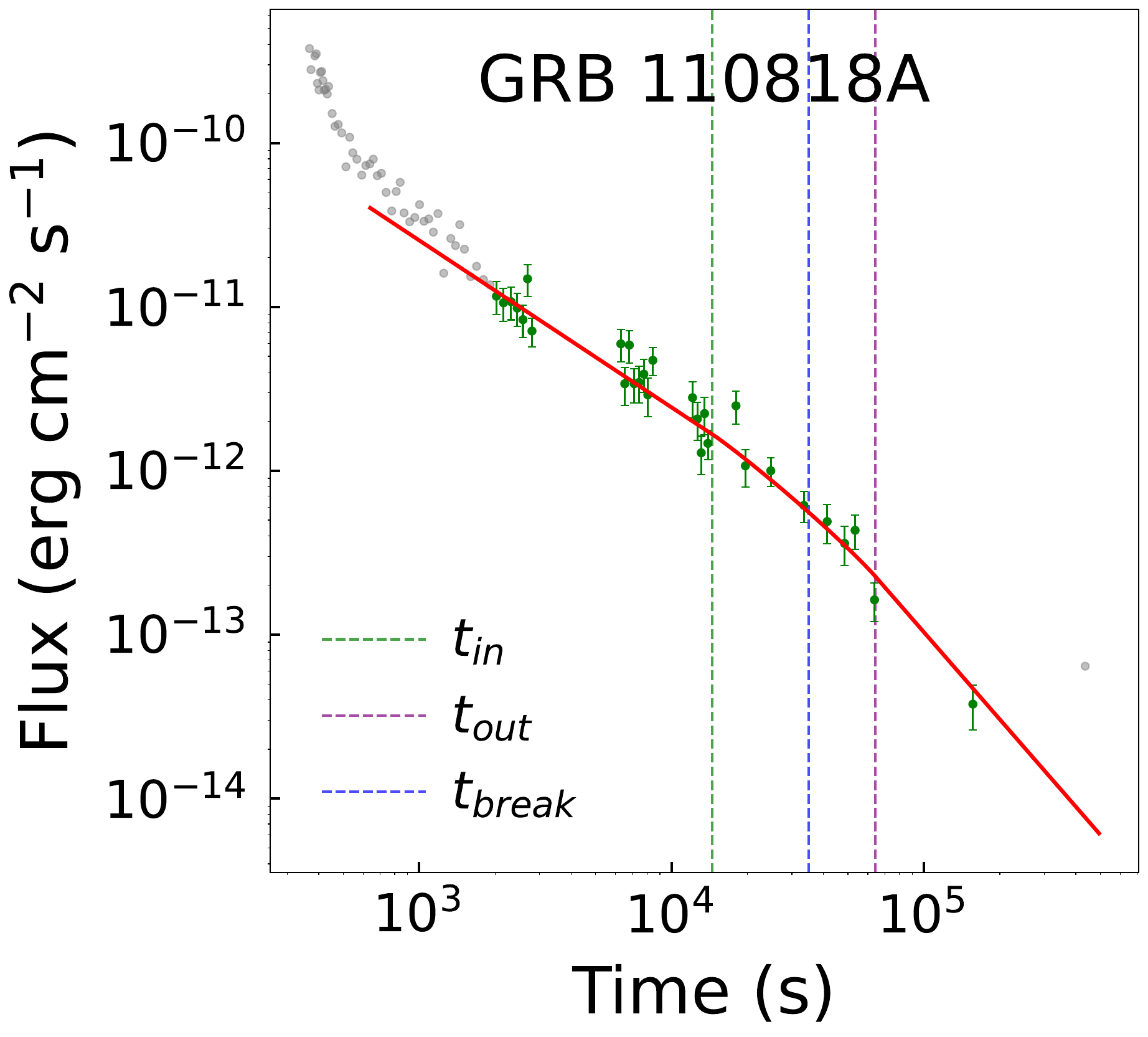}
  \end{subfigure}%
  \hspace{0.01\textwidth}
  \begin{subfigure}[b]{0.23\textwidth}
    \includegraphics[width=\linewidth]{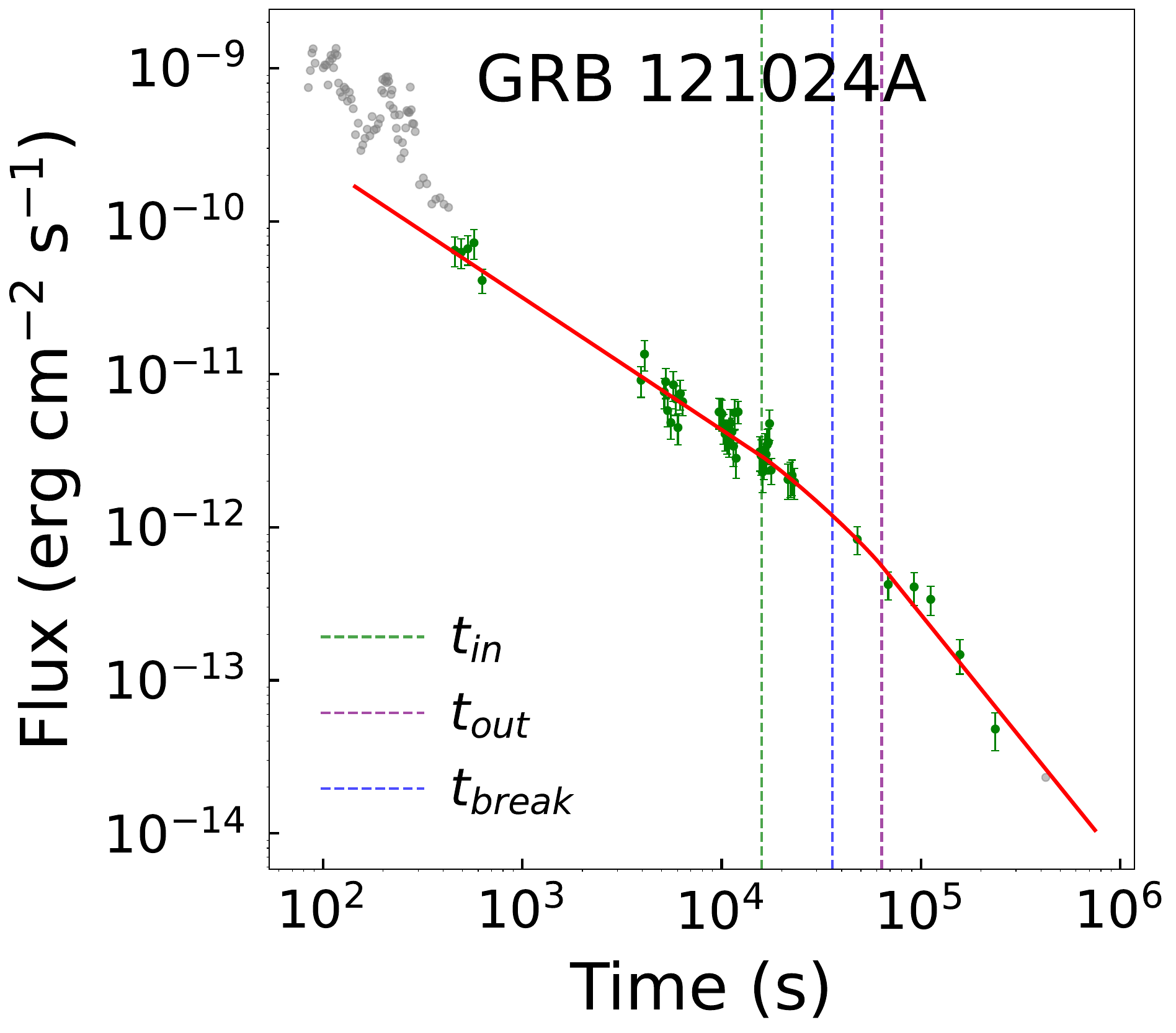}
  \end{subfigure}%
  \hspace{0.01\textwidth}
  \begin{subfigure}[b]{0.23\textwidth}
    \includegraphics[width=\linewidth]{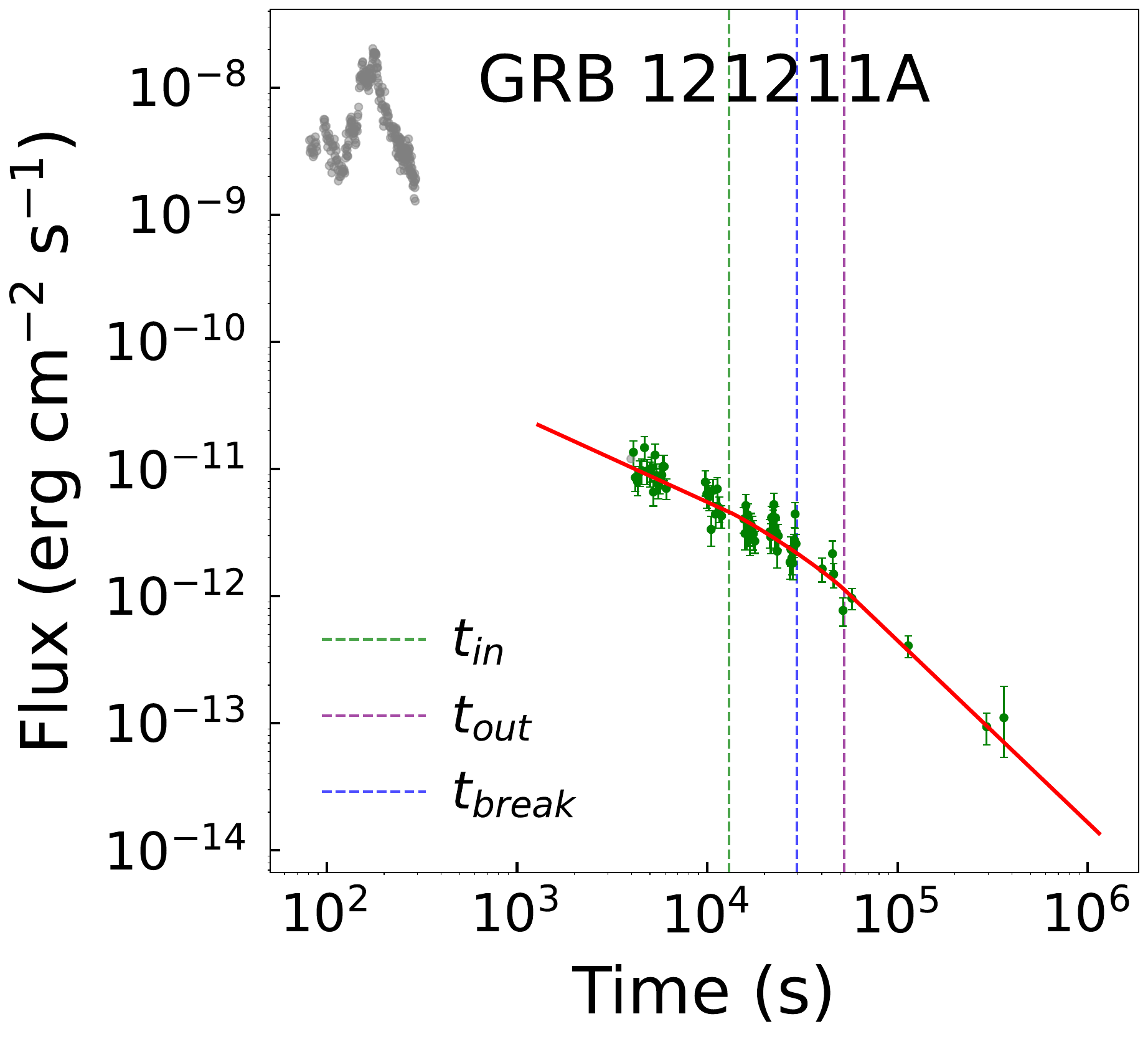}
  \end{subfigure}%
  \hspace{0.01\textwidth}
  \begin{subfigure}[b]{0.23\textwidth}
    \includegraphics[width=\linewidth]{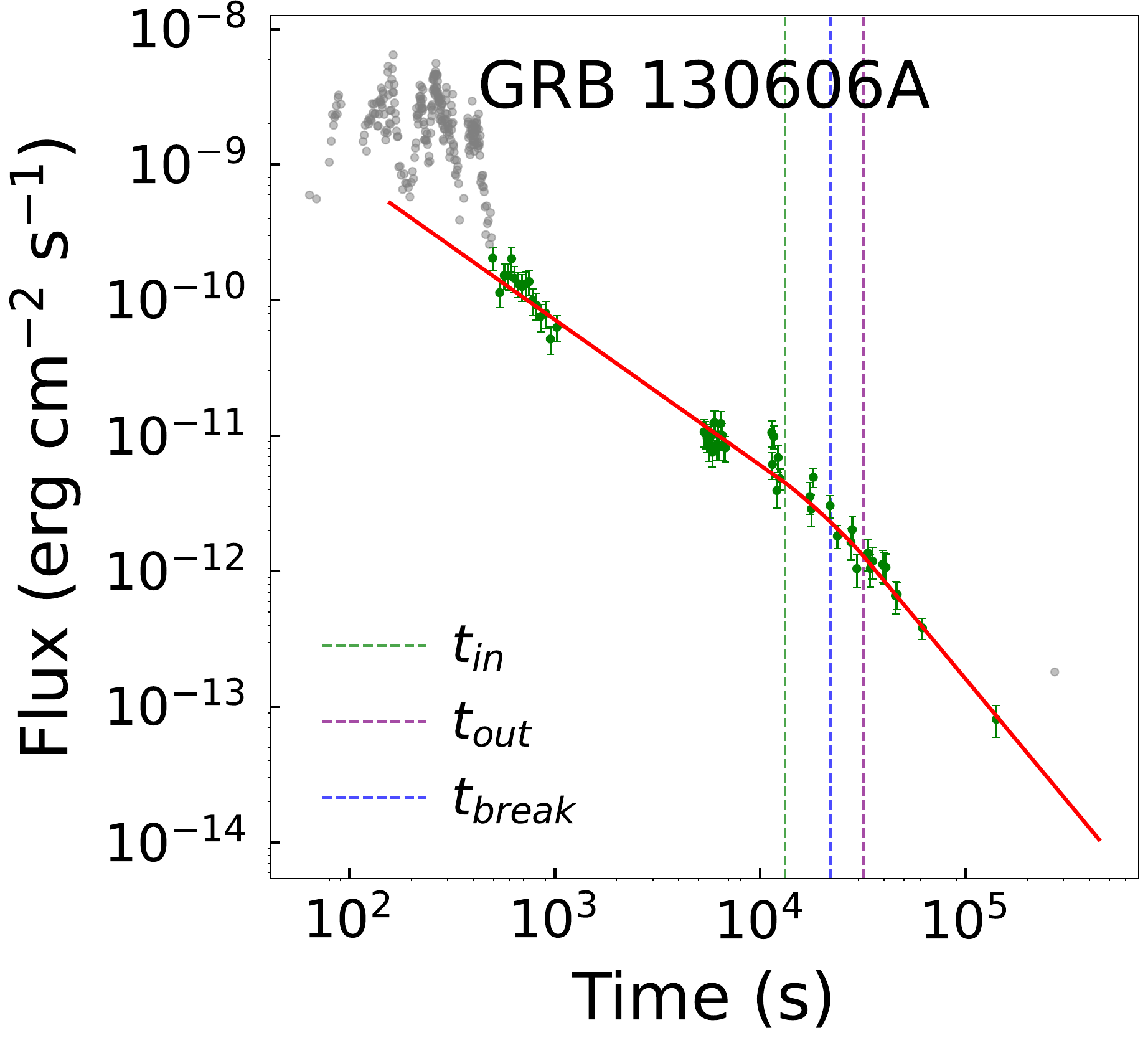}
  \end{subfigure}%
  \\%
  \begin{subfigure}[b]{0.23\textwidth}
    \includegraphics[width=\linewidth]{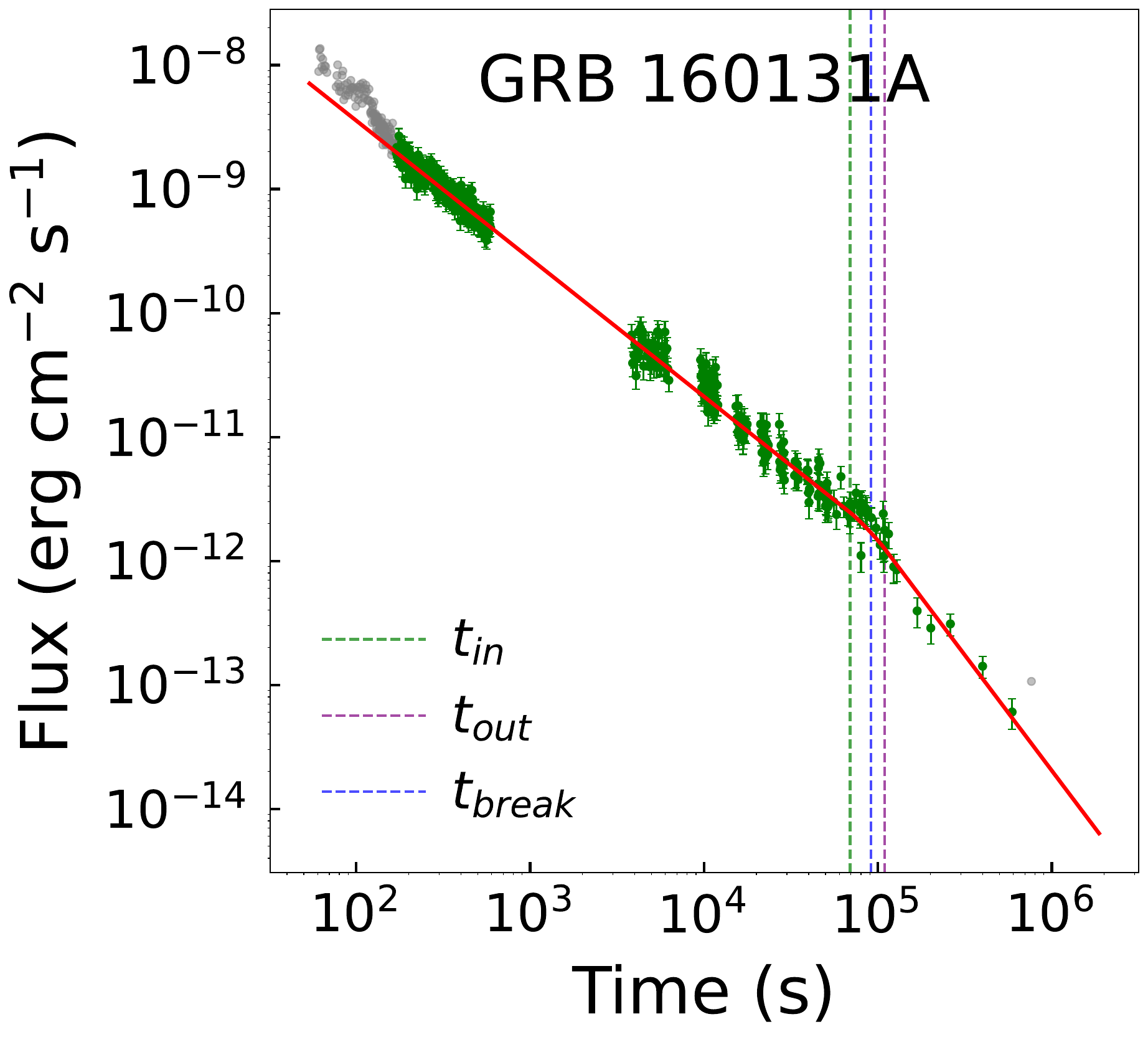}
  \end{subfigure}%
  \hspace{0.01\textwidth}
  \begin{subfigure}[b]{0.23\textwidth}
    \includegraphics[width=\linewidth]{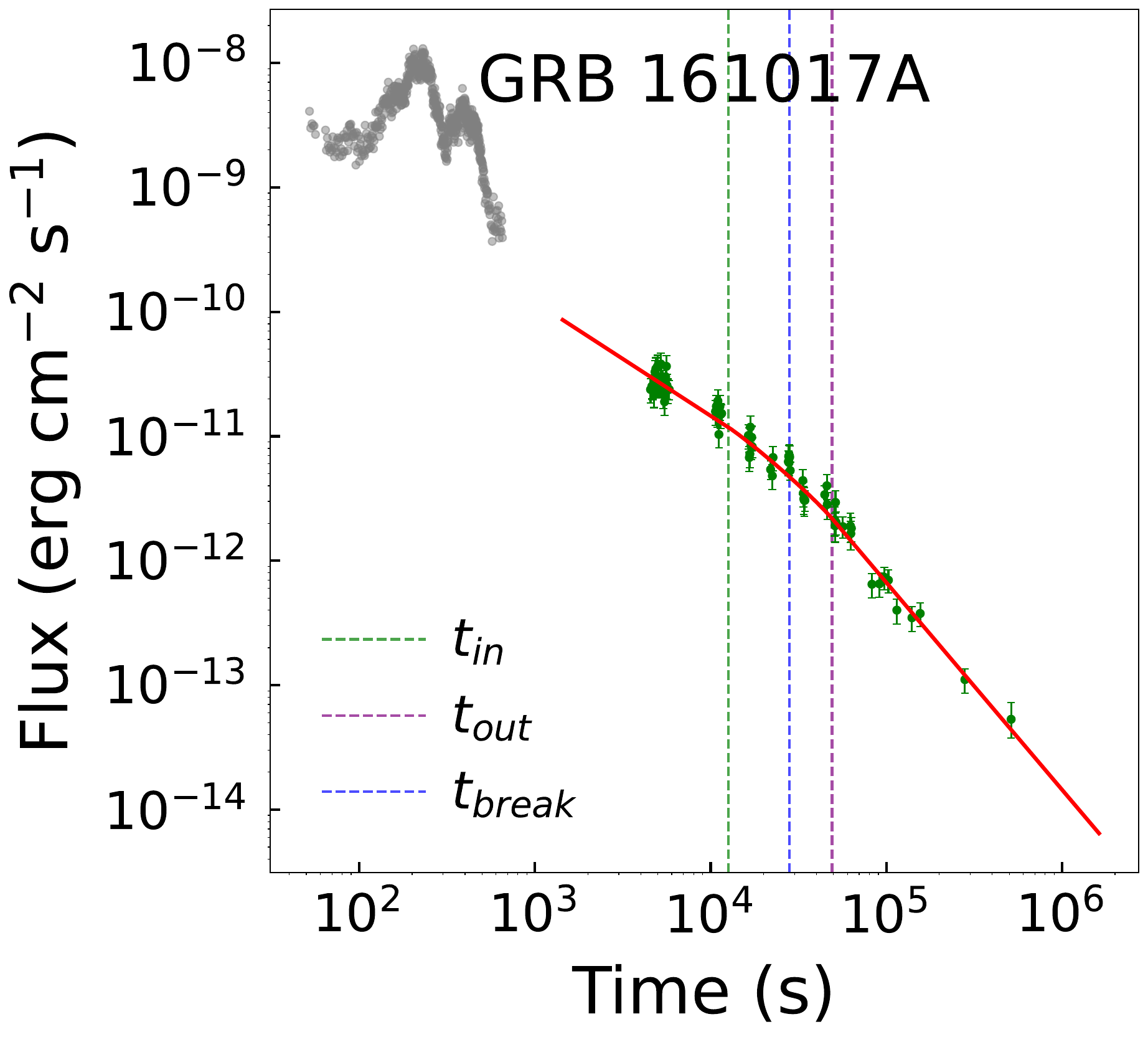}
  \end{subfigure}%
  \hspace{0.01\textwidth}
  \begin{subfigure}[b]{0.23\textwidth}
    \includegraphics[width=\linewidth]{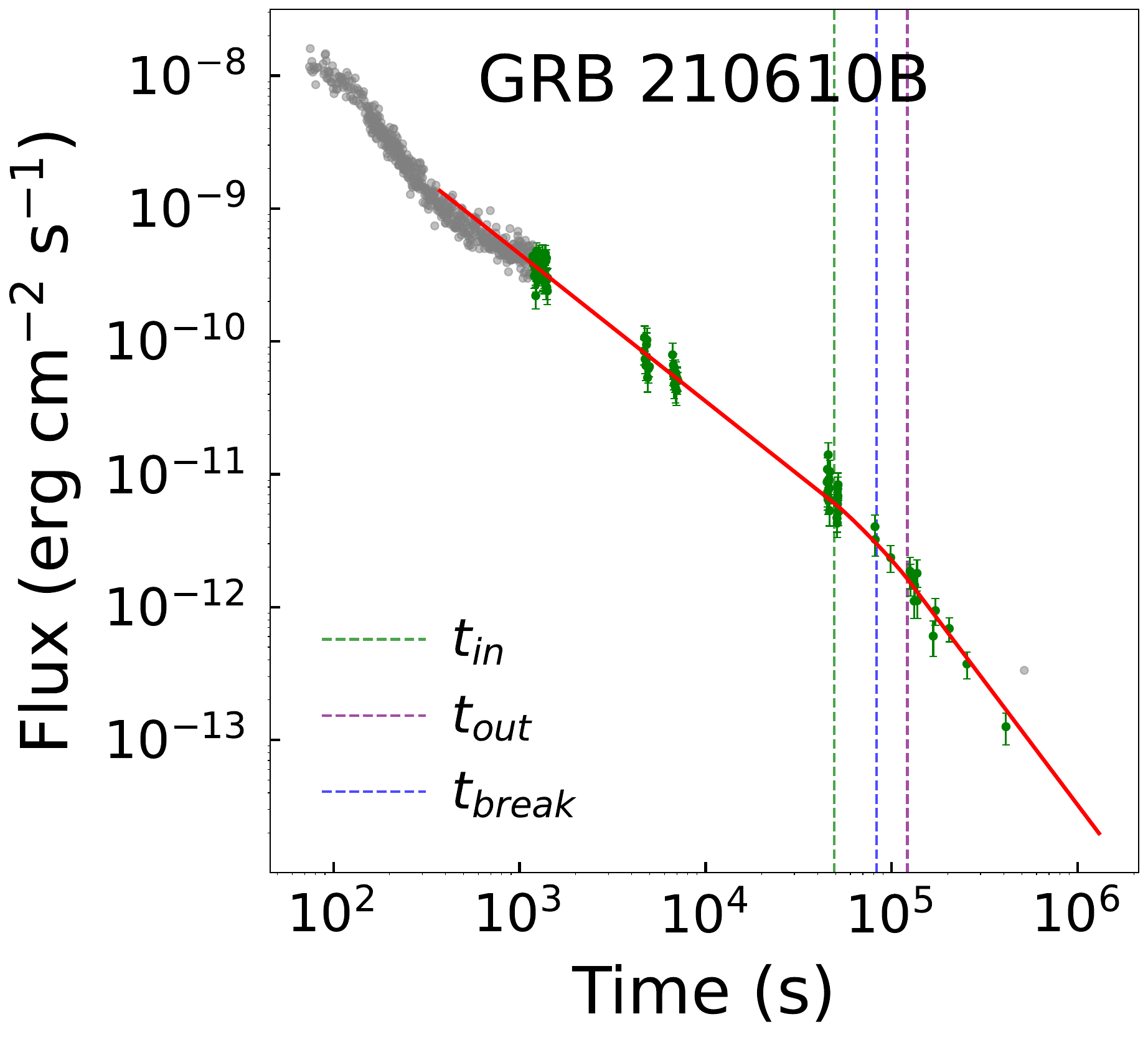}
  \end{subfigure}%
  \hspace{0.01\textwidth}
  \begin{subfigure}[b]{0.23\textwidth}
    \includegraphics[width=\linewidth]{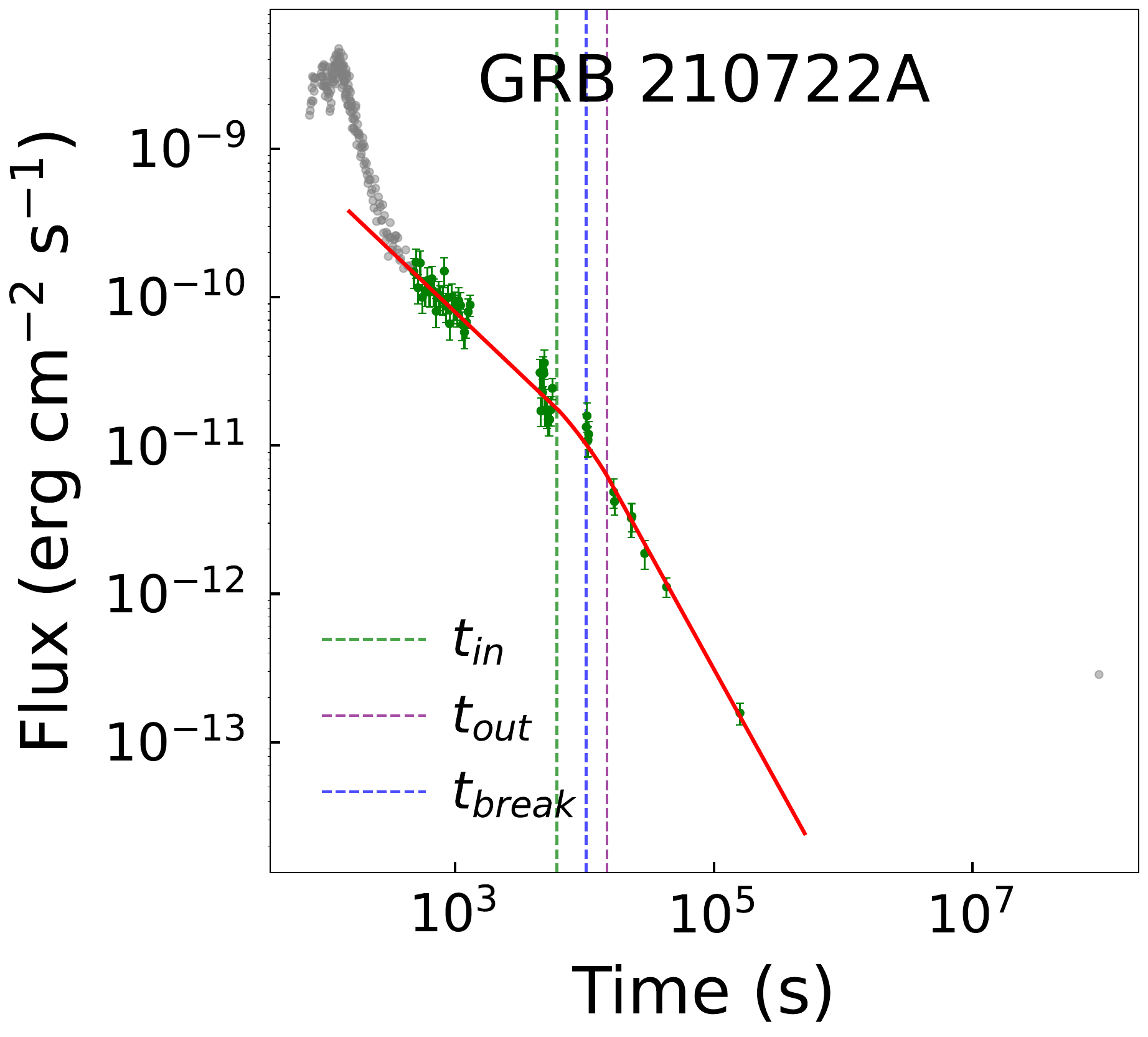}
  \end{subfigure}%
  \\%
  \caption{X-ray afterglow light curve fitting results for model~1. Each panel shows the observed flux (data points) and the best-fit model (solid line) for individual gamma-ray bursts in the ISM environment.}
  \label{fig:ISM_model1_light_curves}
\end{figure}

\begin{figure}[htbp]
\centering
  \begin{subfigure}[b]{0.23\textwidth}
    \includegraphics[width=\linewidth]{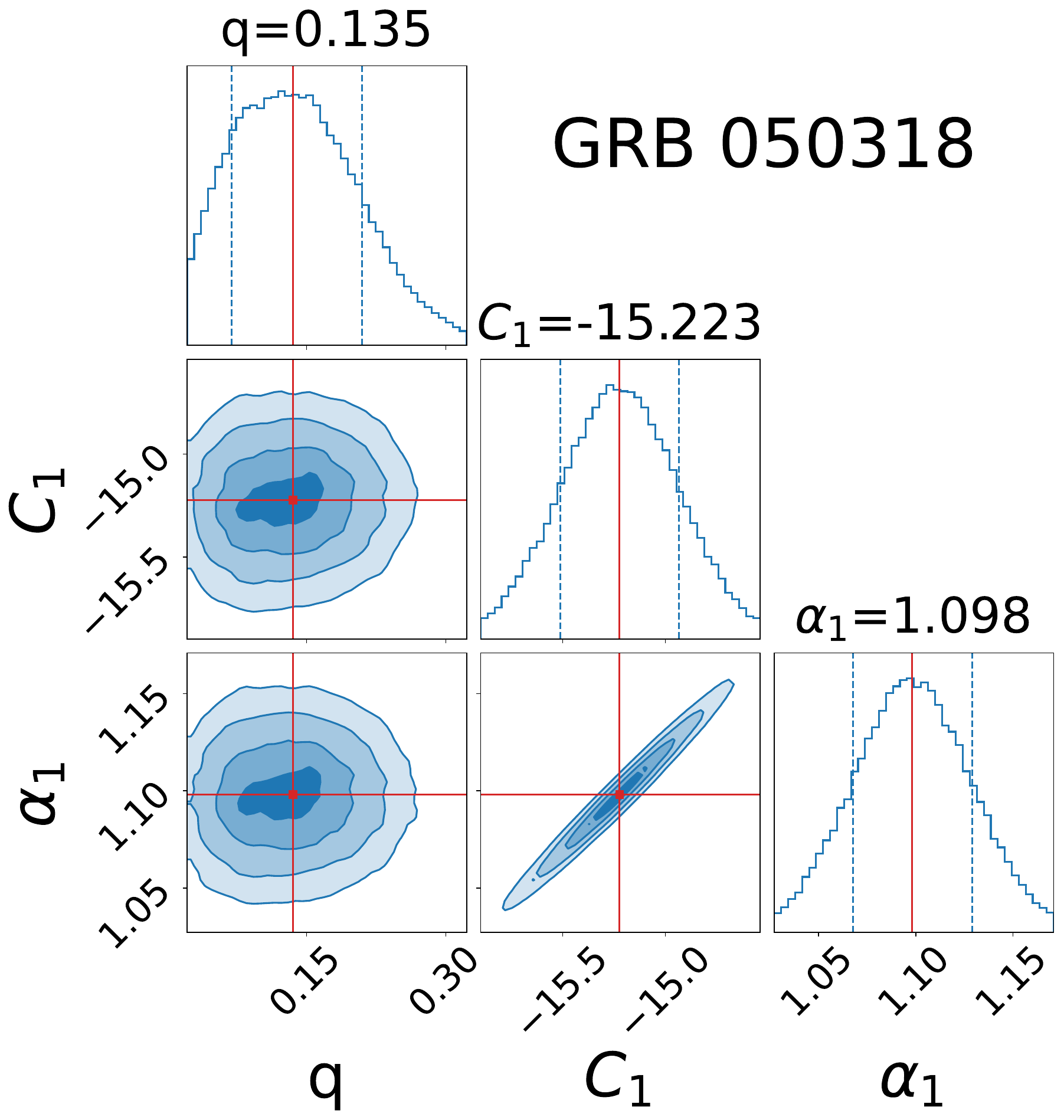}
  \end{subfigure}%
  \hspace{0.01\textwidth}
  \begin{subfigure}[b]{0.23\textwidth}
    \includegraphics[width=\linewidth]{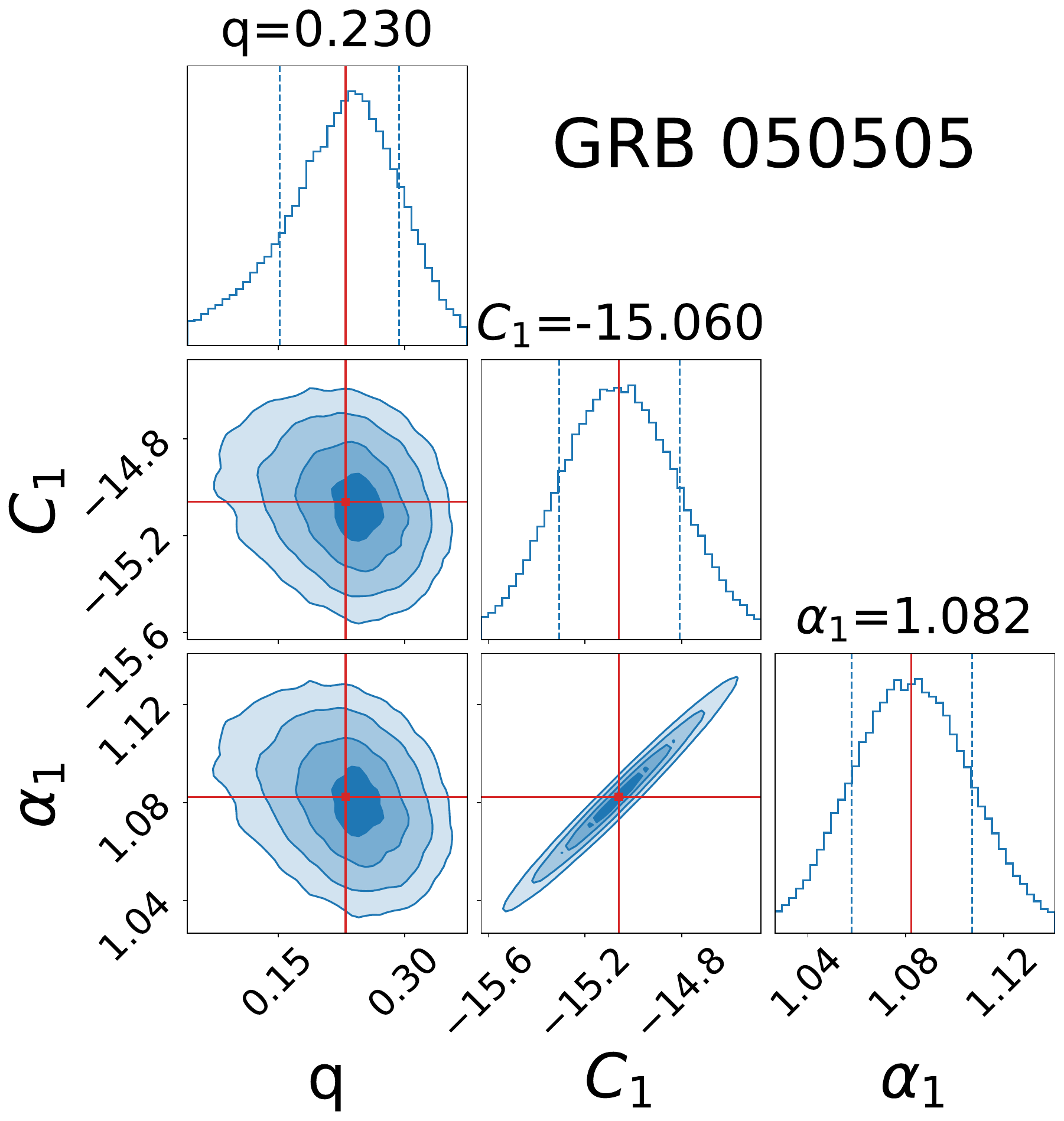}
  \end{subfigure}%
  \hspace{0.01\textwidth}
  \begin{subfigure}[b]{0.23\textwidth}
    \includegraphics[width=\linewidth]{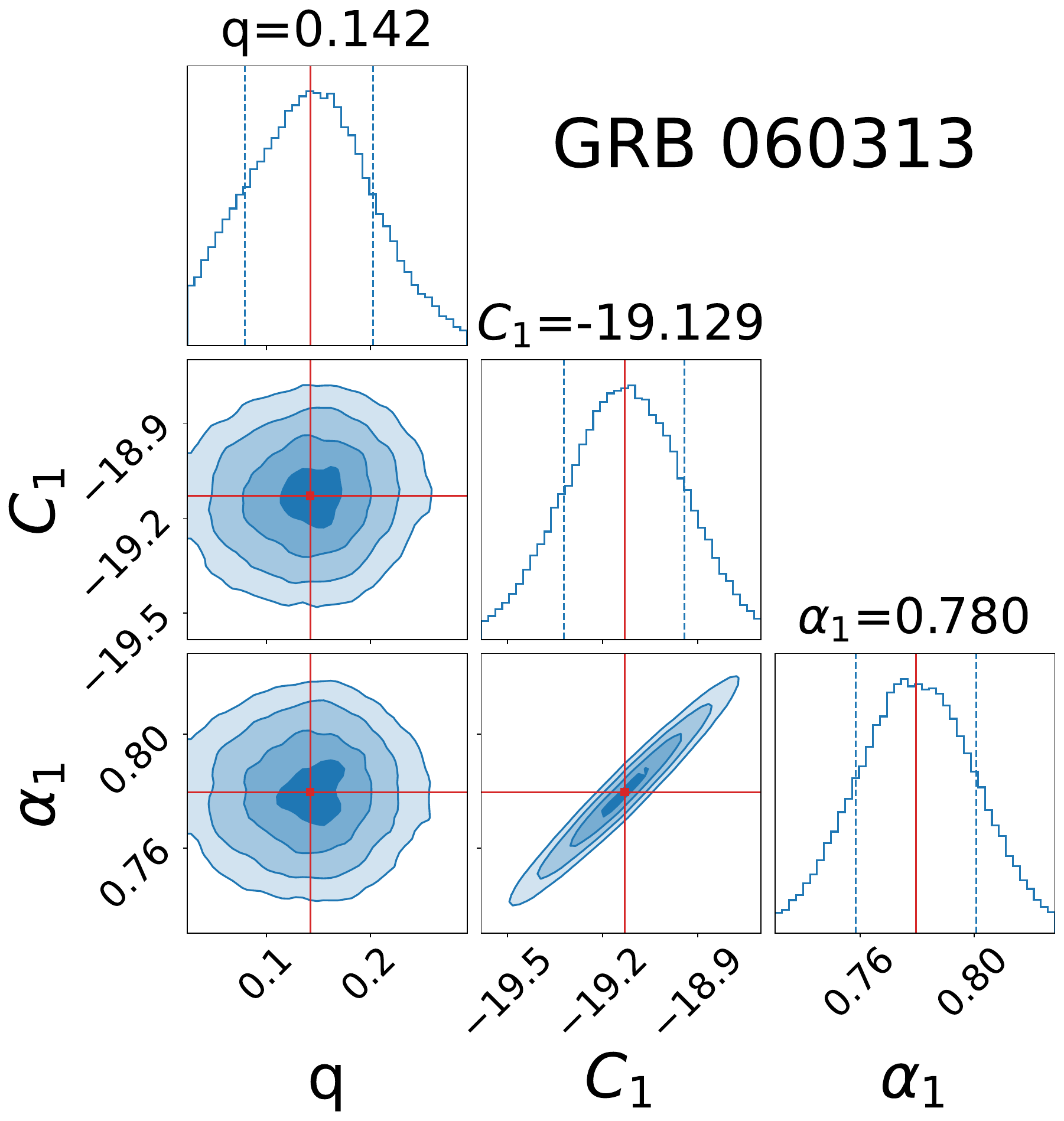}
  \end{subfigure}%
  \hspace{0.01\textwidth}
  \begin{subfigure}[b]{0.23\textwidth}
    \includegraphics[width=\linewidth]{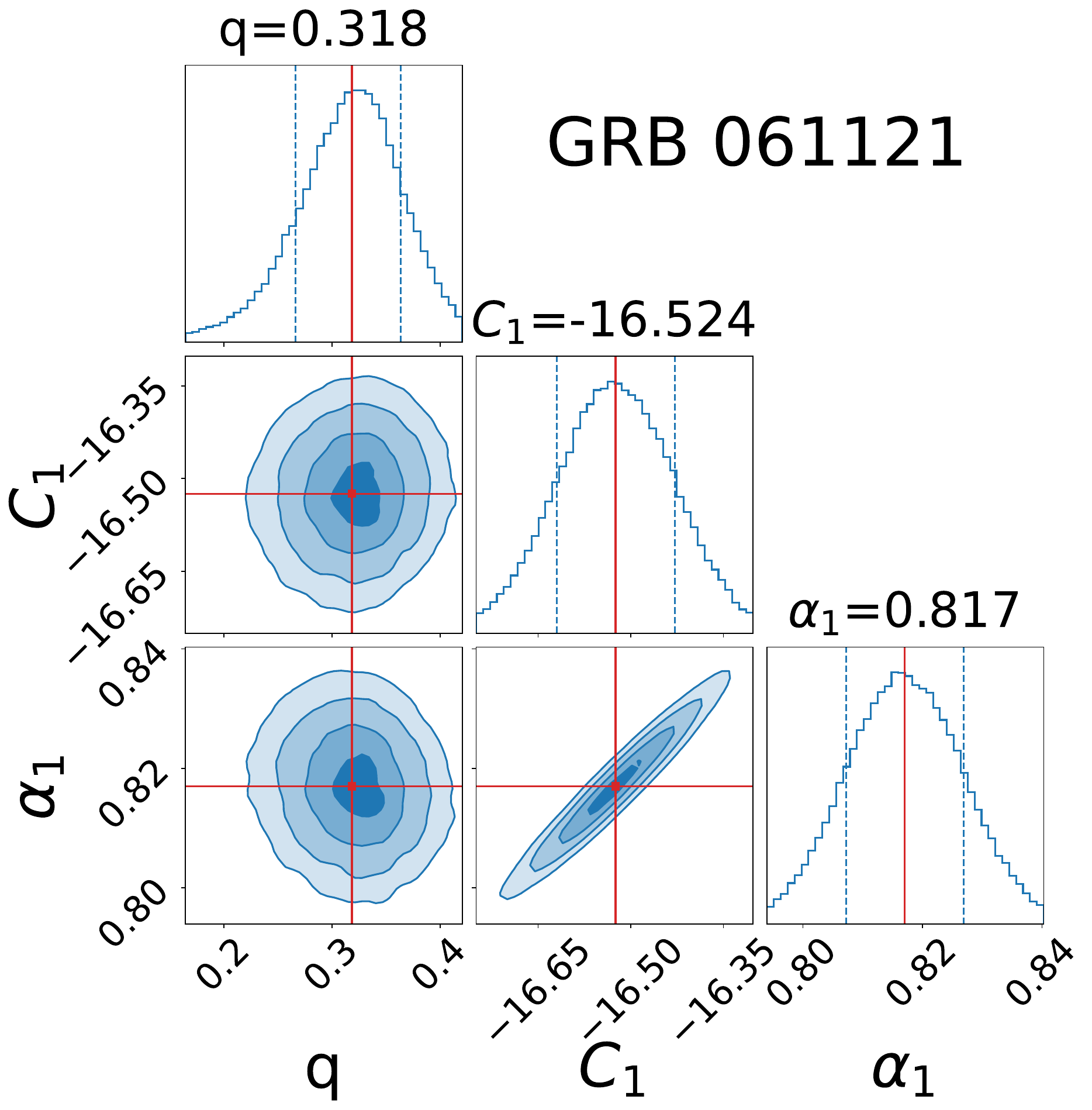}
  \end{subfigure}%
  \\%
  \begin{subfigure}[b]{0.23\textwidth}
    \includegraphics[width=\linewidth]{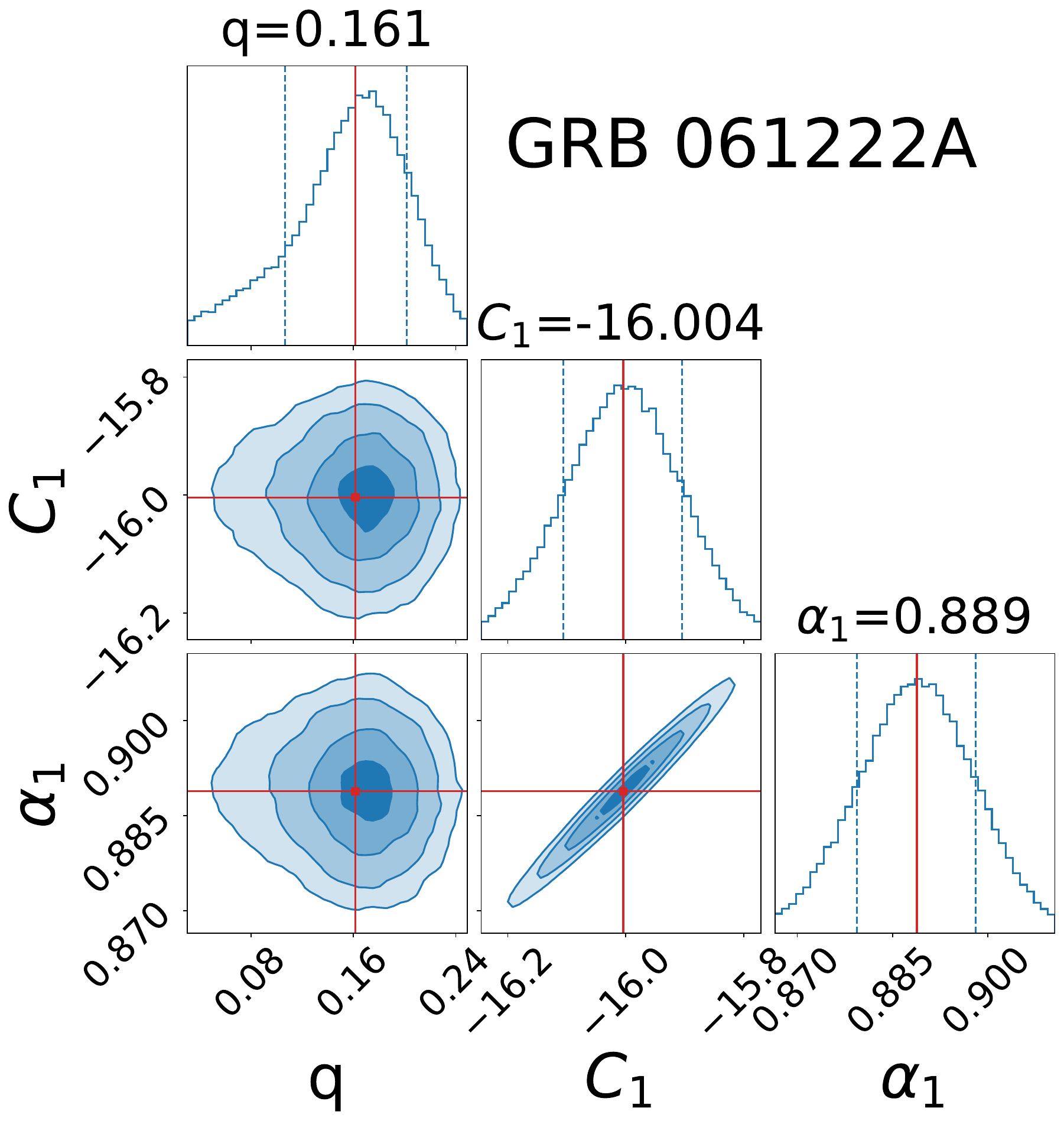}
  \end{subfigure}%
  \hspace{0.01\textwidth}
  \begin{subfigure}[b]{0.23\textwidth}
    \includegraphics[width=\linewidth]{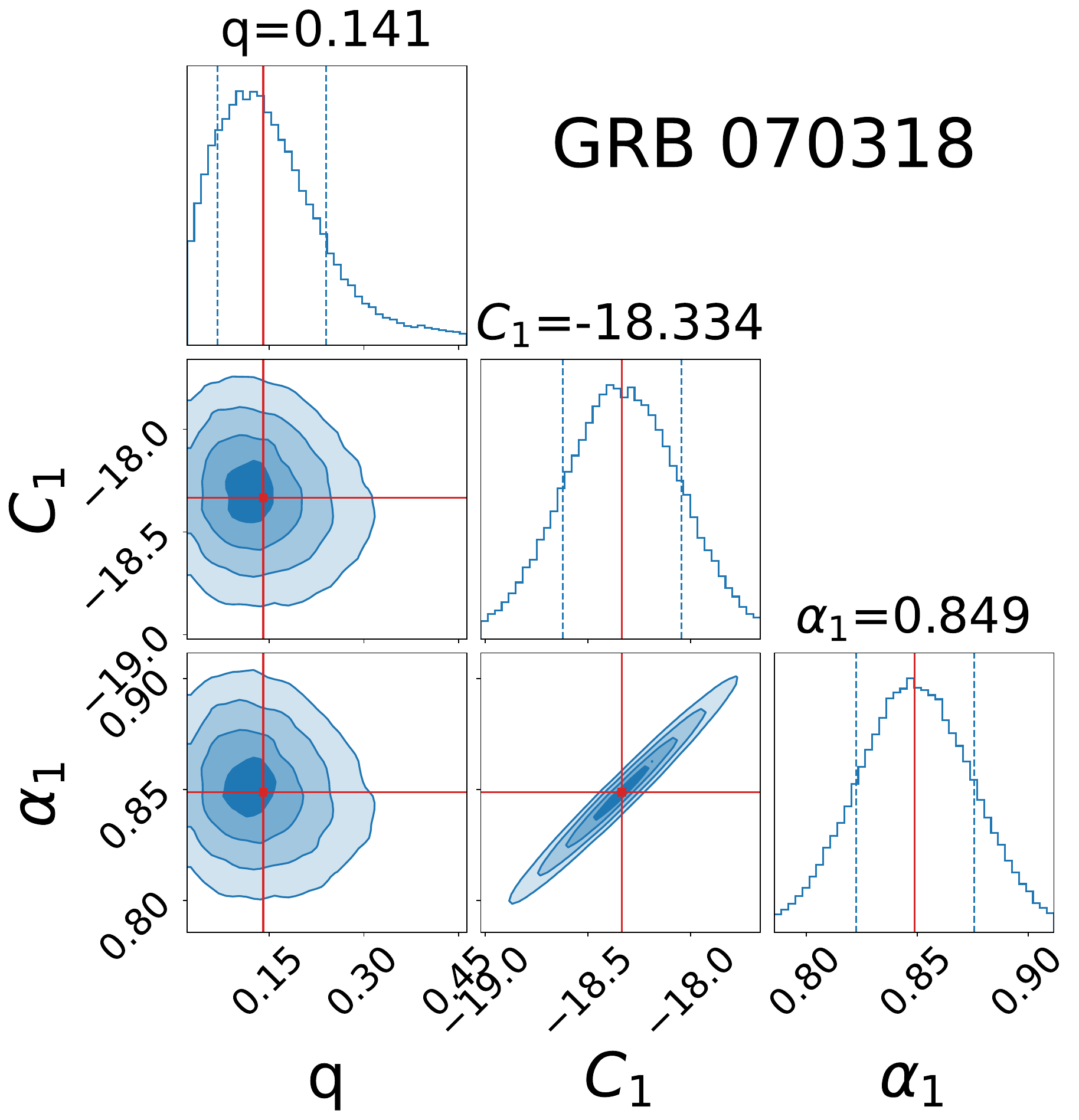}
  \end{subfigure}%
  \hspace{0.01\textwidth}
  \begin{subfigure}[b]{0.23\textwidth}
    \includegraphics[width=\linewidth]{result_070420_10_parameter_distributions.pdf}
  \end{subfigure}%
  \hspace{0.01\textwidth}
  \begin{subfigure}[b]{0.23\textwidth}
    \includegraphics[width=\linewidth]{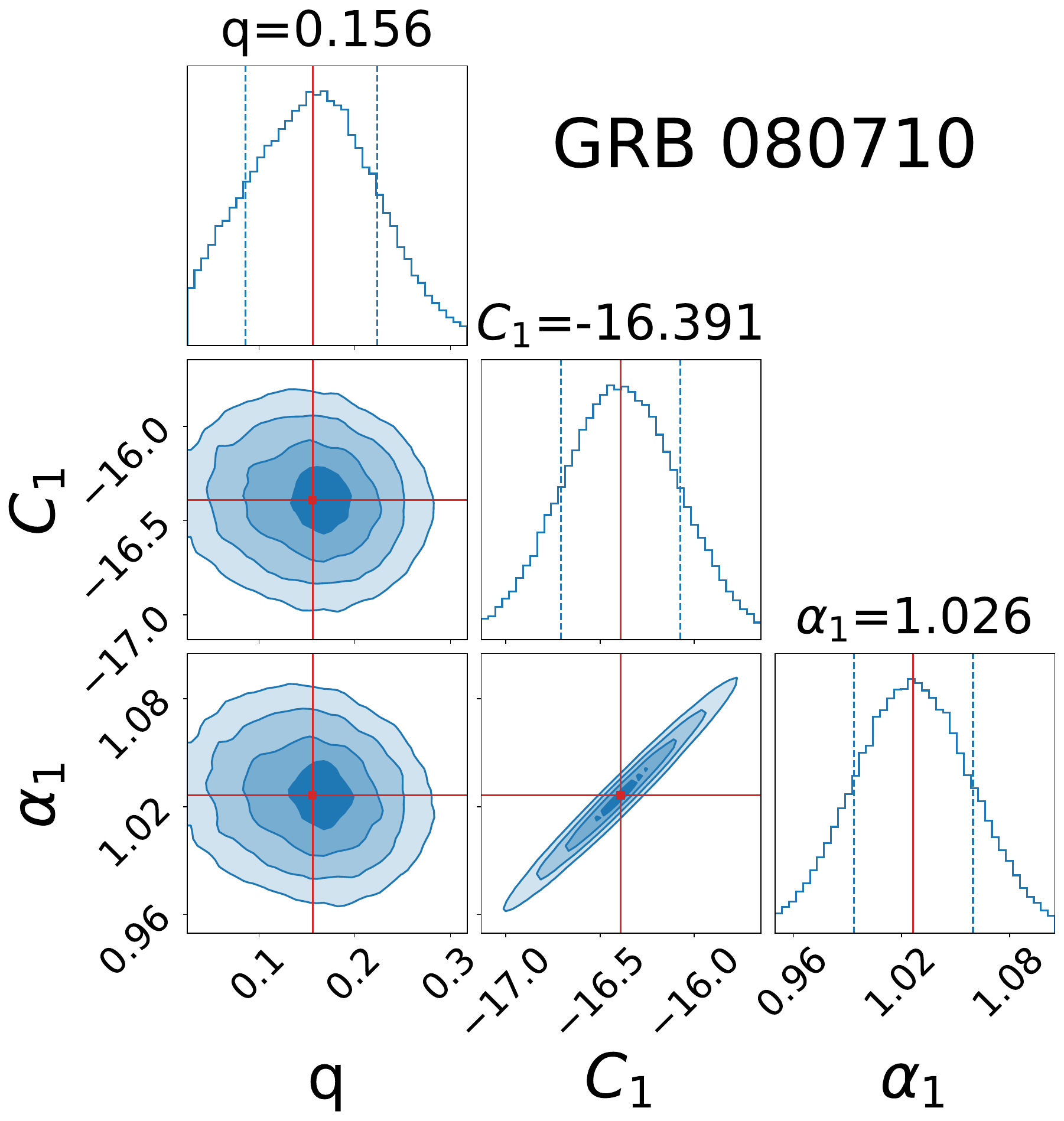}
  \end{subfigure}%
  \\%
  \begin{subfigure}[b]{0.23\textwidth}
    \includegraphics[width=\linewidth]{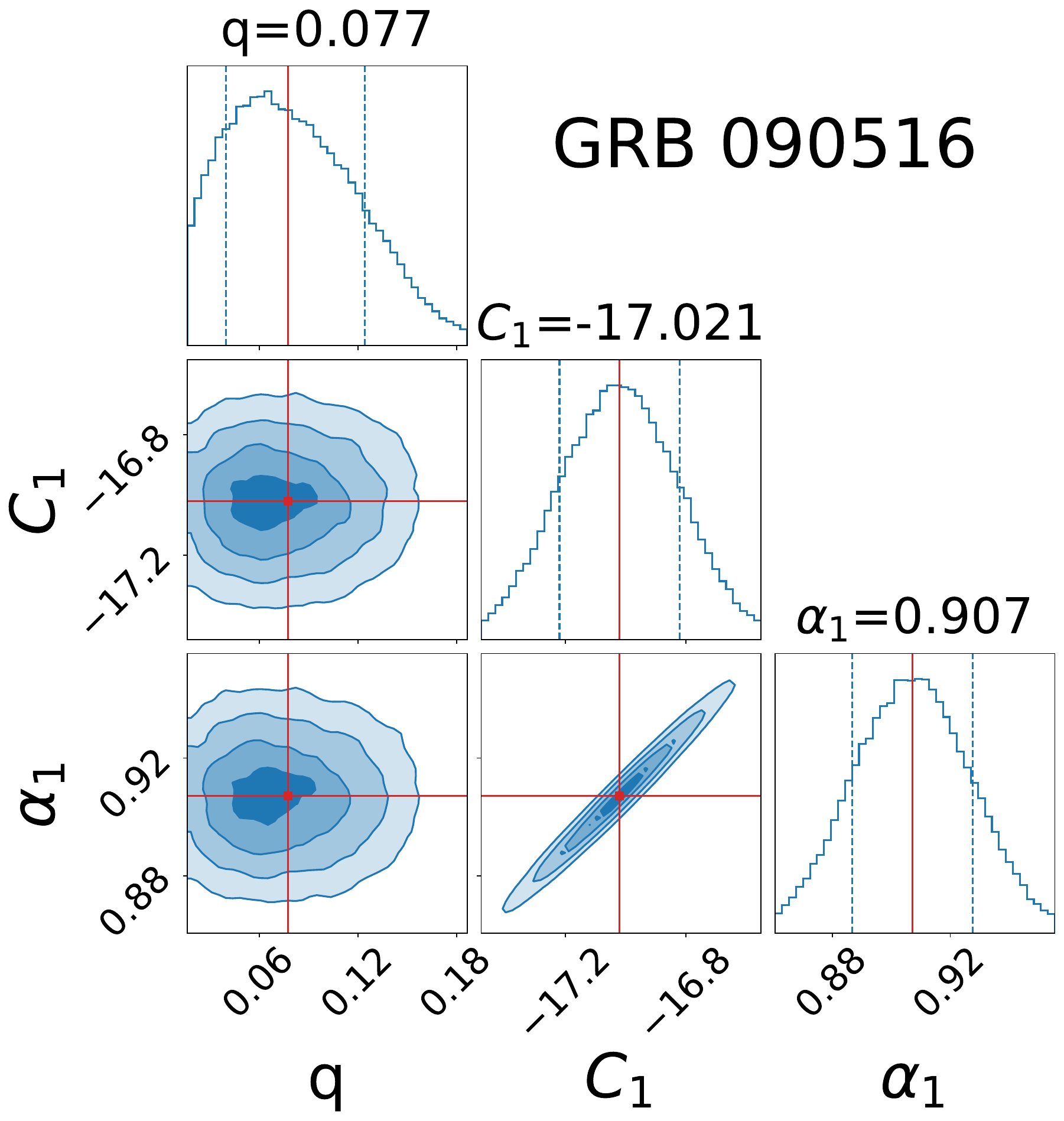}
  \end{subfigure}%
  \hspace{0.01\textwidth}
  \begin{subfigure}[b]{0.23\textwidth}
    \includegraphics[width=\linewidth]{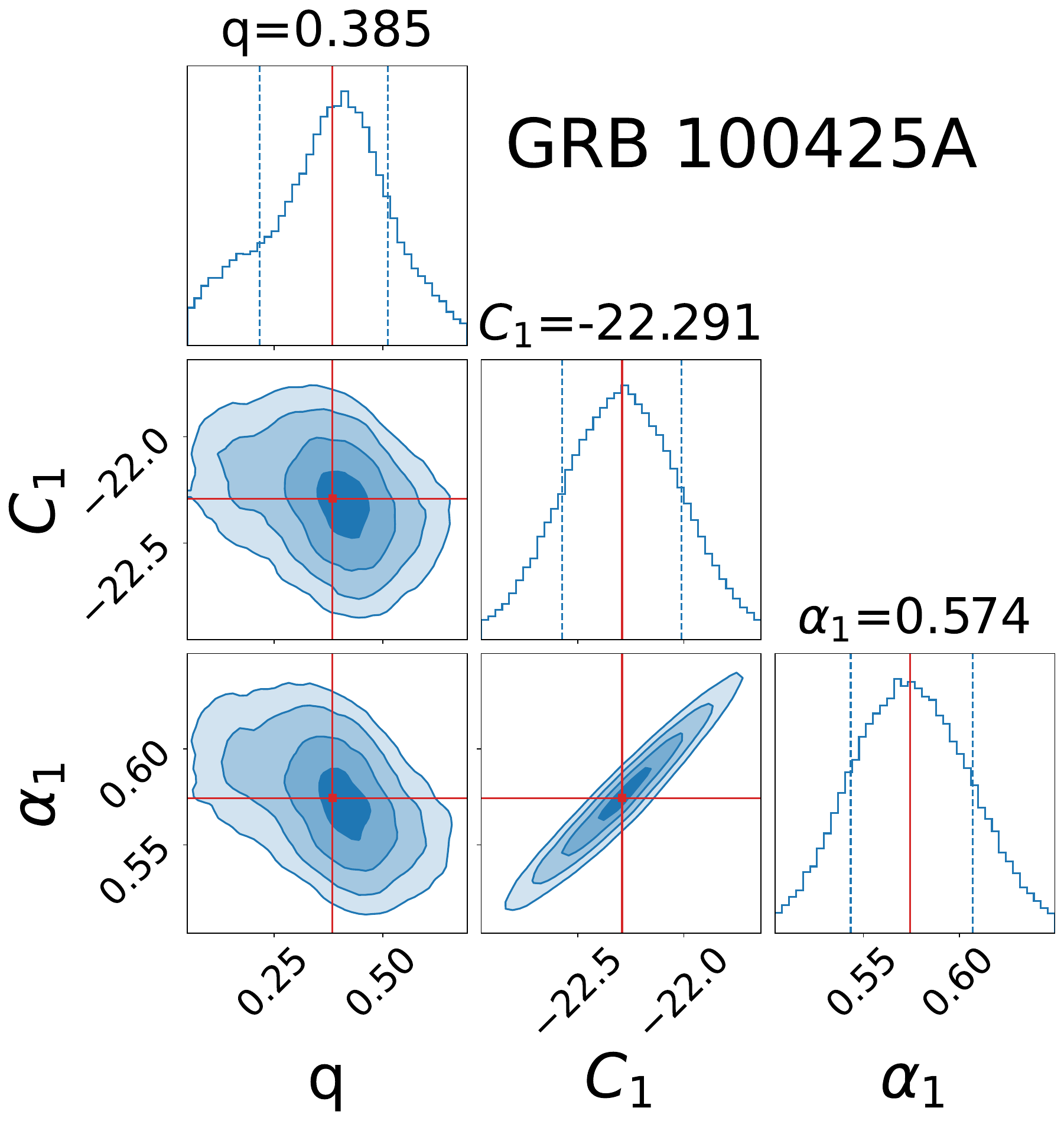}
  \end{subfigure}%
  \hspace{0.01\textwidth}
  \begin{subfigure}[b]{0.23\textwidth}
    \includegraphics[width=\linewidth]{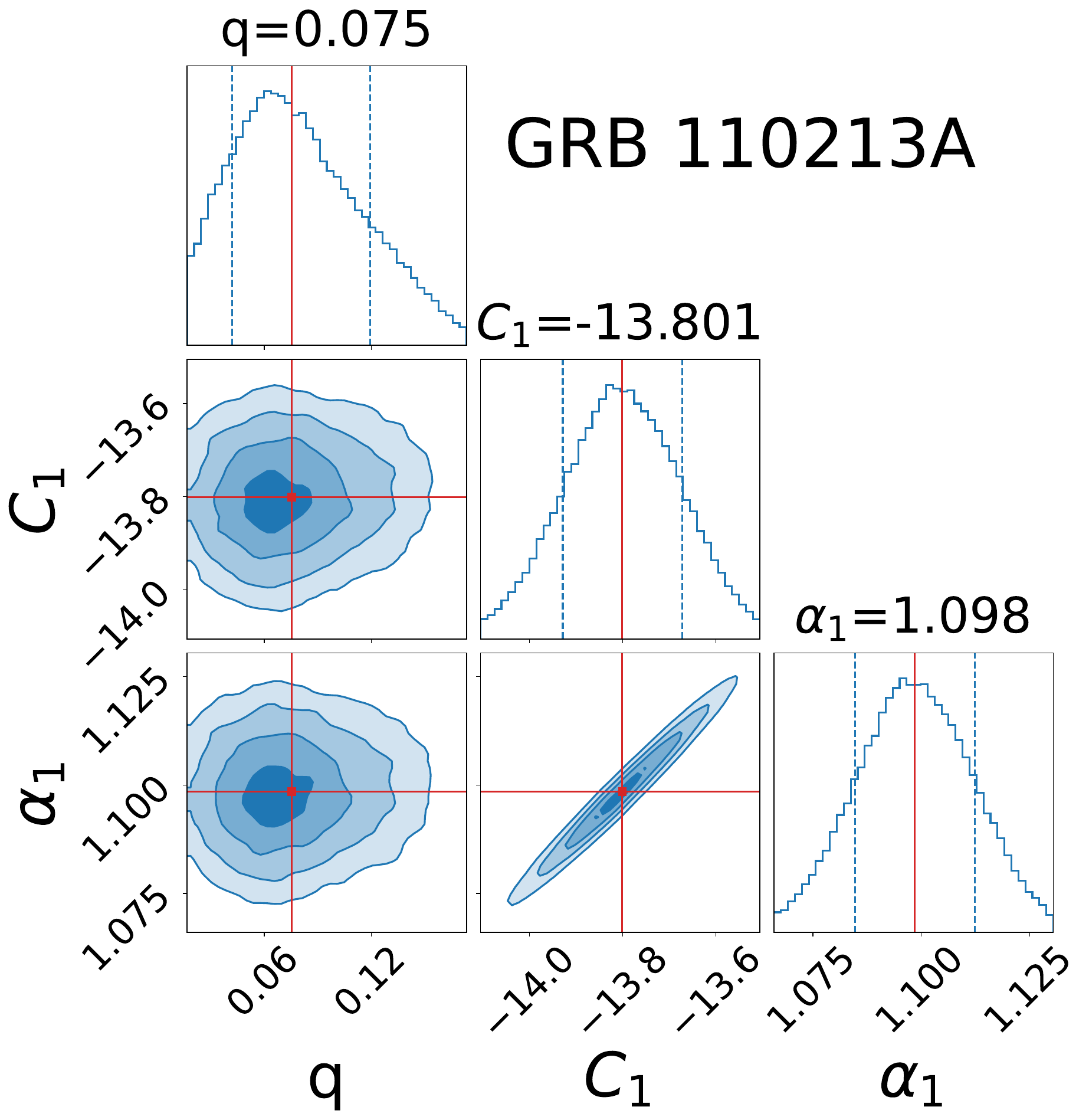}
  \end{subfigure}%
  \hspace{0.01\textwidth}
  \begin{subfigure}[b]{0.23\textwidth}
    \includegraphics[width=\linewidth]{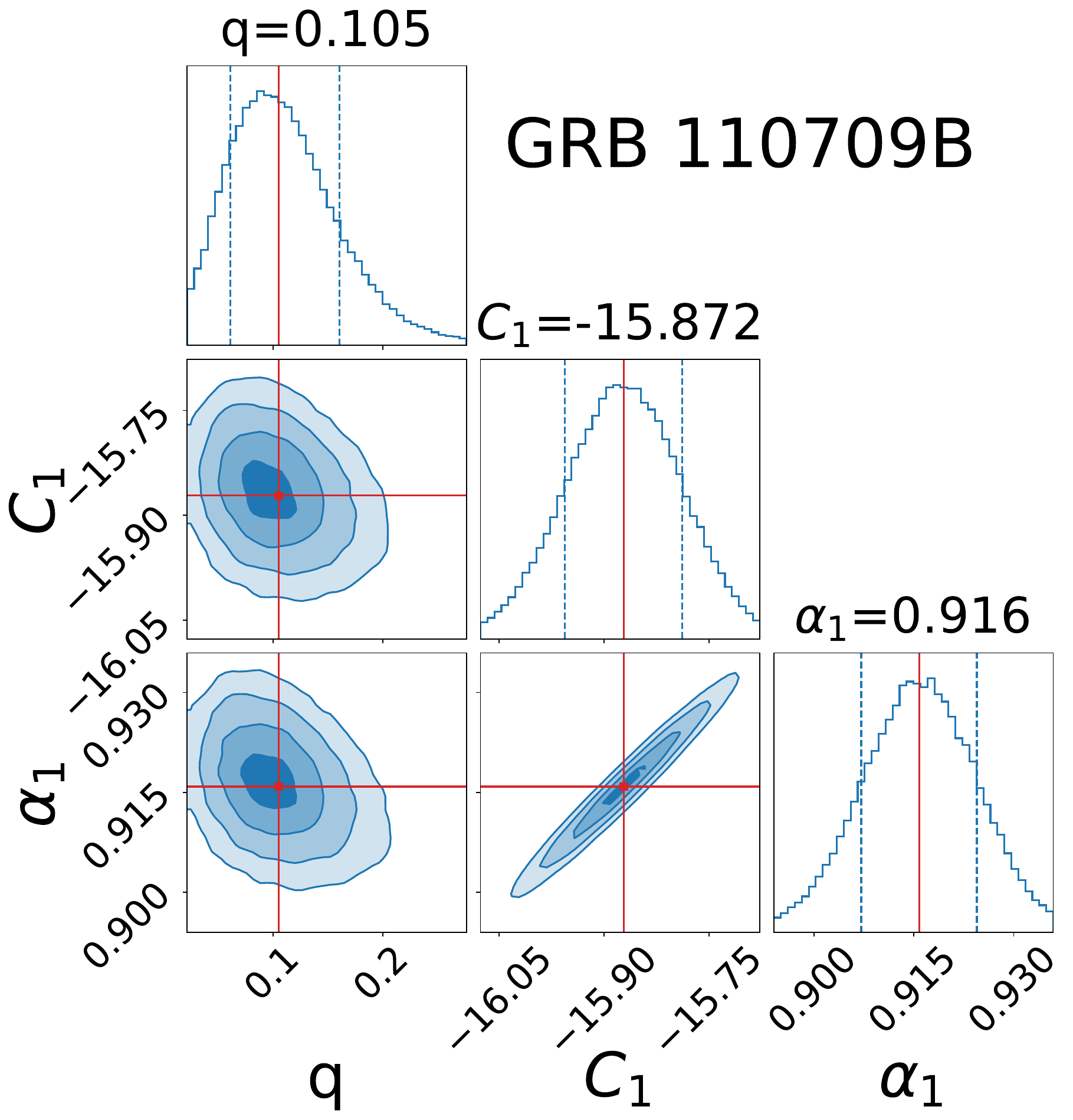}
  \end{subfigure}%
  \\%
  \begin{subfigure}[b]{0.23\textwidth}
    \includegraphics[width=\linewidth]{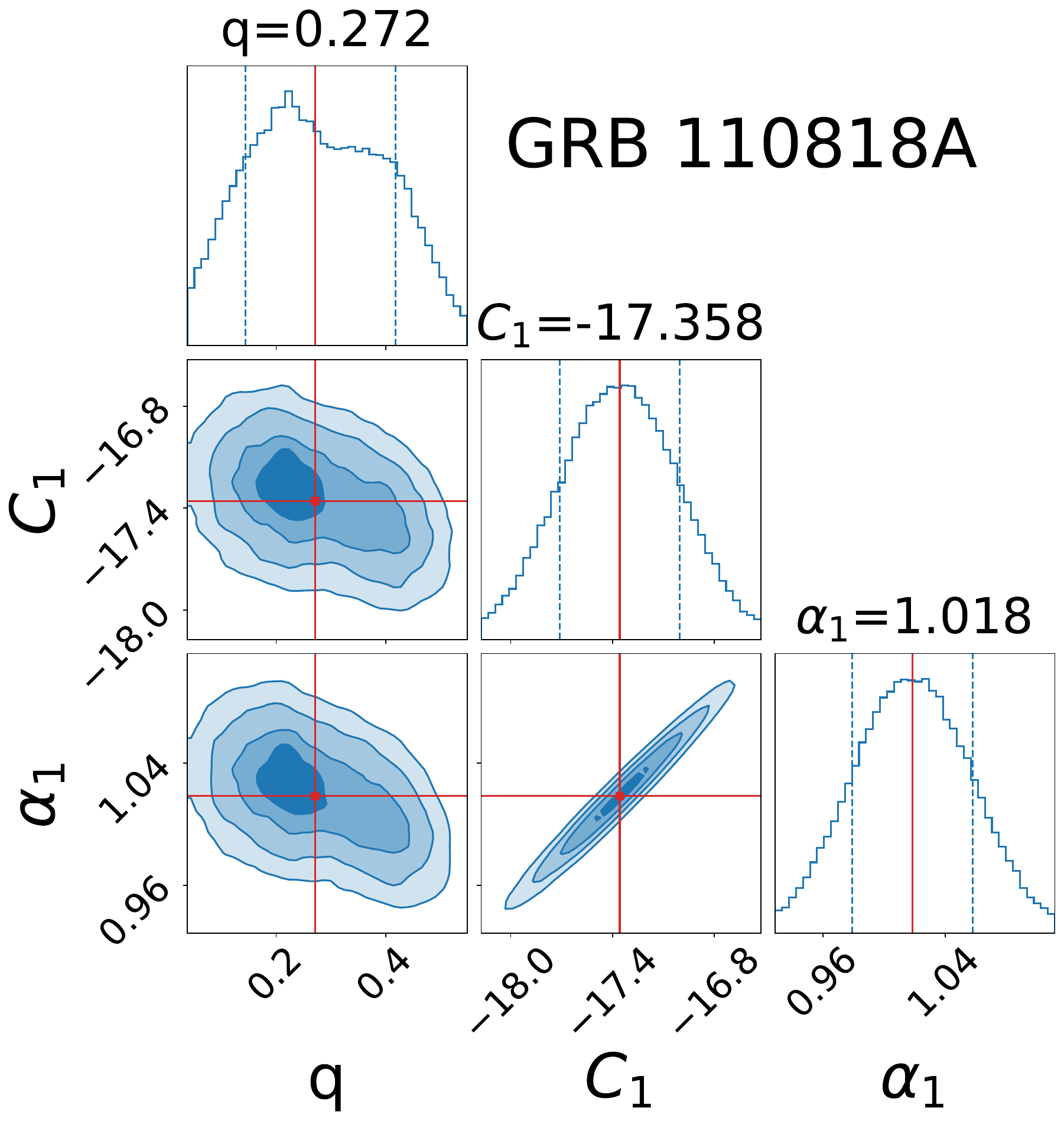}
  \end{subfigure}%
  \hspace{0.01\textwidth}
  \begin{subfigure}[b]{0.23\textwidth}
    \includegraphics[width=\linewidth]{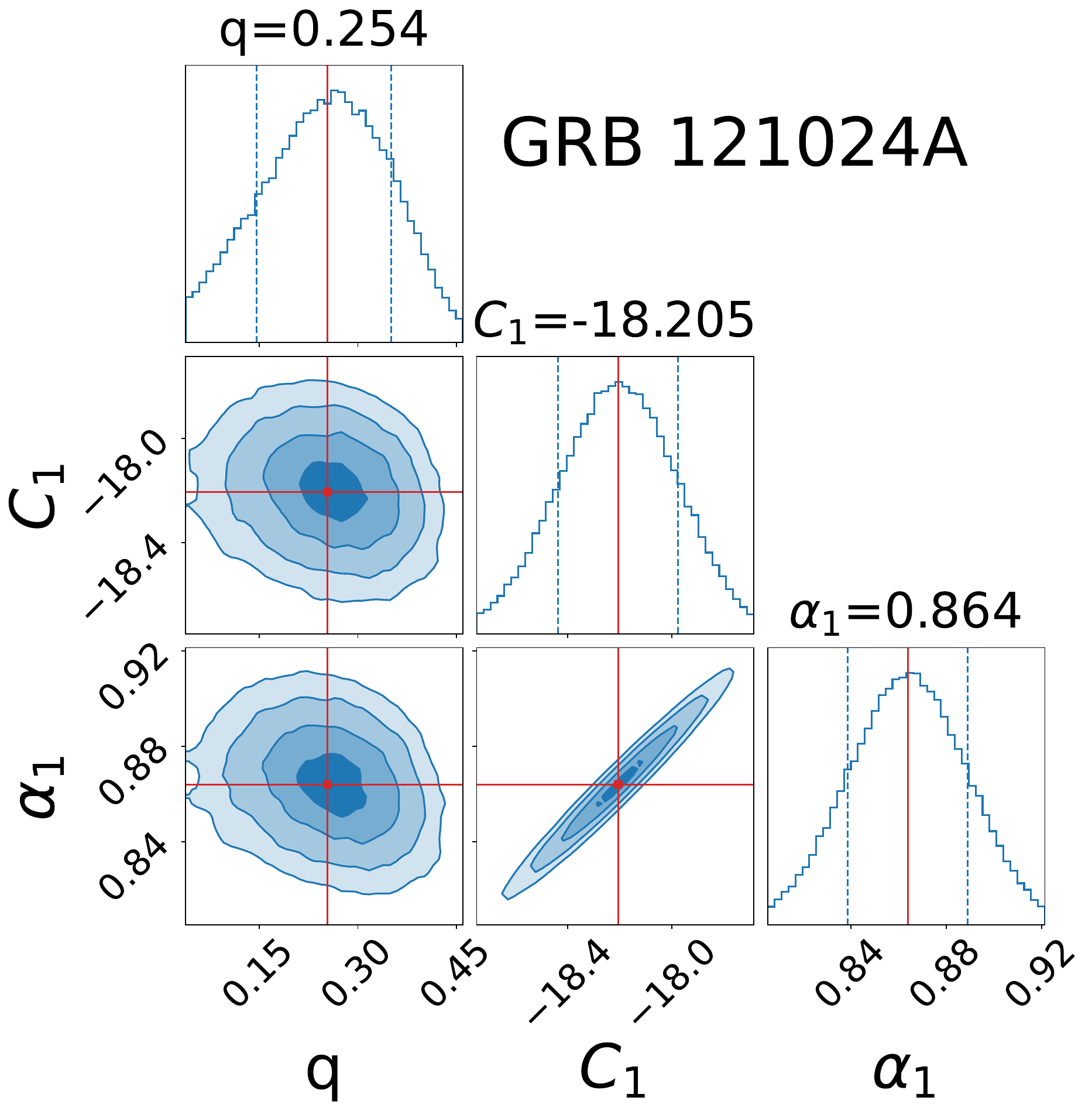}
  \end{subfigure}%
  \hspace{0.01\textwidth}
  \begin{subfigure}[b]{0.23\textwidth}
    \includegraphics[width=\linewidth]{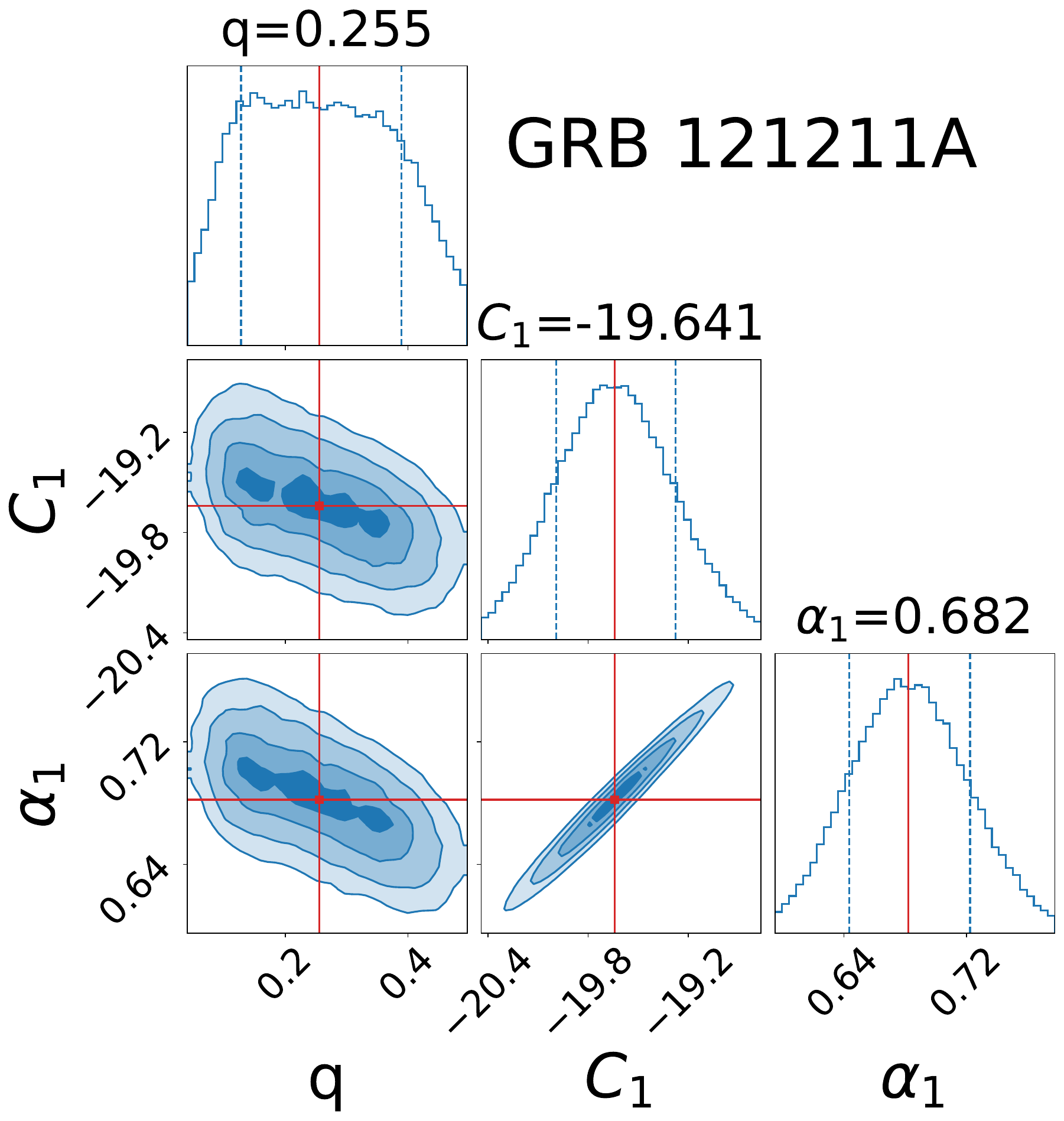}
  \end{subfigure}%
  \hspace{0.01\textwidth}
  \begin{subfigure}[b]{0.23\textwidth}
    \includegraphics[width=\linewidth]{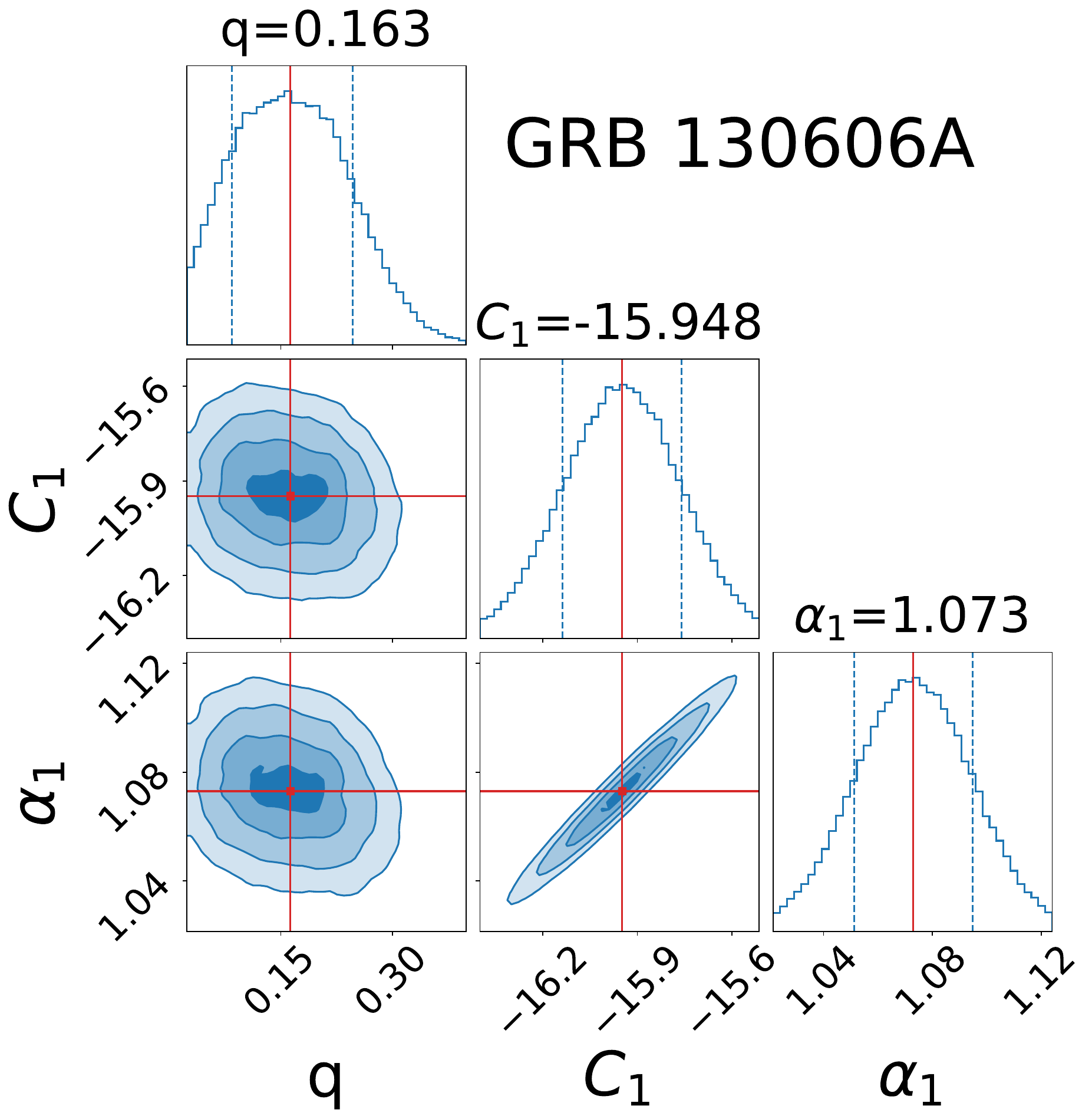}
  \end{subfigure}%
  \\%
  \begin{subfigure}[b]{0.23\textwidth}
    \includegraphics[width=\linewidth]{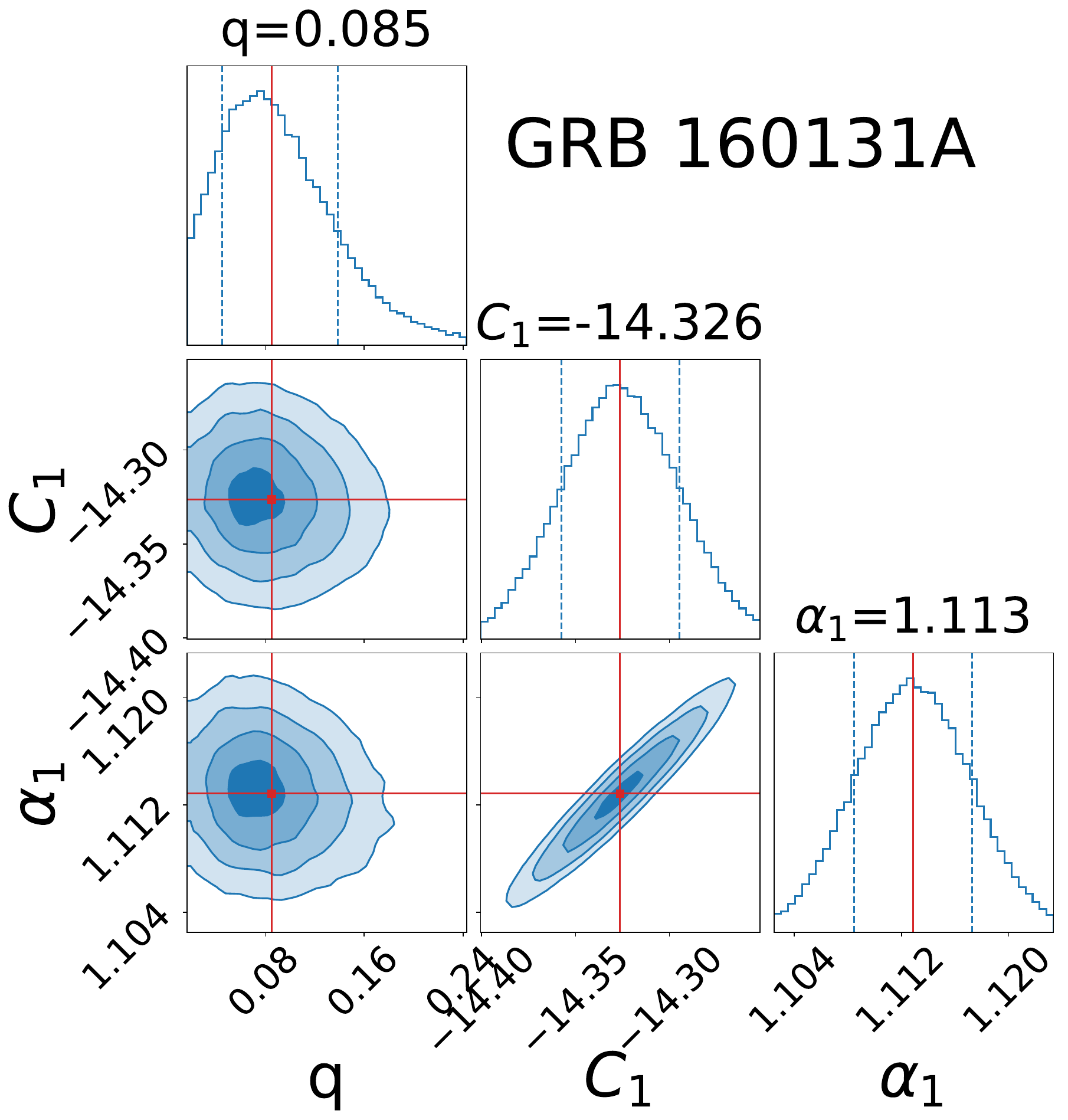}
  \end{subfigure}%
  \hspace{0.01\textwidth}
  \begin{subfigure}[b]{0.23\textwidth}
    \includegraphics[width=\linewidth]{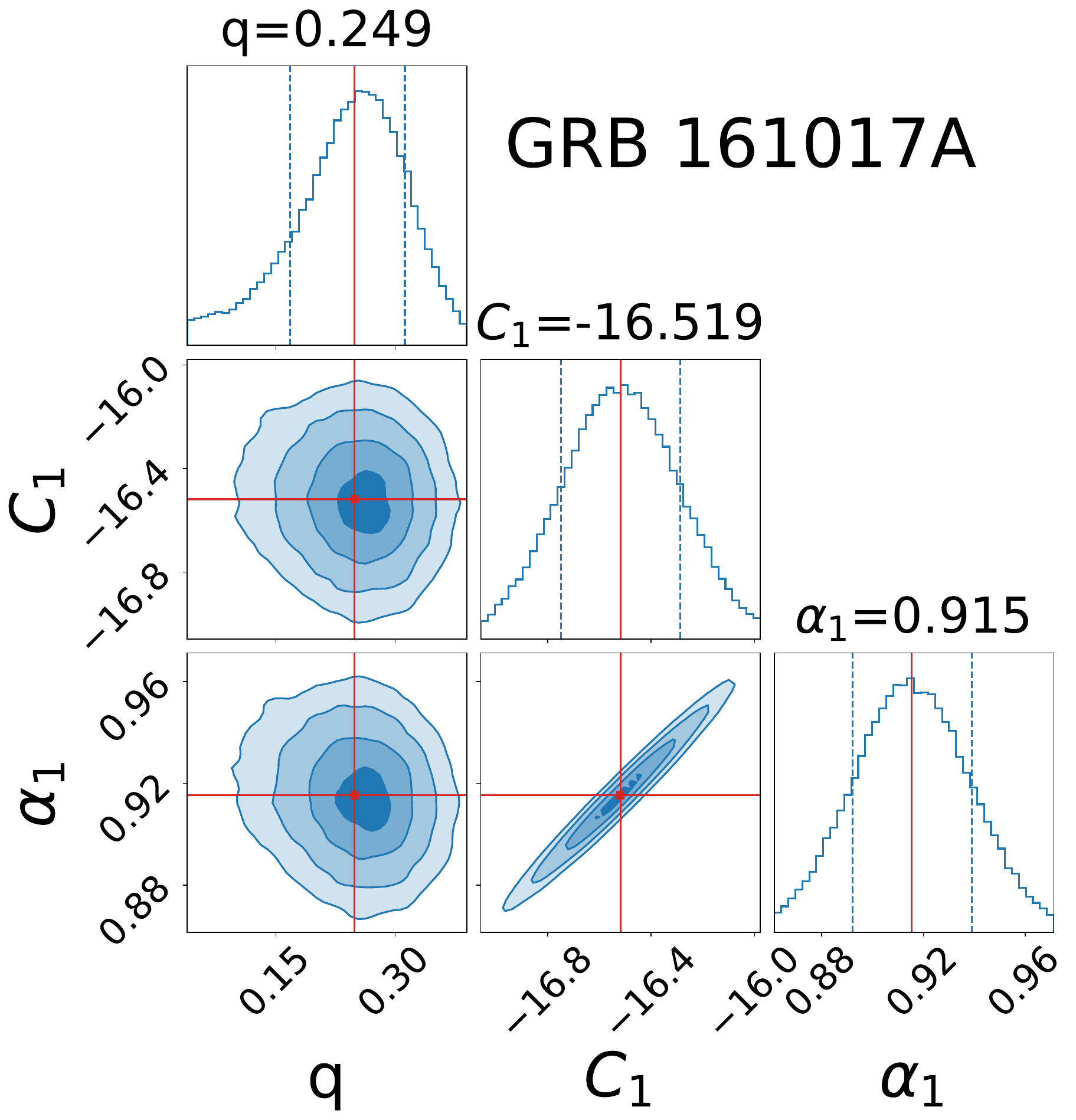}
  \end{subfigure}%
  \hspace{0.01\textwidth}
  \begin{subfigure}[b]{0.23\textwidth}
    \includegraphics[width=\linewidth]{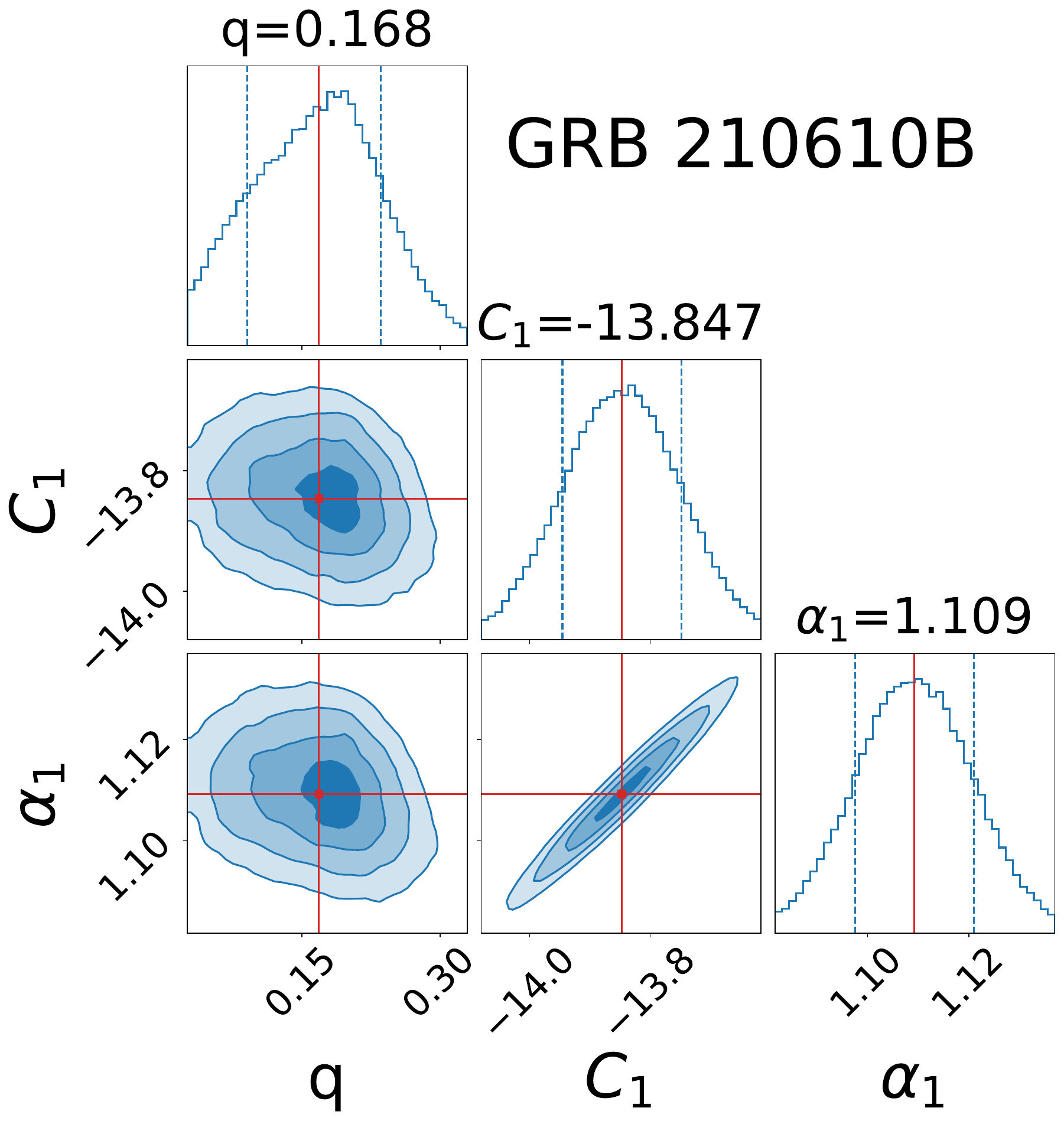}
  \end{subfigure}%
  \hspace{0.01\textwidth}
  \begin{subfigure}[b]{0.23\textwidth}
    \includegraphics[width=\linewidth]{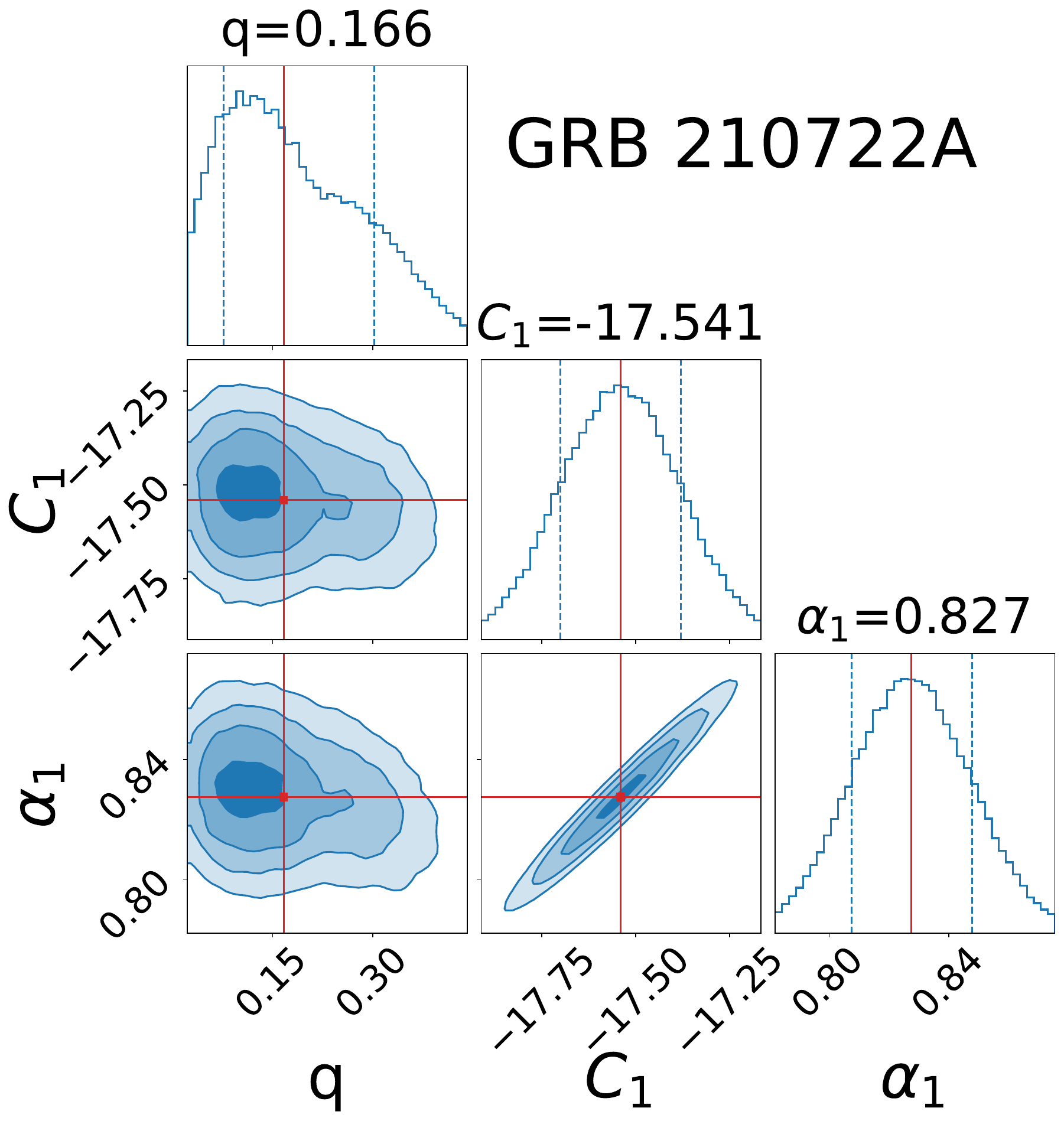}
  \end{subfigure}%
  \\%
  \caption{Posterior parameter distributions for model~1. Each panel displays the MCMC sampling results for key parameters including the off-axis ratio $q = \theta_{\text{obs}}/\theta_{\text{jet}}$, decay index $\alpha_1$, and normalization factor.}
  \label{fig:ISM_model1_parameter_distributions}
\end{figure}

\begin{figure}[htbp]
\centering
  \begin{subfigure}[b]{0.23\textwidth}
    \includegraphics[width=\linewidth]{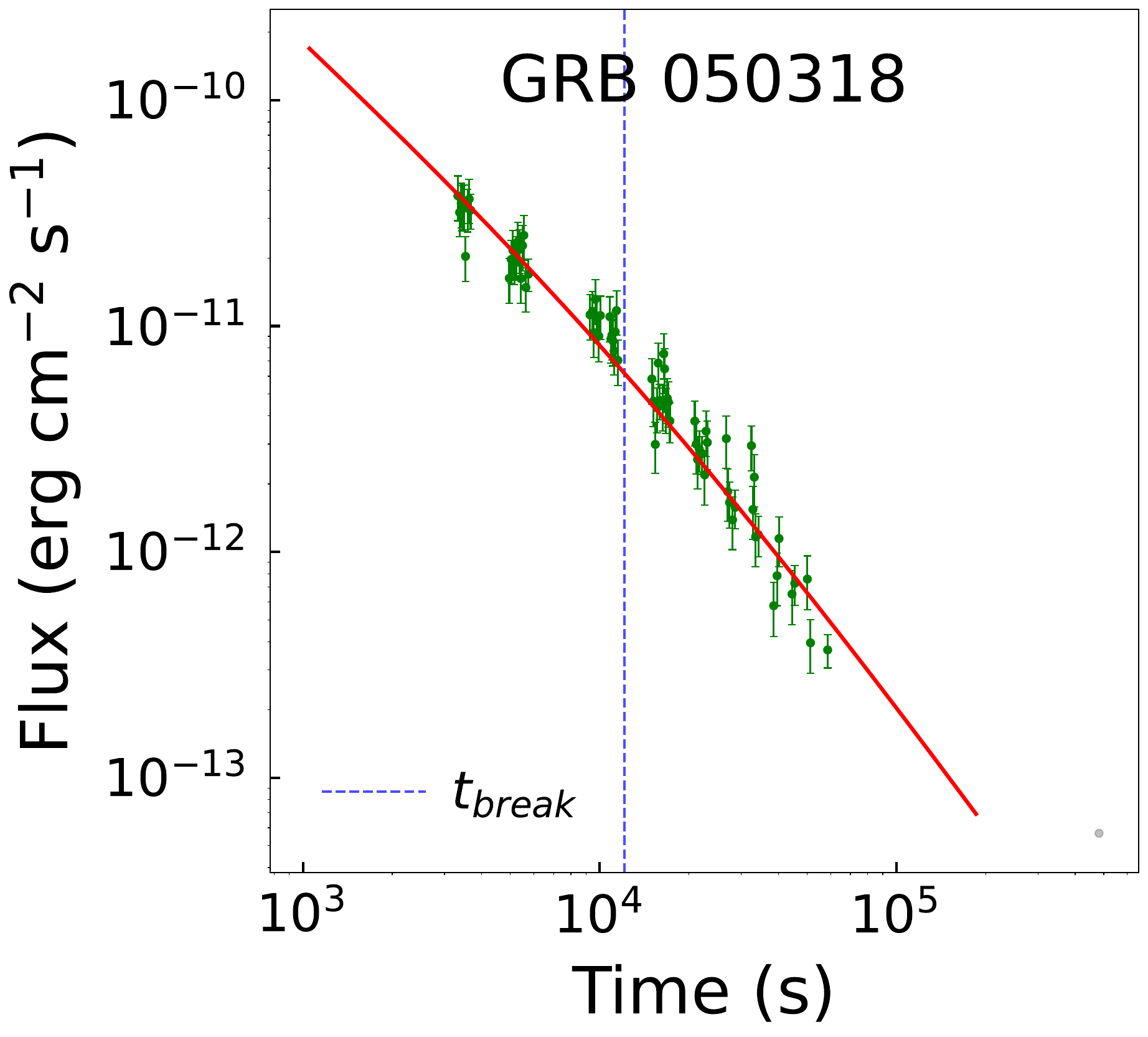}
  \end{subfigure}%
  \hspace{0.01\textwidth}
  \begin{subfigure}[b]{0.23\textwidth}
    \includegraphics[width=\linewidth]{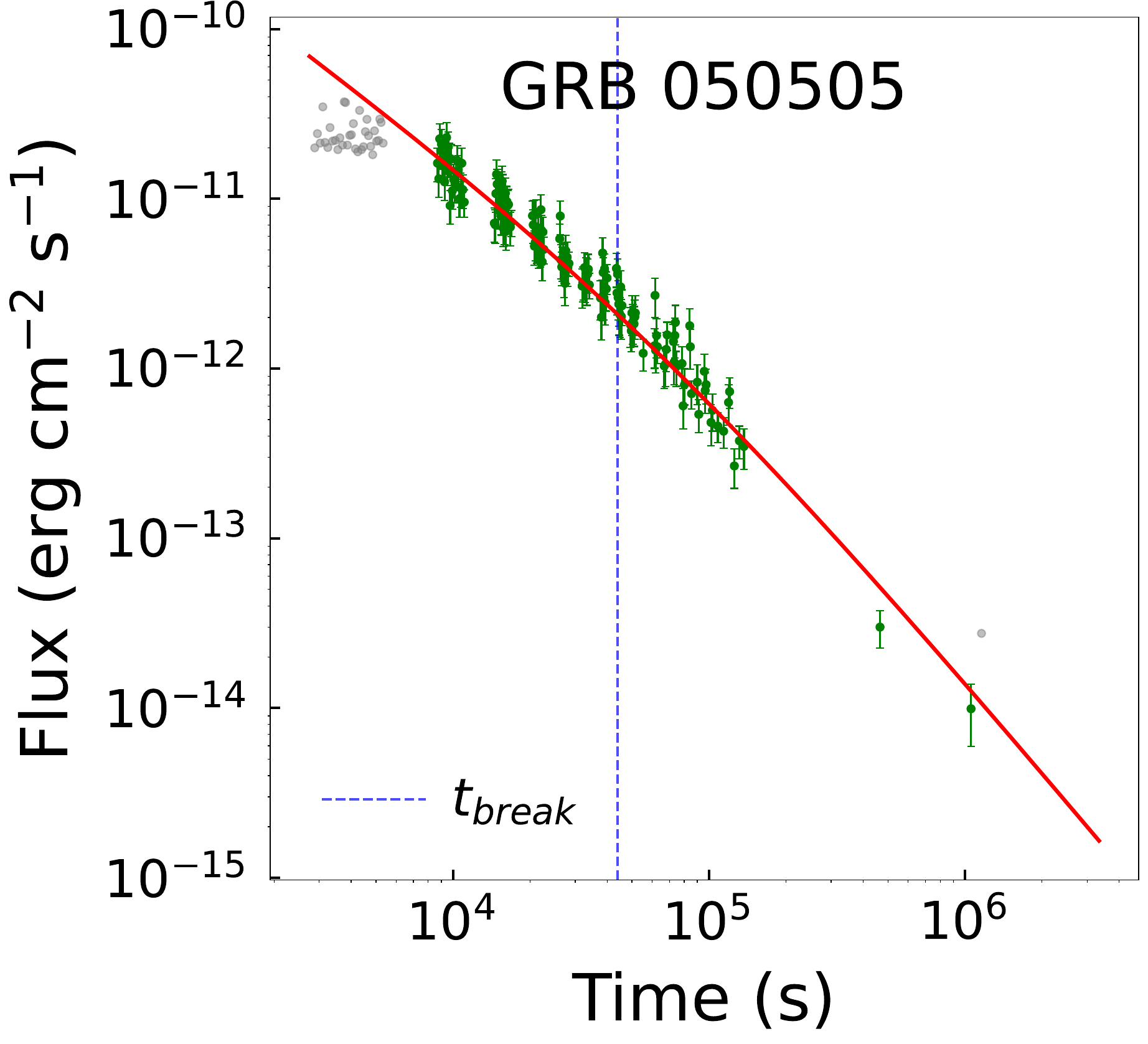}
  \end{subfigure}%
  \hspace{0.01\textwidth}
  \begin{subfigure}[b]{0.23\textwidth}
    \includegraphics[width=\linewidth]{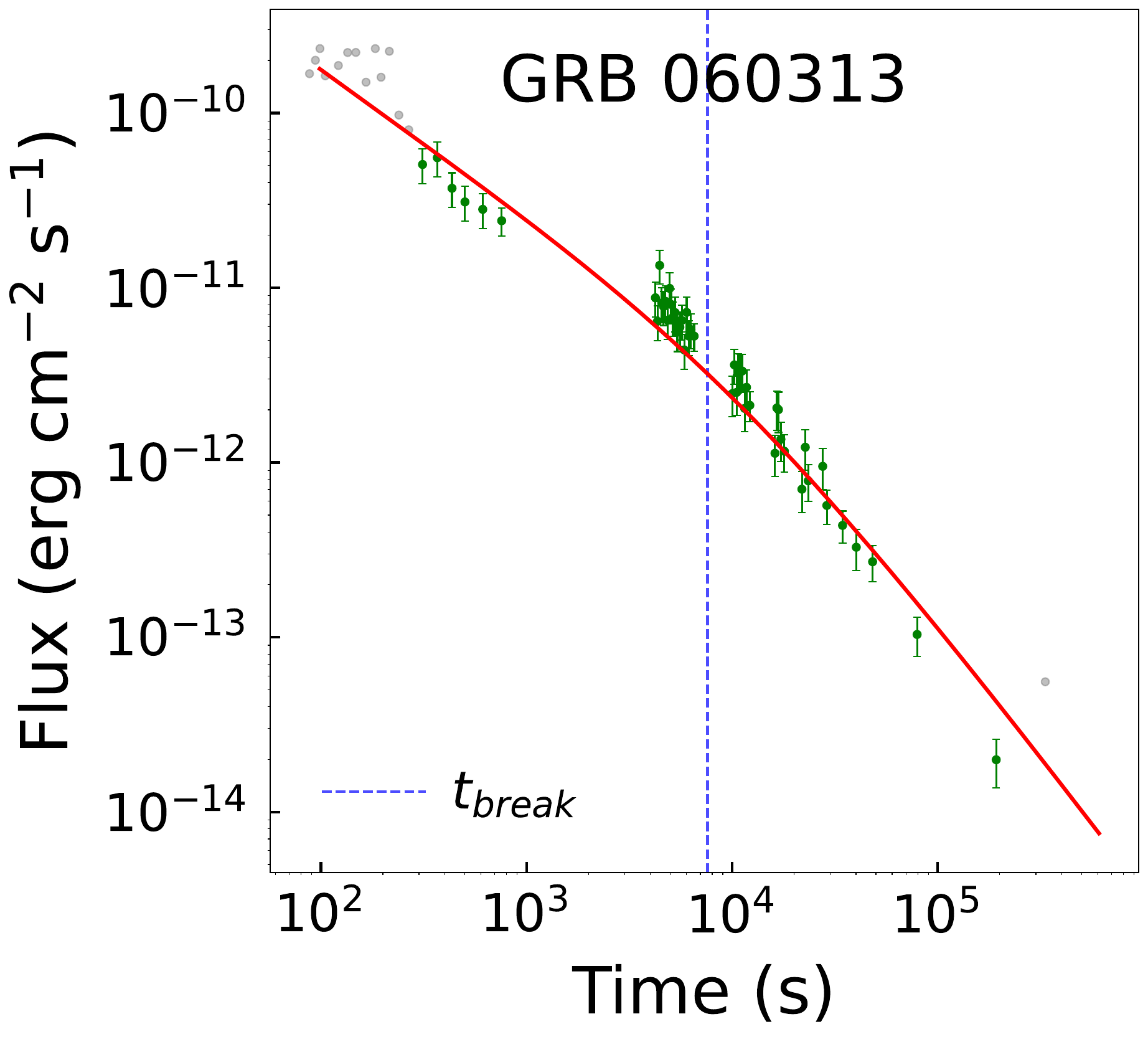}
  \end{subfigure}%
  \hspace{0.01\textwidth}
  \begin{subfigure}[b]{0.23\textwidth}
    \includegraphics[width=\linewidth]{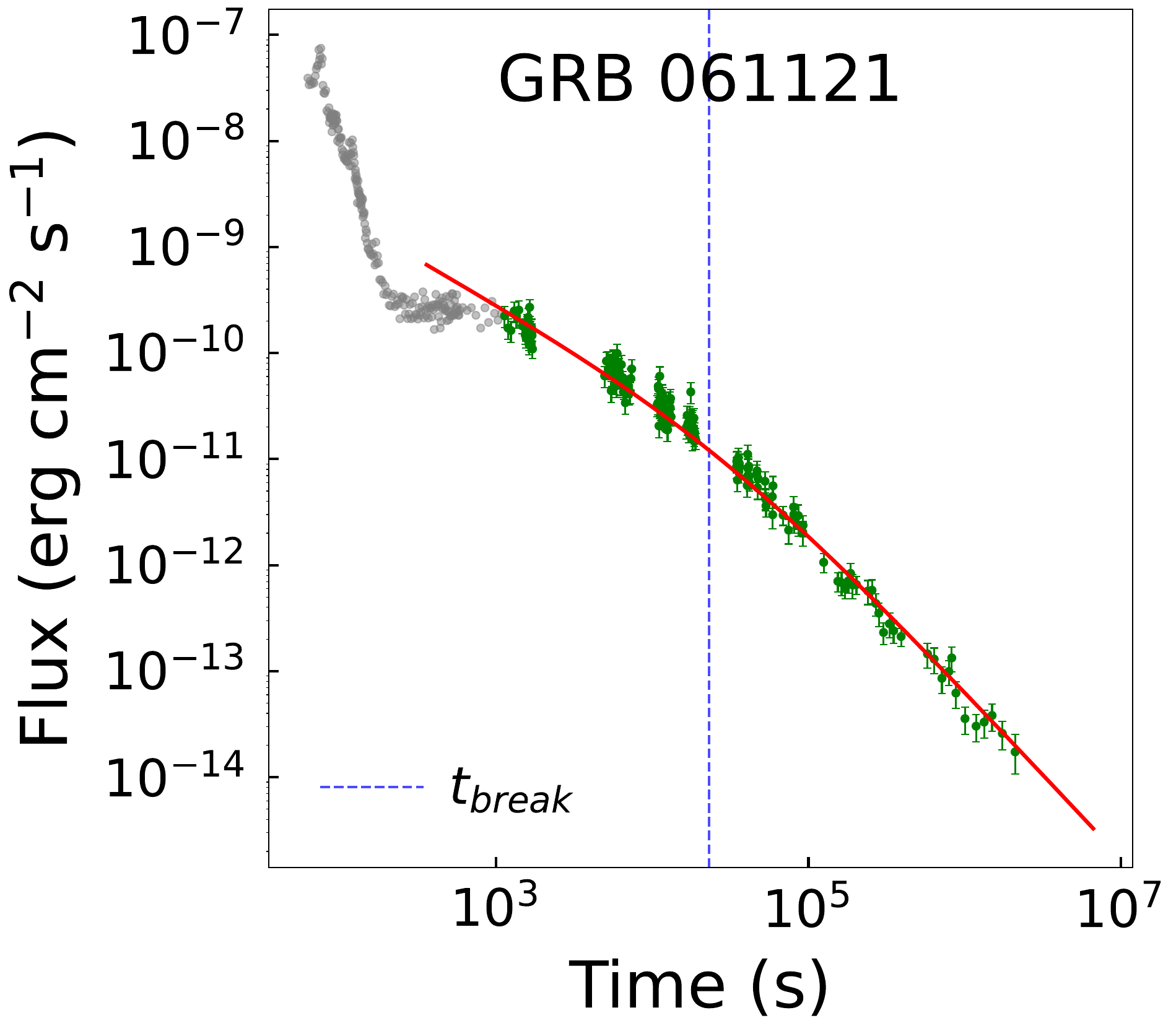}
  \end{subfigure}%
  \\%
  \begin{subfigure}[b]{0.23\textwidth}
    \includegraphics[width=\linewidth]{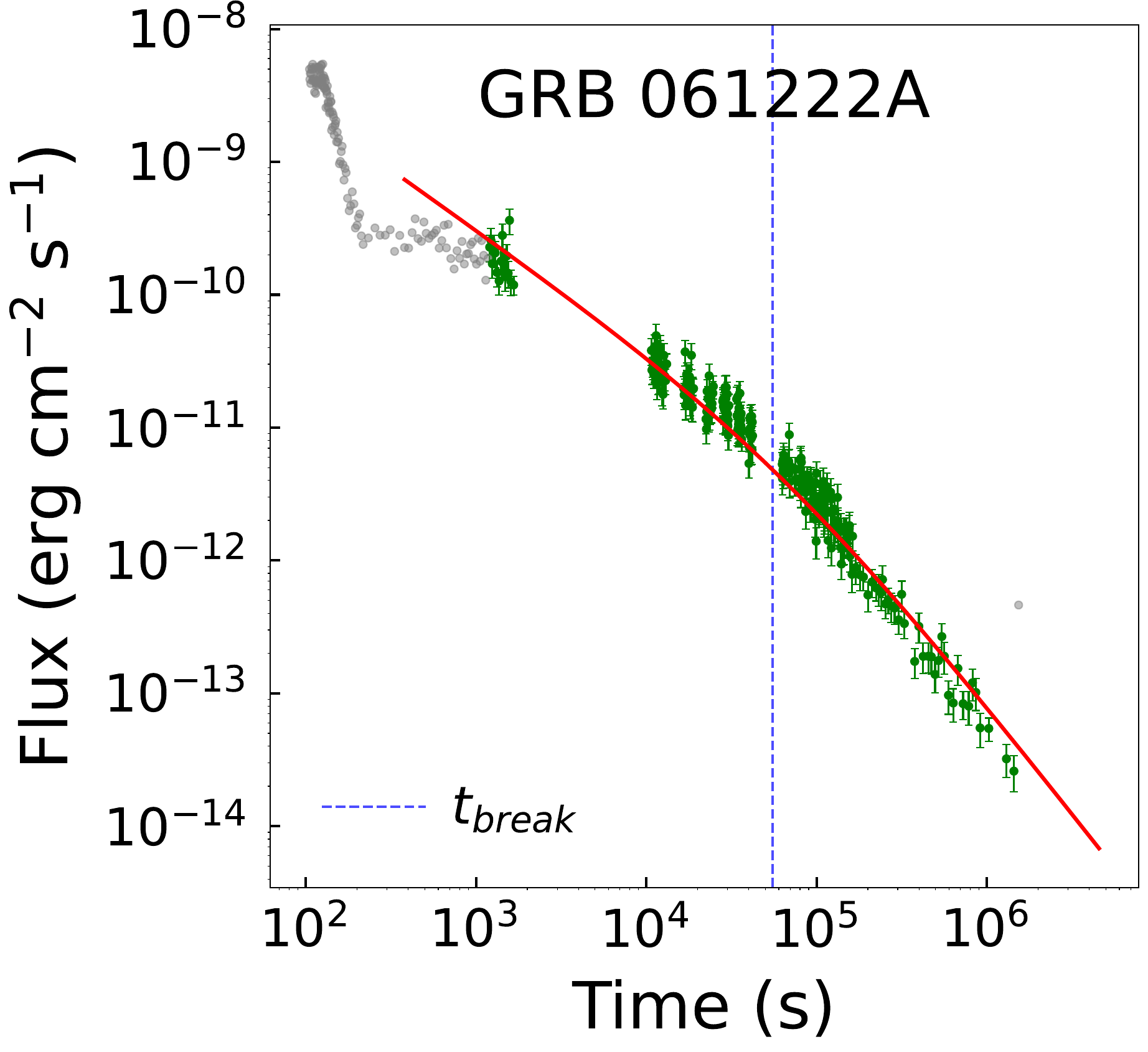}
  \end{subfigure}%
  \hspace{0.01\textwidth}
  \begin{subfigure}[b]{0.23\textwidth}
    \includegraphics[width=\linewidth]{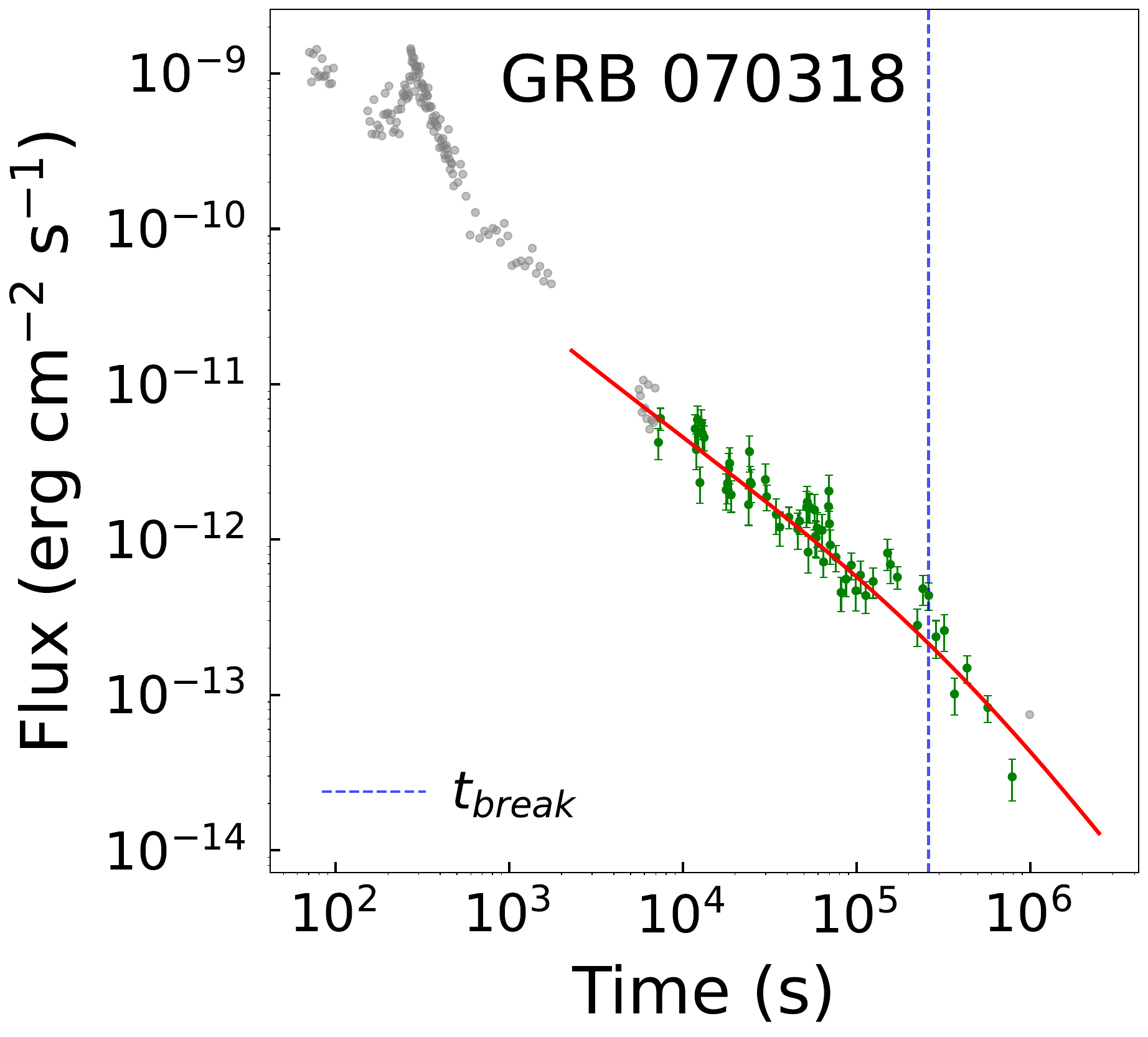}
  \end{subfigure}%
  \hspace{0.01\textwidth}
  \begin{subfigure}[b]{0.23\textwidth}
    \includegraphics[width=\linewidth]{result_070420_20_flux_vs_time.pdf}
  \end{subfigure}%
  \hspace{0.01\textwidth}
  \begin{subfigure}[b]{0.23\textwidth}
    \includegraphics[width=\linewidth]{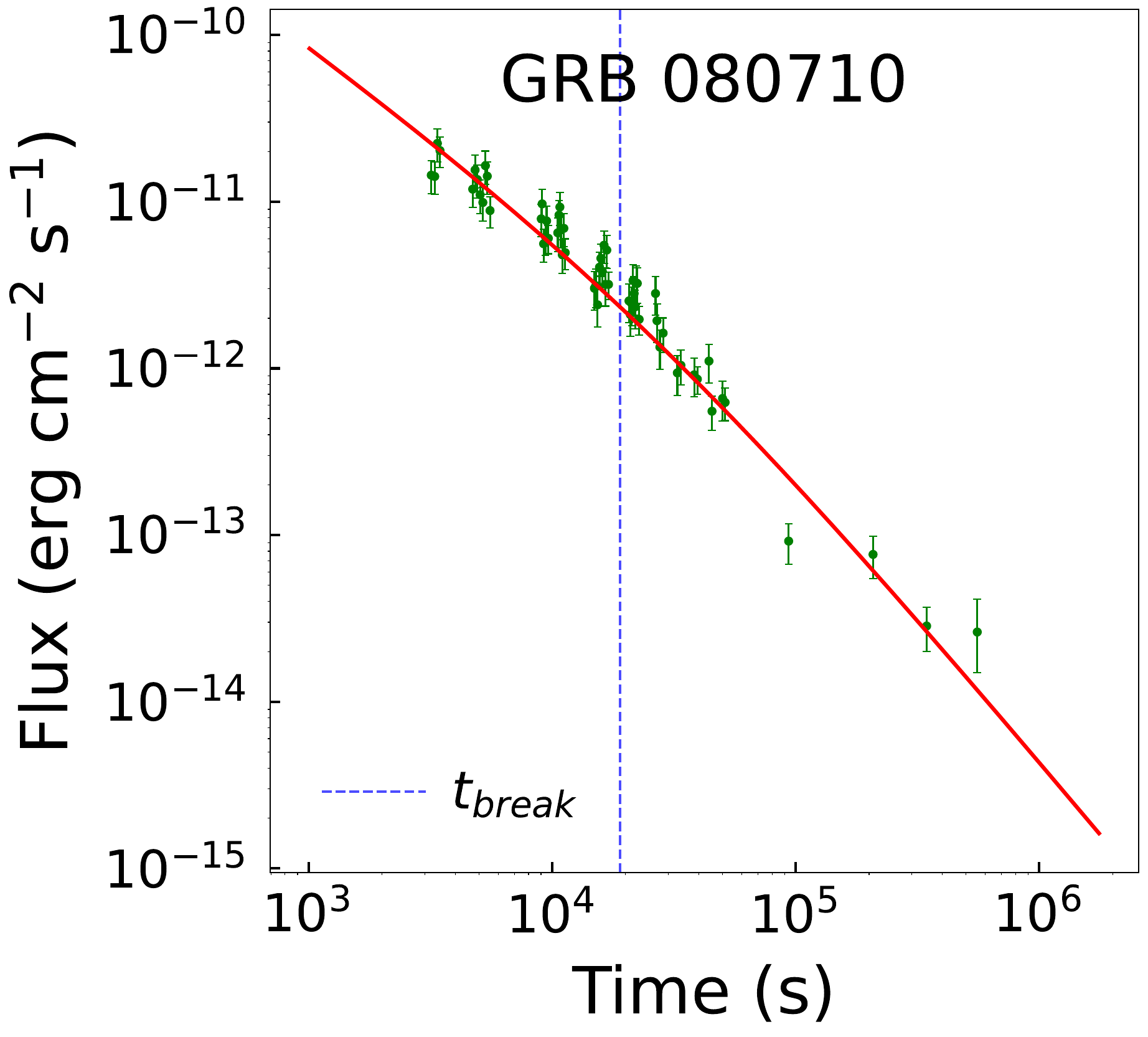}
  \end{subfigure}%
  \\%
  \begin{subfigure}[b]{0.23\textwidth}
    \includegraphics[width=\linewidth]{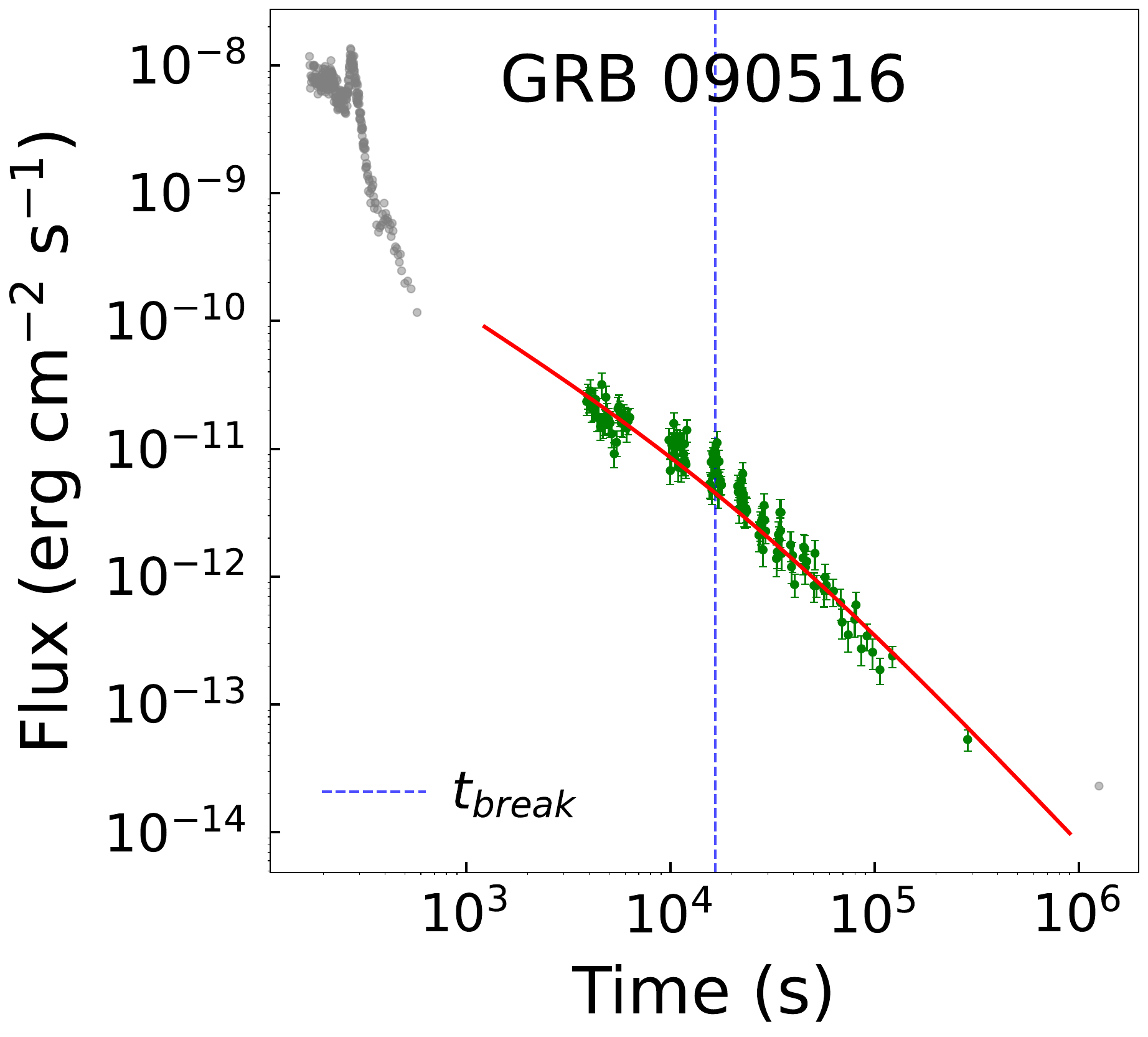}
  \end{subfigure}%
  \hspace{0.01\textwidth}
  \begin{subfigure}[b]{0.23\textwidth}
    \includegraphics[width=\linewidth]{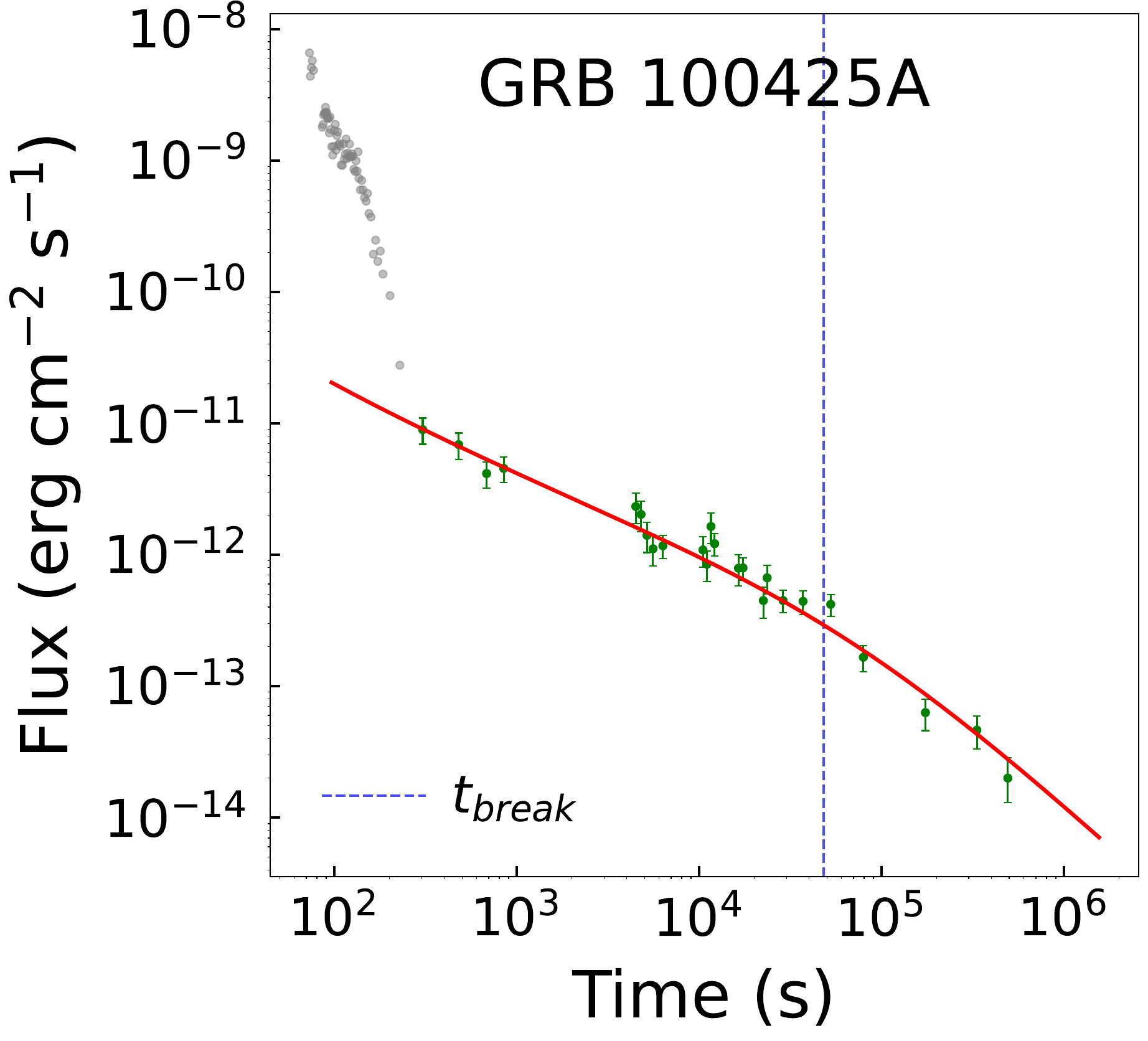}
  \end{subfigure}%
  \hspace{0.01\textwidth}
  \begin{subfigure}[b]{0.23\textwidth}
    \includegraphics[width=\linewidth]{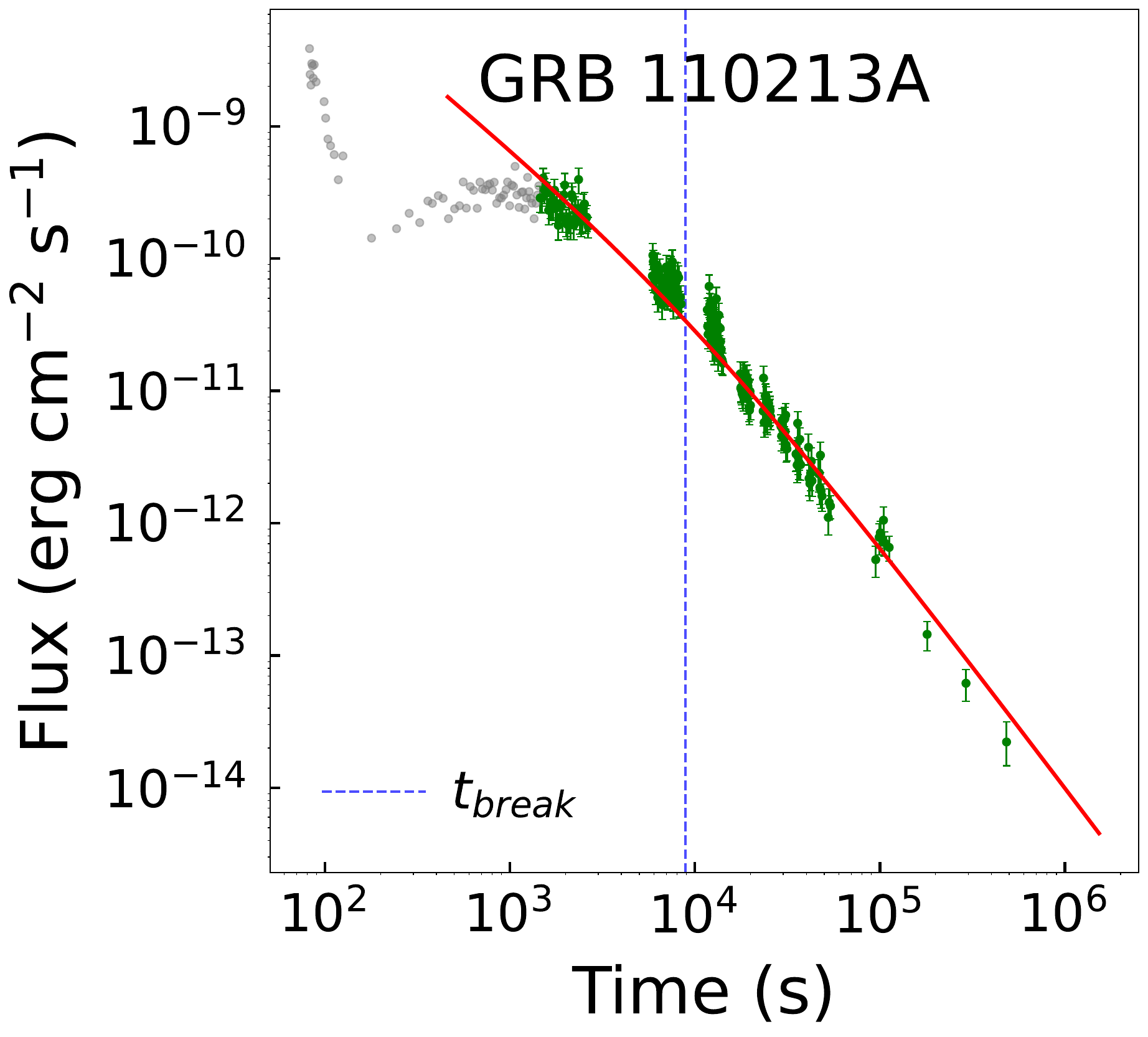}
  \end{subfigure}%
  \hspace{0.01\textwidth}
  \begin{subfigure}[b]{0.23\textwidth}
    \includegraphics[width=\linewidth]{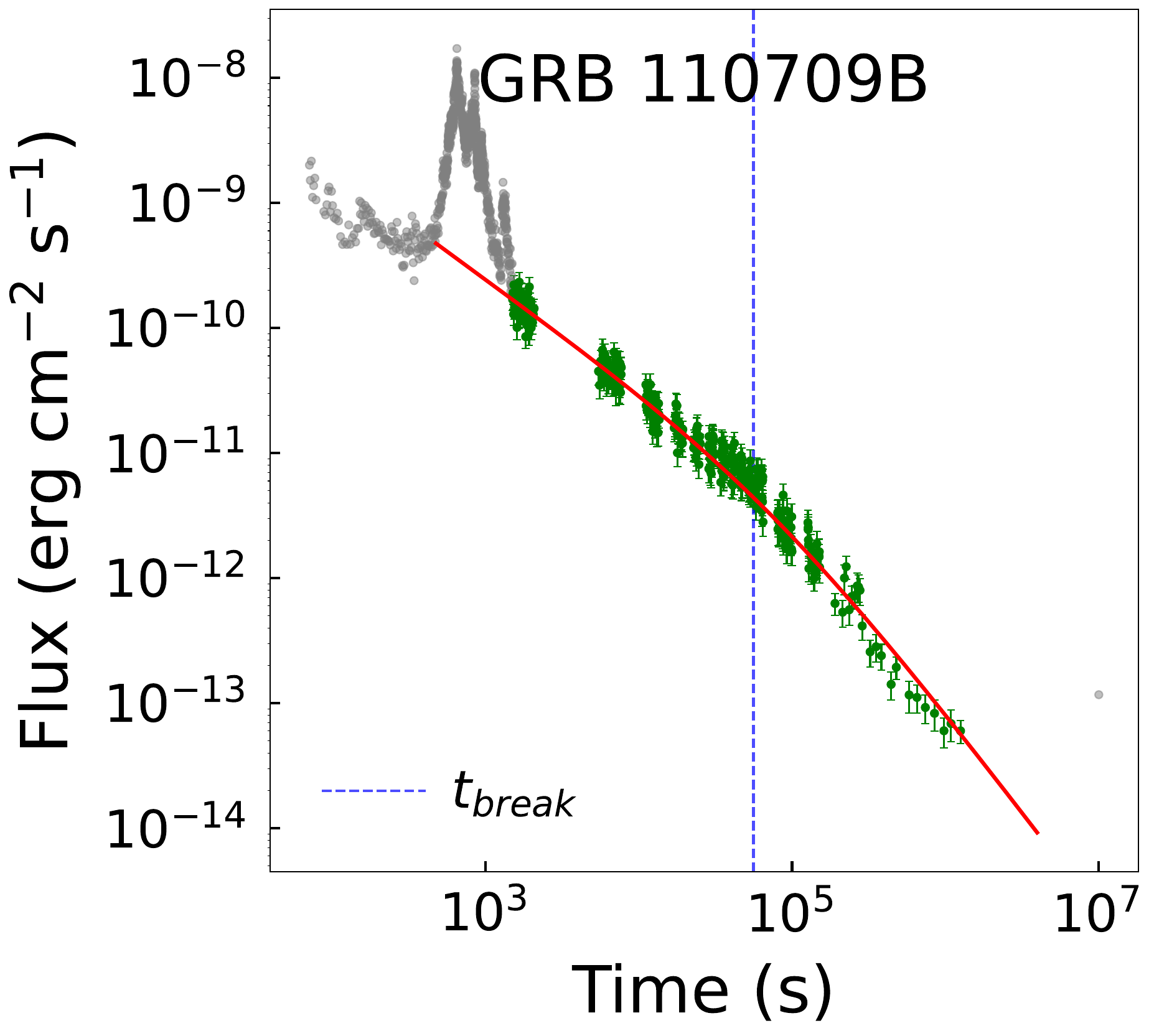}
  \end{subfigure}%
  \\%
  \begin{subfigure}[b]{0.23\textwidth}
    \includegraphics[width=\linewidth]{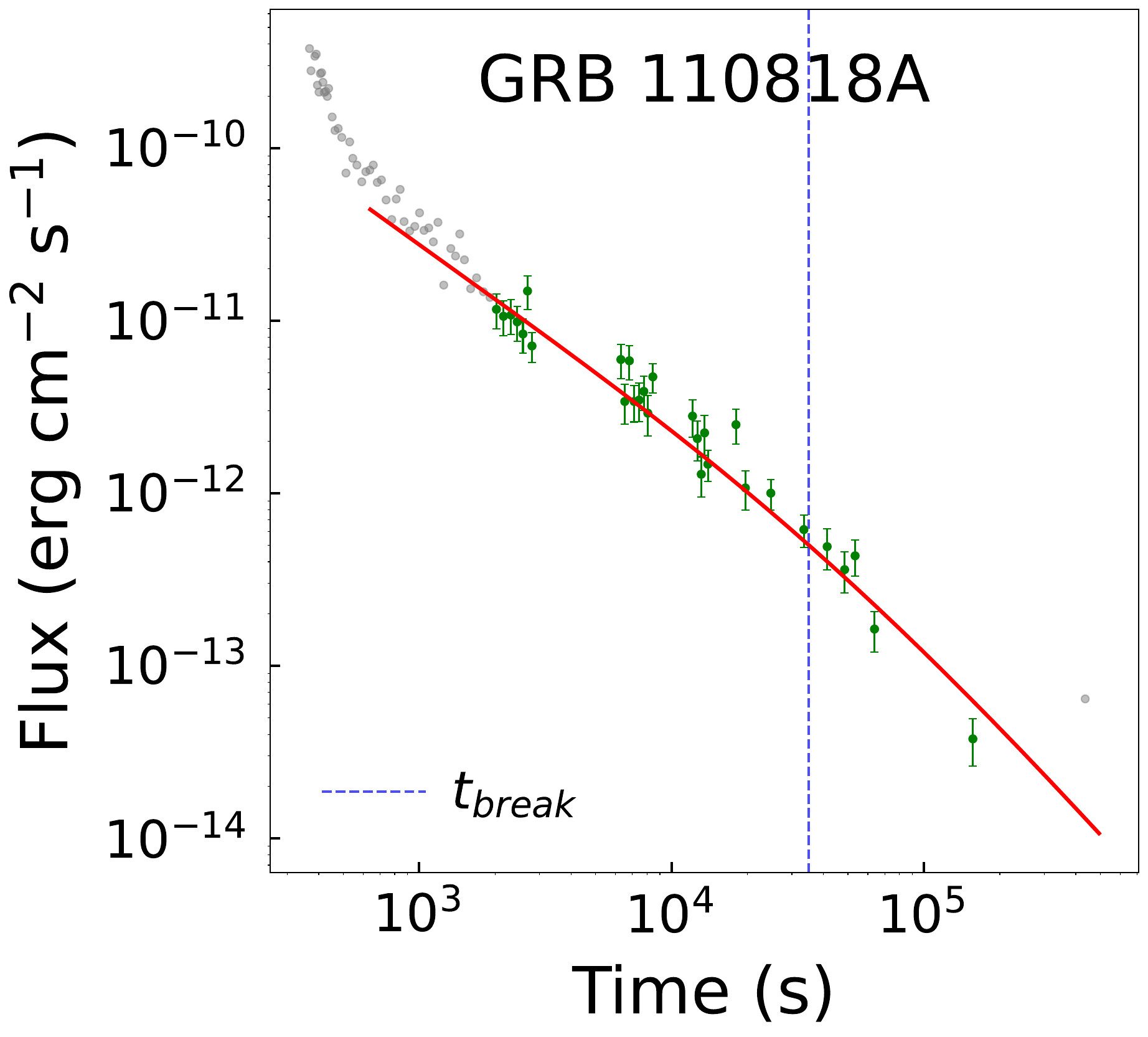}
  \end{subfigure}%
  \hspace{0.01\textwidth}
  \begin{subfigure}[b]{0.23\textwidth}
    \includegraphics[width=\linewidth]{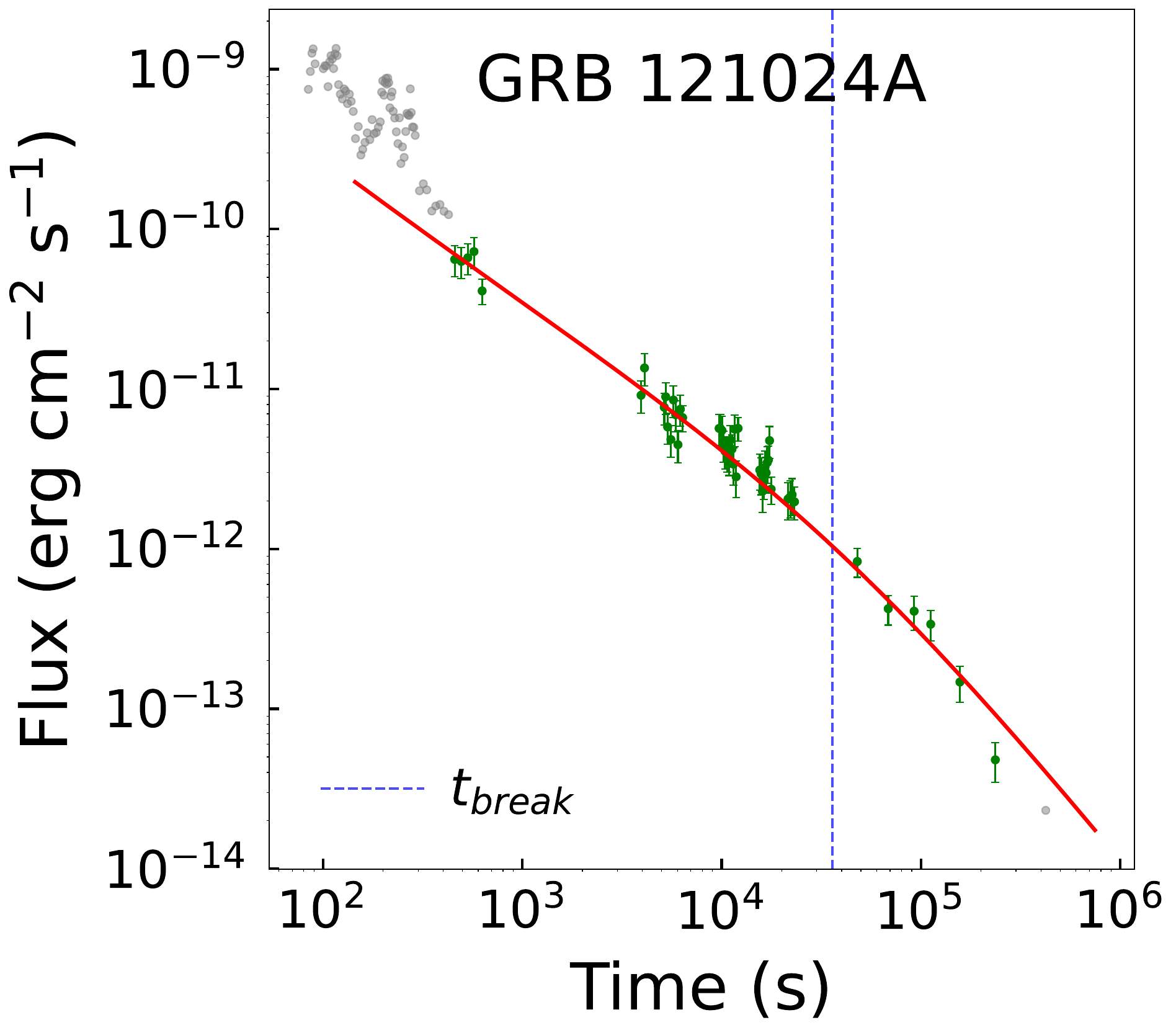}
  \end{subfigure}%
  \hspace{0.01\textwidth}
  \begin{subfigure}[b]{0.23\textwidth}
    \includegraphics[width=\linewidth]{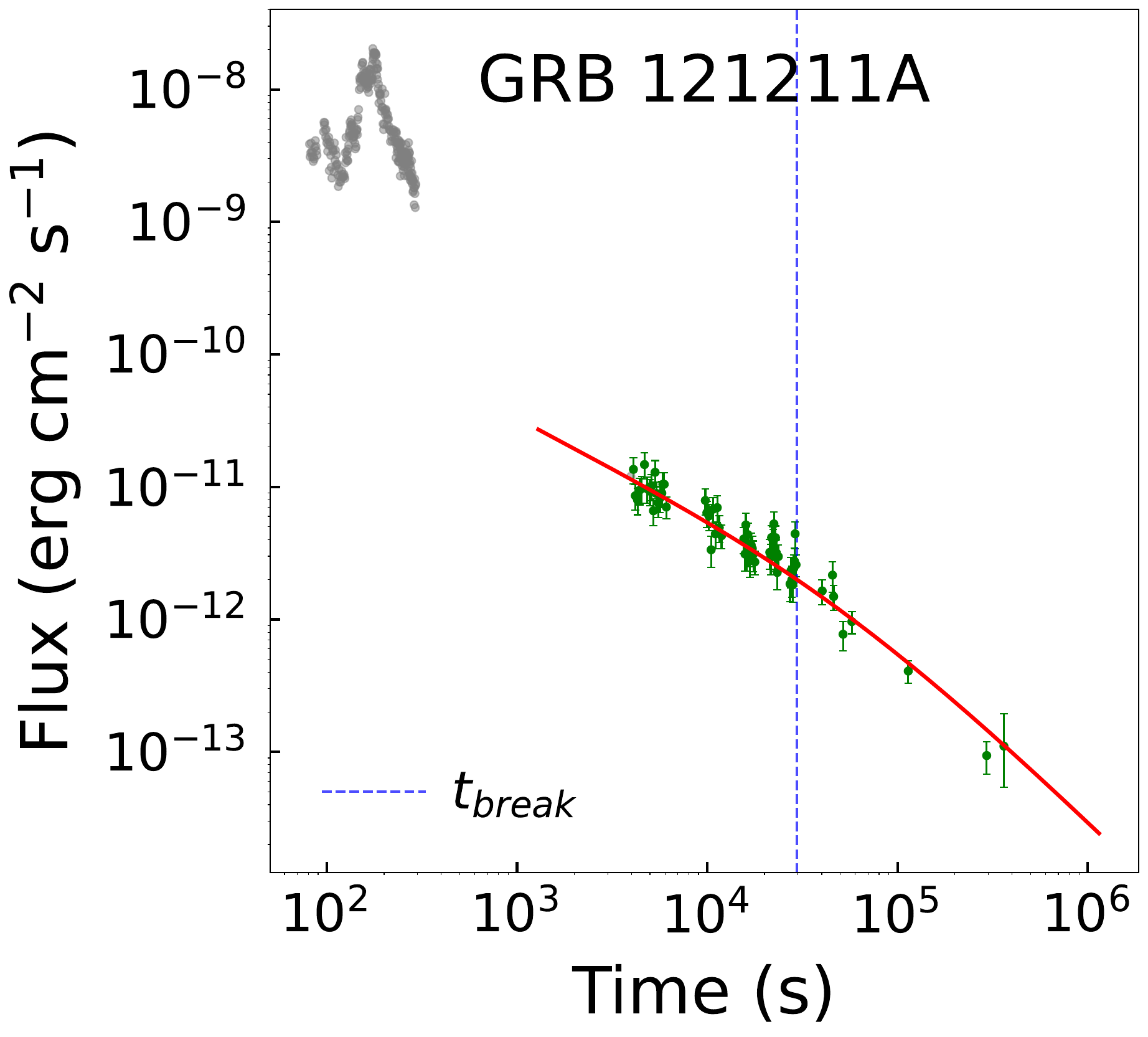}
  \end{subfigure}%
  \hspace{0.01\textwidth}
  \begin{subfigure}[b]{0.23\textwidth}
    \includegraphics[width=\linewidth]{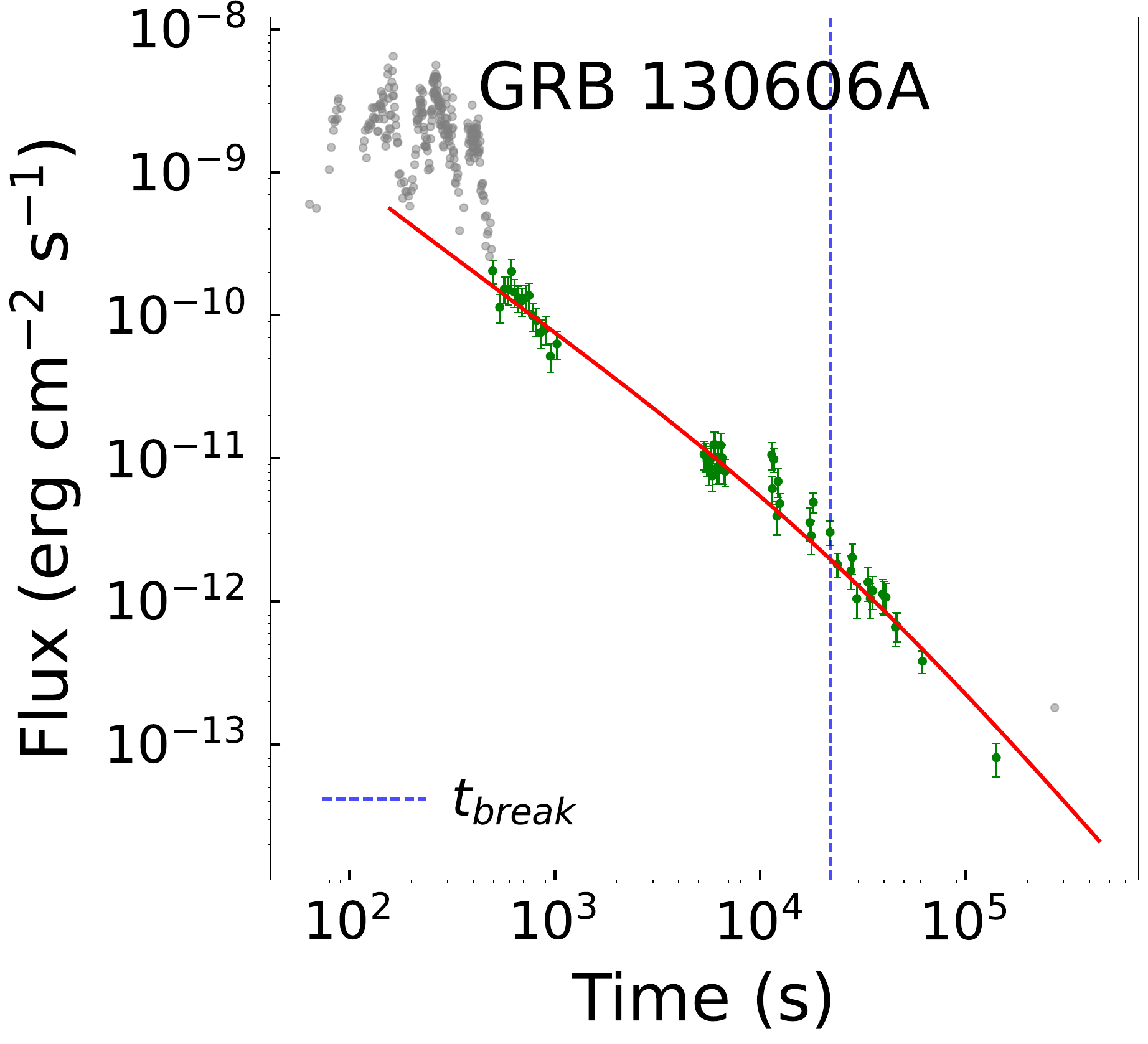}
  \end{subfigure}%
  \\%
  \begin{subfigure}[b]{0.23\textwidth}
    \includegraphics[width=\linewidth]{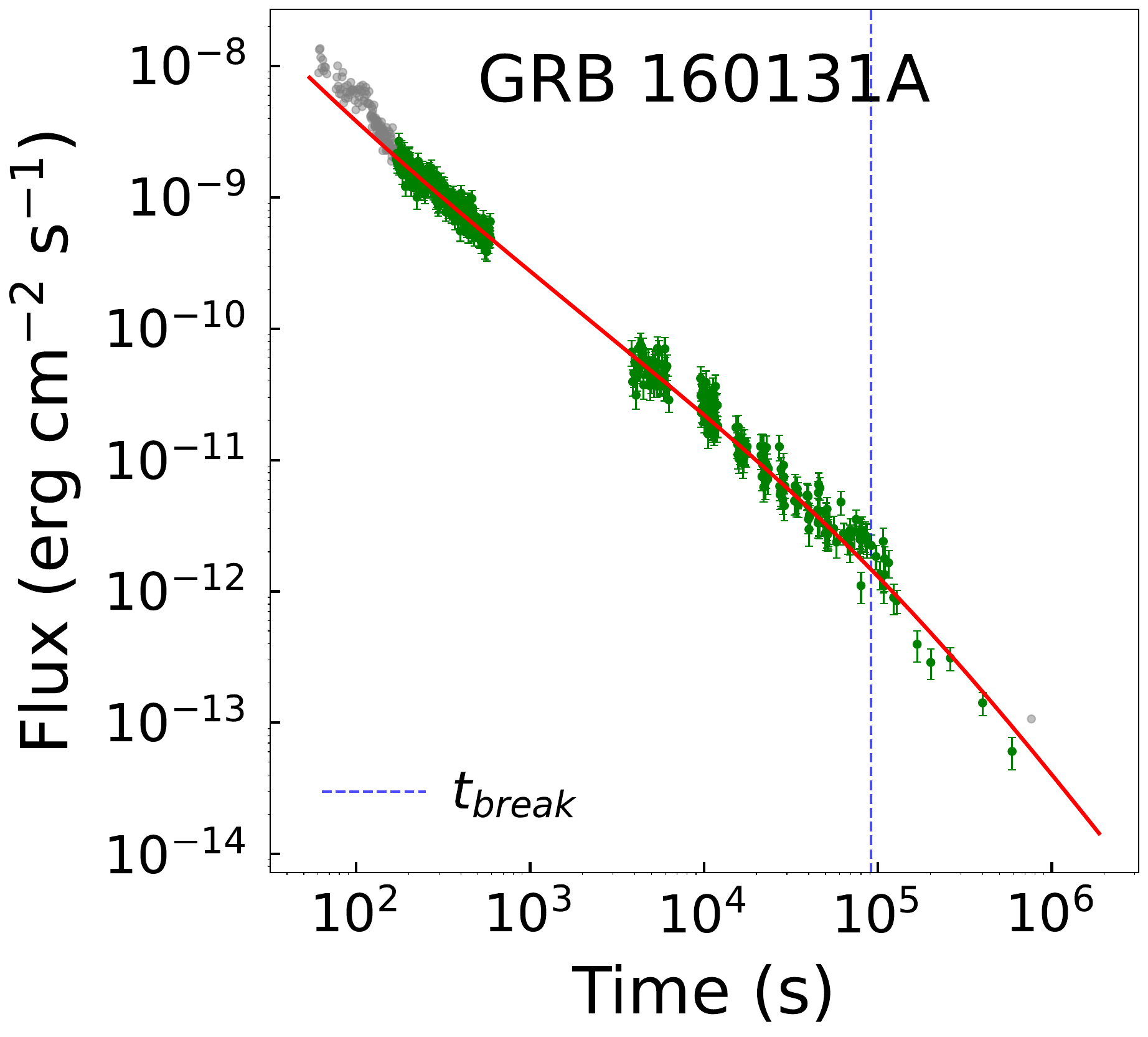}
  \end{subfigure}%
  \hspace{0.01\textwidth}
  \begin{subfigure}[b]{0.23\textwidth}
    \includegraphics[width=\linewidth]{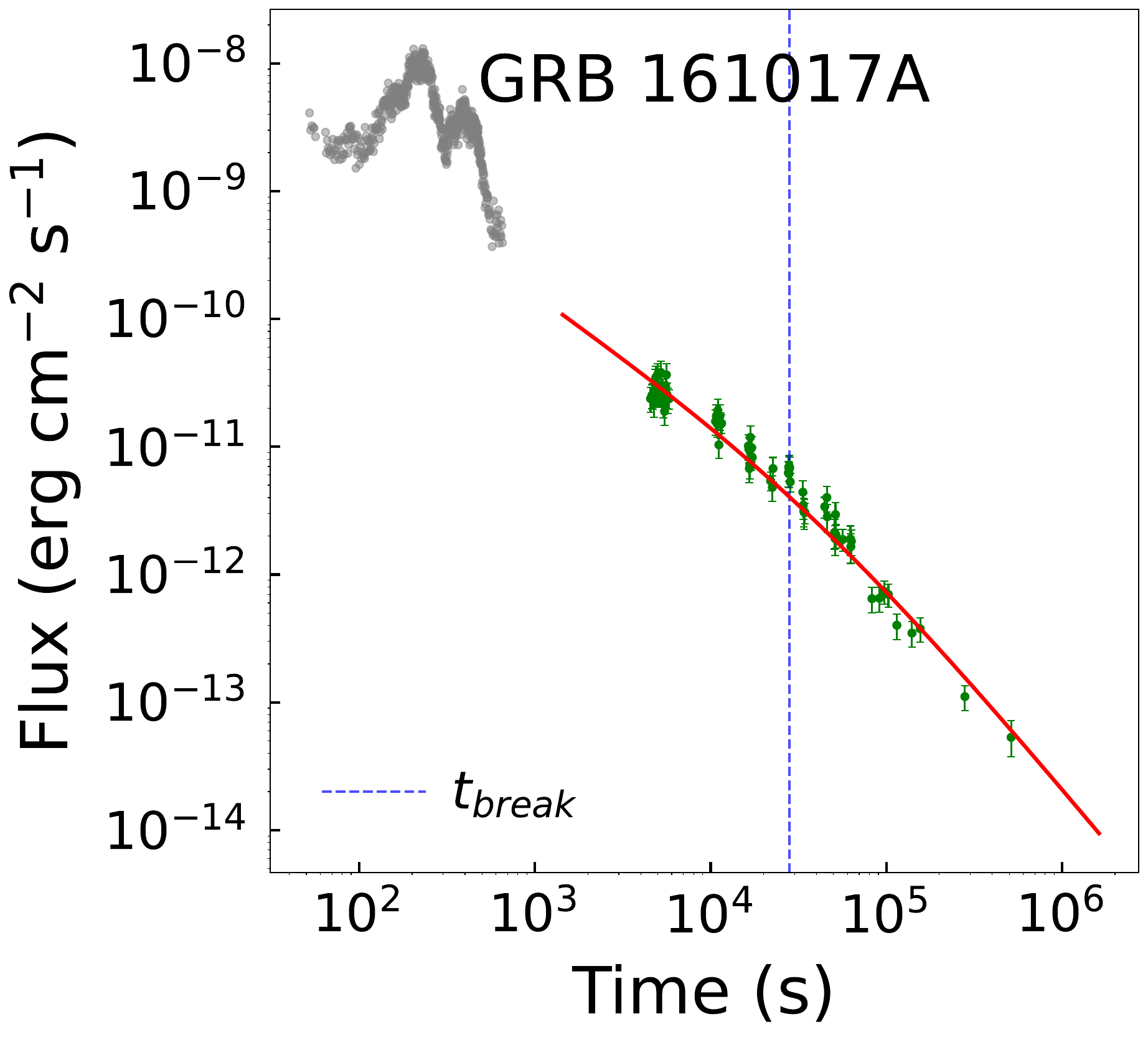}
  \end{subfigure}%
  \hspace{0.01\textwidth}
  \begin{subfigure}[b]{0.23\textwidth}
    \includegraphics[width=\linewidth]{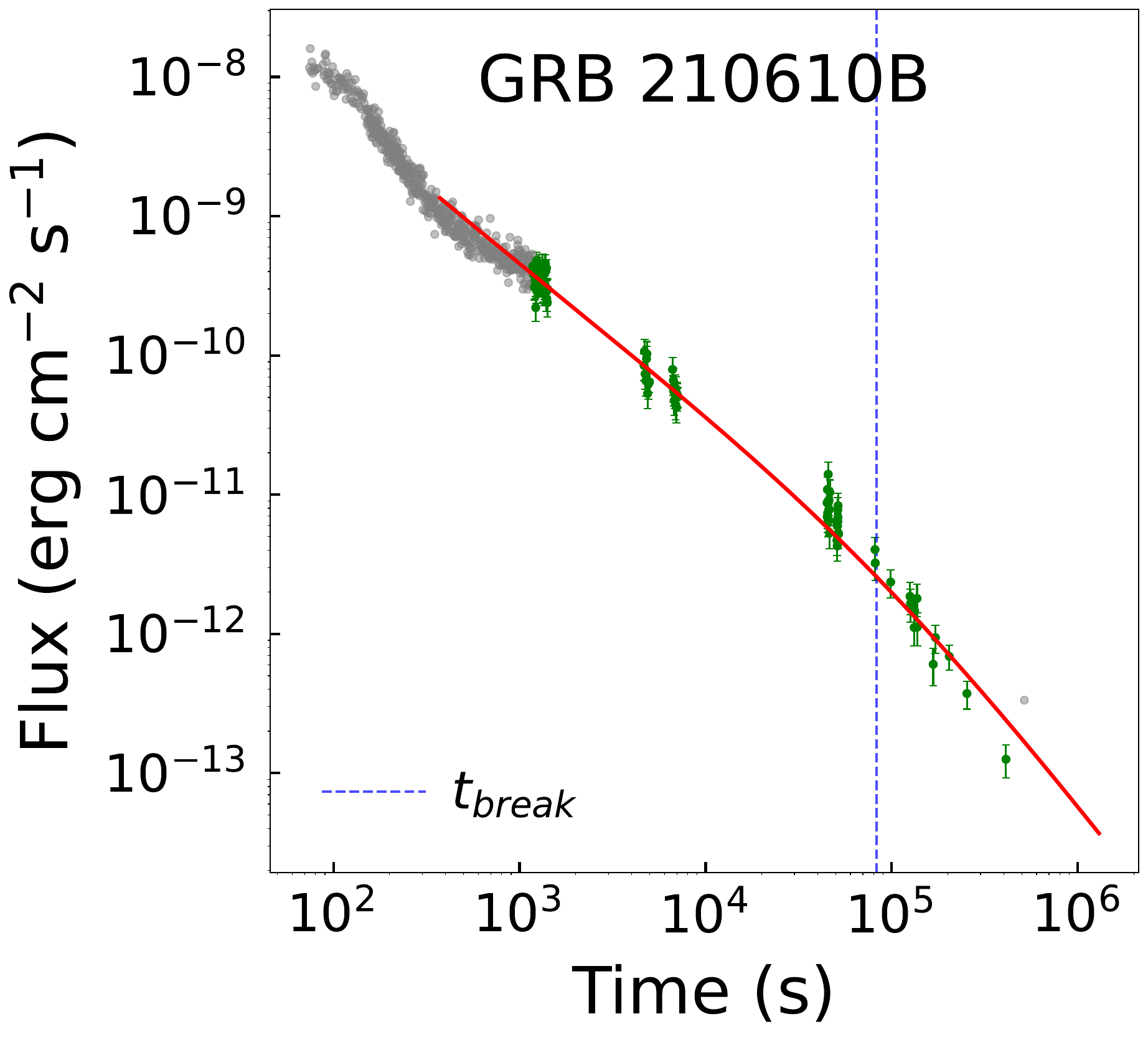}
  \end{subfigure}%
  \hspace{0.01\textwidth}
  \begin{subfigure}[b]{0.23\textwidth}
    \includegraphics[width=\linewidth]{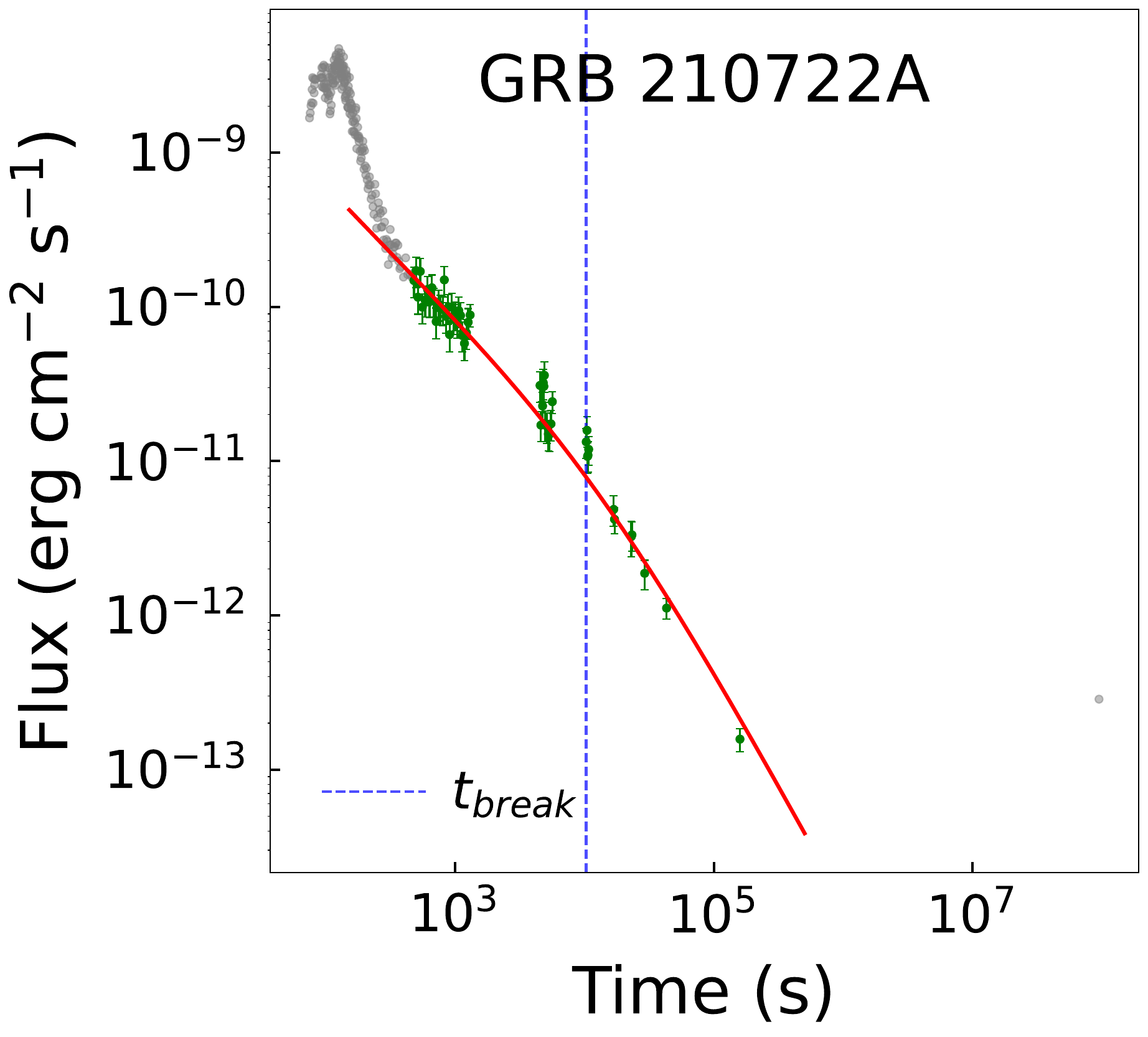}
  \end{subfigure}%
  \\%
  \caption{X-ray afterglow light curve fitting results for model~2 with high-latitude emission. Each panel shows the observed flux (data points) and the best-fit model (solid line) for individual gamma-ray bursts in the ISM environment.}
  \label{fig:ISM_model2_light_curves}
\end{figure}

\begin{figure}[htbp]
\centering
  \begin{subfigure}[b]{0.23\textwidth}
    \includegraphics[width=\linewidth]{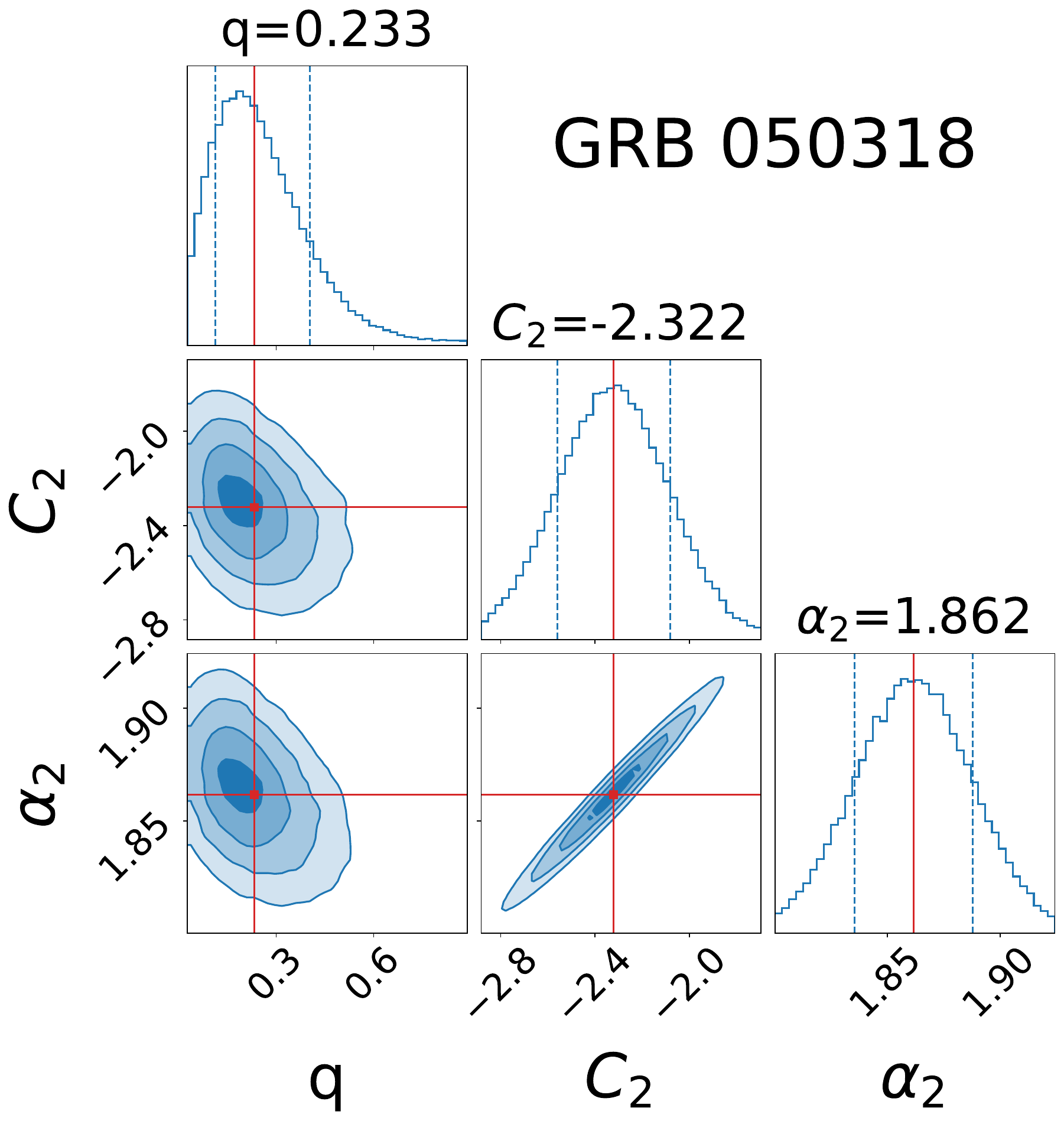}
  \end{subfigure}%
  \hspace{0.01\textwidth}
  \begin{subfigure}[b]{0.23\textwidth}
    \includegraphics[width=\linewidth]{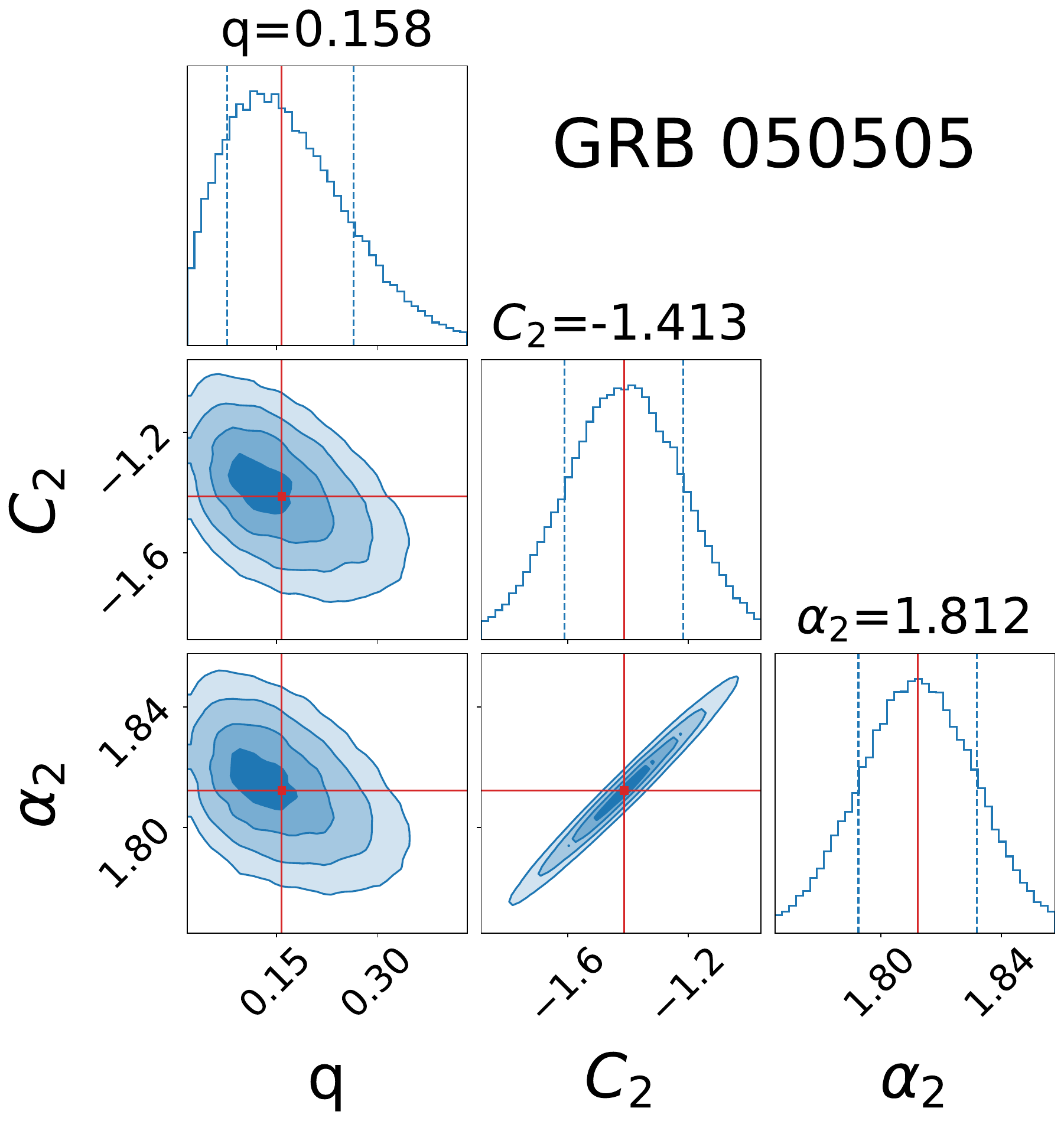}
  \end{subfigure}%
  \hspace{0.01\textwidth}
  \begin{subfigure}[b]{0.23\textwidth}
    \includegraphics[width=\linewidth]{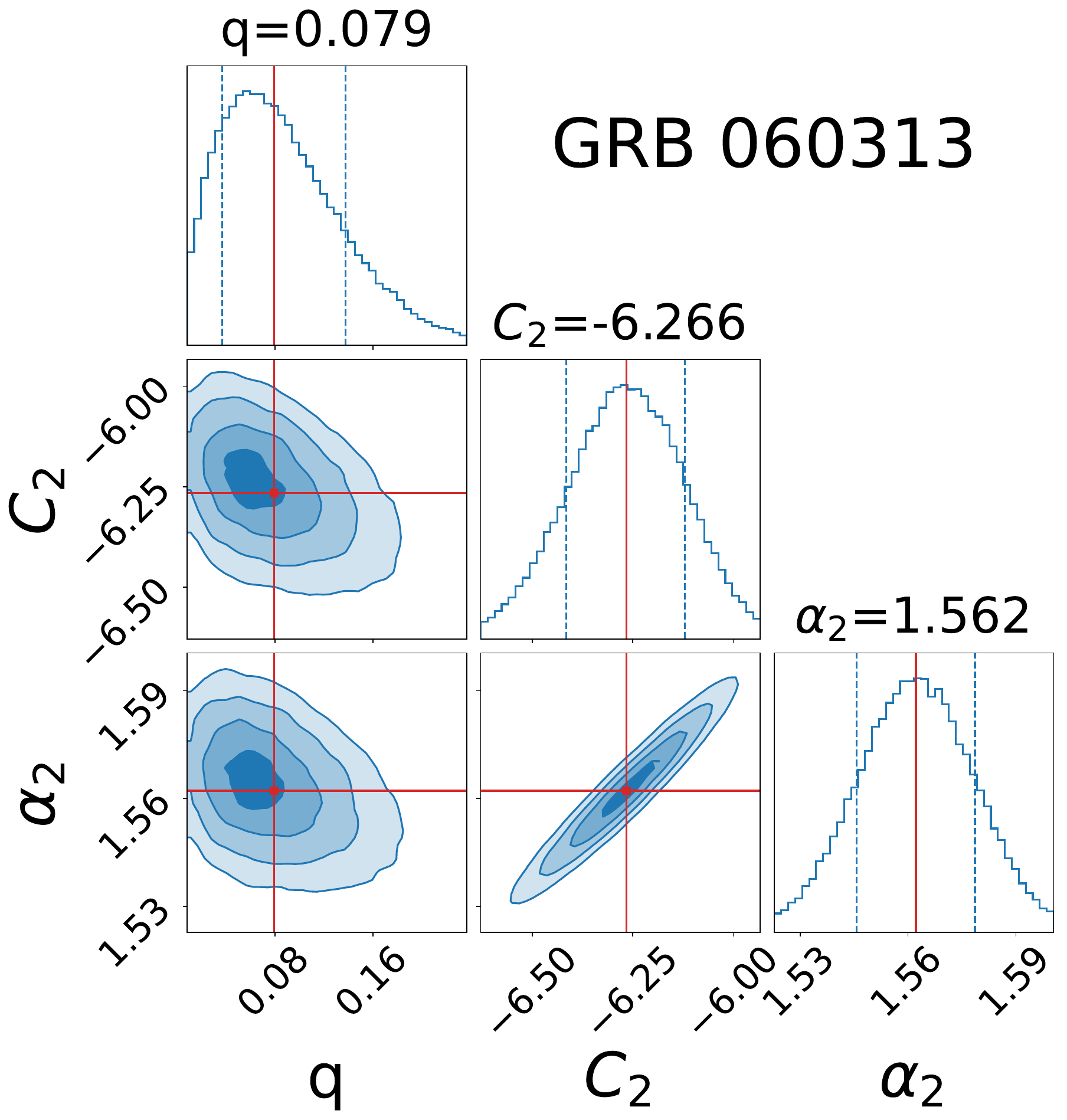}
  \end{subfigure}%
  \hspace{0.01\textwidth}
  \begin{subfigure}[b]{0.23\textwidth}
    \includegraphics[width=\linewidth]{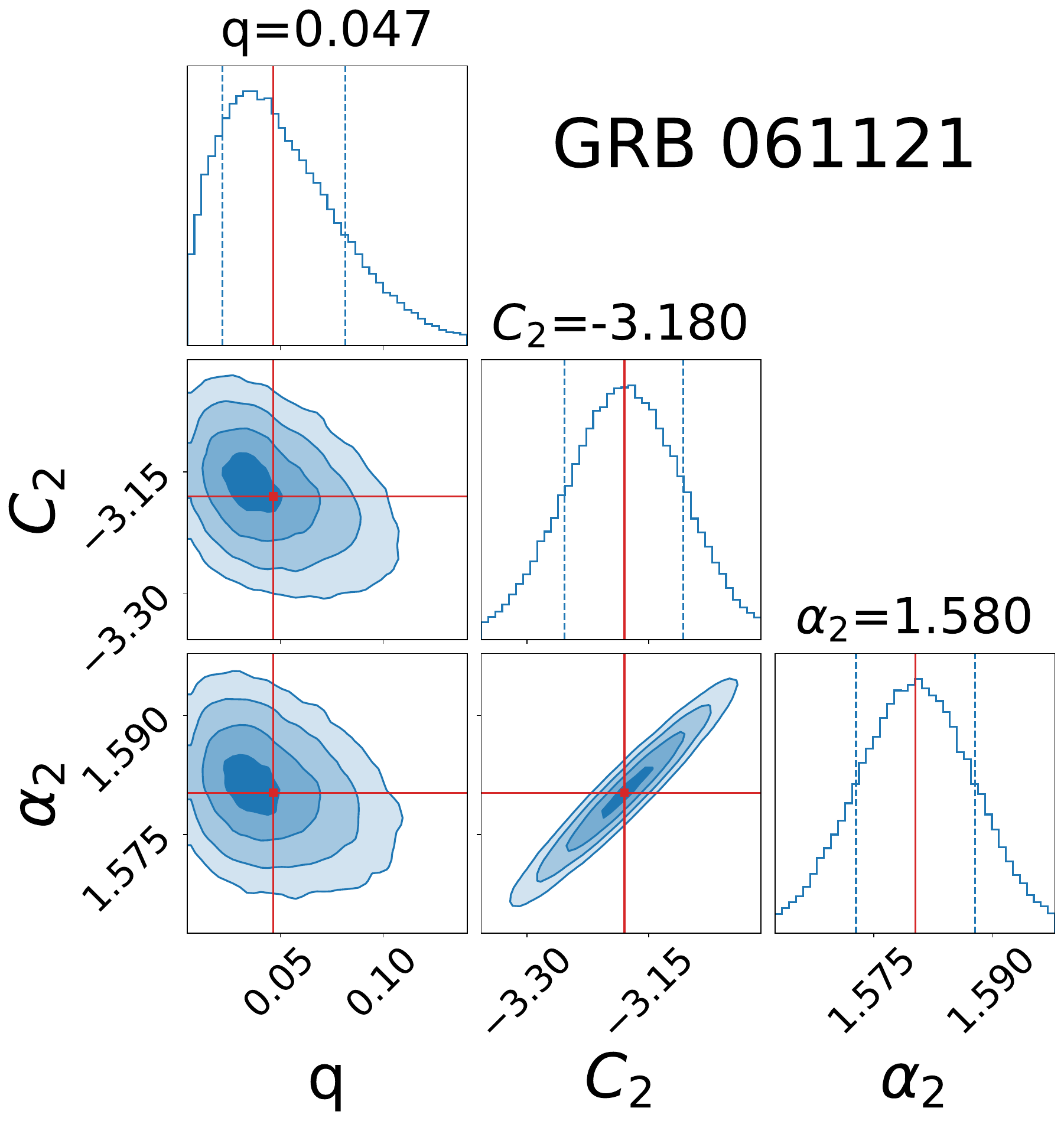}
  \end{subfigure}%
  \\%
  \begin{subfigure}[b]{0.23\textwidth}
    \includegraphics[width=\linewidth]{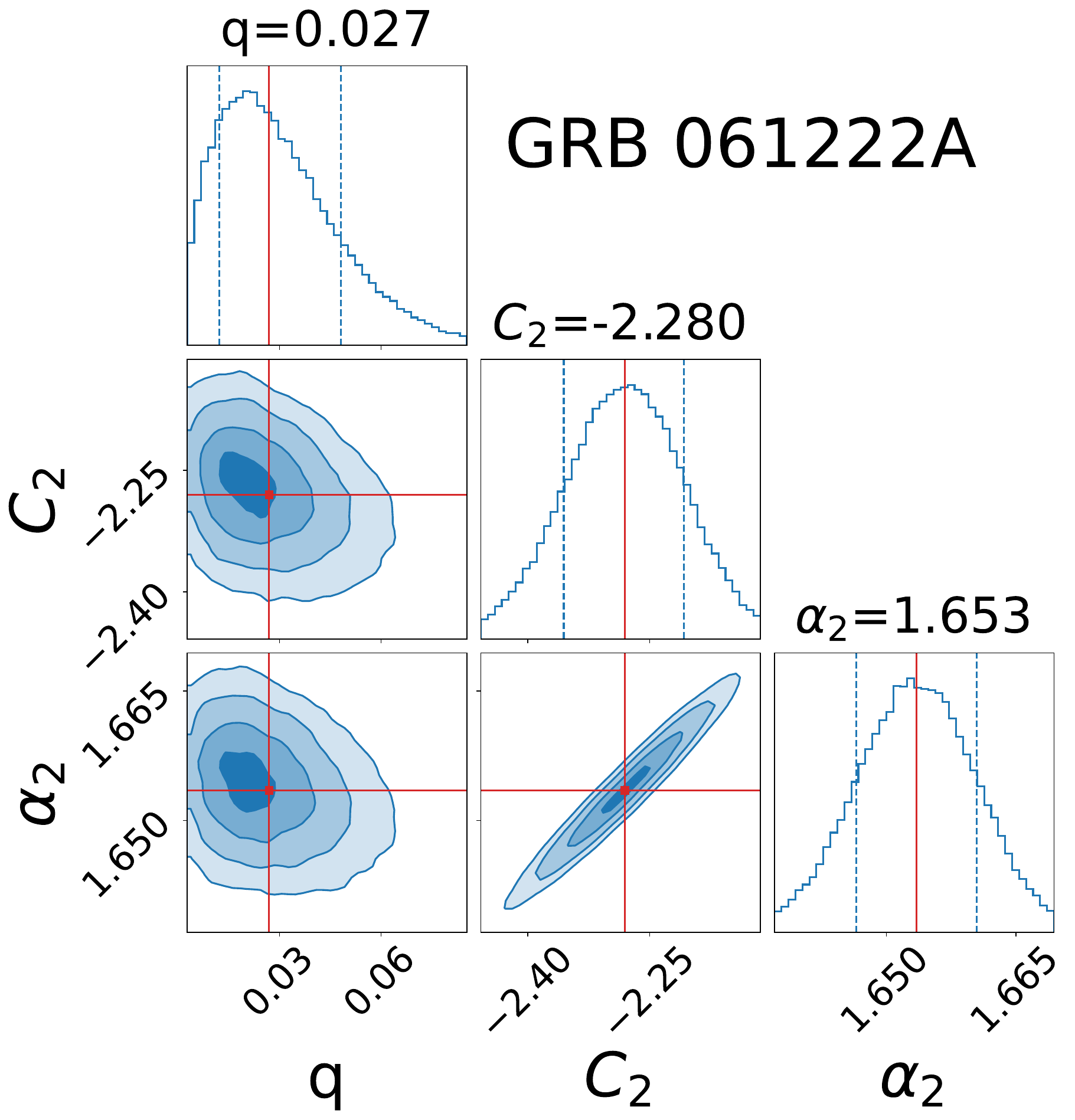}
  \end{subfigure}%
  \hspace{0.01\textwidth}
  \begin{subfigure}[b]{0.23\textwidth}
    \includegraphics[width=\linewidth]{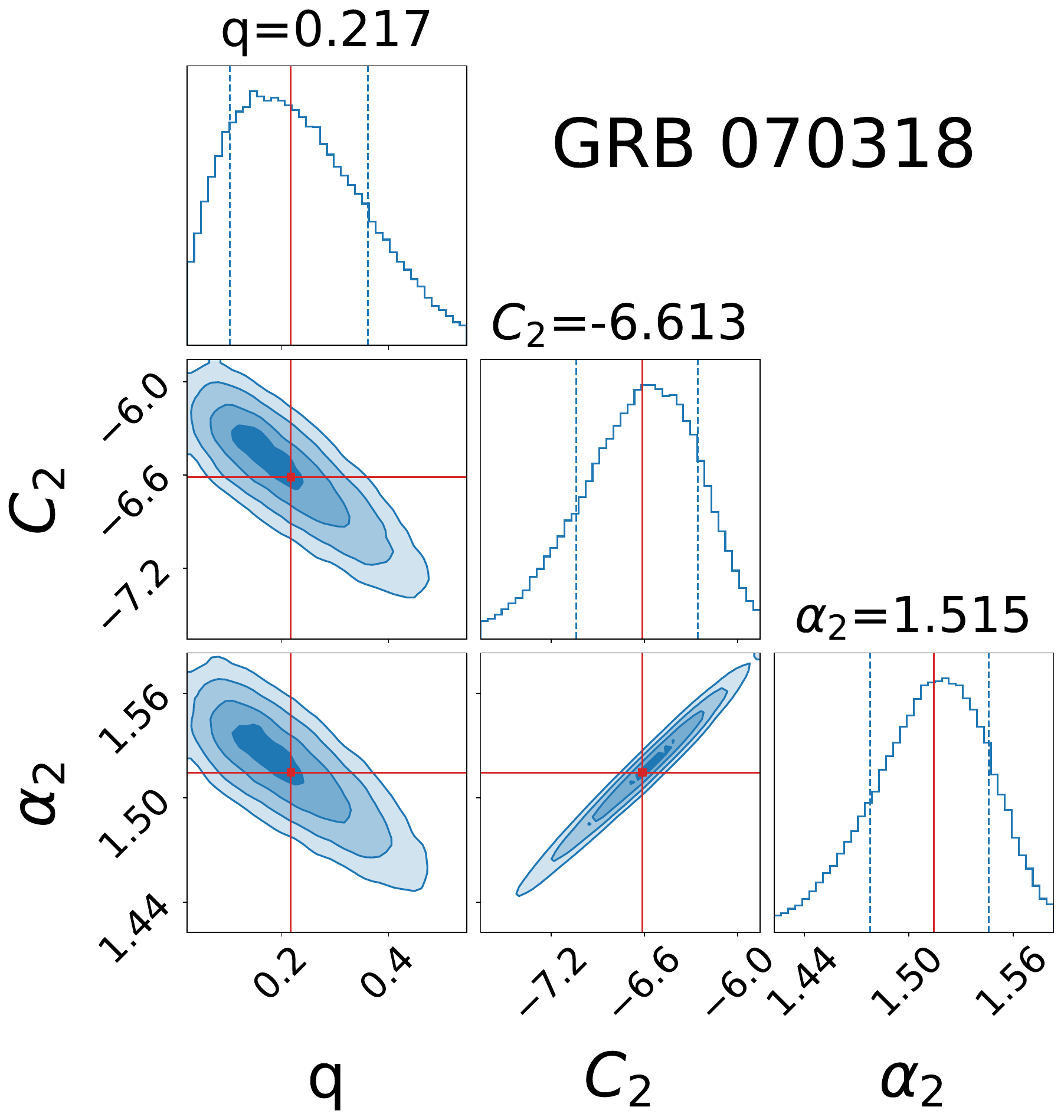}
  \end{subfigure}%
  \hspace{0.01\textwidth}
  \begin{subfigure}[b]{0.23\textwidth}
    \includegraphics[width=\linewidth]{result_070420_20_parameter_distributions.pdf}
  \end{subfigure}%
  \hspace{0.01\textwidth}
  \begin{subfigure}[b]{0.23\textwidth}
    \includegraphics[width=\linewidth]{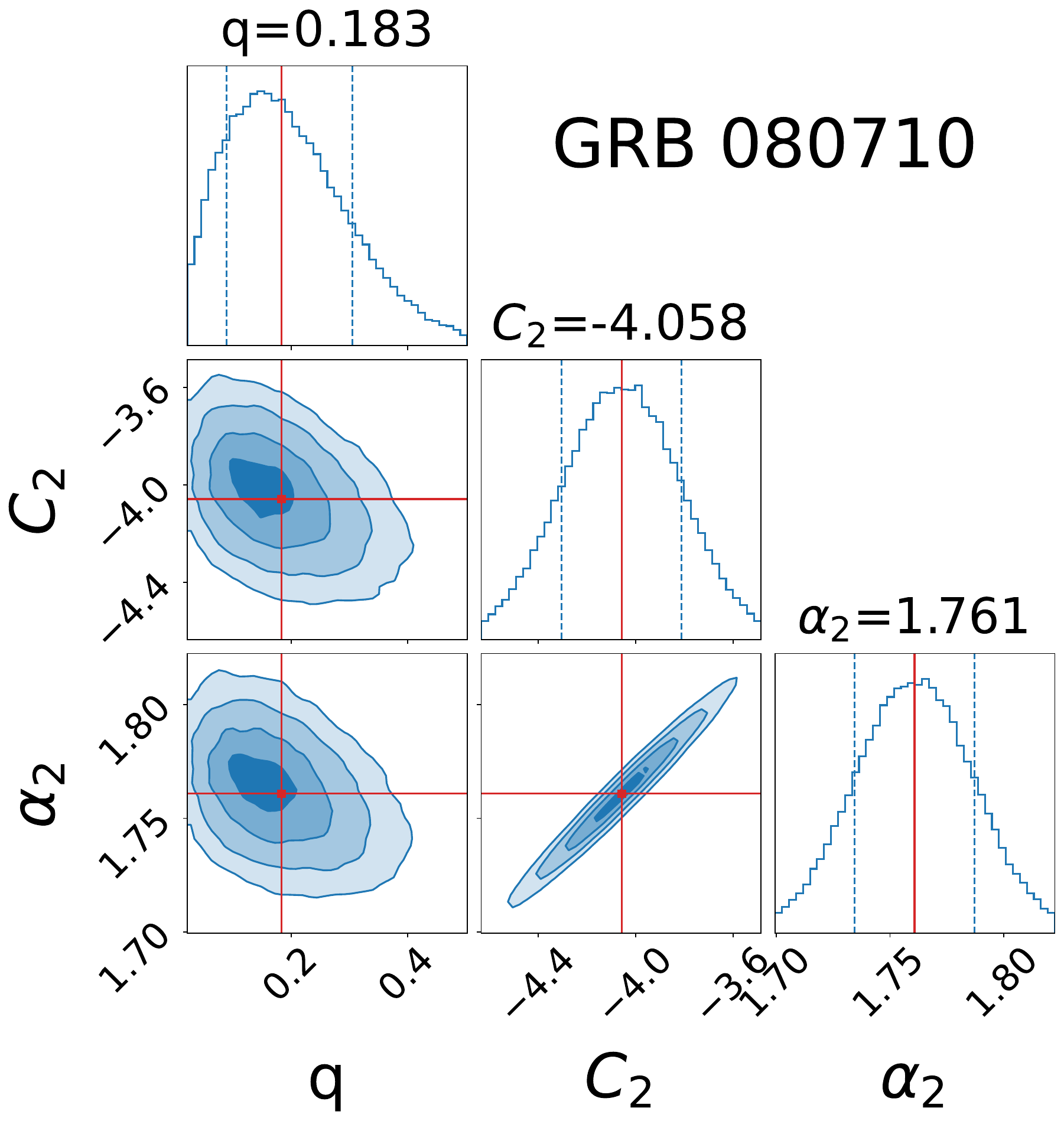}
  \end{subfigure}%
  \\%
  \begin{subfigure}[b]{0.23\textwidth}
    \includegraphics[width=\linewidth]{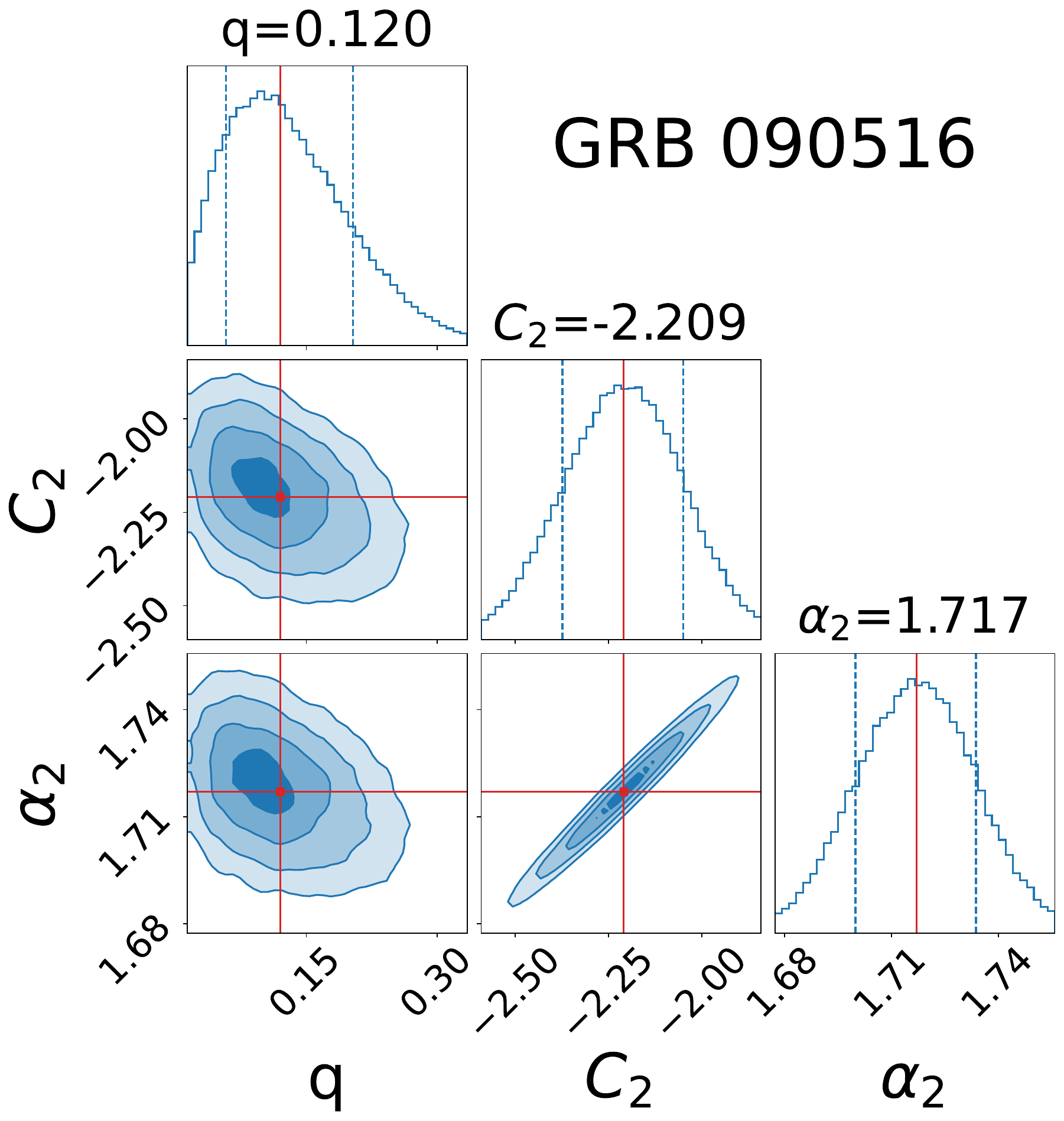}
  \end{subfigure}%
  \hspace{0.01\textwidth}
  \begin{subfigure}[b]{0.23\textwidth}
    \includegraphics[width=\linewidth]{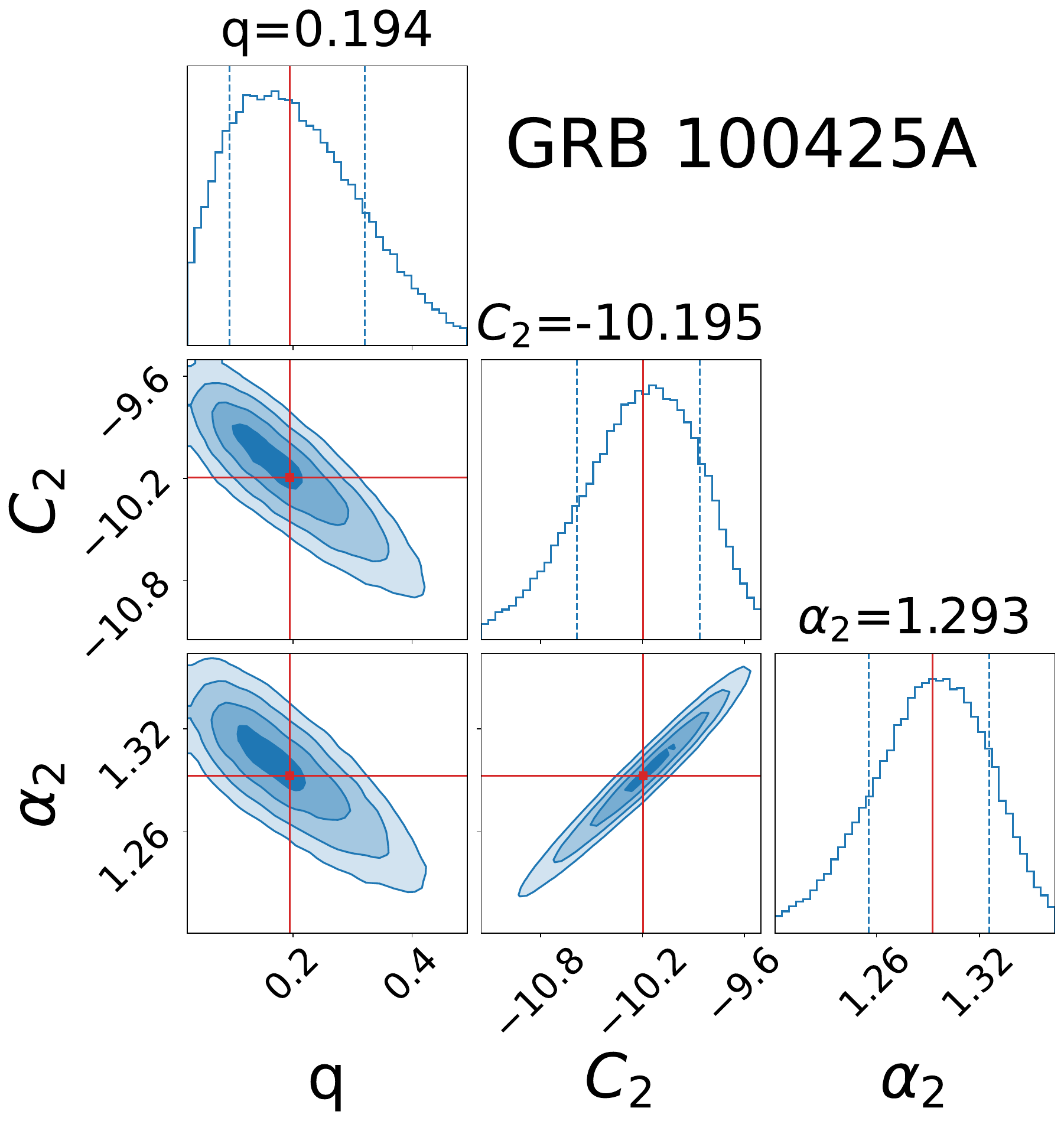}
  \end{subfigure}%
  \hspace{0.01\textwidth}
  \begin{subfigure}[b]{0.23\textwidth}
    \includegraphics[width=\linewidth]{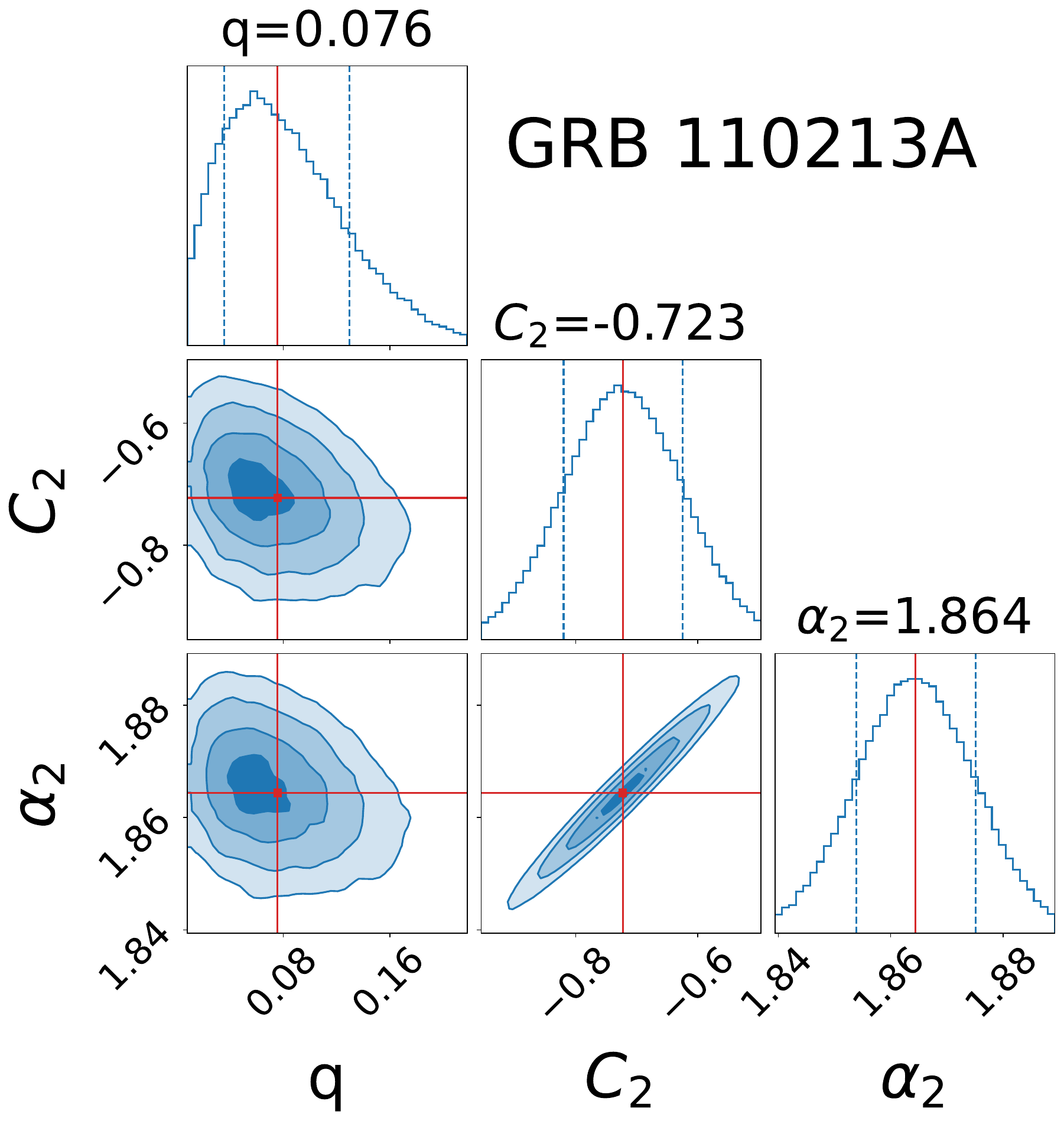}
  \end{subfigure}%
  \hspace{0.01\textwidth}
  \begin{subfigure}[b]{0.23\textwidth}
    \includegraphics[width=\linewidth]{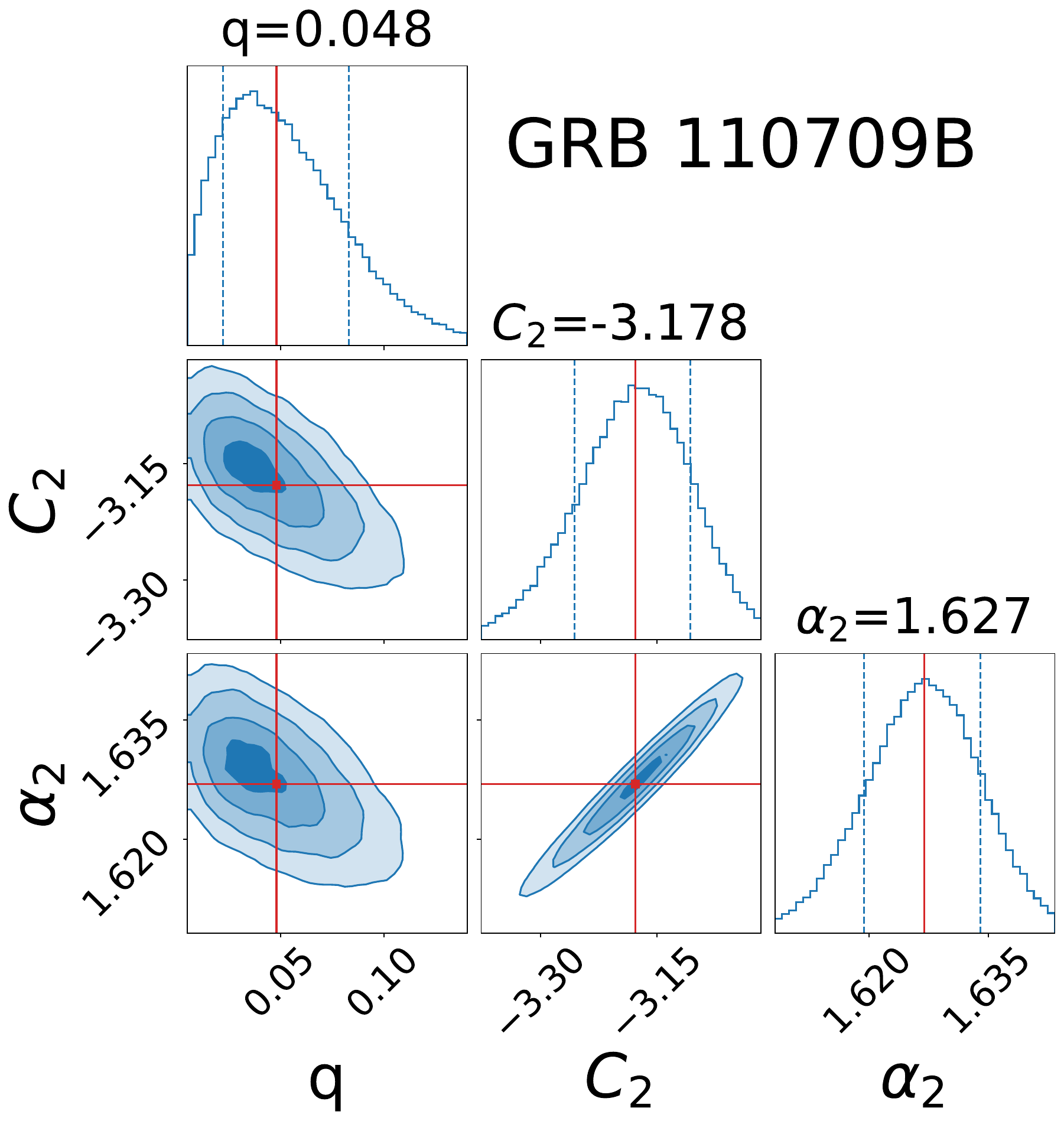}
  \end{subfigure}%
  \\%
  \begin{subfigure}[b]{0.23\textwidth}
    \includegraphics[width=\linewidth]{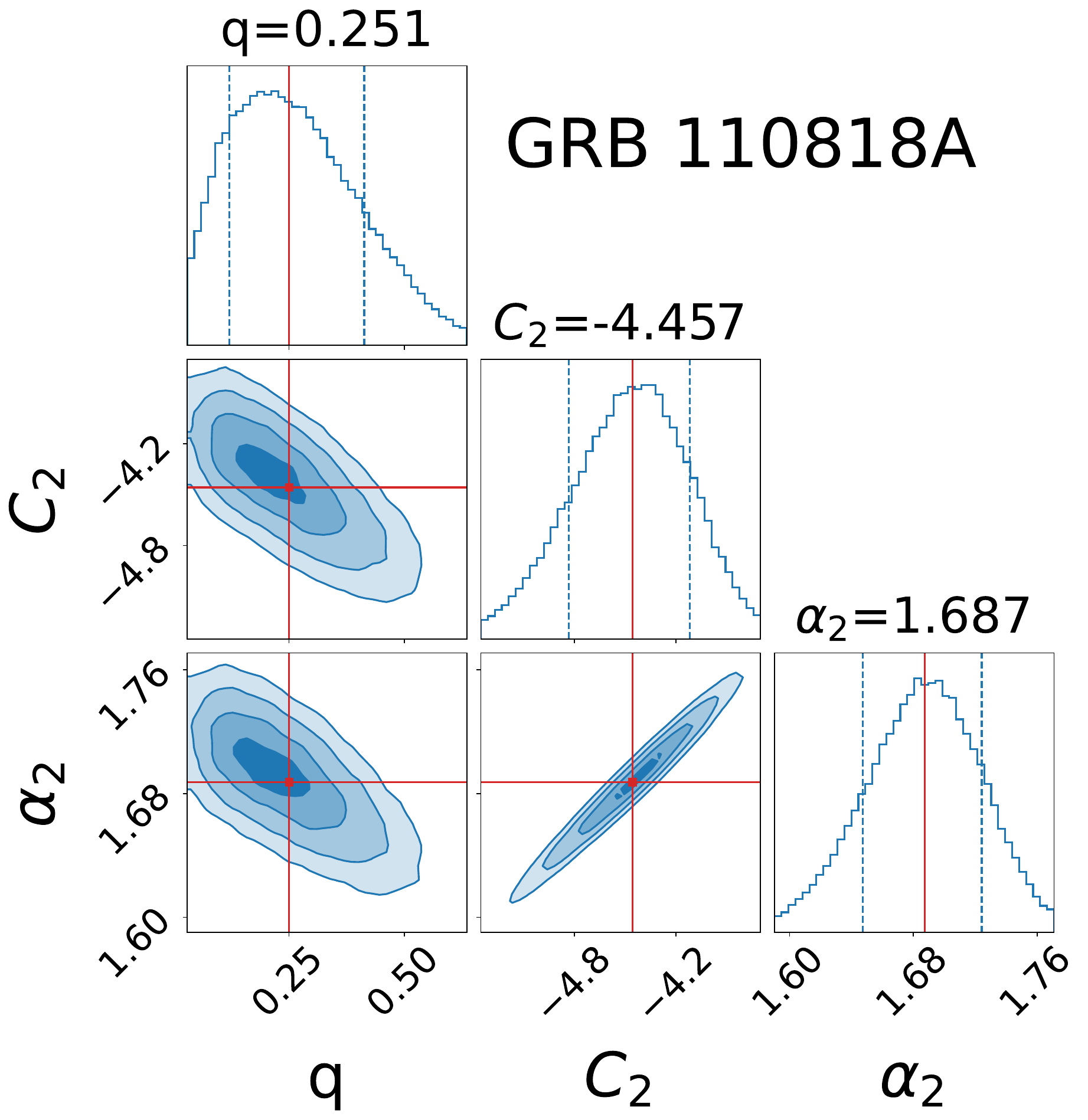}
  \end{subfigure}%
  \hspace{0.01\textwidth}
  \begin{subfigure}[b]{0.23\textwidth}
    \includegraphics[width=\linewidth]{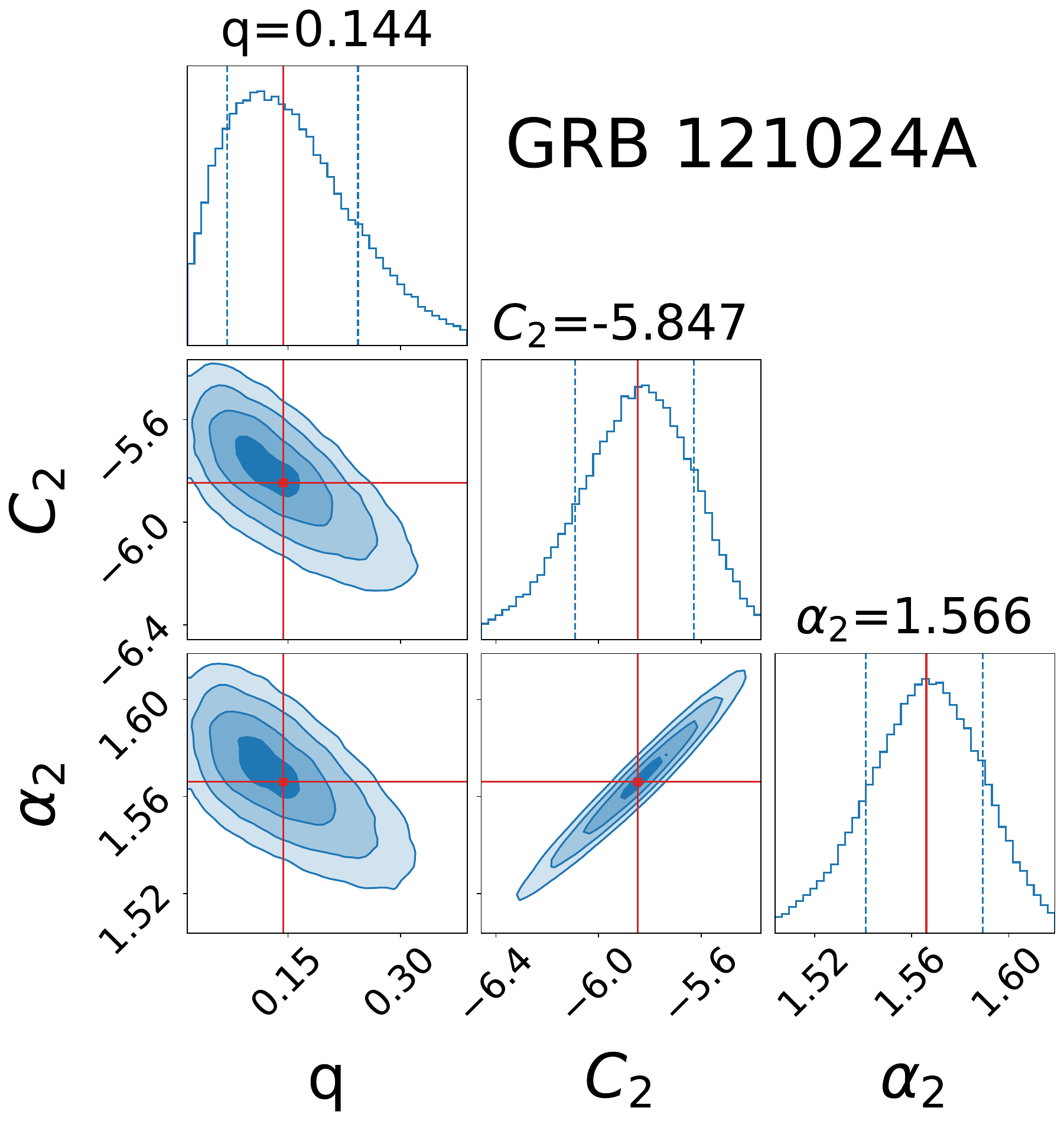}
  \end{subfigure}%
  \hspace{0.01\textwidth}
  \begin{subfigure}[b]{0.23\textwidth}
    \includegraphics[width=\linewidth]{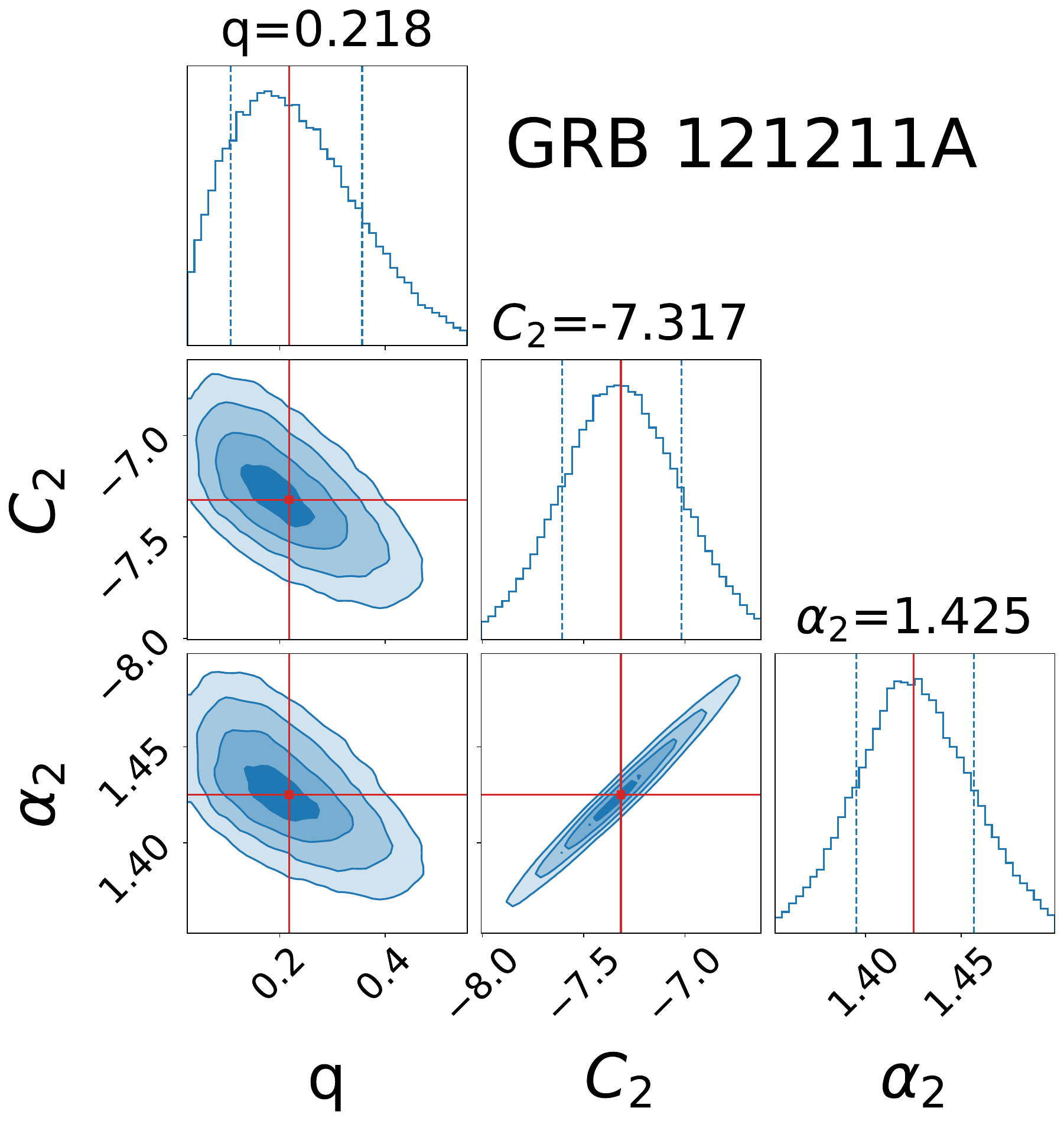}
  \end{subfigure}%
  \hspace{0.01\textwidth}
  \begin{subfigure}[b]{0.23\textwidth}
    \includegraphics[width=\linewidth]{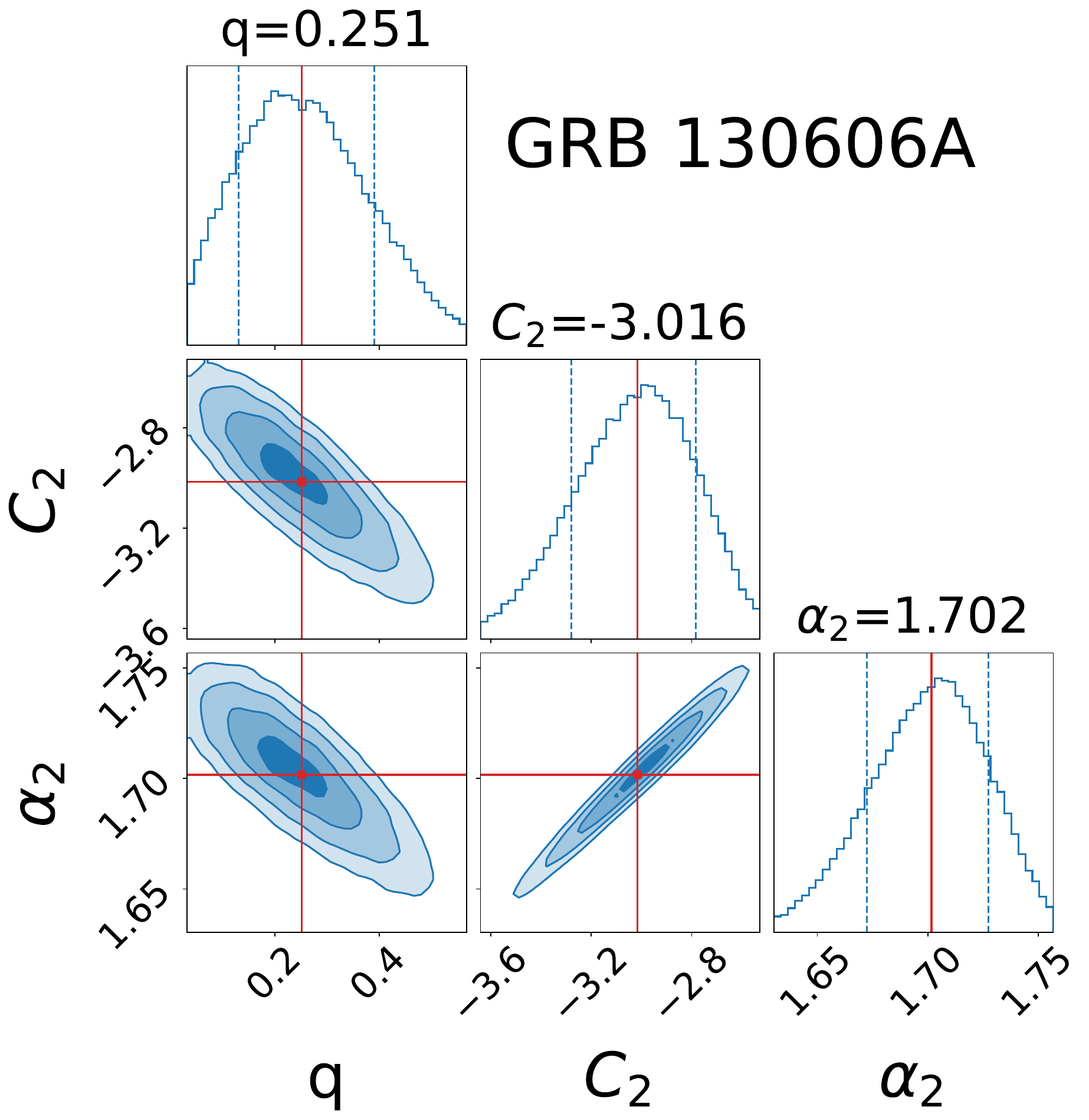}
  \end{subfigure}%
  \\%
  \begin{subfigure}[b]{0.23\textwidth}
    \includegraphics[width=\linewidth]{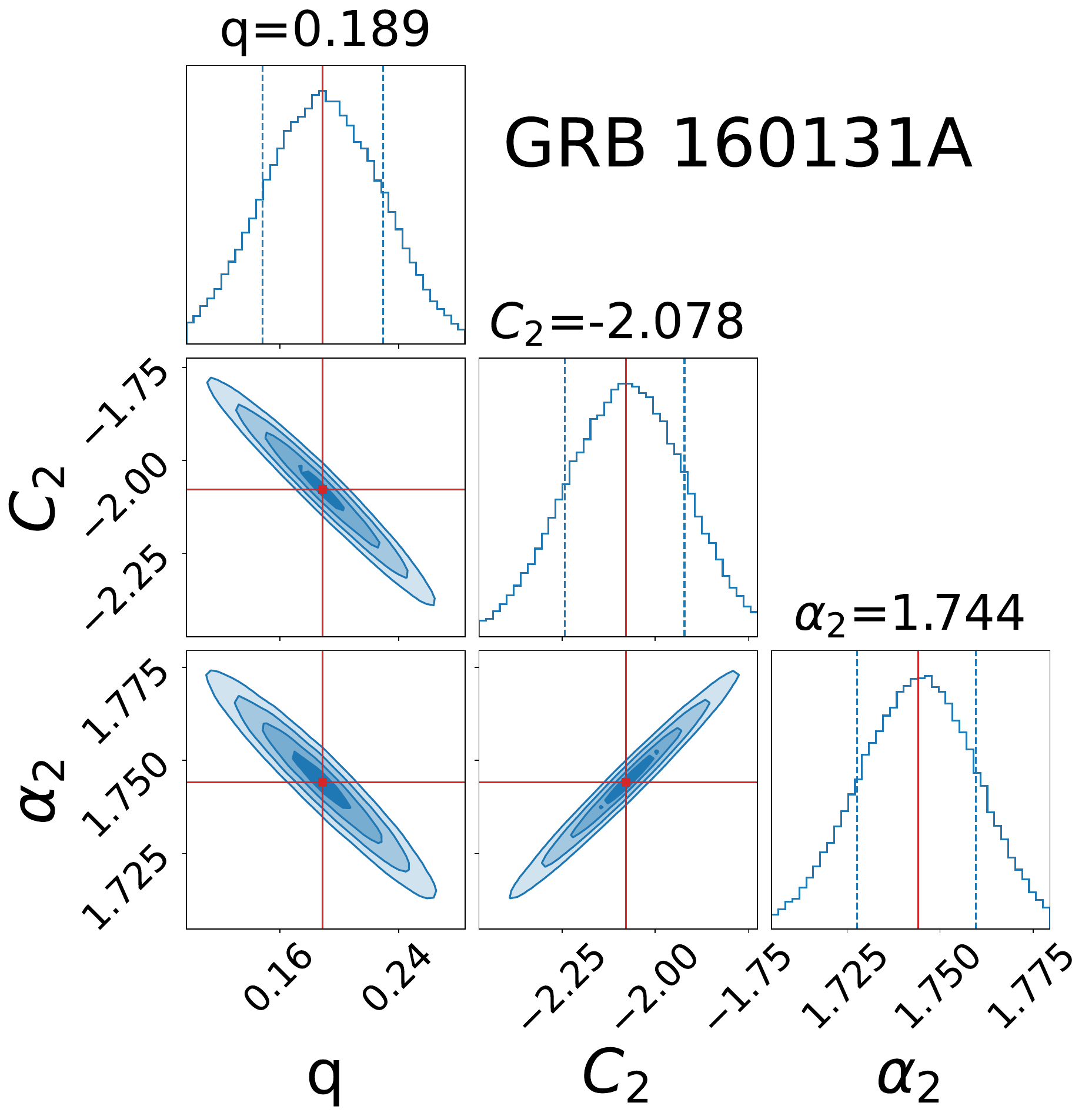}
  \end{subfigure}%
  \hspace{0.01\textwidth}
  \begin{subfigure}[b]{0.23\textwidth}
    \includegraphics[width=\linewidth]{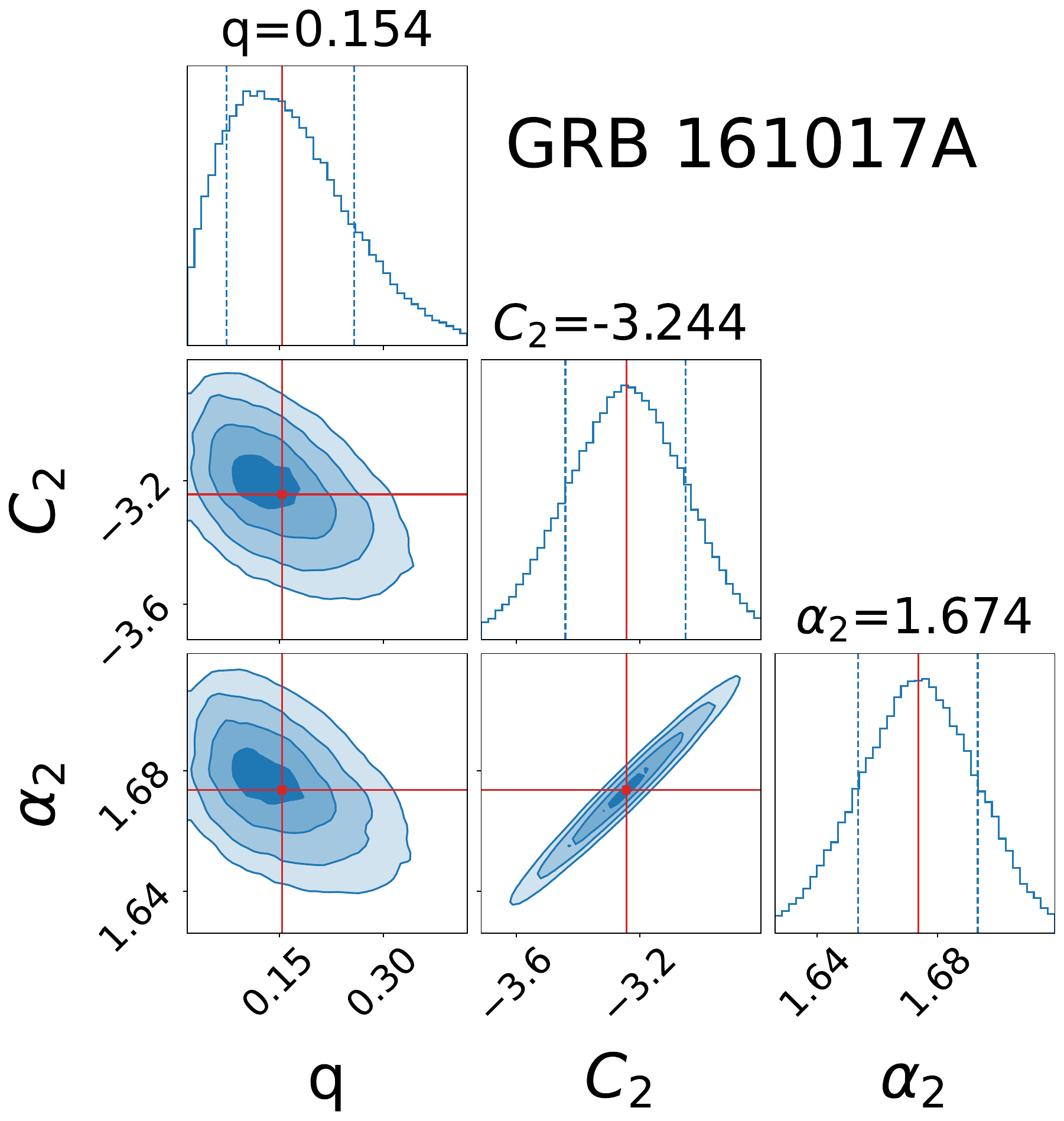}
  \end{subfigure}%
  \hspace{0.01\textwidth}
  \begin{subfigure}[b]{0.23\textwidth}
    \includegraphics[width=\linewidth]{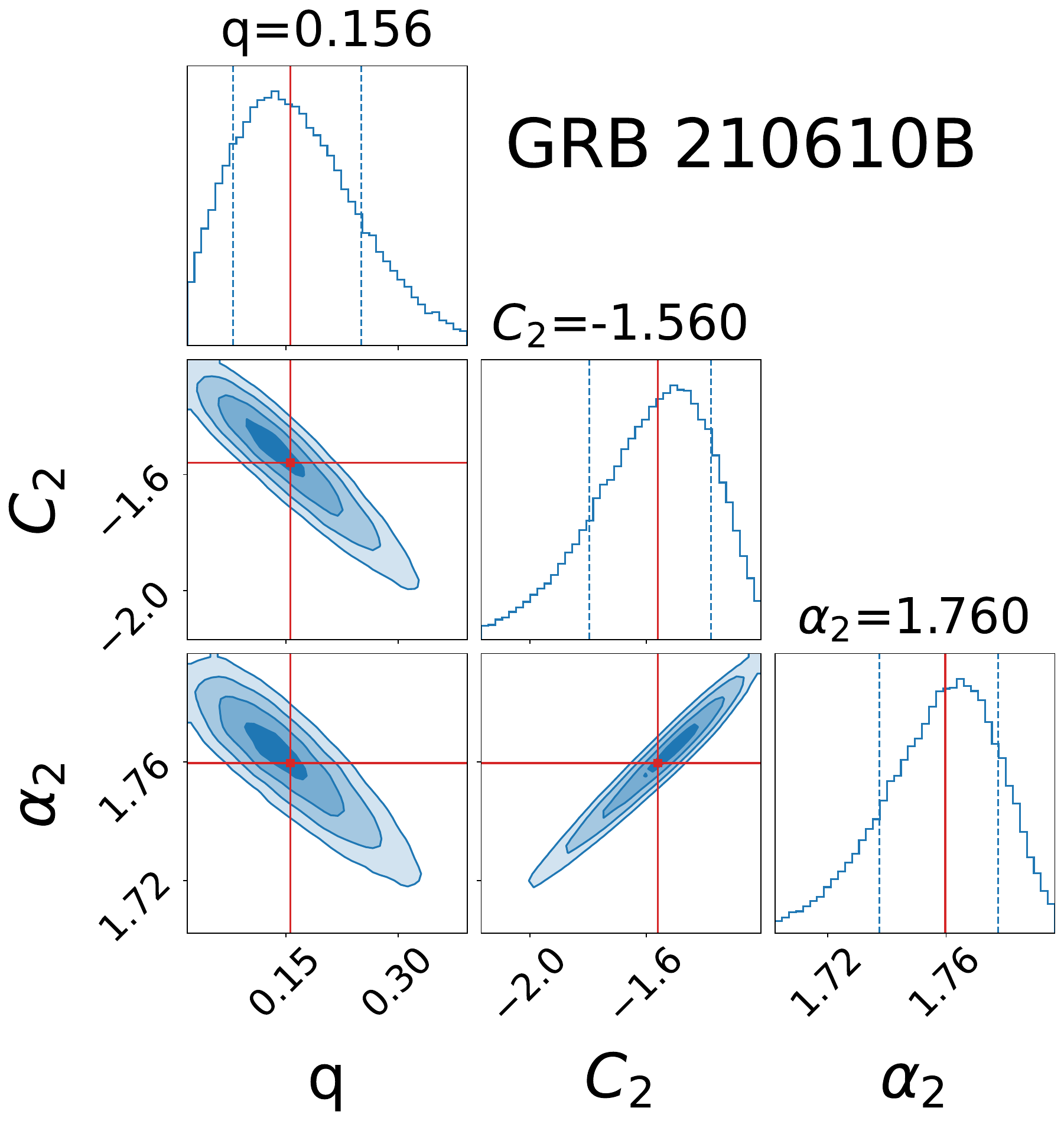}
  \end{subfigure}%
  \hspace{0.01\textwidth}
  \begin{subfigure}[b]{0.23\textwidth}
    \includegraphics[width=\linewidth]{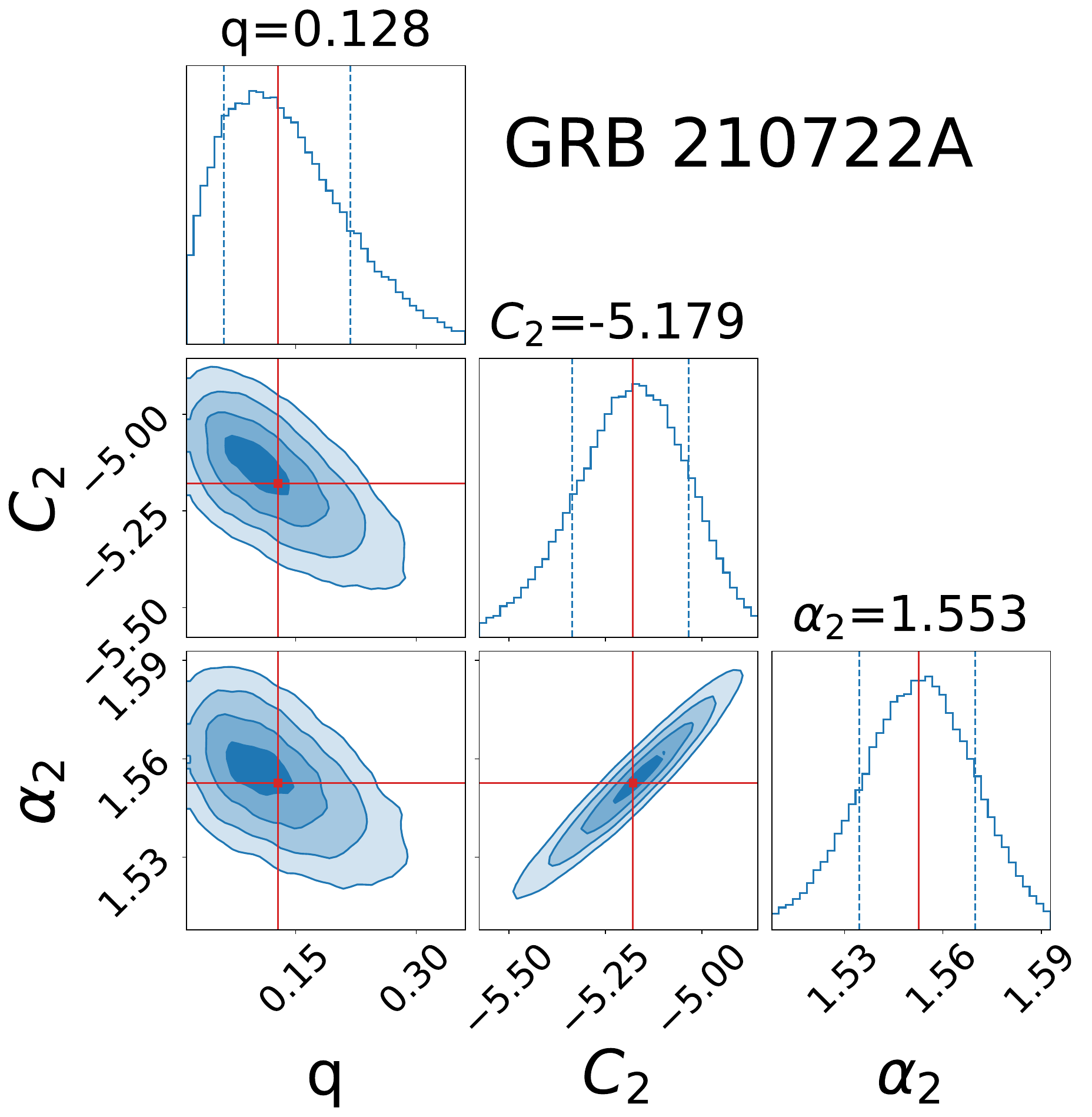}
  \end{subfigure}%
  \\%
  \caption{Posterior parameter distributions for model~2 with full consideration of high-latitude emission. Each panel displays the MCMC sampling results for key parameters including the off-axis ratio $q = \theta_{\text{obs}}/\theta_{\text{jet}}$, decay index $\alpha_2$, and normalization factor.}
  \label{fig:ISM_model2_parameter_distributions}
\end{figure}

\begin{figure}[htbp]
\centering
  \begin{subfigure}[b]{0.23\textwidth}
    \includegraphics[width=\linewidth]{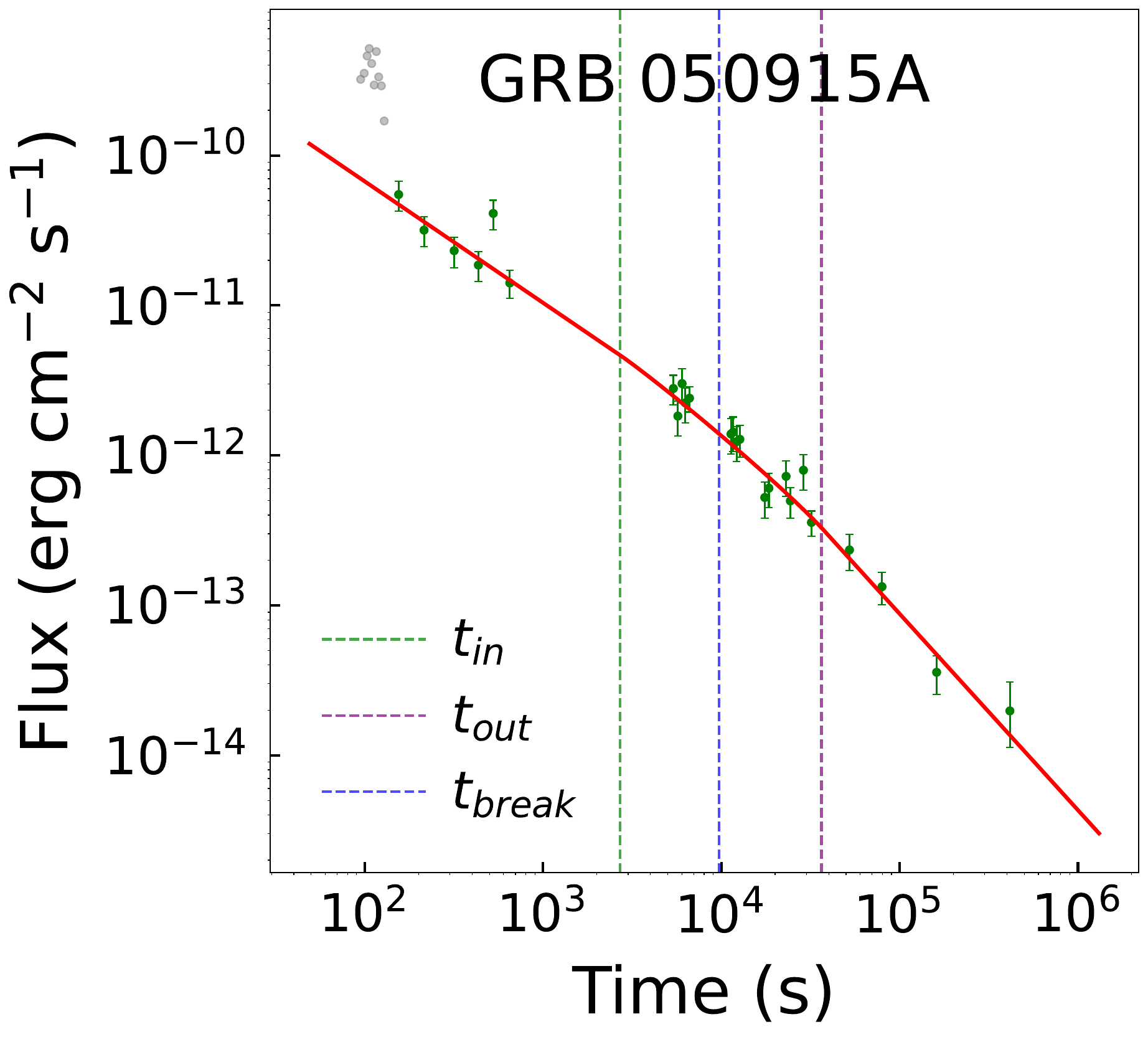}
  \end{subfigure}%
  \hspace{0.01\textwidth}
  \begin{subfigure}[b]{0.23\textwidth}
    \includegraphics[width=\linewidth]{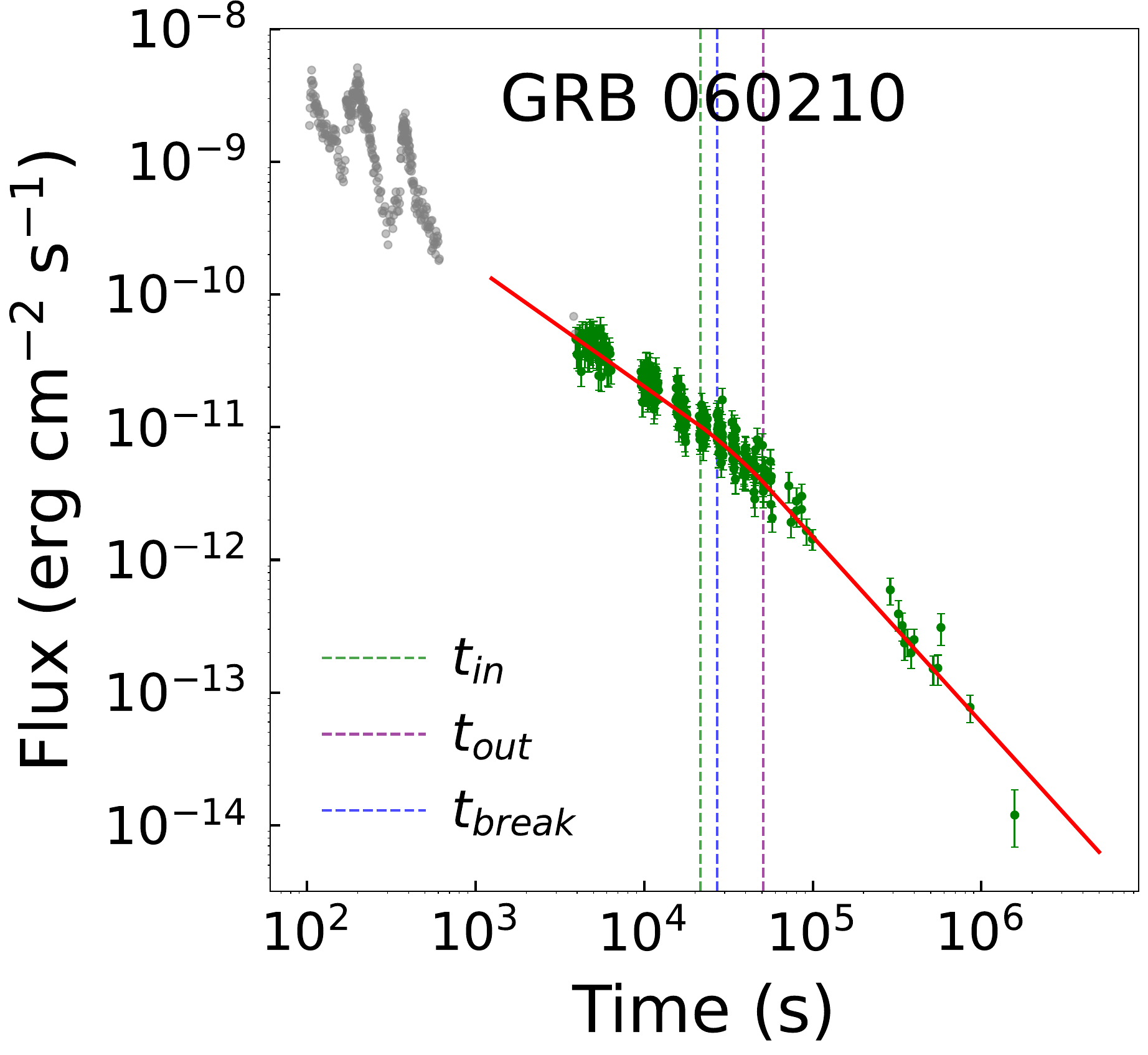}
  \end{subfigure}%
  \hspace{0.01\textwidth}
  \begin{subfigure}[b]{0.23\textwidth}
    \includegraphics[width=\linewidth]{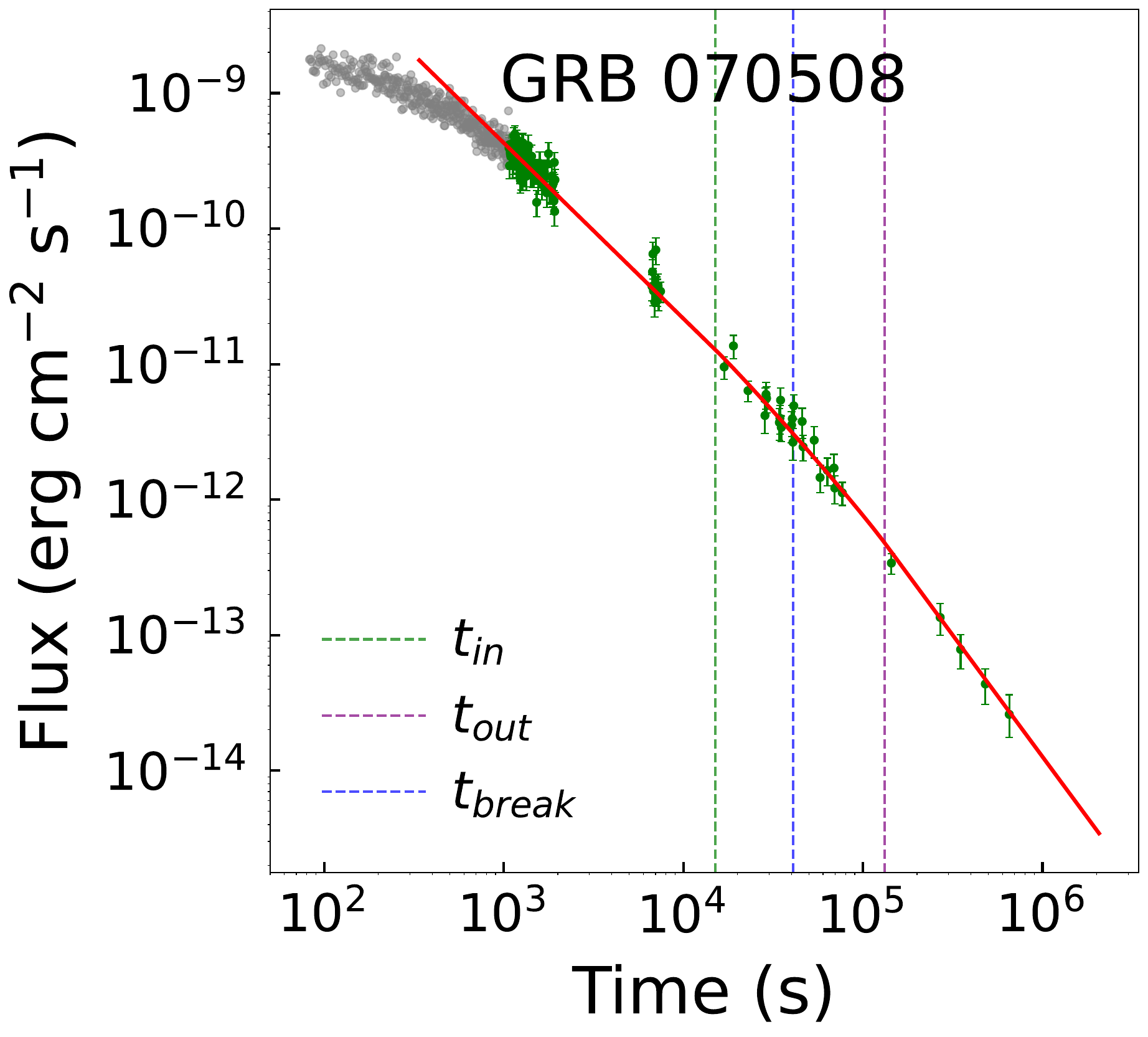}
  \end{subfigure}%
  \hspace{0.01\textwidth}
  \begin{subfigure}[b]{0.23\textwidth}
    \includegraphics[width=\linewidth]{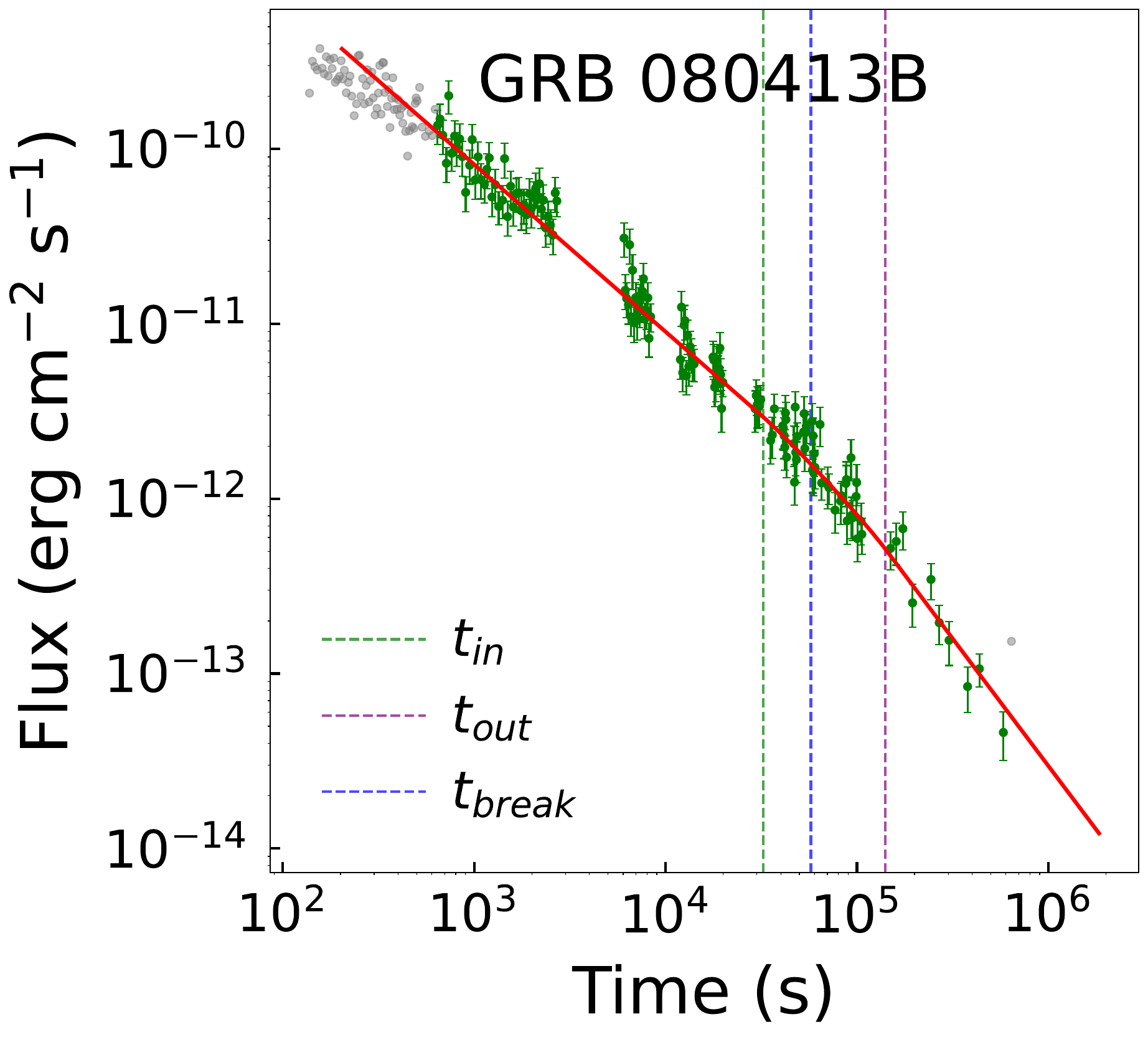}
  \end{subfigure}%
  \\%
  \begin{subfigure}[b]{0.23\textwidth}
    \includegraphics[width=\linewidth]{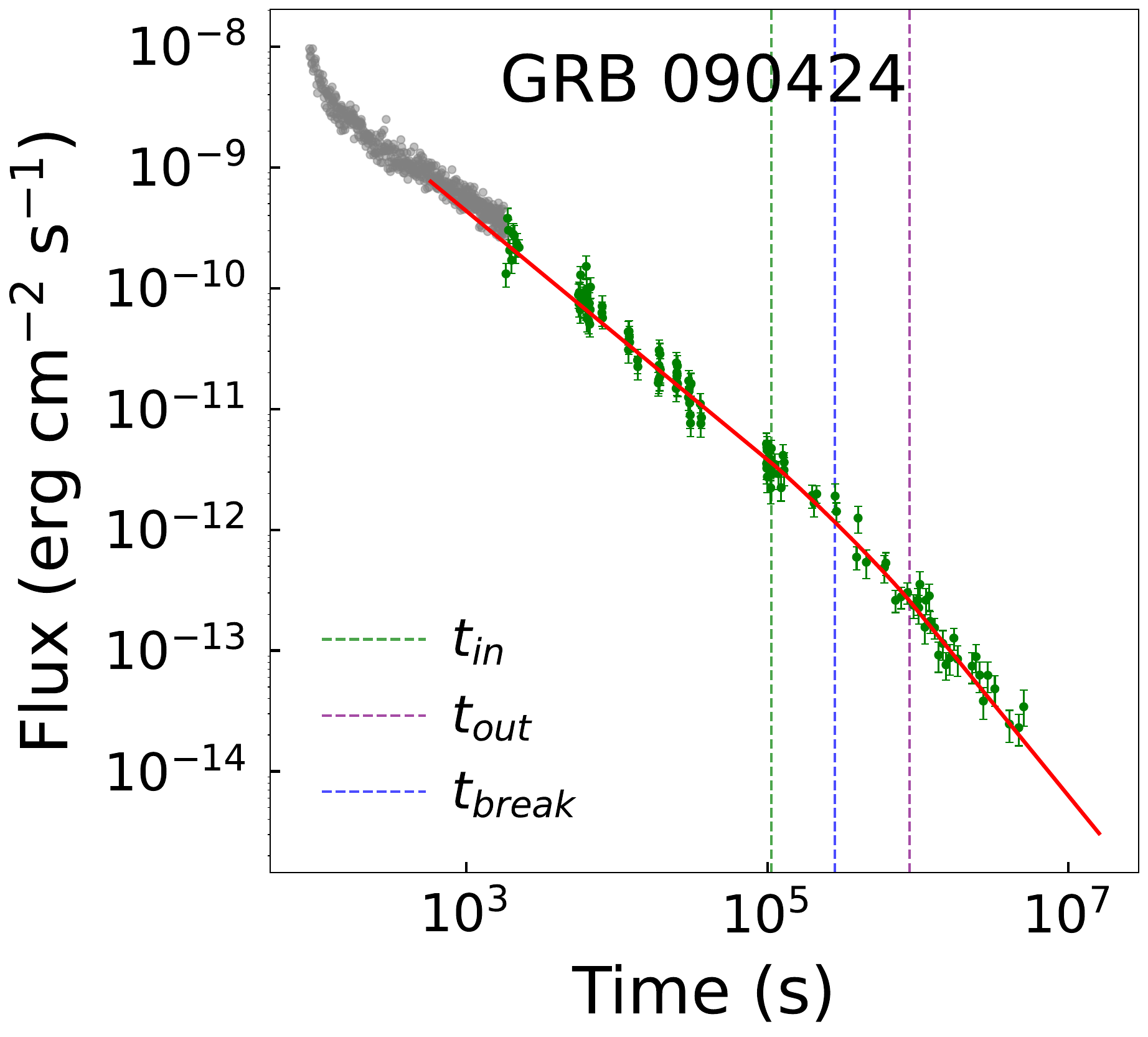}
  \end{subfigure}%
  \hspace{0.01\textwidth}
  \begin{subfigure}[b]{0.23\textwidth}
    \includegraphics[width=\linewidth]{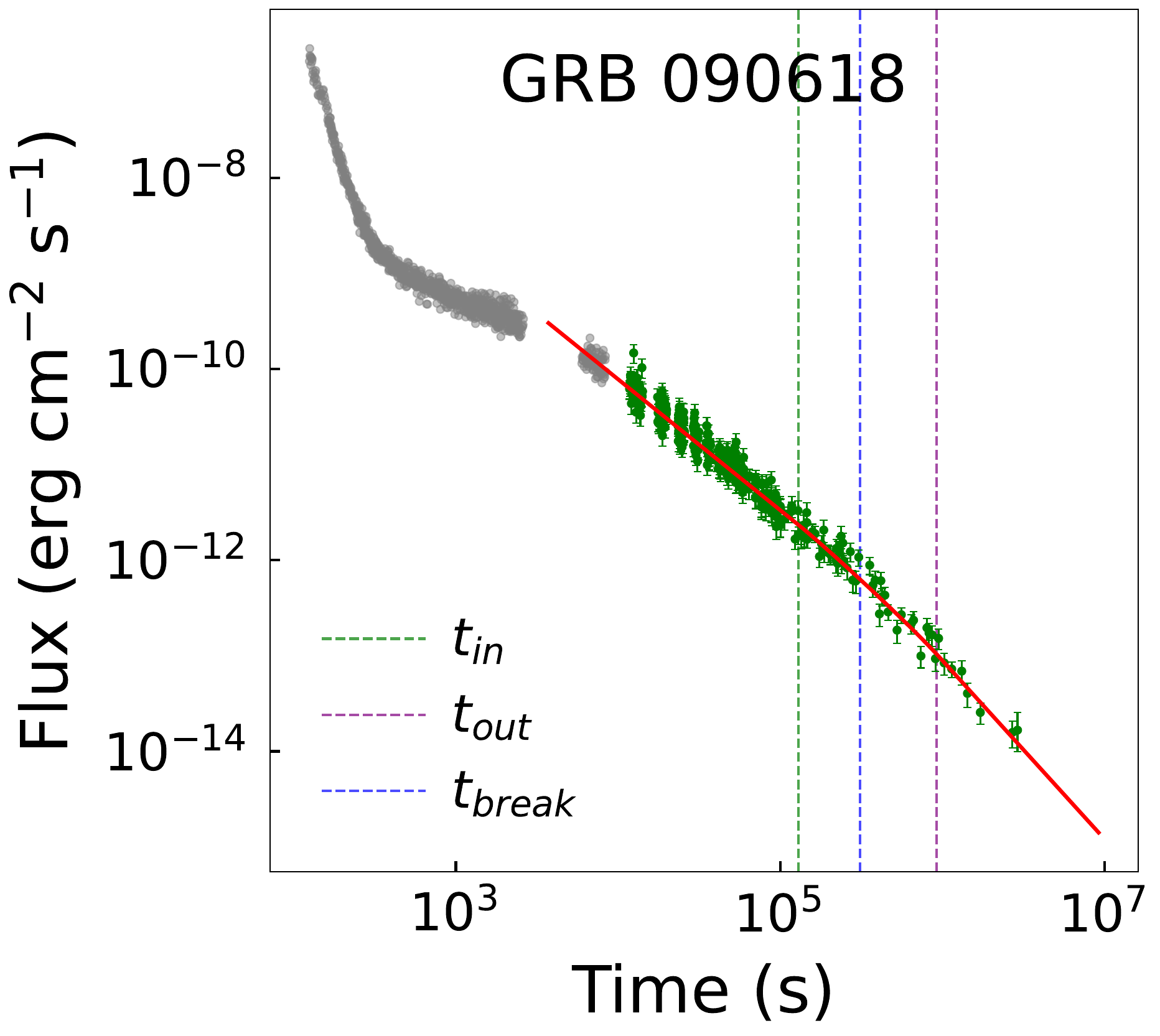}
  \end{subfigure}%
  \hspace{0.01\textwidth}
  \begin{subfigure}[b]{0.23\textwidth}
    \includegraphics[width=\linewidth]{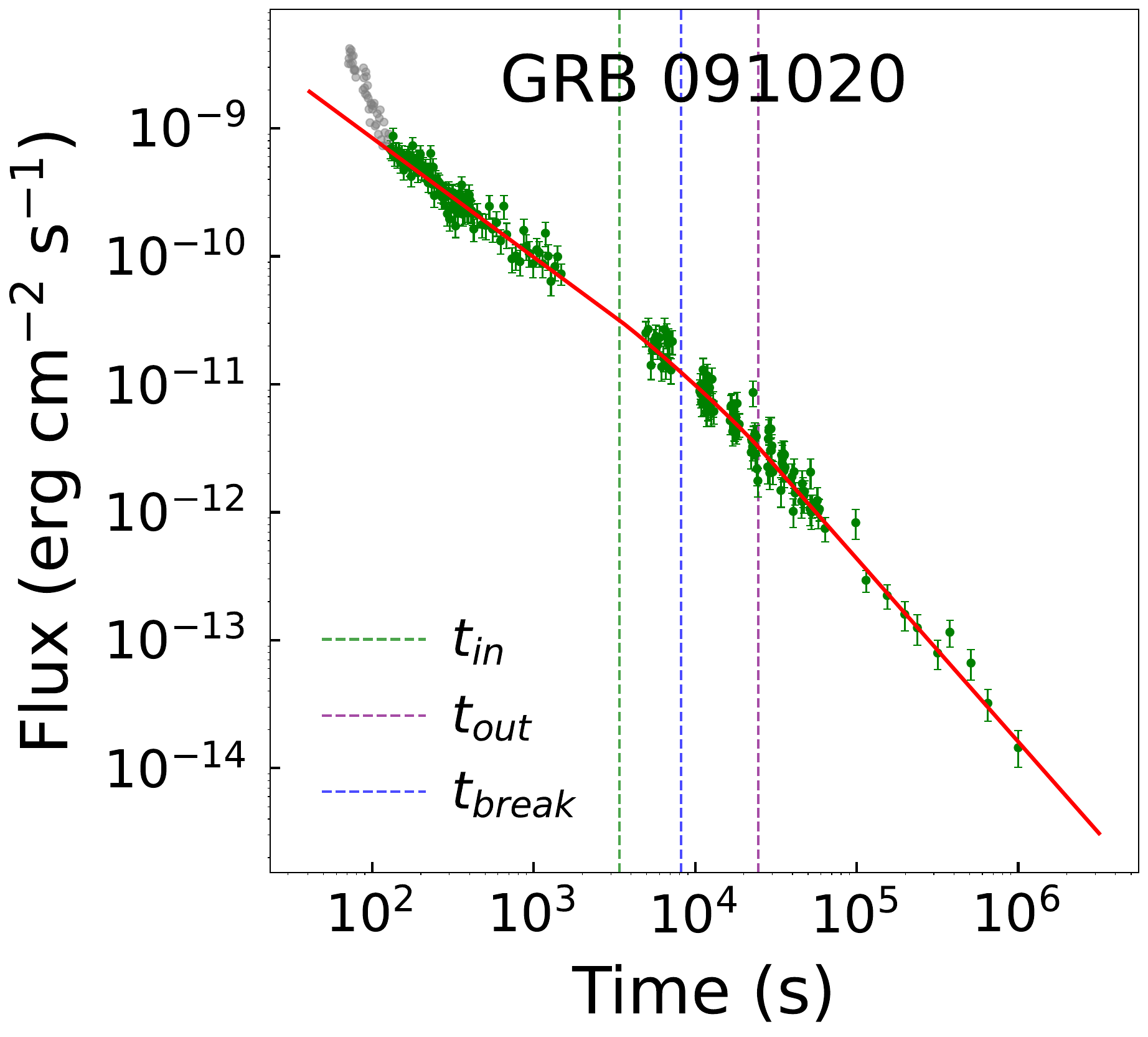}
  \end{subfigure}%
  \hspace{0.01\textwidth}
  \begin{subfigure}[b]{0.23\textwidth}
    \includegraphics[width=\linewidth]{result_091029_10_flux_vs_time.pdf}
  \end{subfigure}%
  \\%
  \begin{subfigure}[b]{0.23\textwidth}
    \includegraphics[width=\linewidth]{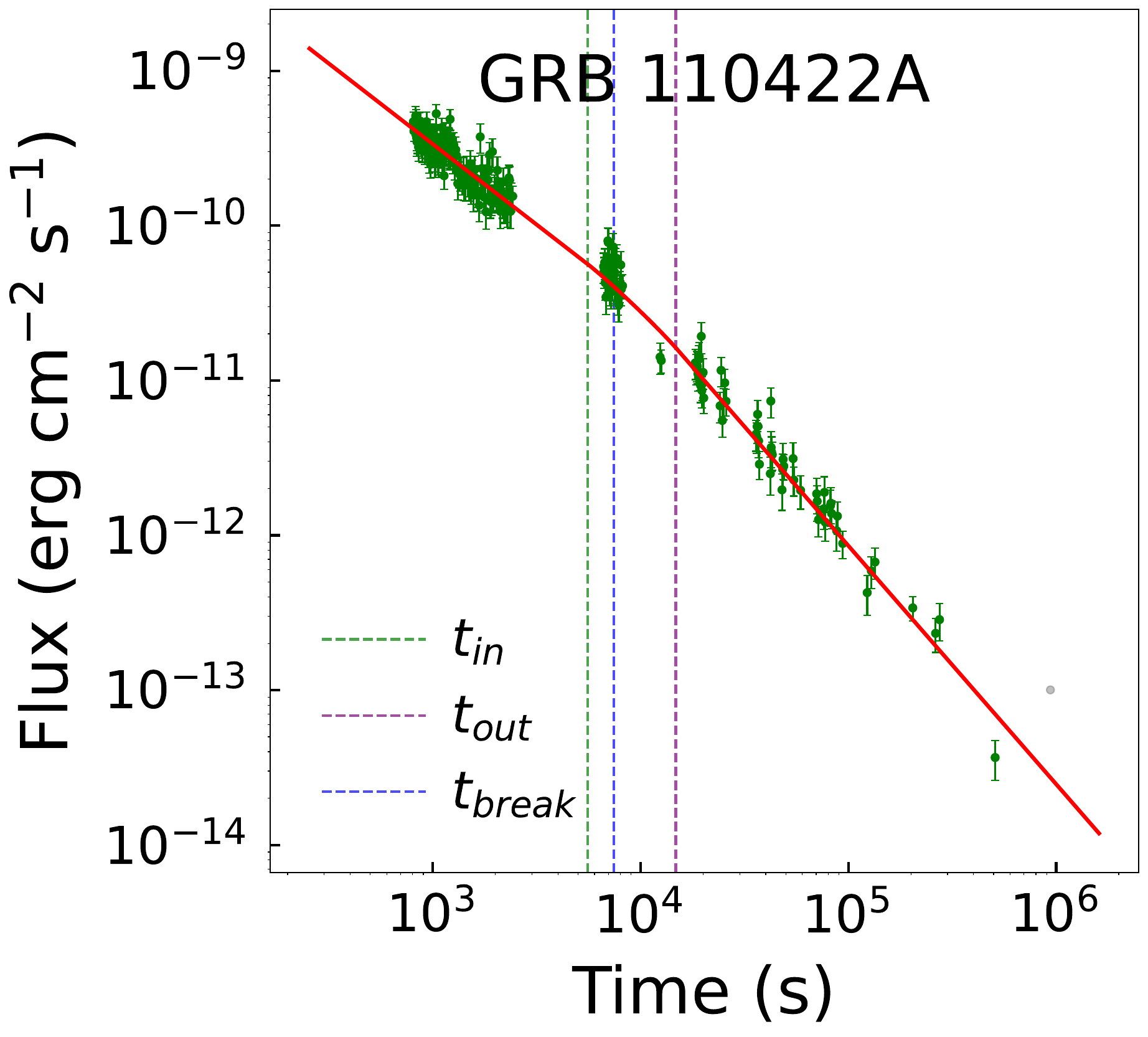}
  \end{subfigure}%
  \hspace{0.01\textwidth}
  \begin{subfigure}[b]{0.23\textwidth}
    \includegraphics[width=\linewidth]{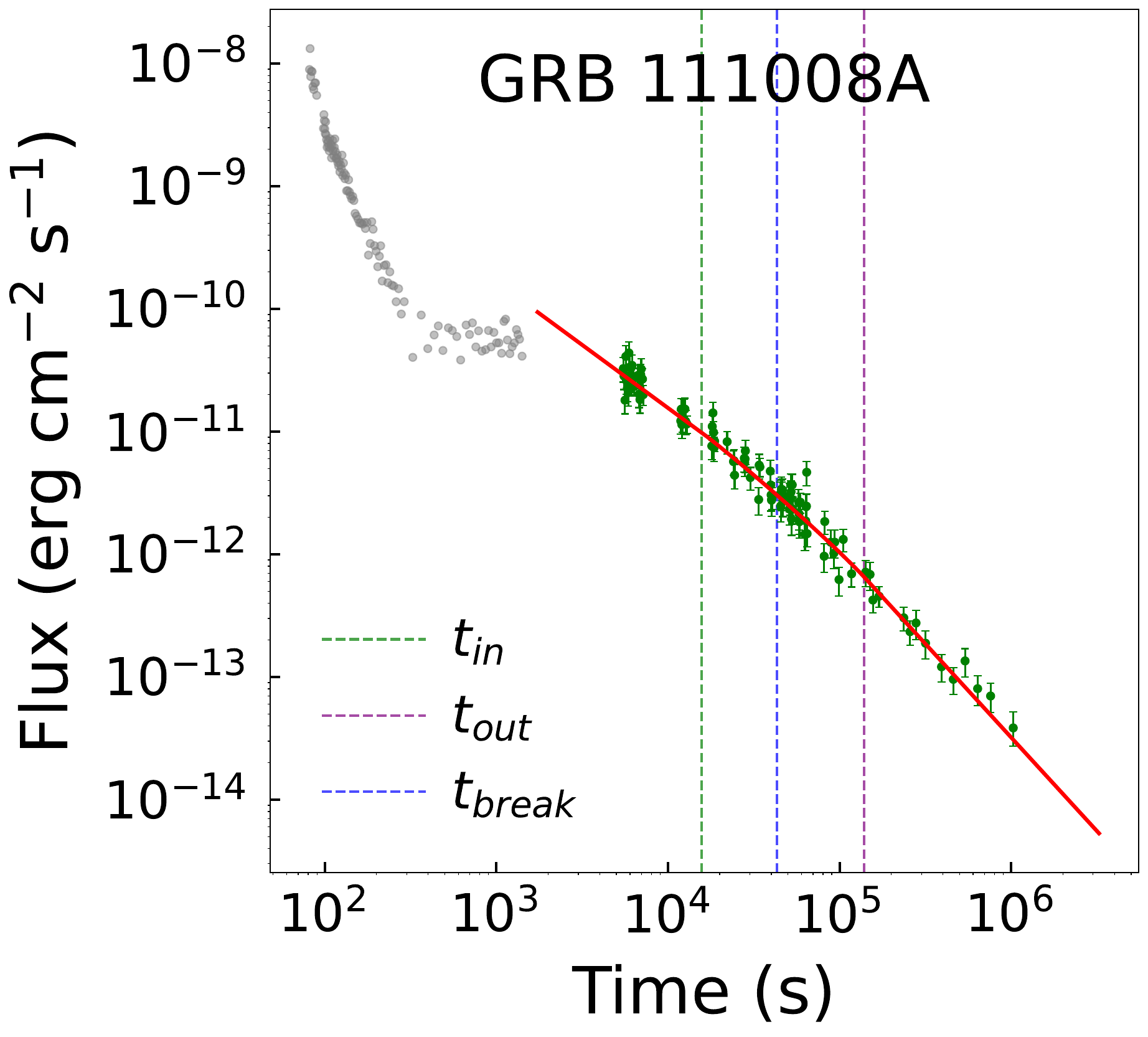}
  \end{subfigure}%
  \hspace{0.01\textwidth}
  \begin{subfigure}[b]{0.23\textwidth}
    \includegraphics[width=\linewidth]{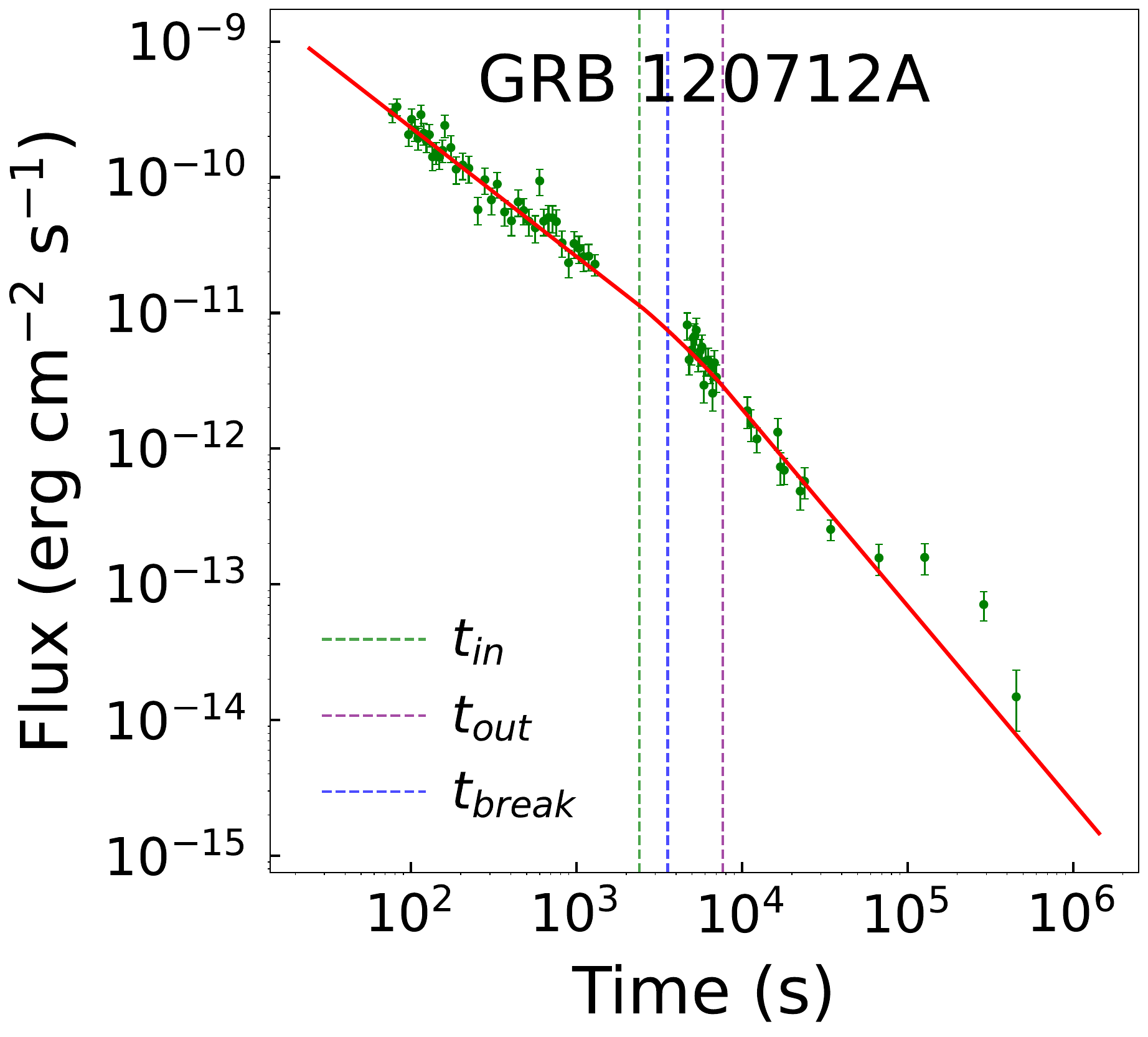}
  \end{subfigure}%
  \hspace{0.01\textwidth}
  \begin{subfigure}[b]{0.23\textwidth}
    \includegraphics[width=\linewidth]{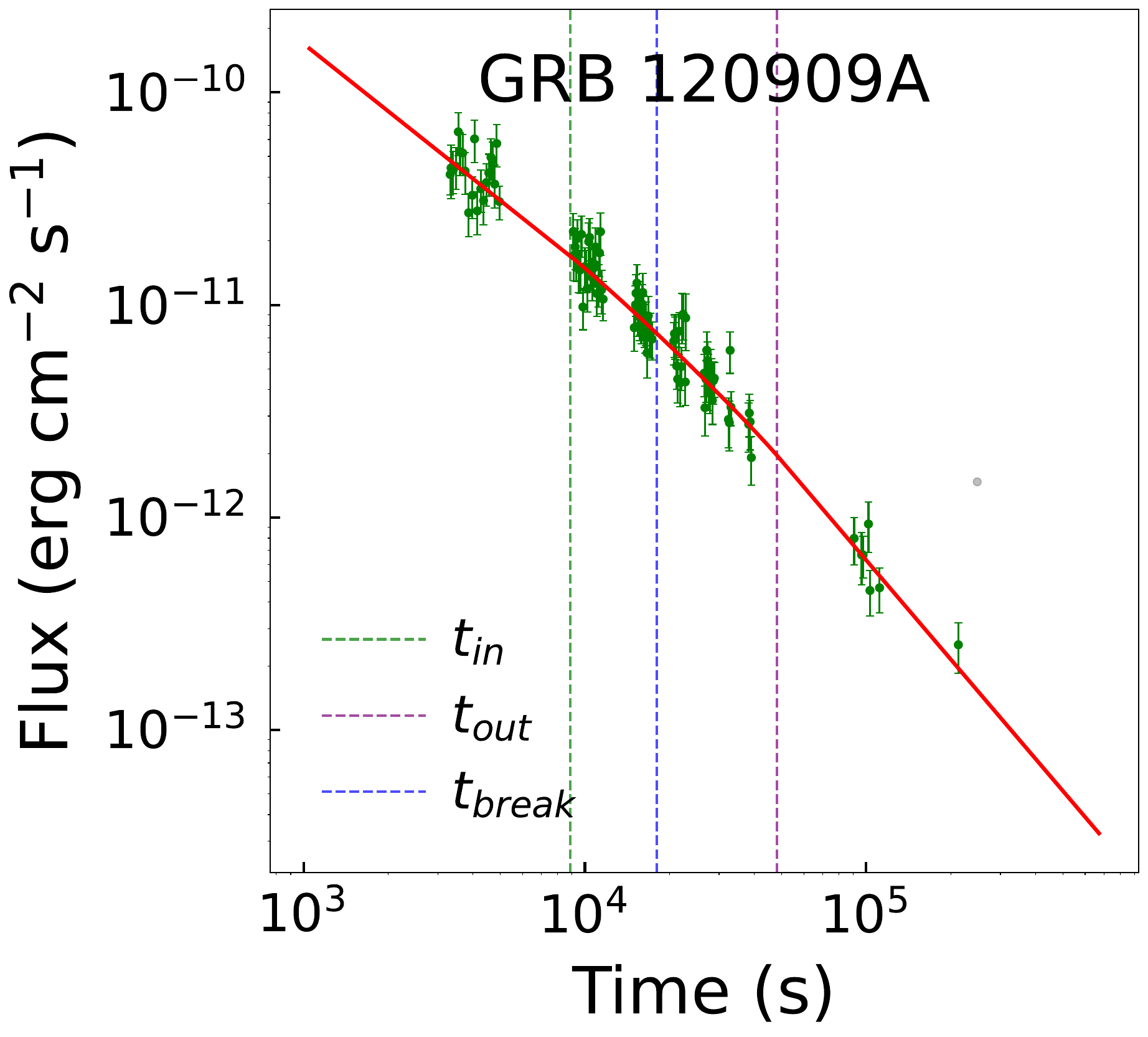}
  \end{subfigure}%
  \\%
  \begin{subfigure}[b]{0.23\textwidth}
    \includegraphics[width=\linewidth]{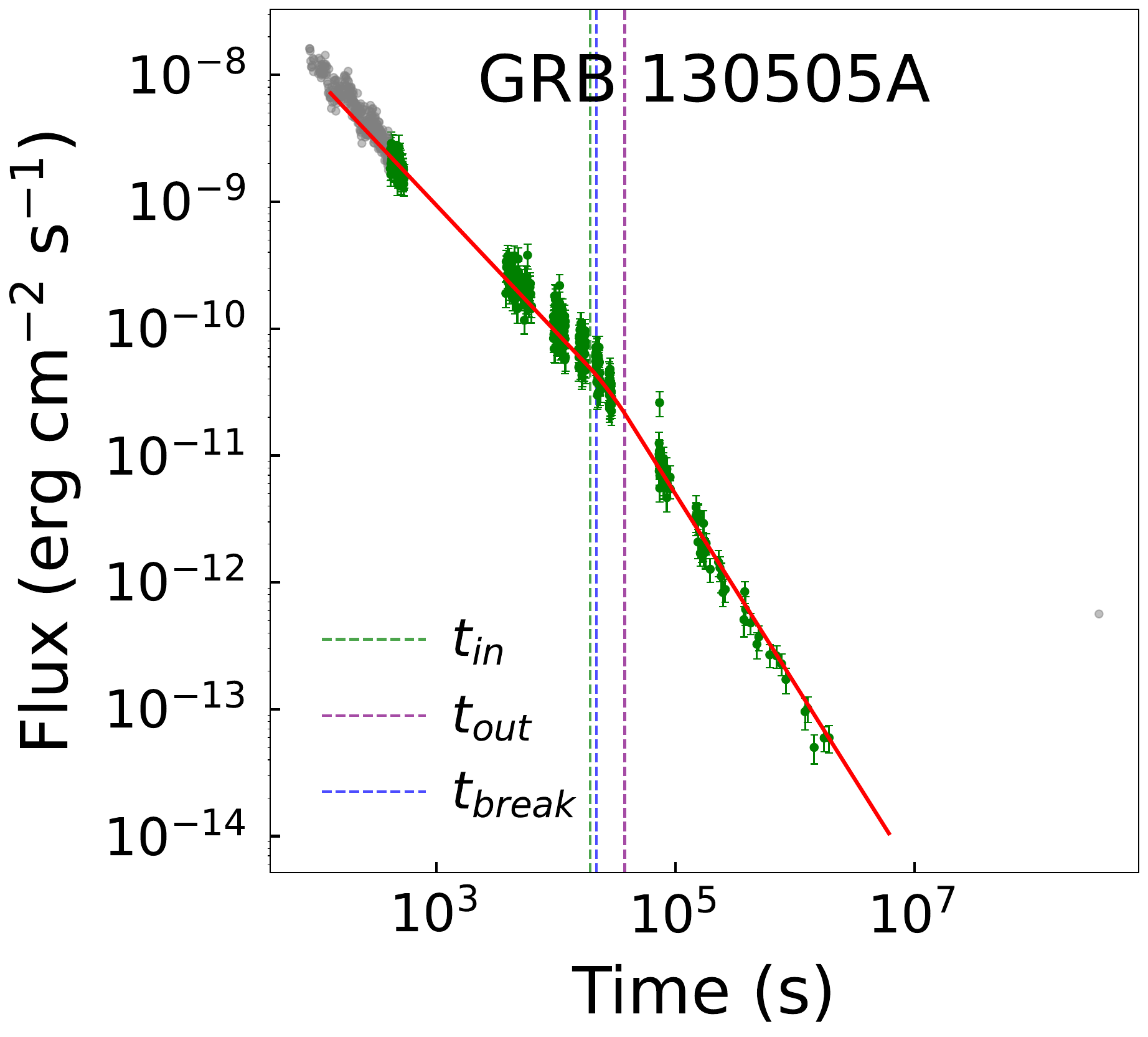}
  \end{subfigure}%
  \hspace{0.01\textwidth}
  \begin{subfigure}[b]{0.23\textwidth}
    \includegraphics[width=\linewidth]{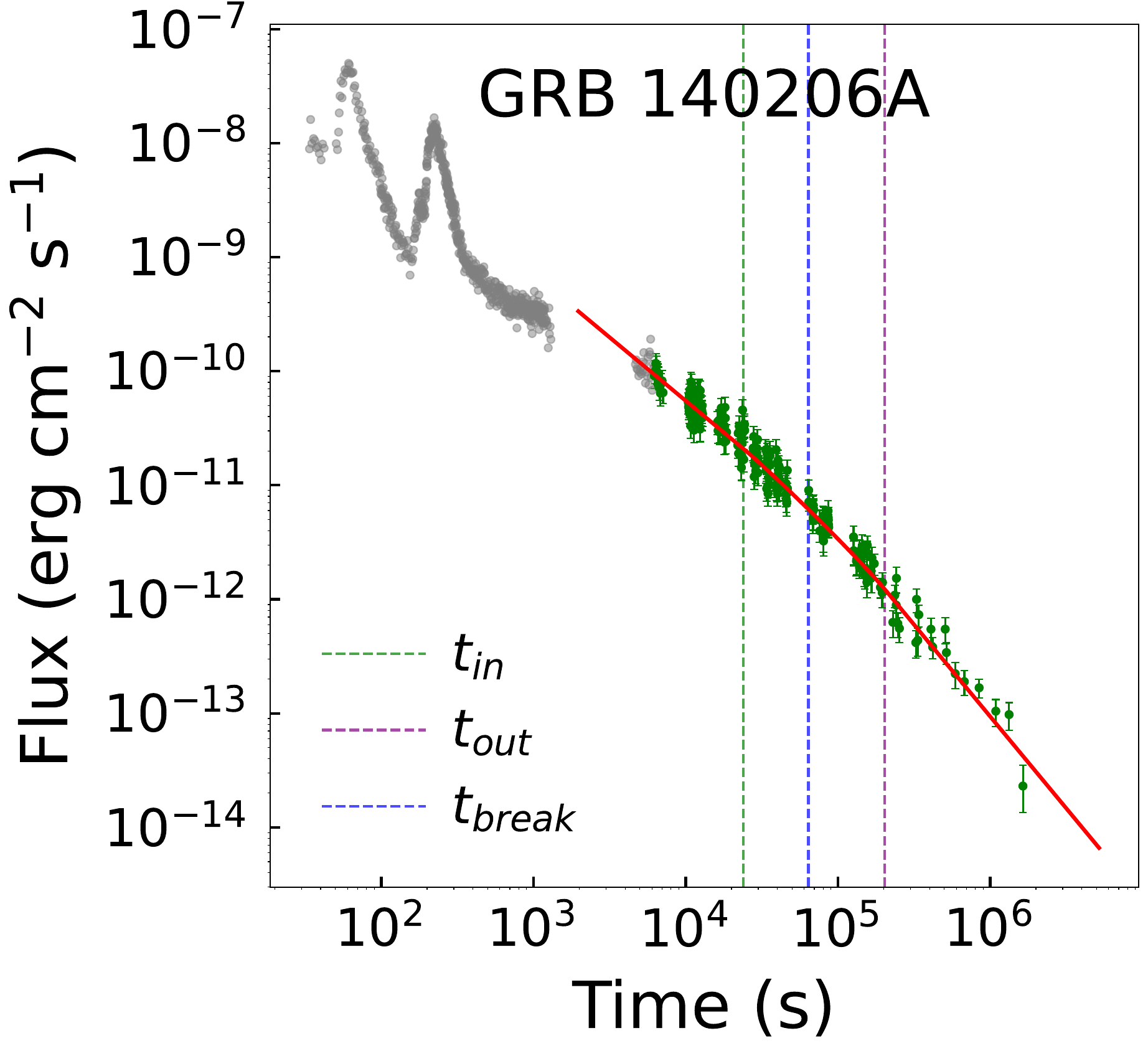}
  \end{subfigure}%
  \hspace{0.01\textwidth}
  \begin{subfigure}[b]{0.23\textwidth}
    \includegraphics[width=\linewidth]{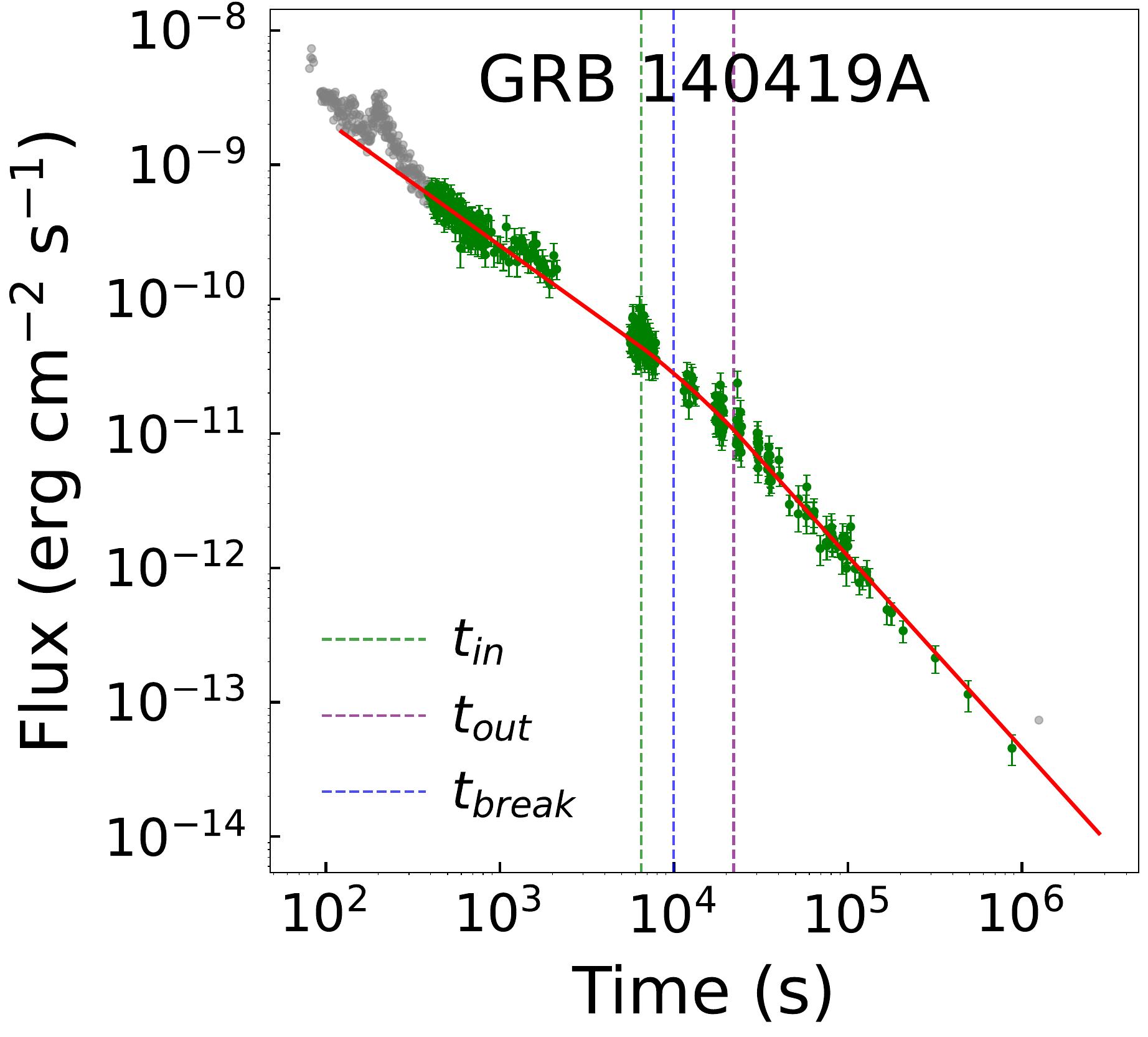}
  \end{subfigure}%
  \hspace{0.01\textwidth}
  \begin{subfigure}[b]{0.23\textwidth}
    \includegraphics[width=\linewidth]{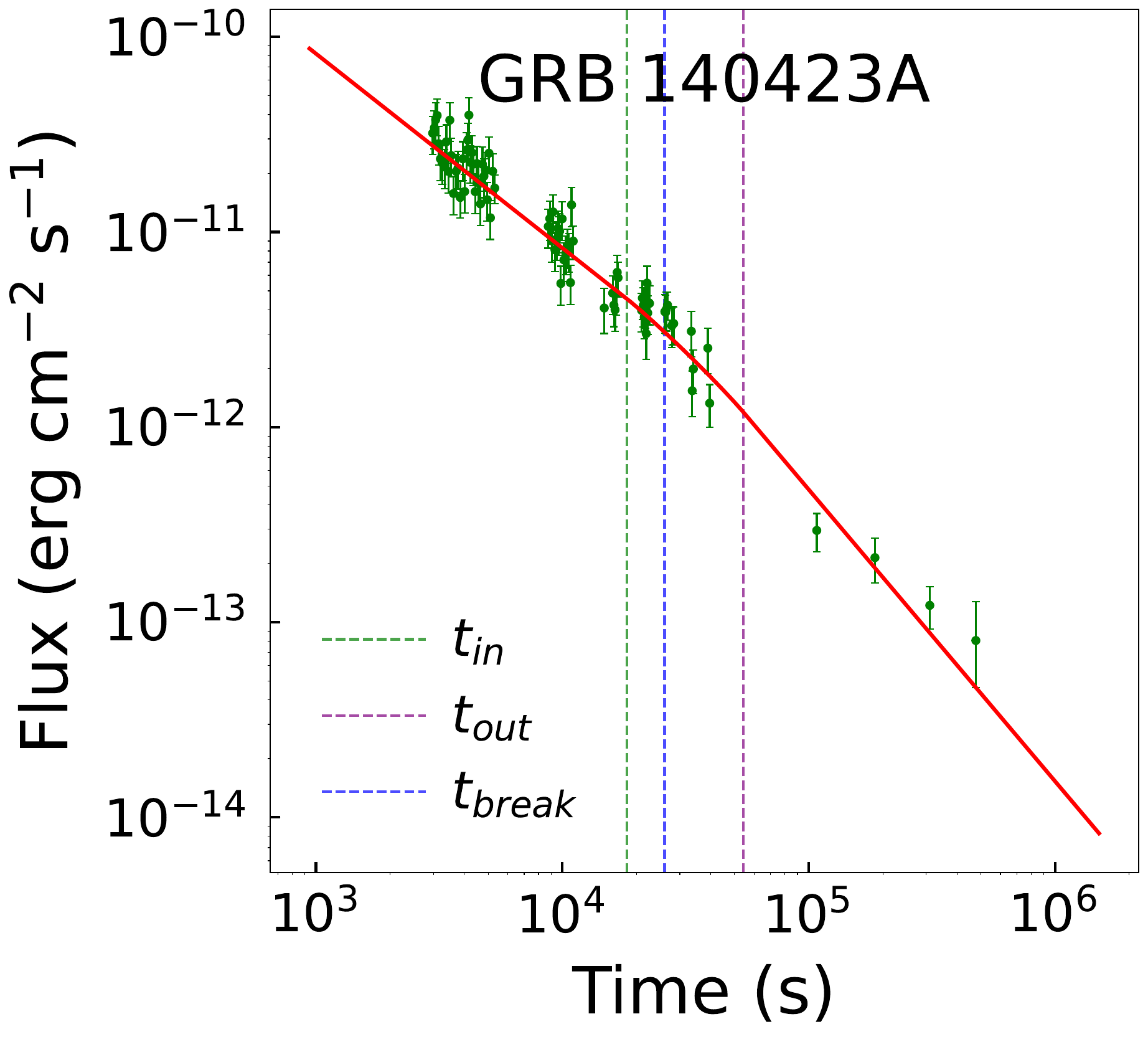}
  \end{subfigure}%
  \\%
  \begin{subfigure}[b]{0.23\textwidth}
    \includegraphics[width=\linewidth]{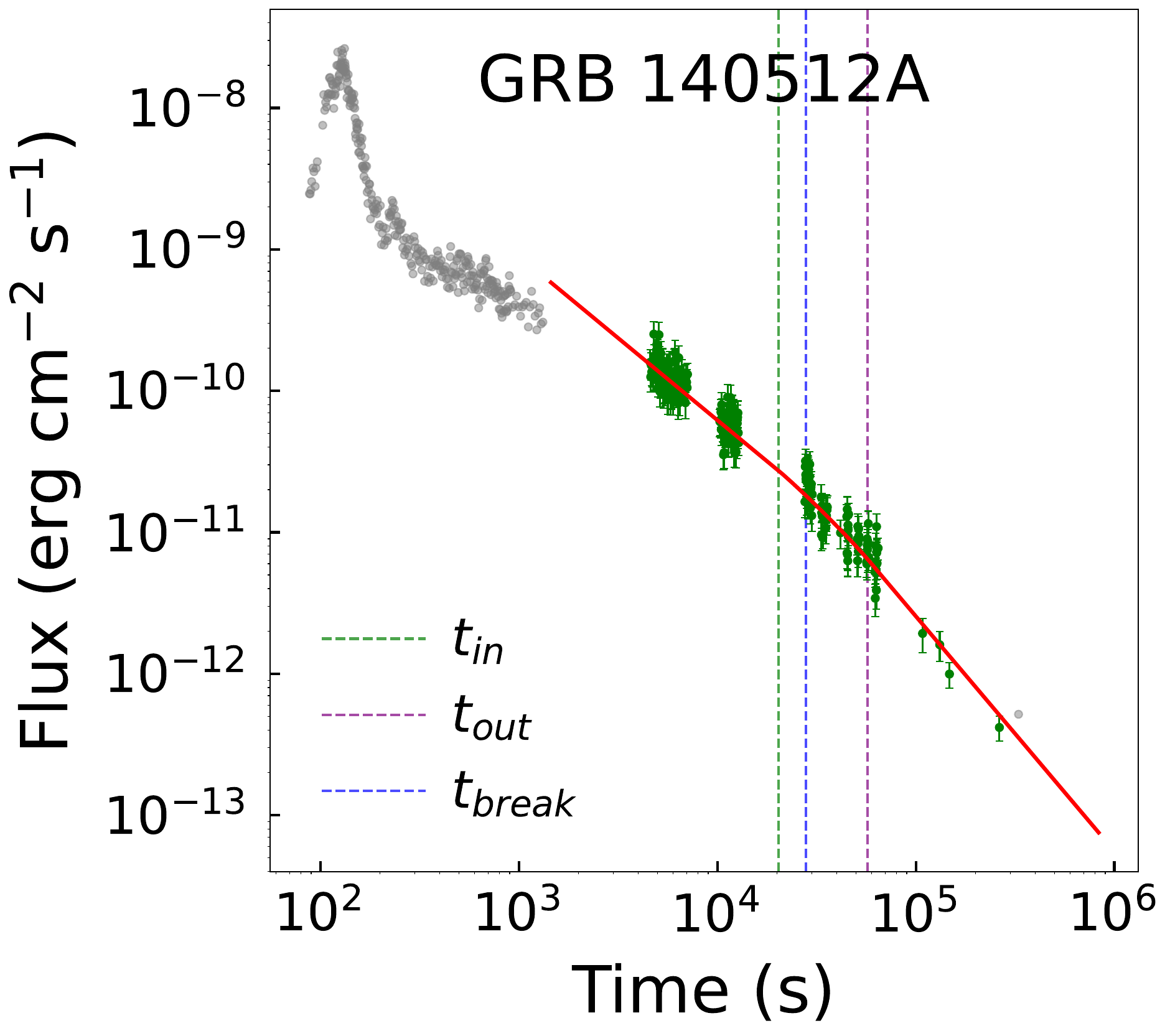}
  \end{subfigure}%
  \hspace{0.01\textwidth}
  \begin{subfigure}[b]{0.23\textwidth}
    \includegraphics[width=\linewidth]{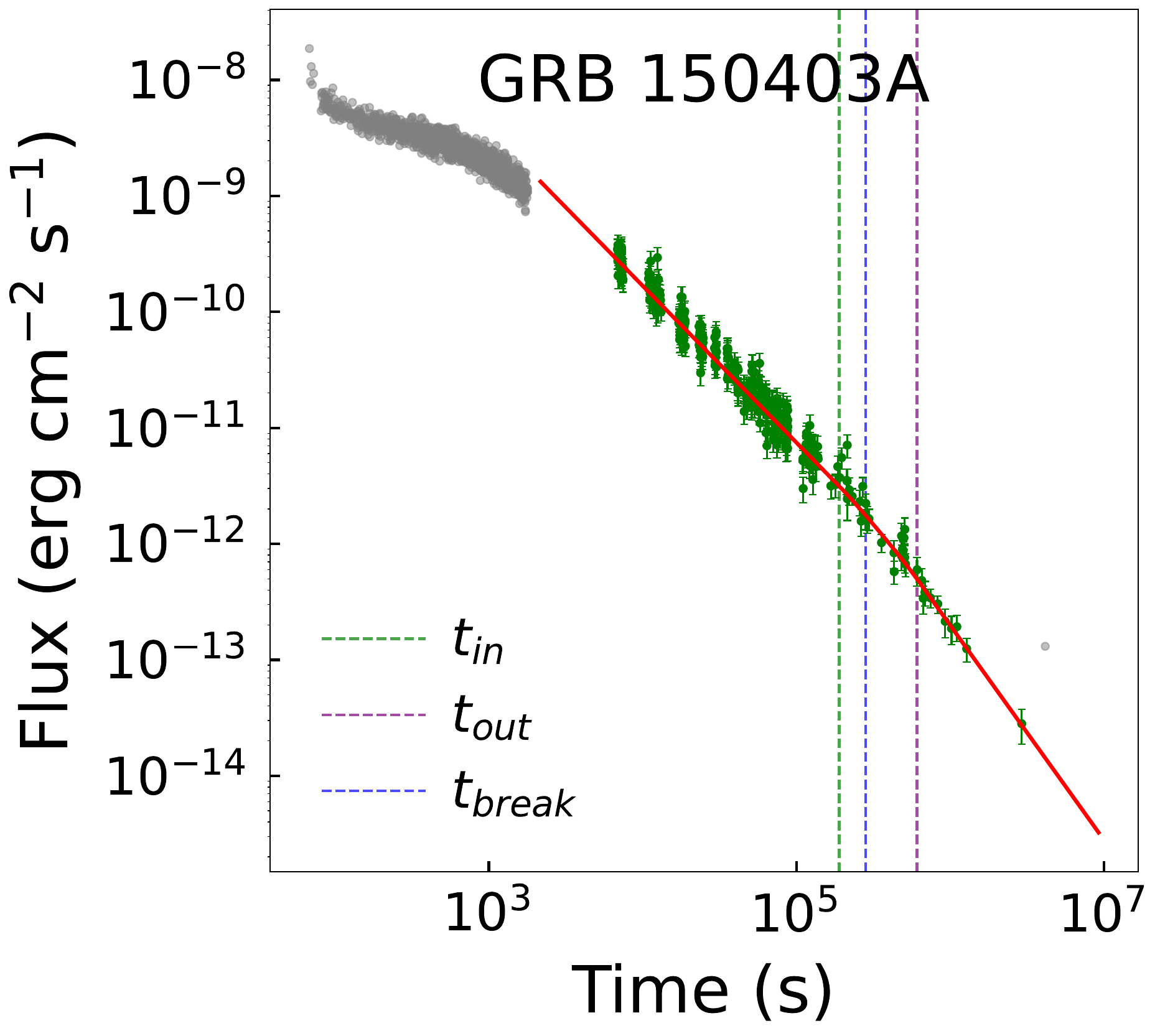}
  \end{subfigure}%
  \hspace{0.01\textwidth}
  \begin{subfigure}[b]{0.23\textwidth}
    \includegraphics[width=\linewidth]{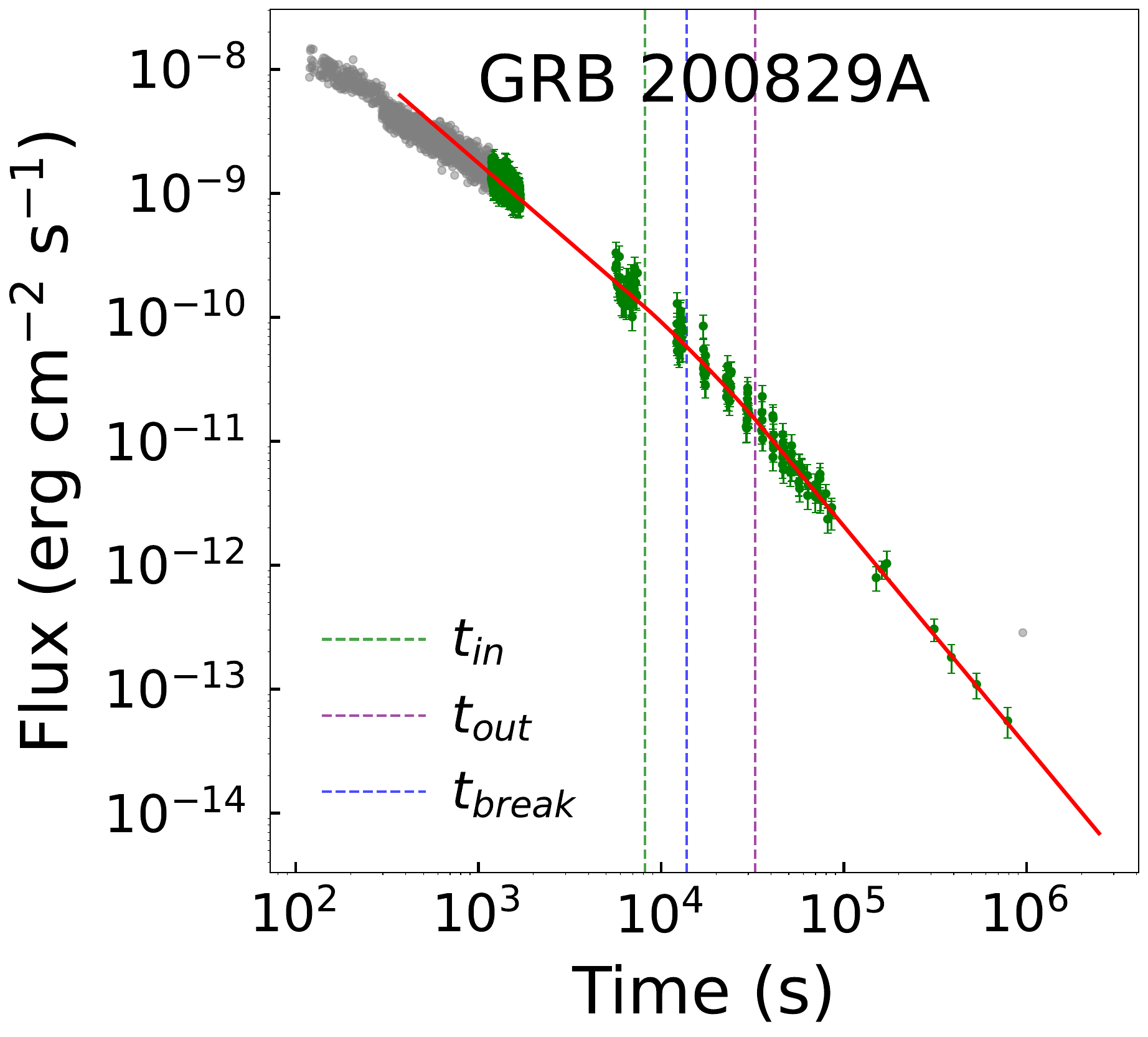}
  \end{subfigure}%
  \hspace{0.01\textwidth}
  \begin{subfigure}[b]{0.23\textwidth}
    \includegraphics[width=\linewidth]{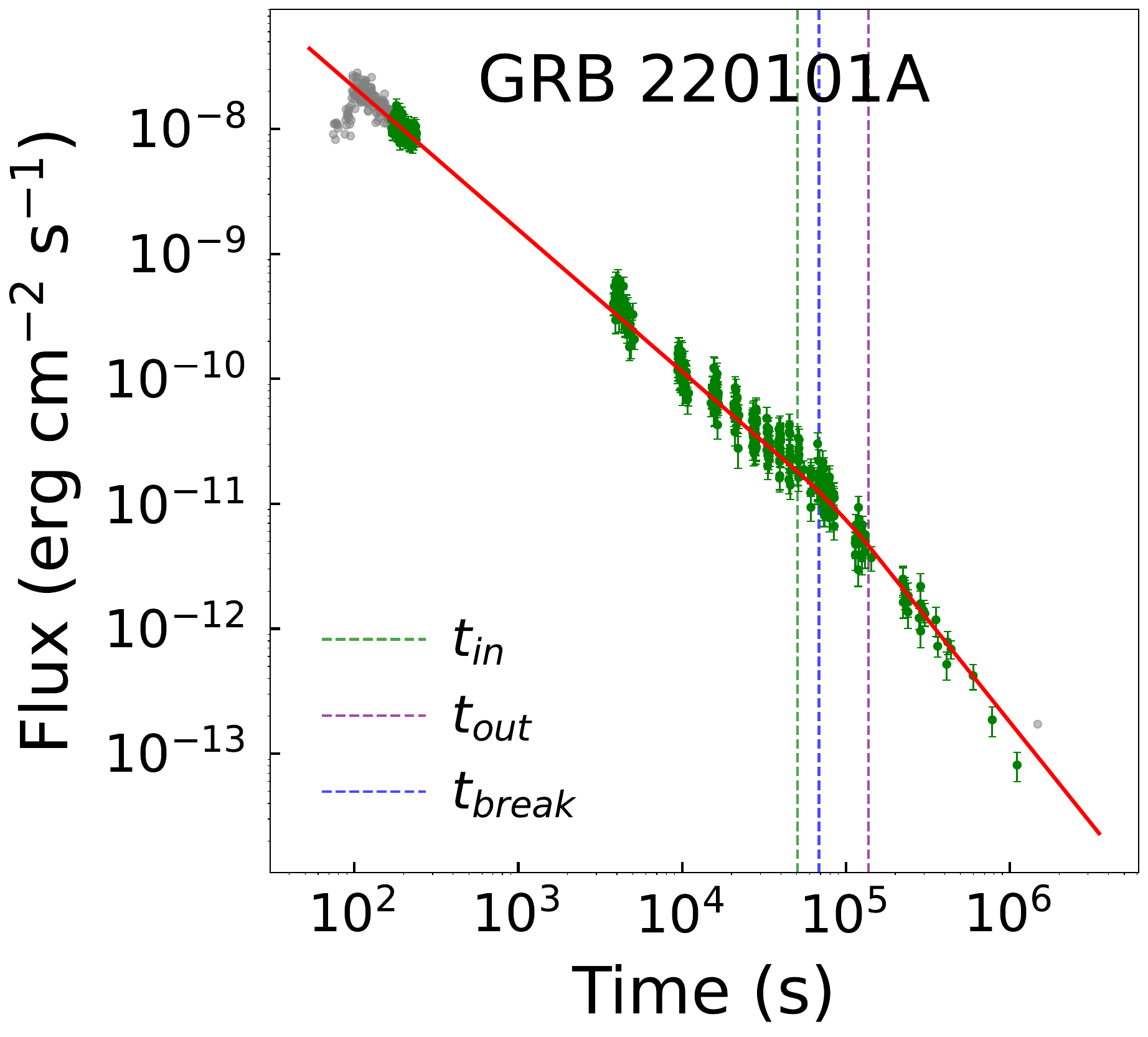}
  \end{subfigure}%
  \\%
  \caption{X-ray afterglow light curve fitting results for model~1. Each panel shows the observed flux (data points) and the best-fit model (solid line) for individual gamma-ray bursts in the stellar wind environment, characterized by density profile $n \propto R^{-2}$.}
  \label{fig:wind_model1_light_curves}
\end{figure}

\begin{figure}[htbp]
\centering
  \begin{subfigure}[b]{0.23\textwidth}
    \includegraphics[width=\linewidth]{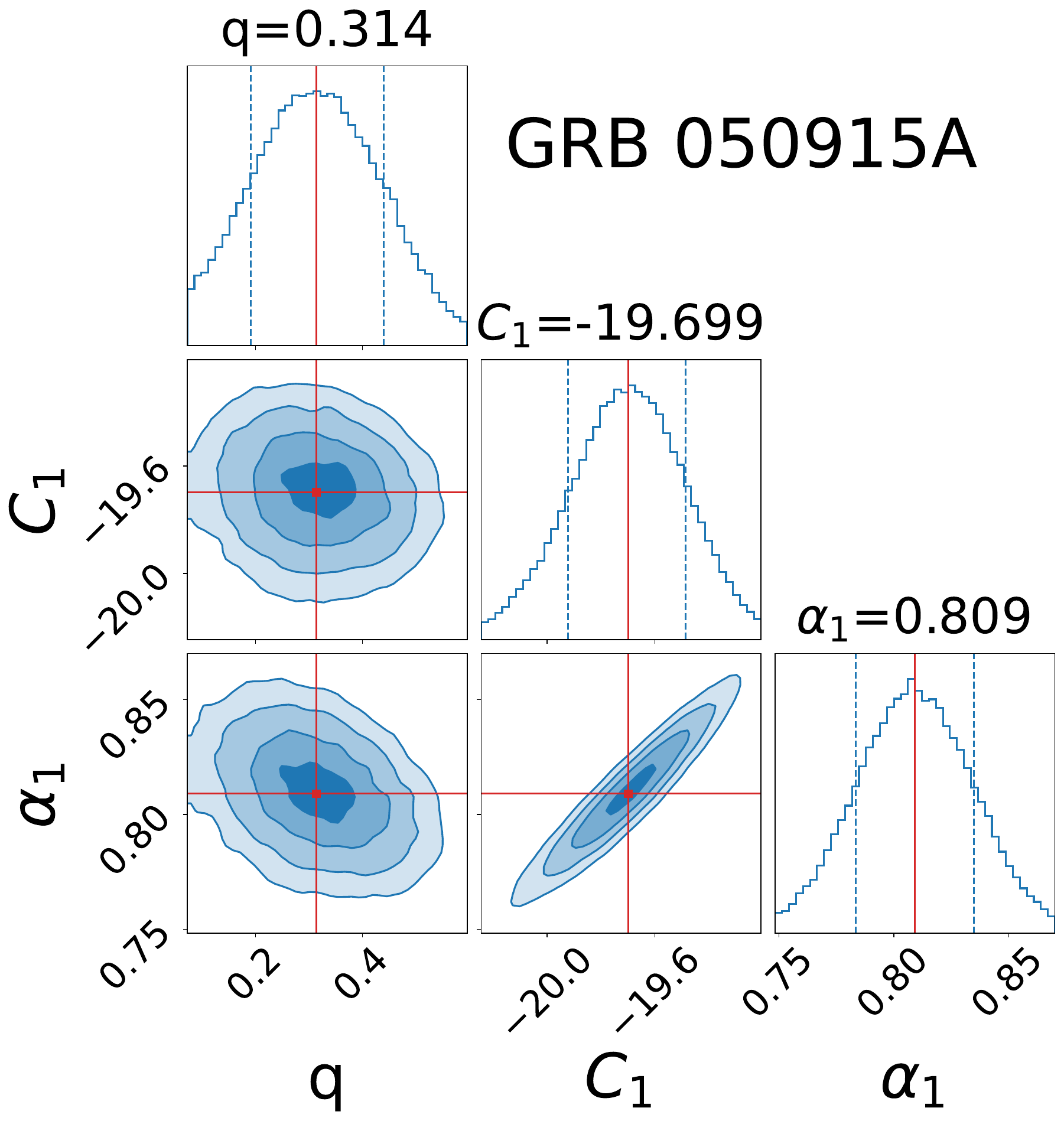}
  \end{subfigure}%
  \hspace{0.01\textwidth}
  \begin{subfigure}[b]{0.23\textwidth}
    \includegraphics[width=\linewidth]{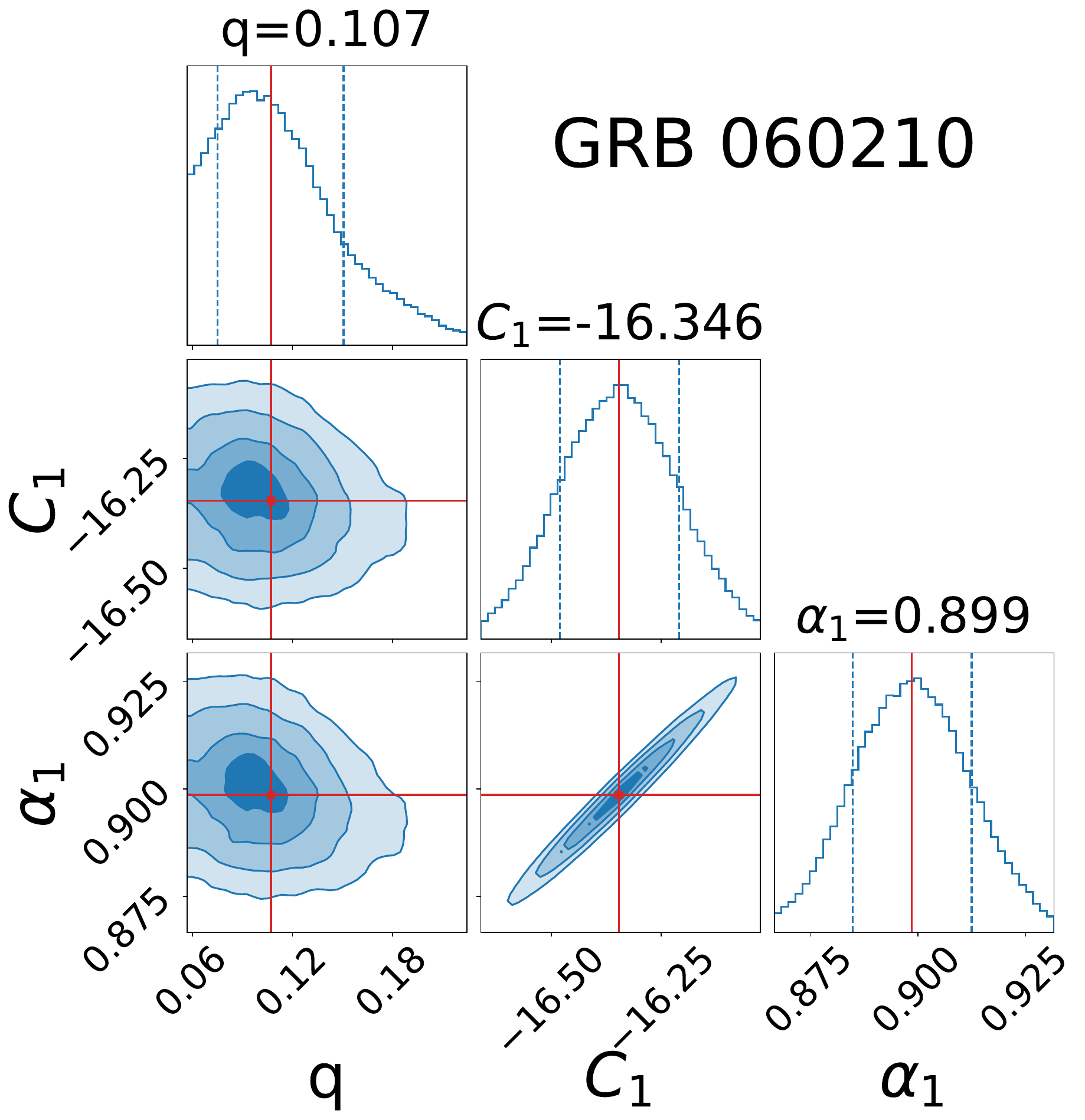}
  \end{subfigure}%
  \hspace{0.01\textwidth}
  \begin{subfigure}[b]{0.23\textwidth}
    \includegraphics[width=\linewidth]{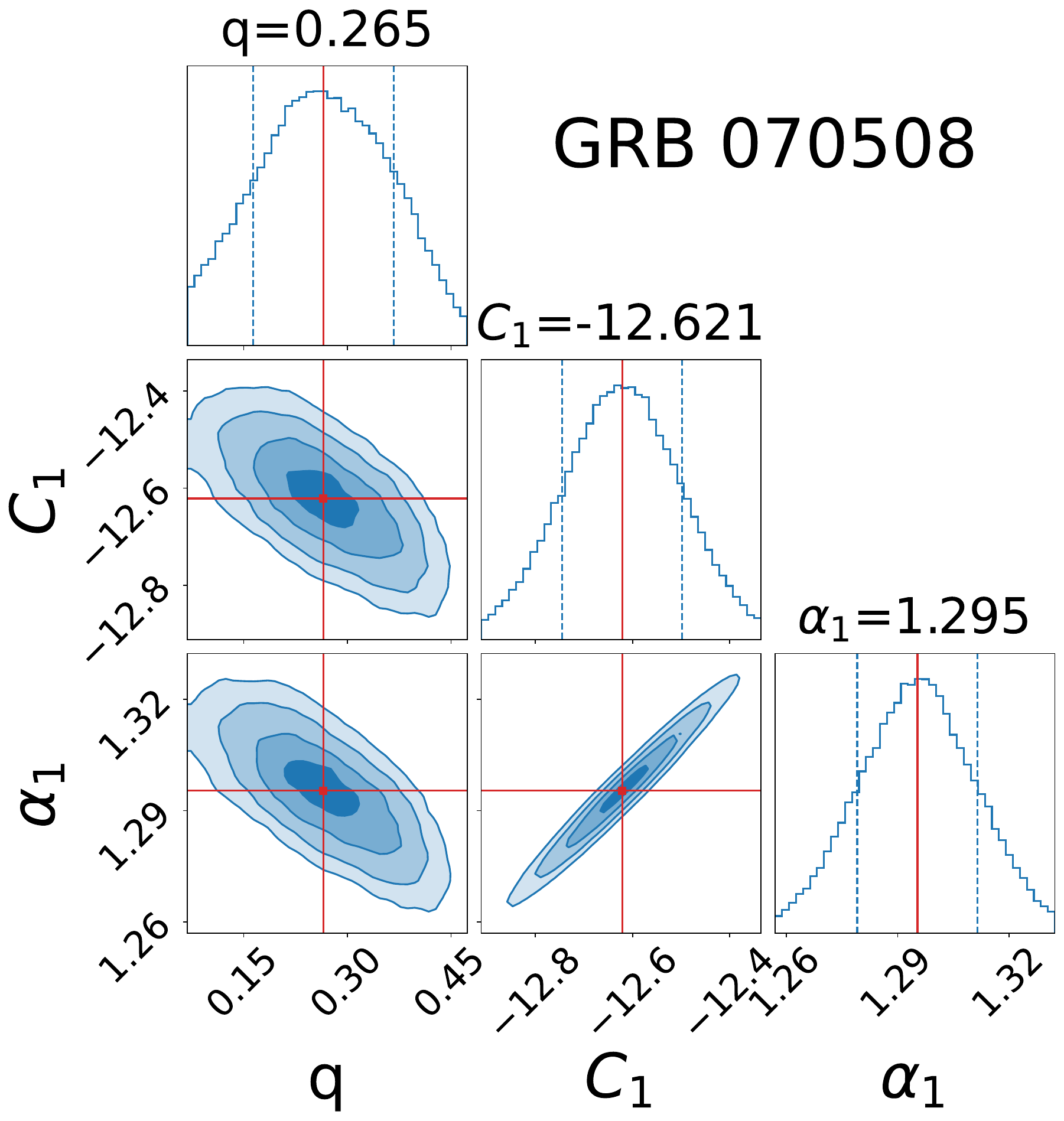}
  \end{subfigure}%
  \hspace{0.01\textwidth}
  \begin{subfigure}[b]{0.23\textwidth}
    \includegraphics[width=\linewidth]{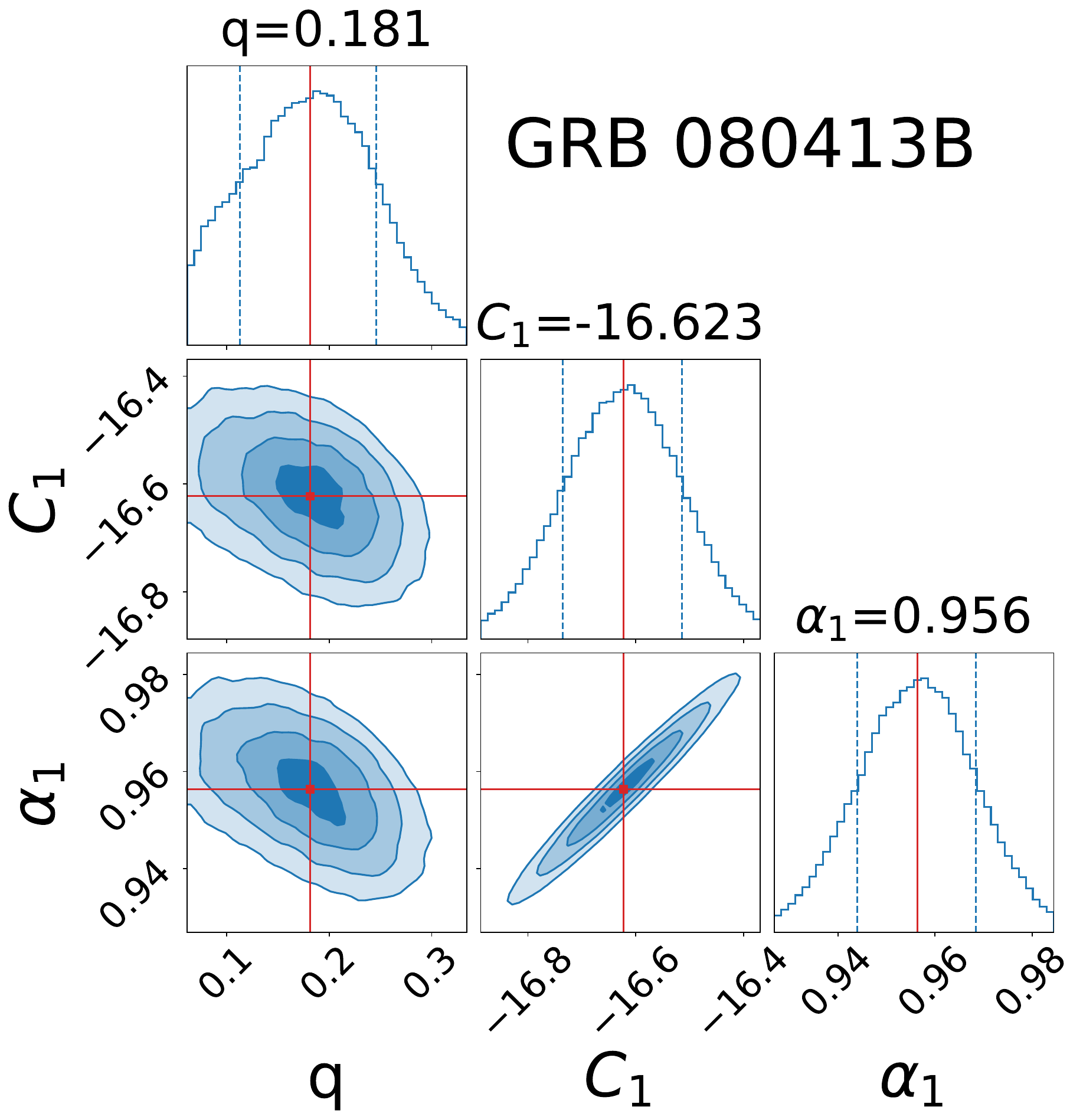}
  \end{subfigure}%
  \\%
  \begin{subfigure}[b]{0.23\textwidth}
    \includegraphics[width=\linewidth]{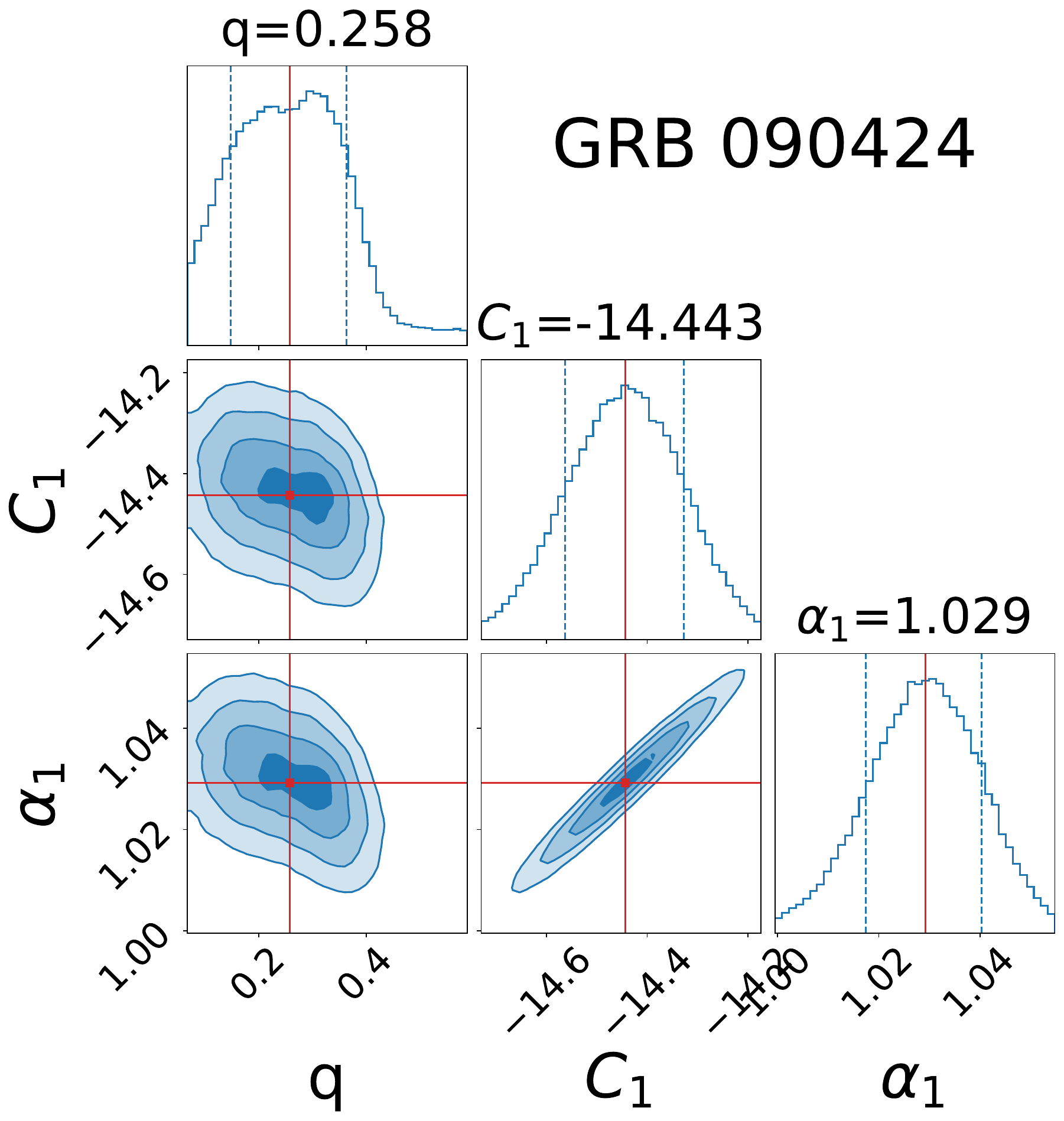}
  \end{subfigure}%
  \hspace{0.01\textwidth}
  \begin{subfigure}[b]{0.23\textwidth}
    \includegraphics[width=\linewidth]{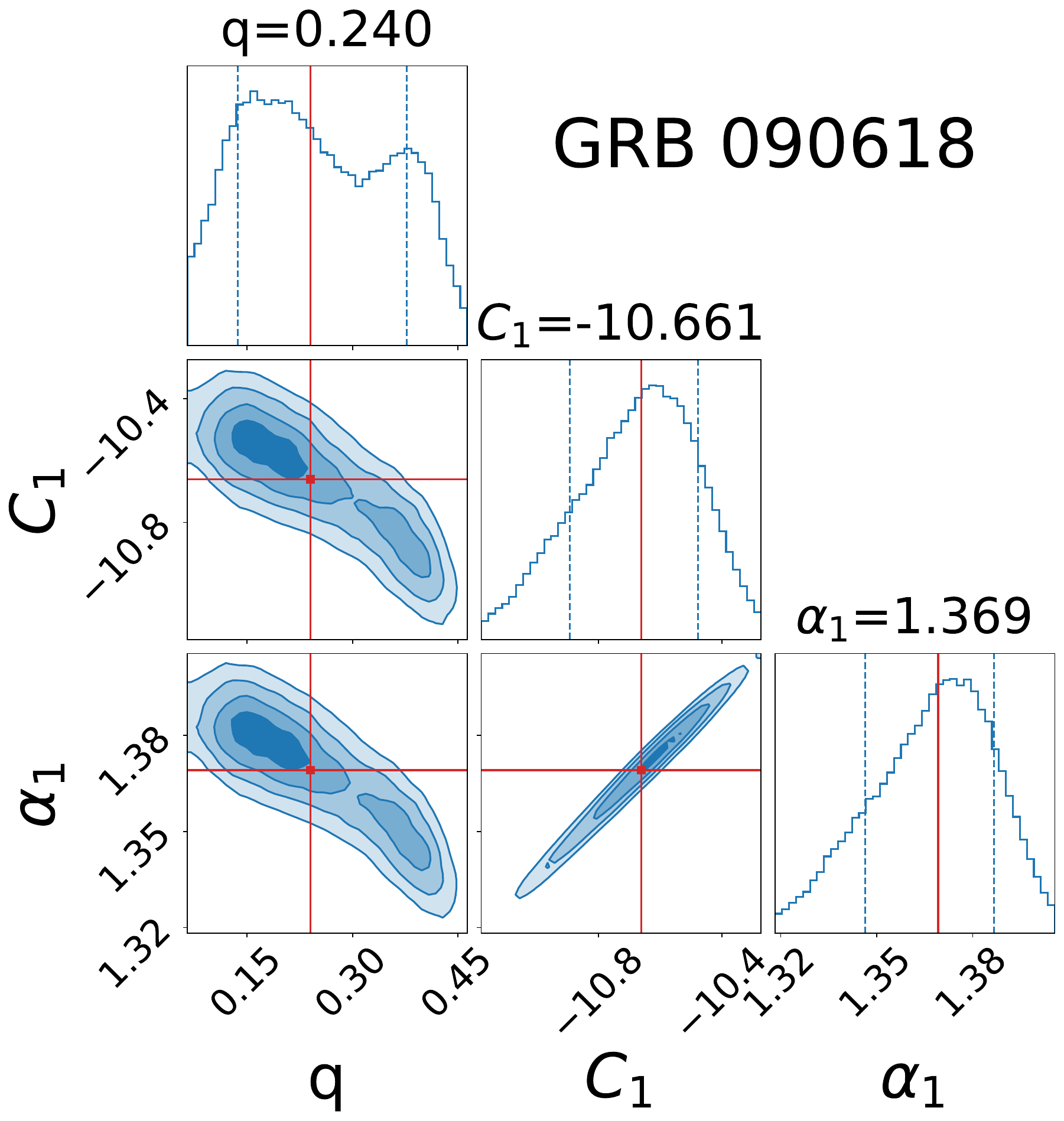}
  \end{subfigure}%
  \hspace{0.01\textwidth}
  \begin{subfigure}[b]{0.23\textwidth}
    \includegraphics[width=\linewidth]{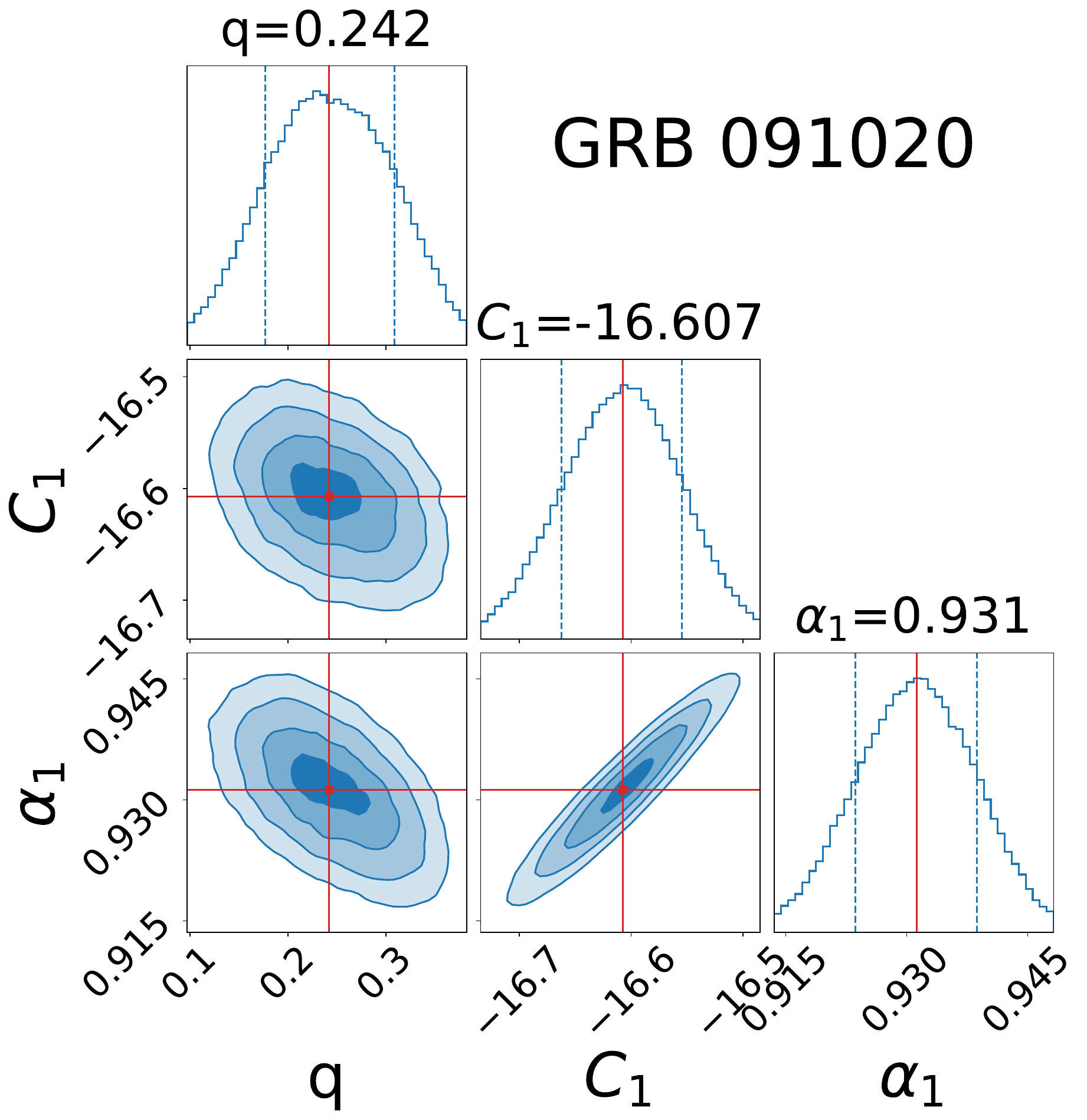}
  \end{subfigure}%
  \hspace{0.01\textwidth}
  \begin{subfigure}[b]{0.23\textwidth}
    \includegraphics[width=\linewidth]{result_091029_10_parameter_distributions.pdf}
  \end{subfigure}%
  \\%
  \begin{subfigure}[b]{0.23\textwidth}
    \includegraphics[width=\linewidth]{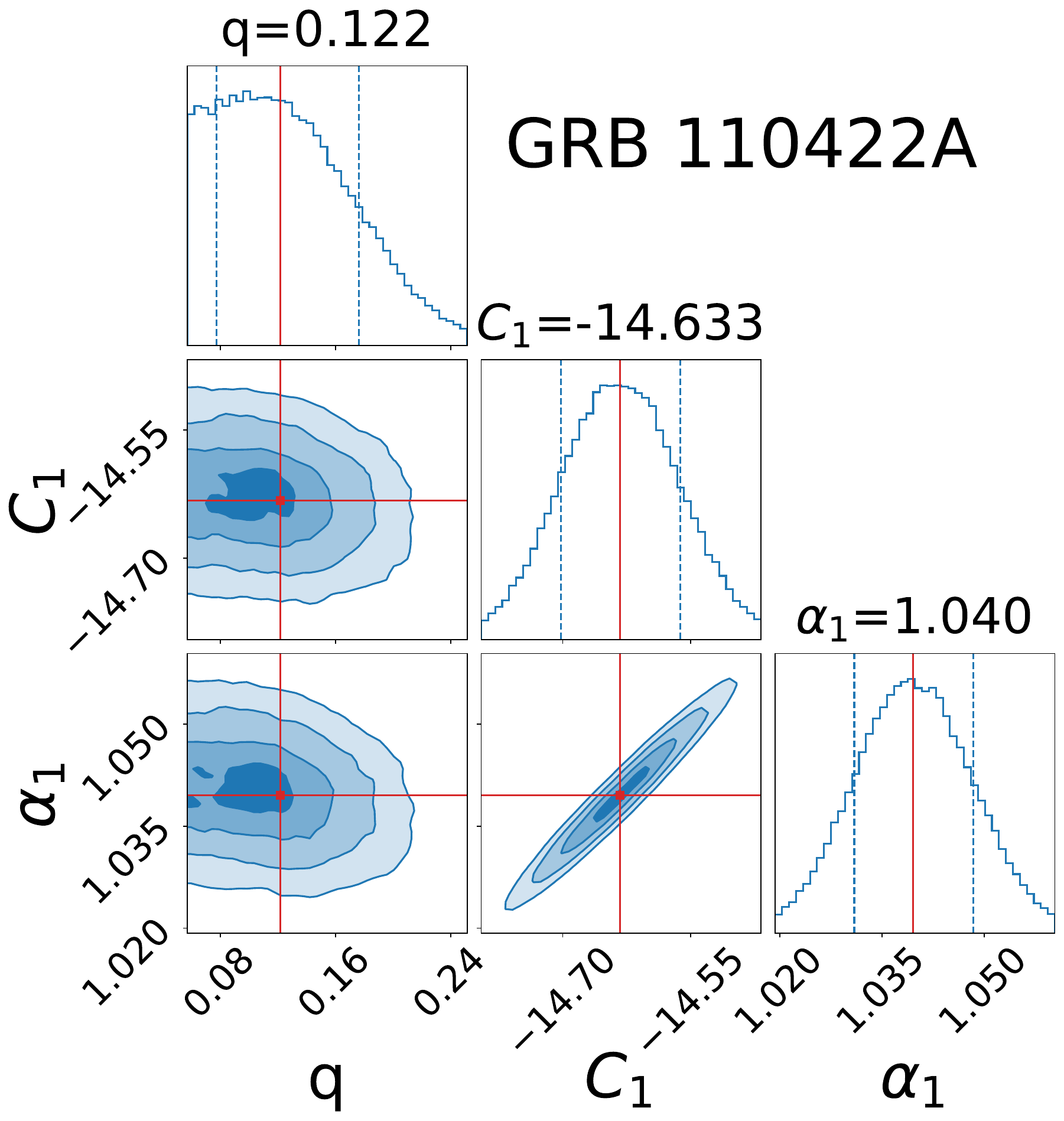}
  \end{subfigure}%
  \hspace{0.01\textwidth}
  \begin{subfigure}[b]{0.23\textwidth}
    \includegraphics[width=\linewidth]{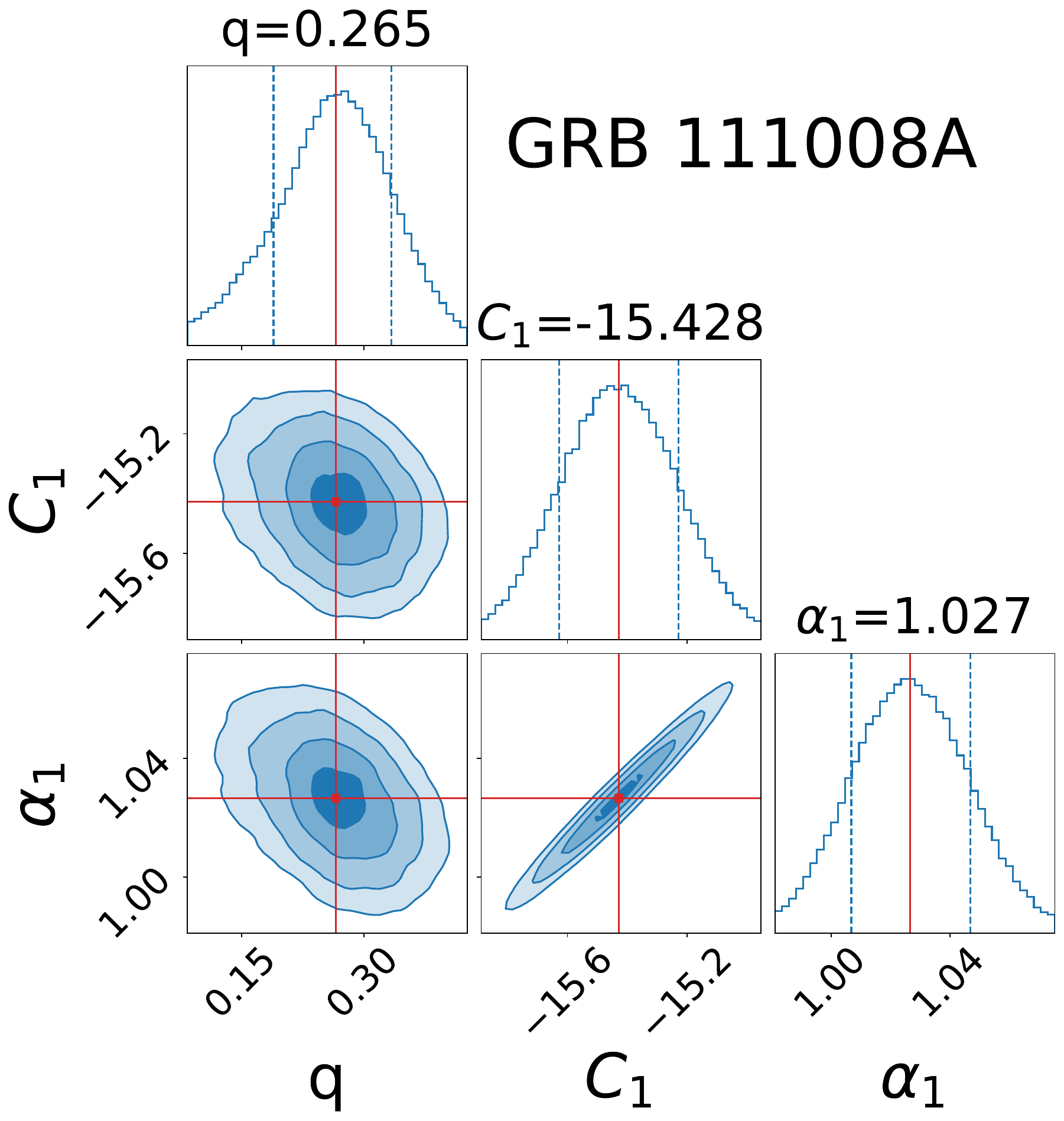}
  \end{subfigure}%
  \hspace{0.01\textwidth}
  \begin{subfigure}[b]{0.23\textwidth}
    \includegraphics[width=\linewidth]{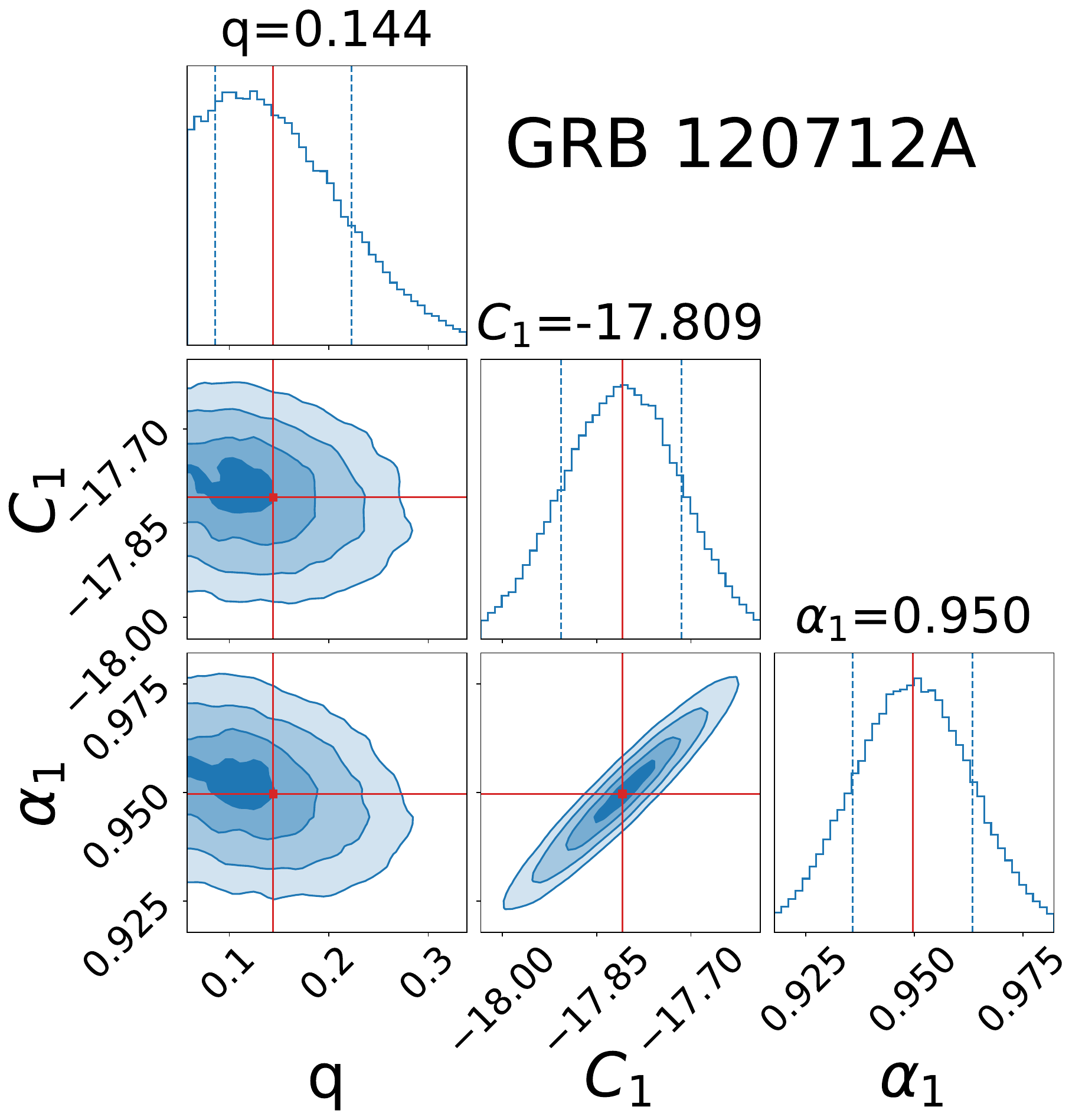}
  \end{subfigure}%
  \hspace{0.01\textwidth}
  \begin{subfigure}[b]{0.23\textwidth}
    \includegraphics[width=\linewidth]{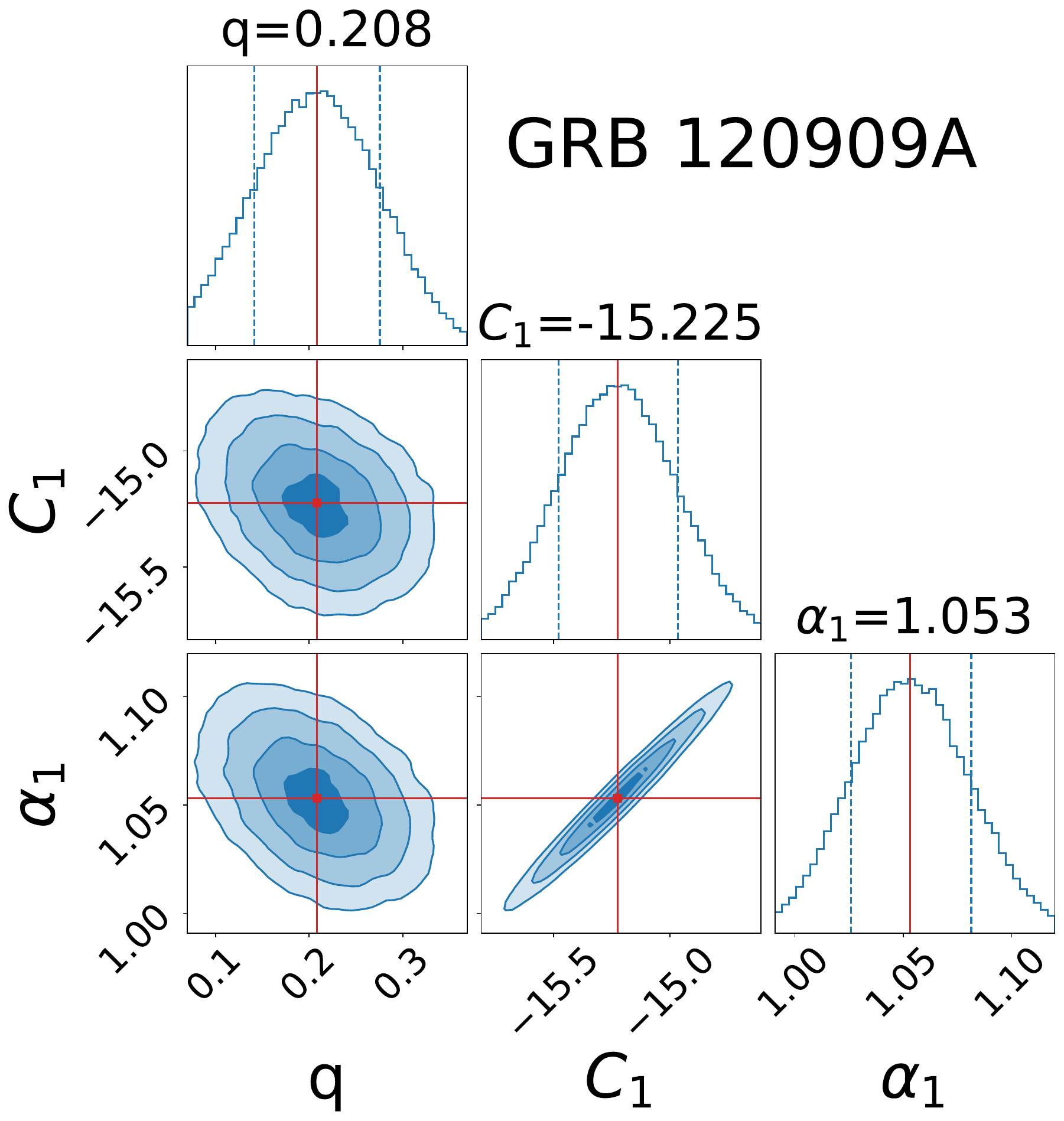}
  \end{subfigure}%
  \\%
  \begin{subfigure}[b]{0.23\textwidth}
    \includegraphics[width=\linewidth]{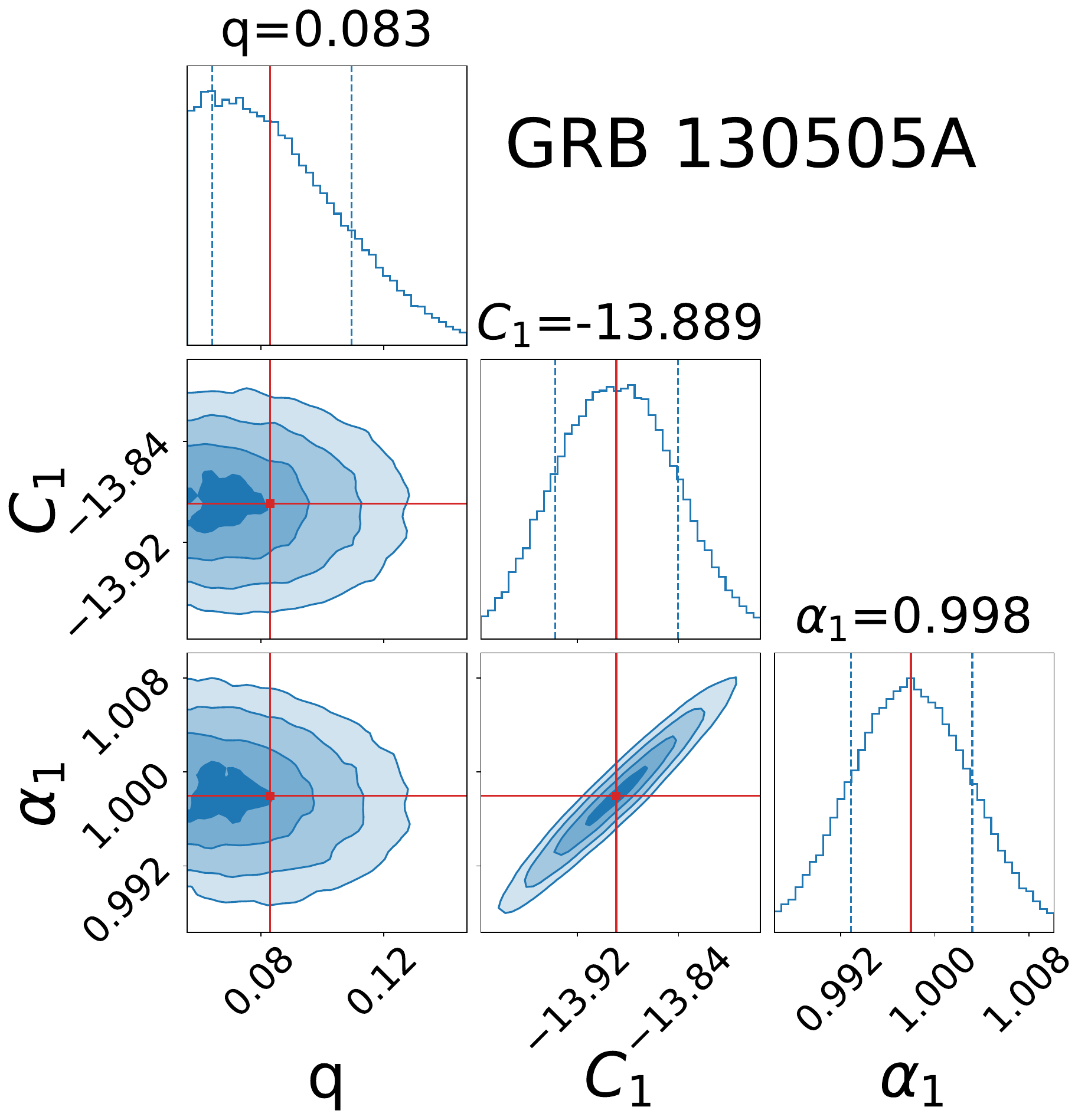}
  \end{subfigure}%
  \hspace{0.01\textwidth}
  \begin{subfigure}[b]{0.23\textwidth}
    \includegraphics[width=\linewidth]{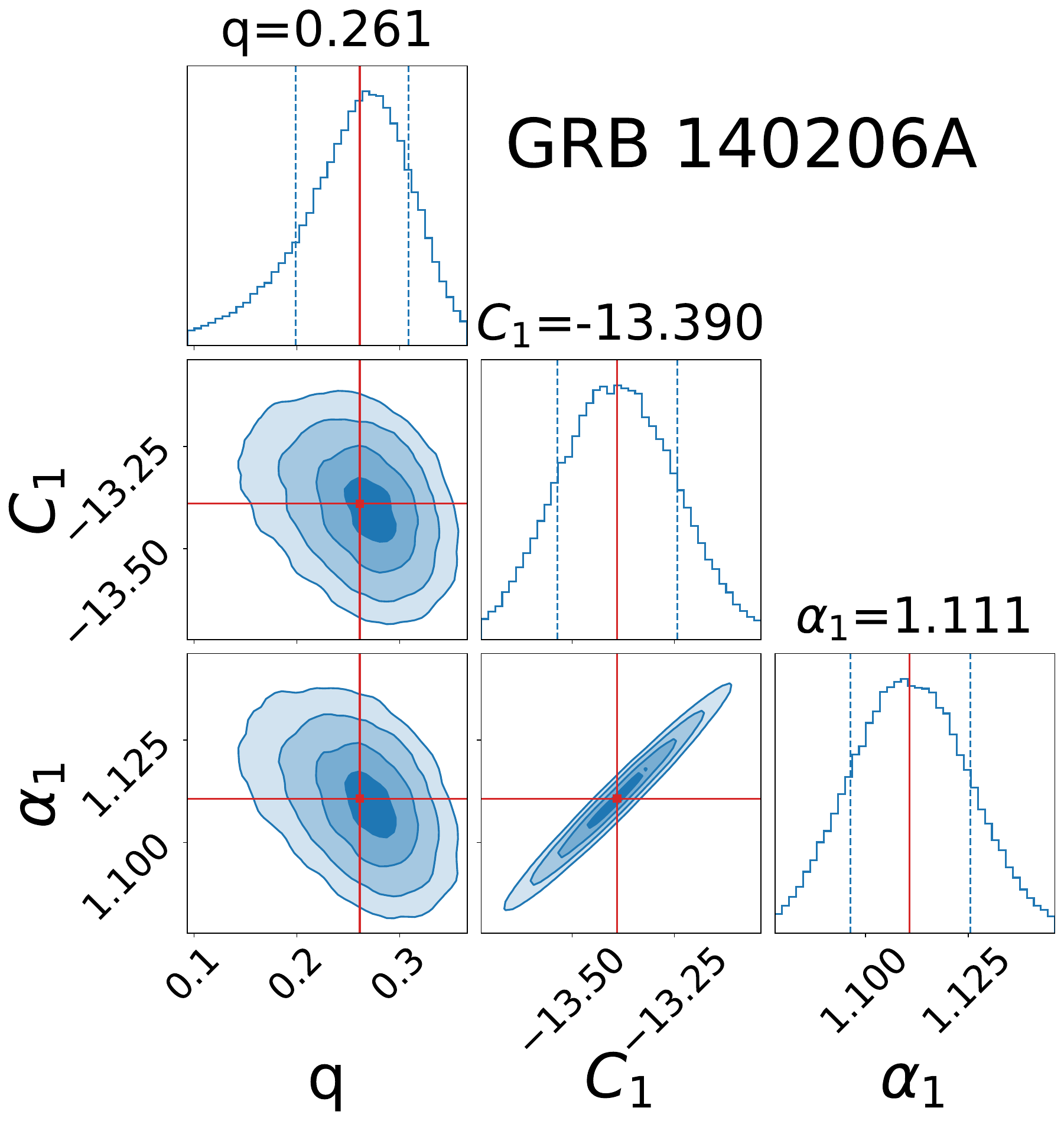}
  \end{subfigure}%
  \hspace{0.01\textwidth}
  \begin{subfigure}[b]{0.23\textwidth}
    \includegraphics[width=\linewidth]{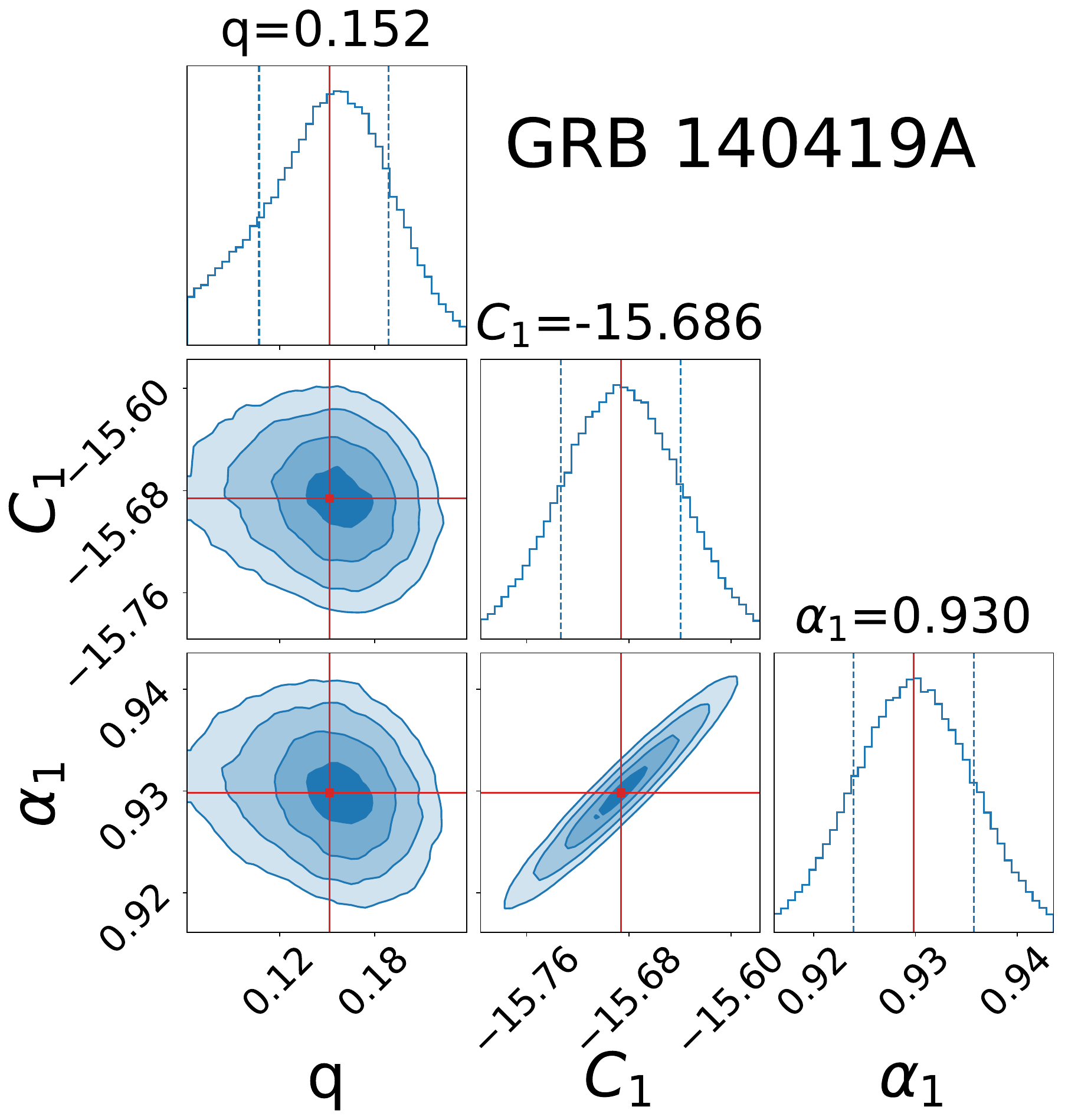}
  \end{subfigure}%
  \hspace{0.01\textwidth}
  \begin{subfigure}[b]{0.23\textwidth}
    \includegraphics[width=\linewidth]{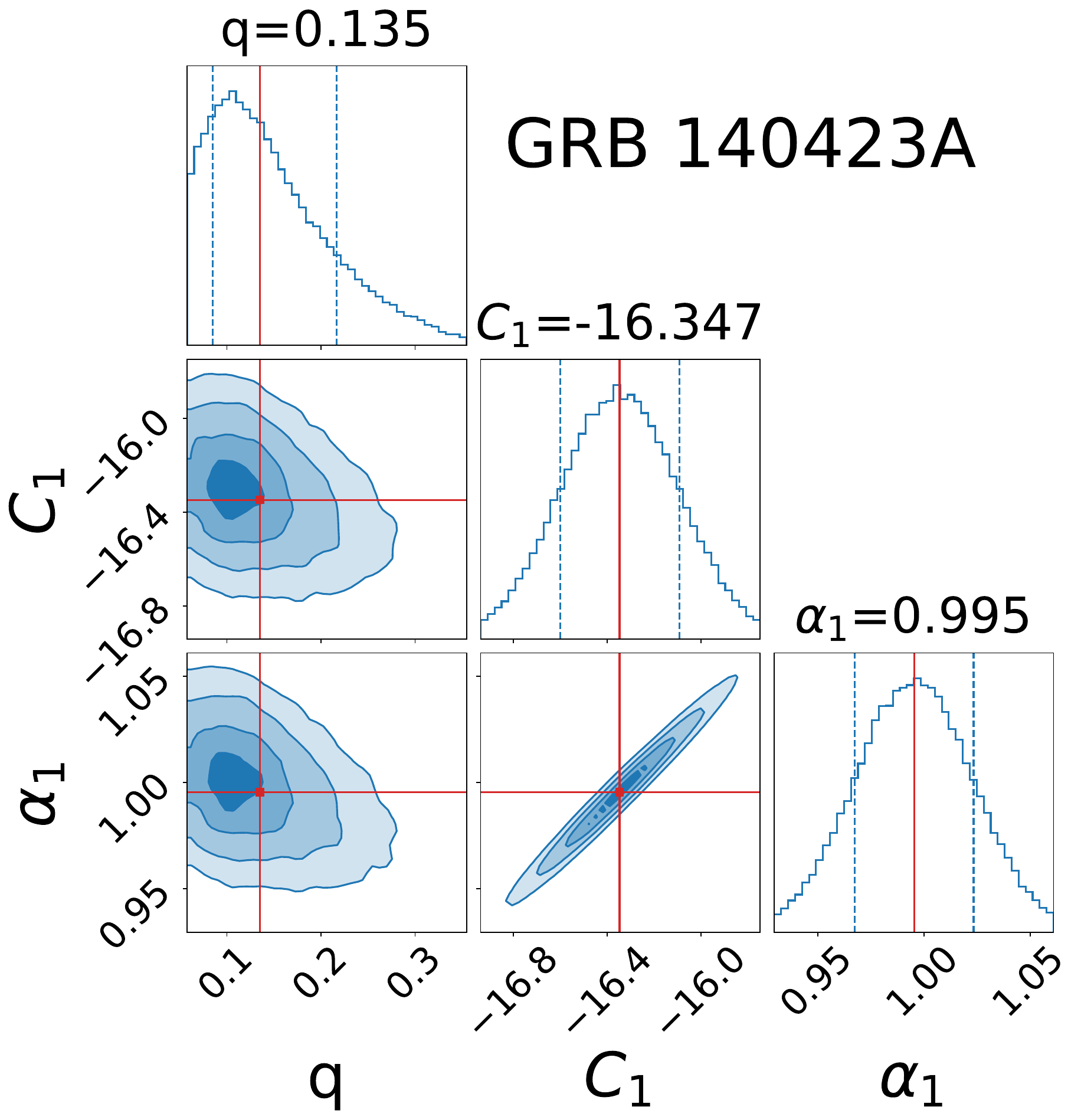}
  \end{subfigure}%
  \\%
  \begin{subfigure}[b]{0.23\textwidth}
    \includegraphics[width=\linewidth]{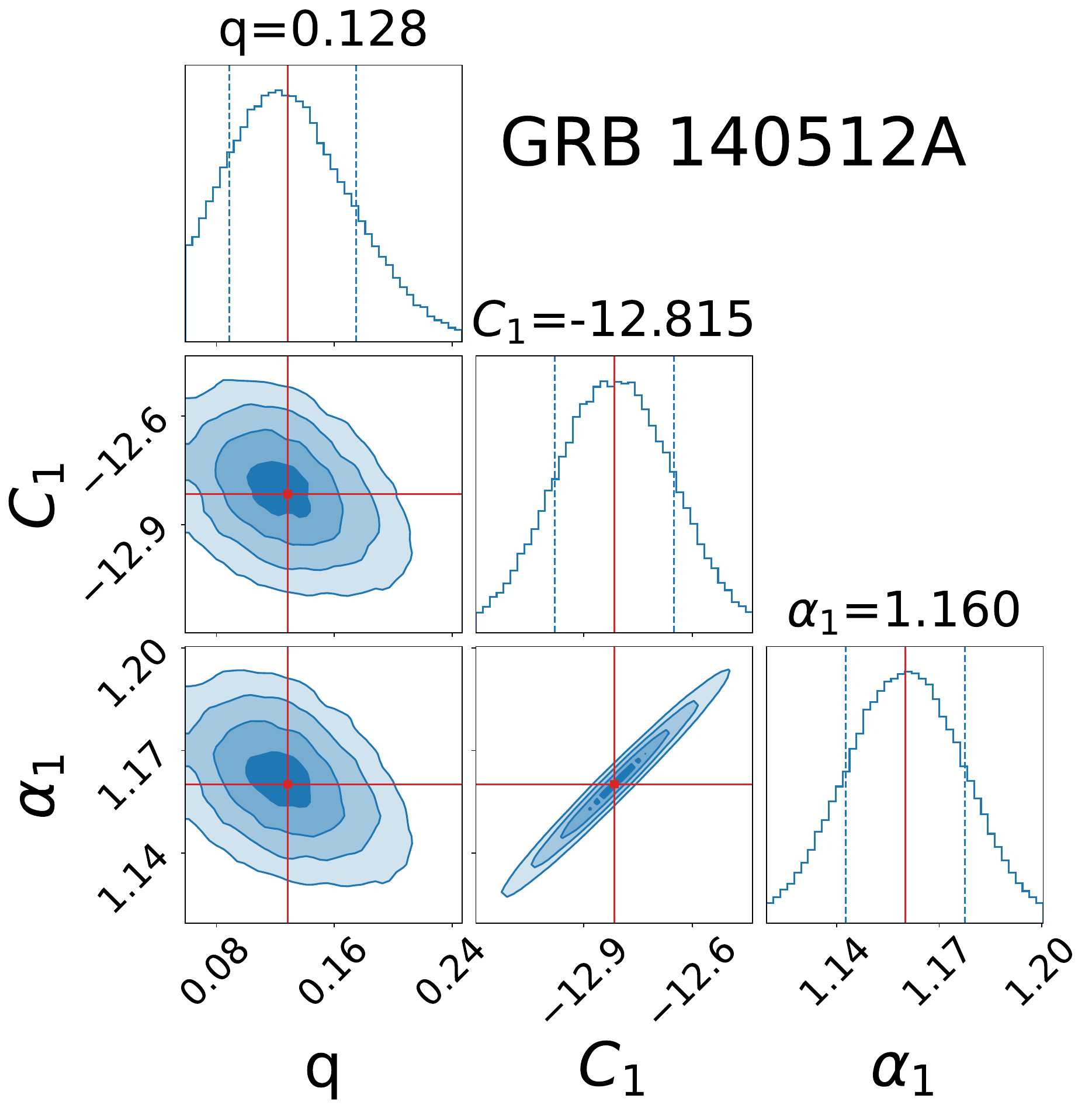}
  \end{subfigure}%
  \hspace{0.01\textwidth}
  \begin{subfigure}[b]{0.23\textwidth}
    \includegraphics[width=\linewidth]{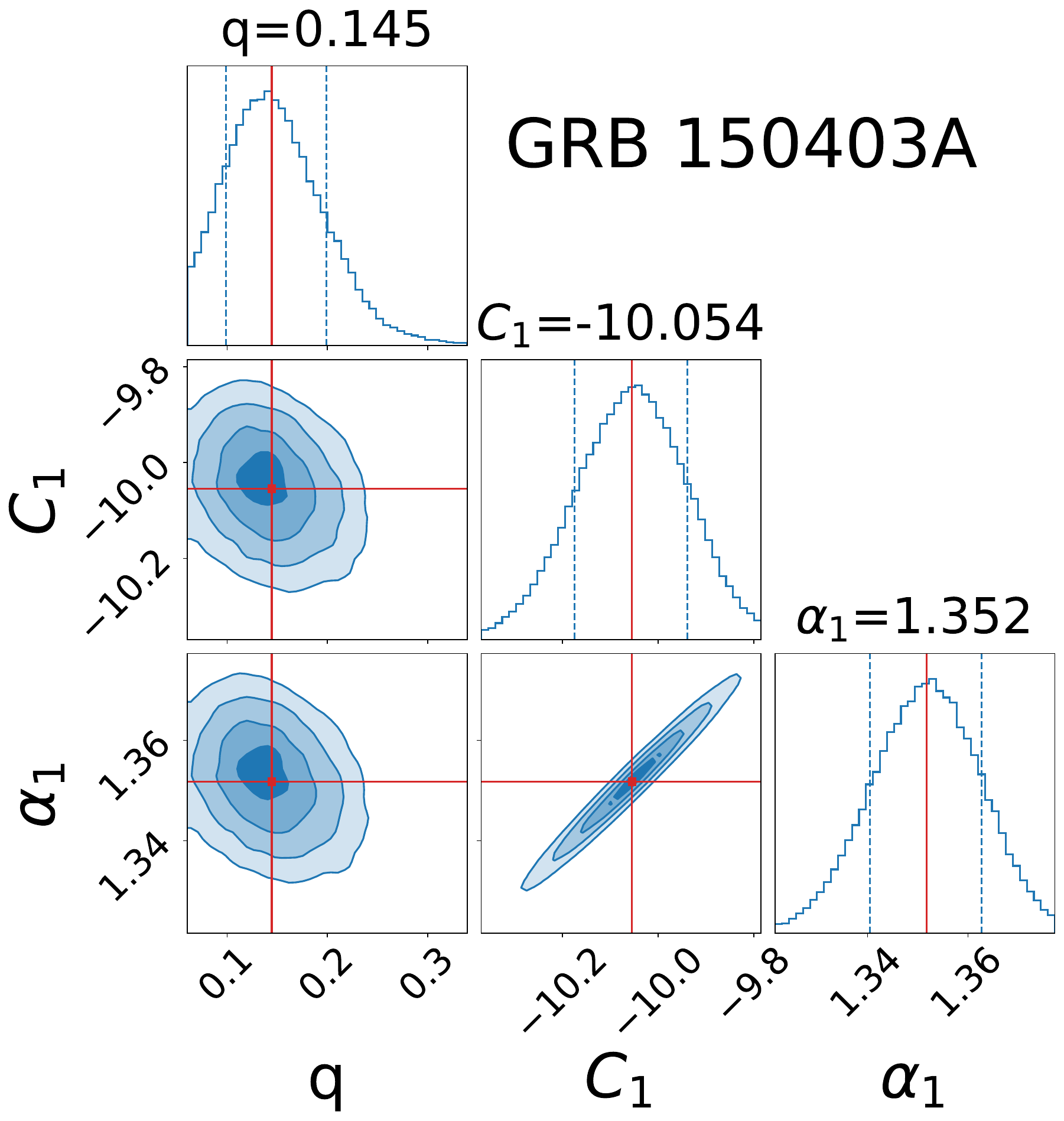}
  \end{subfigure}%
  \hspace{0.01\textwidth}
  \begin{subfigure}[b]{0.23\textwidth}
    \includegraphics[width=\linewidth]{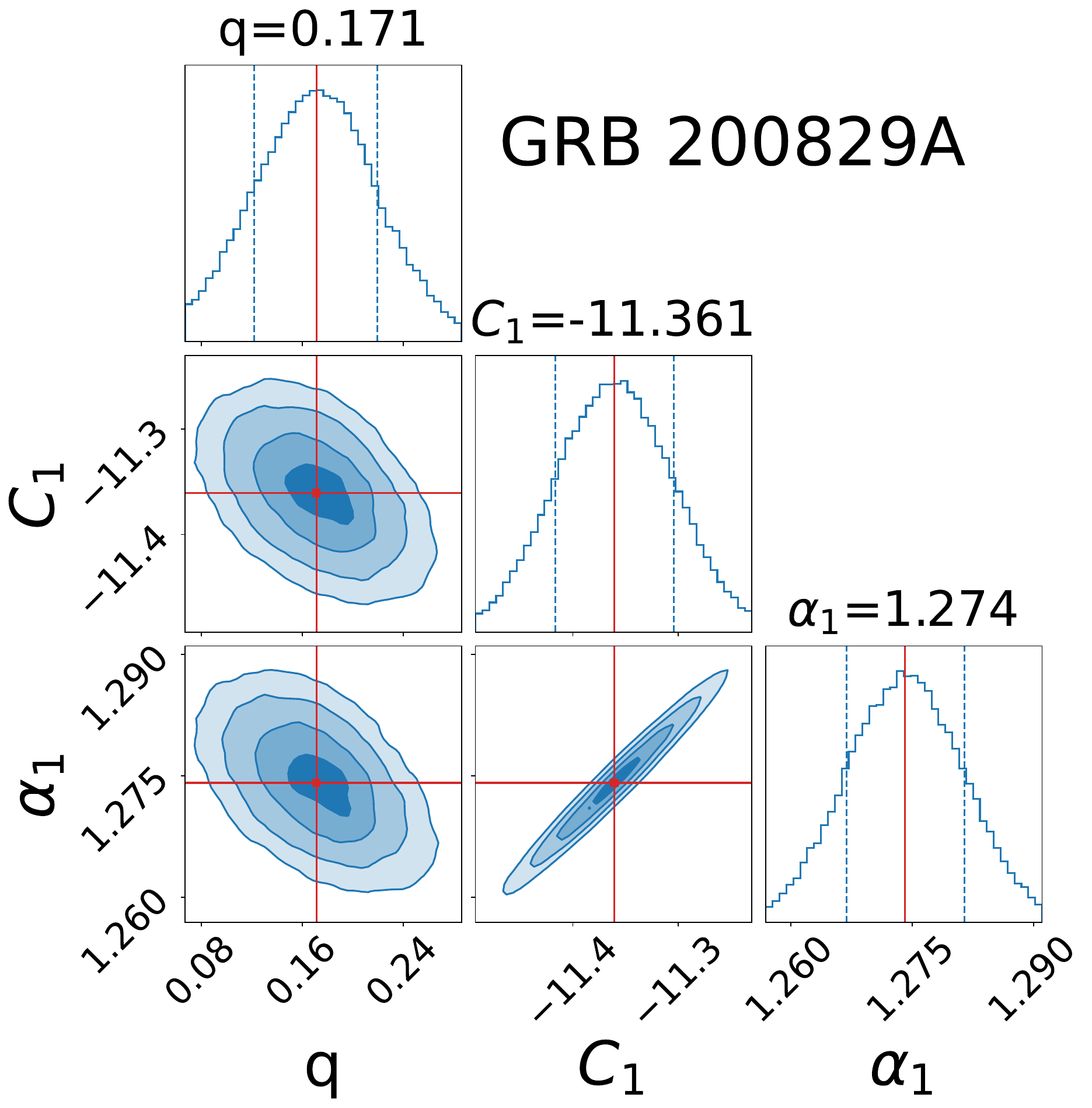}
  \end{subfigure}%
  \hspace{0.01\textwidth}
  \begin{subfigure}[b]{0.23\textwidth}
    \includegraphics[width=\linewidth]{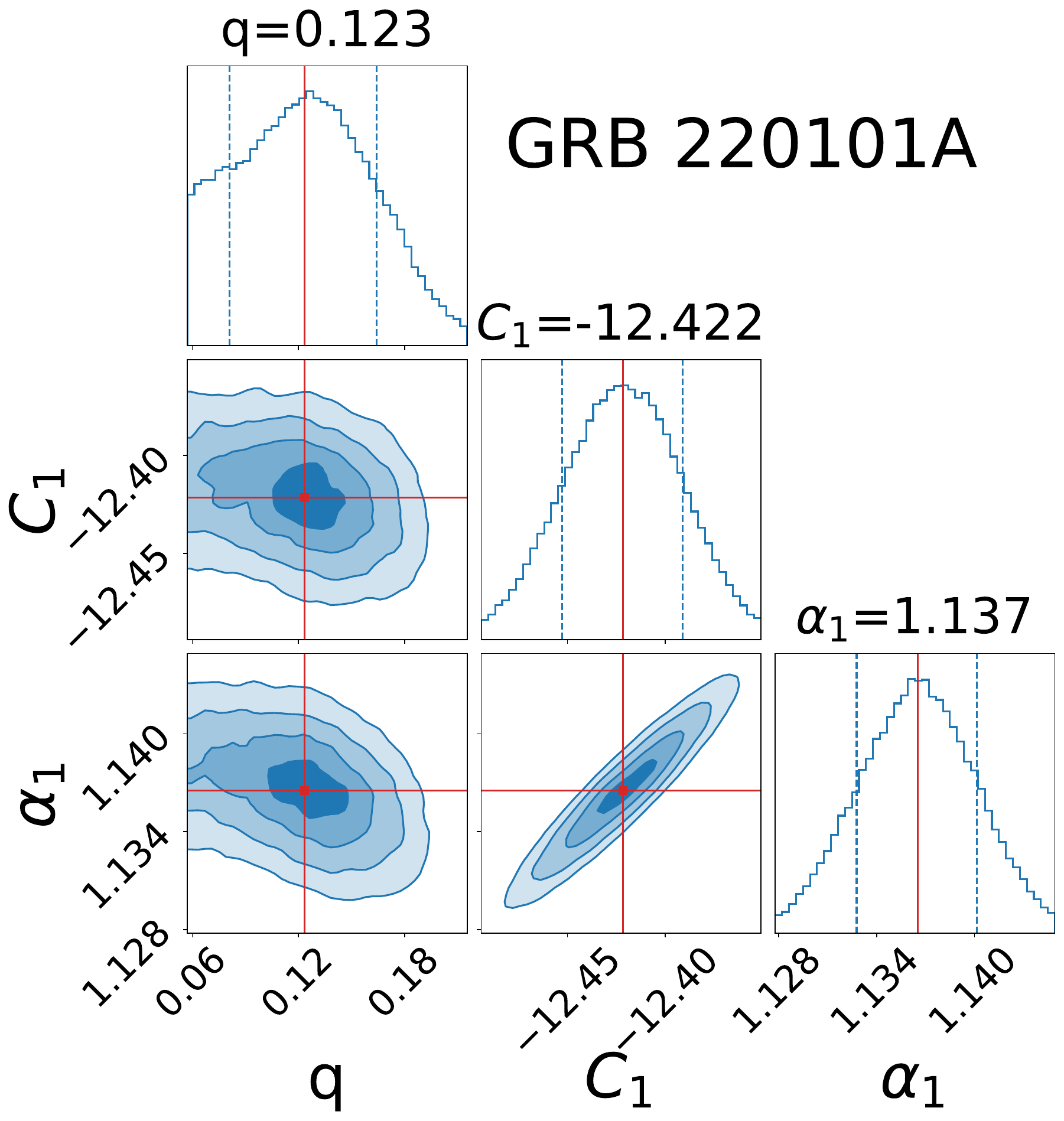}
  \end{subfigure}%
  \\%
  \caption{Posterior parameter distributions for model~1. Each panel displays the MCMC sampling results for key parameters including the off-axis ratio $q = \theta_{\text{obs}}/\theta_{\text{jet}}$, decay index $\alpha_1$, and normalization factor in the stellar wind environment.}
  \label{fig:wind_model1_parameter_distributions}
\end{figure}

\begin{figure}[htbp]
\centering
  \begin{subfigure}[b]{0.23\textwidth}
    \includegraphics[width=\linewidth]{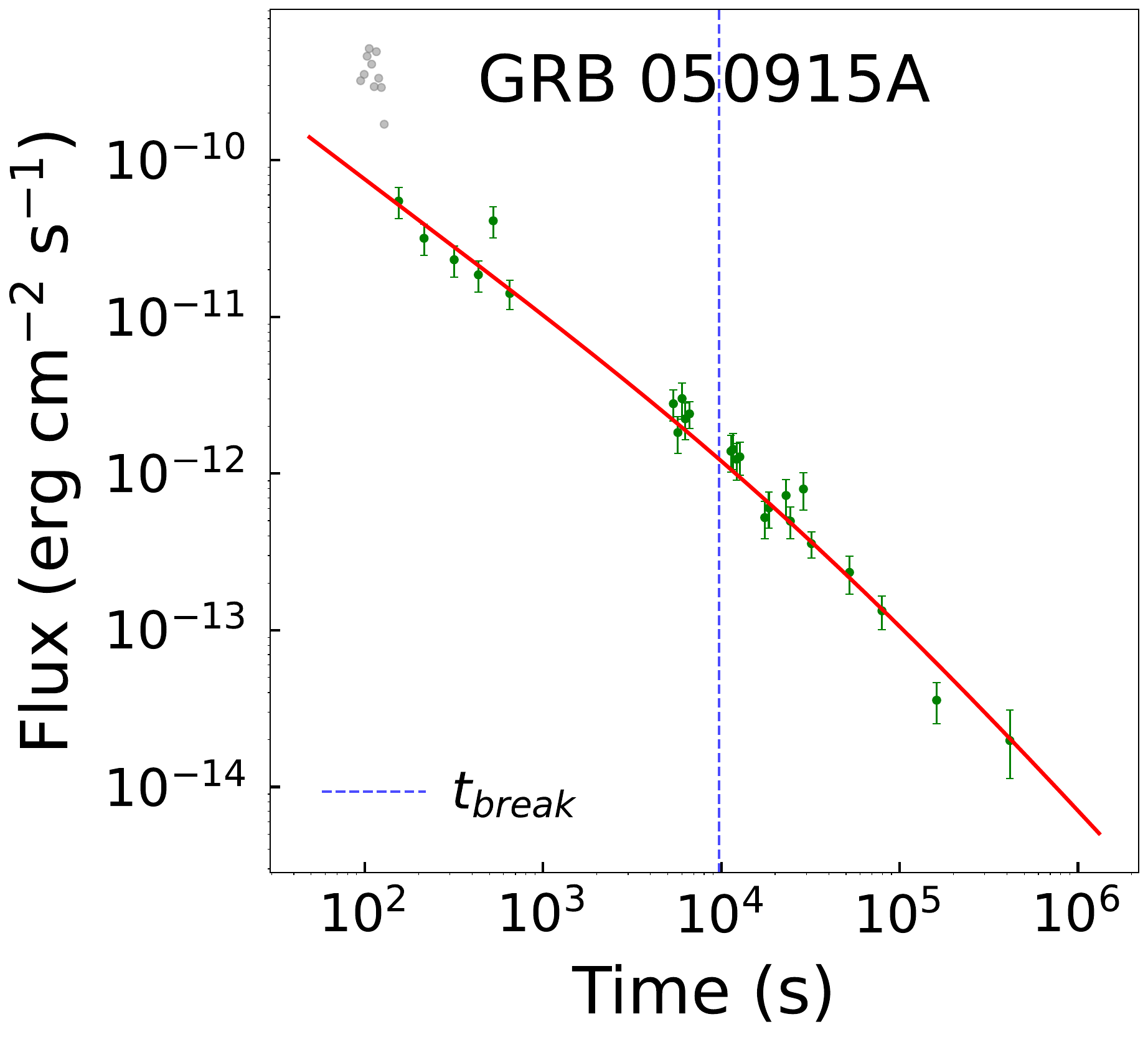}
  \end{subfigure}%
  \hspace{0.01\textwidth}
  \begin{subfigure}[b]{0.23\textwidth}
    \includegraphics[width=\linewidth]{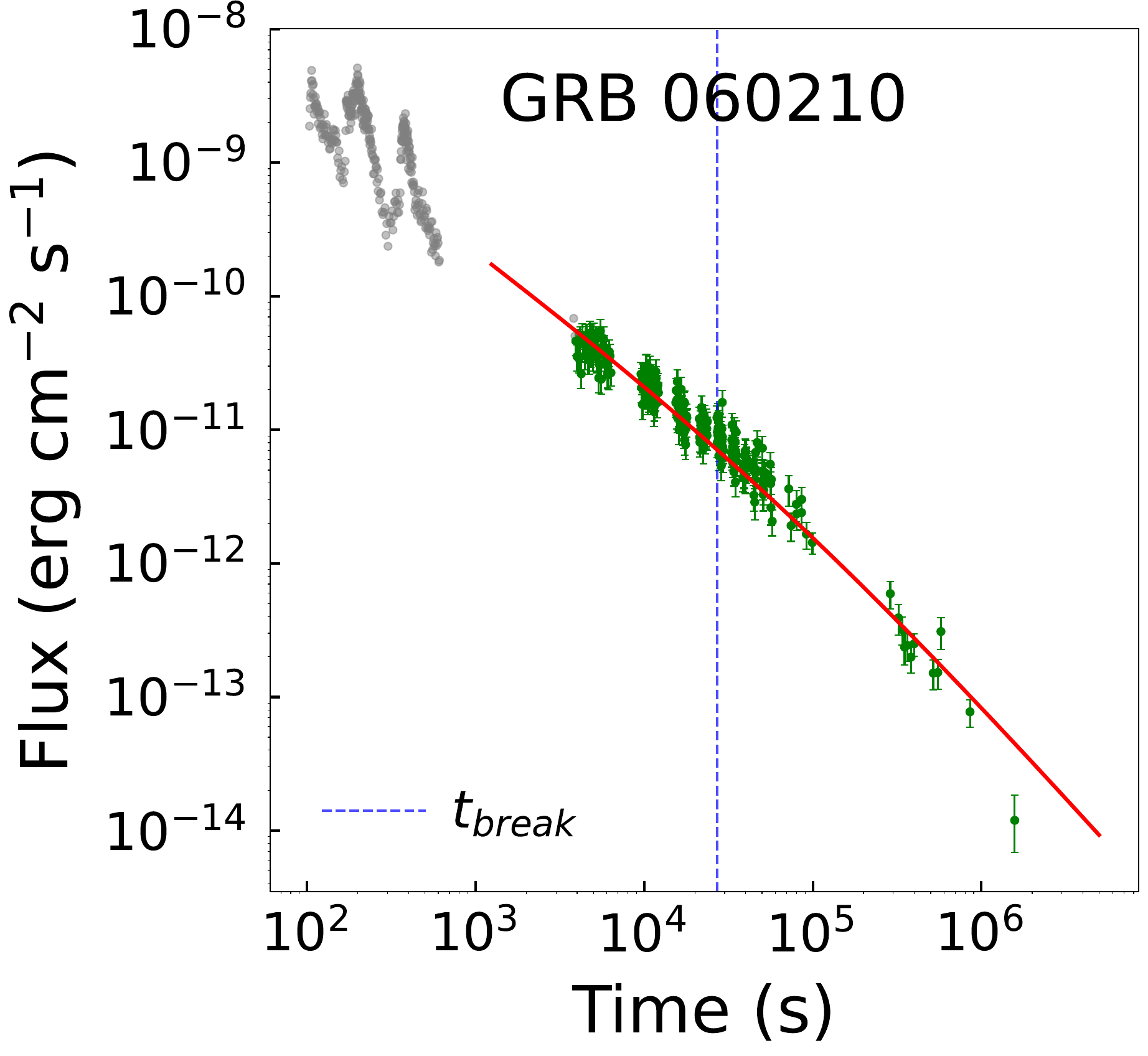}
  \end{subfigure}%
  \hspace{0.01\textwidth}
  \begin{subfigure}[b]{0.23\textwidth}
    \includegraphics[width=\linewidth]{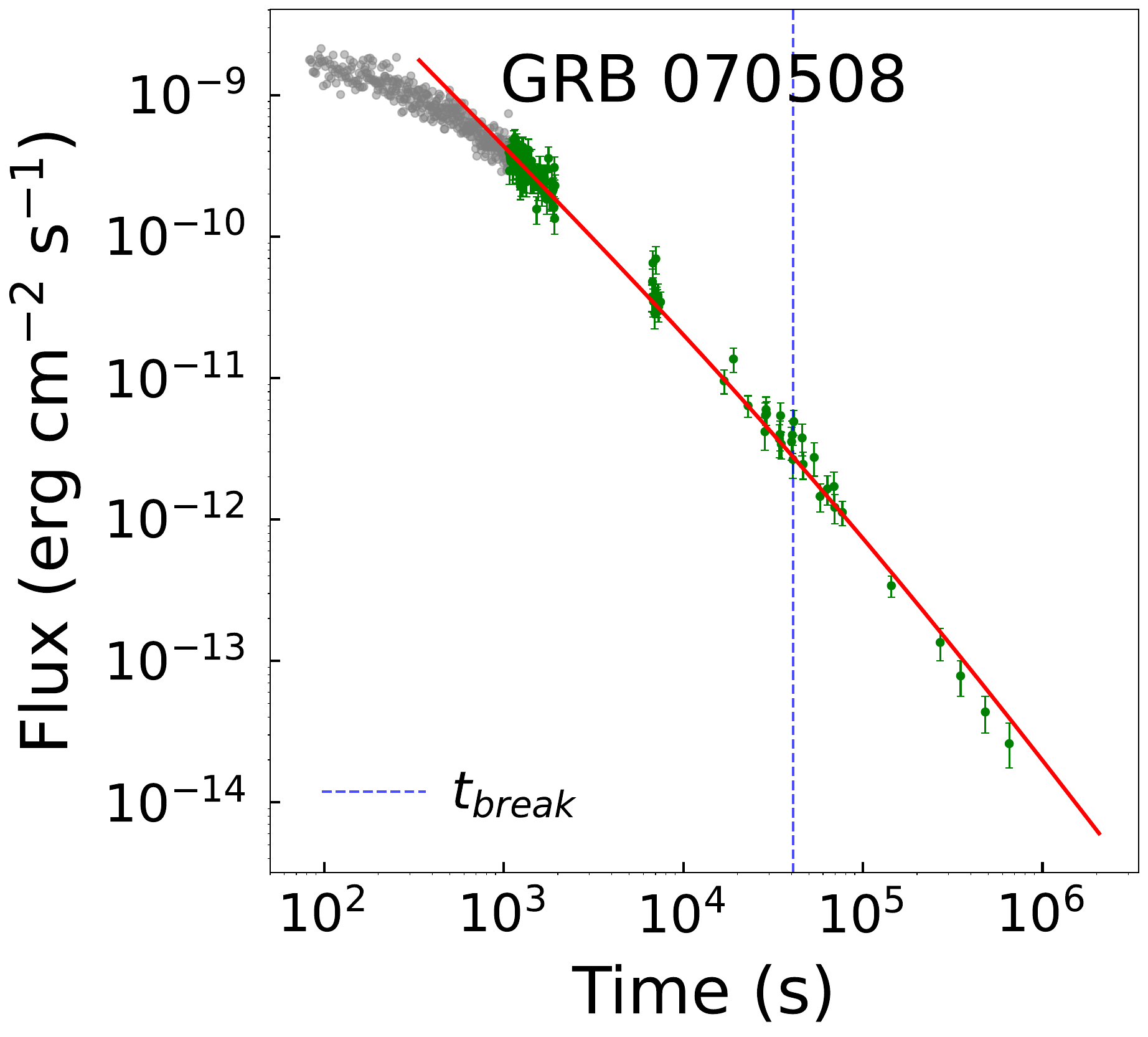}
  \end{subfigure}%
  \hspace{0.01\textwidth}
  \begin{subfigure}[b]{0.23\textwidth}
    \includegraphics[width=\linewidth]{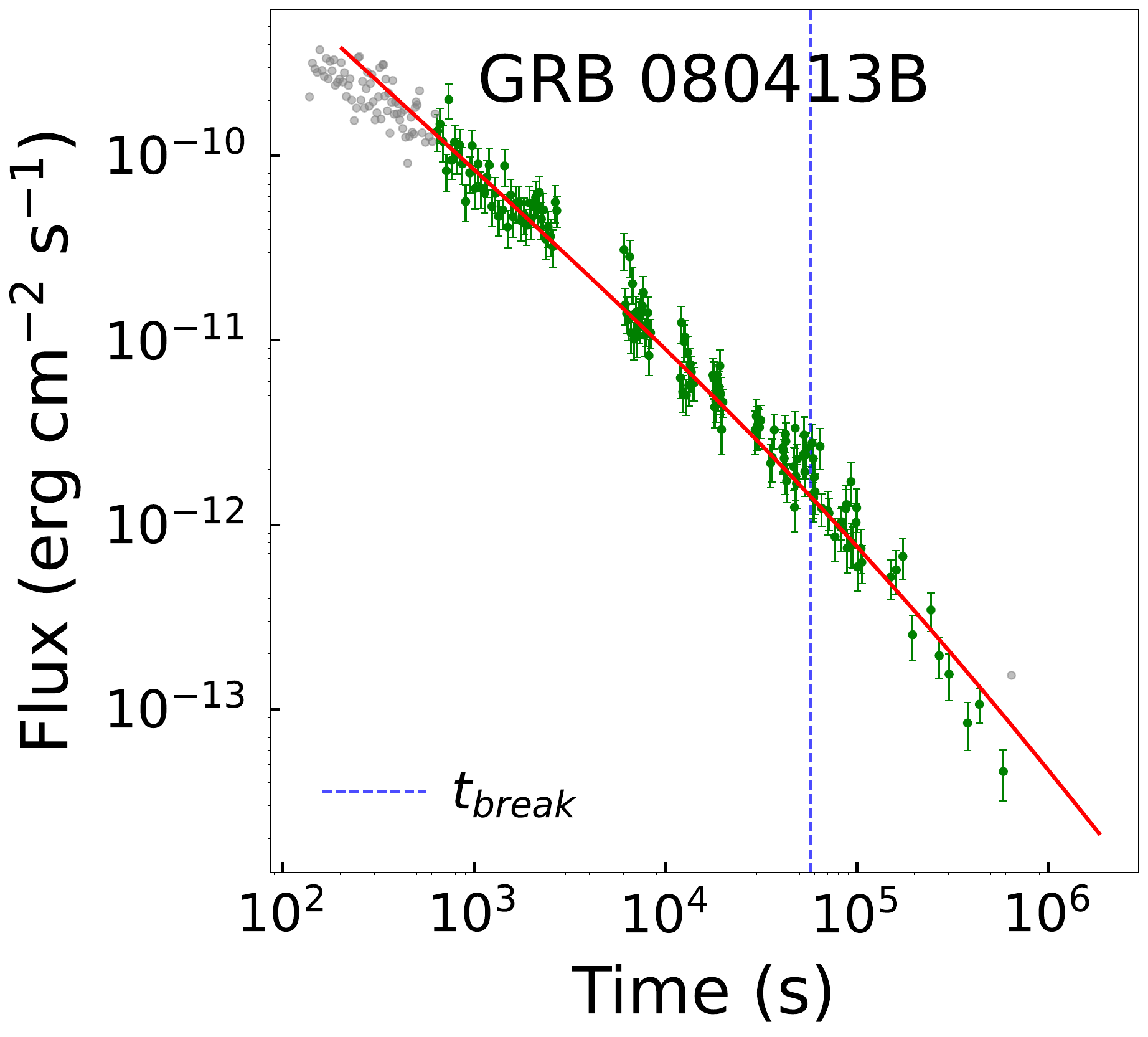}
  \end{subfigure}%
  \\%
  \begin{subfigure}[b]{0.23\textwidth}
    \includegraphics[width=\linewidth]{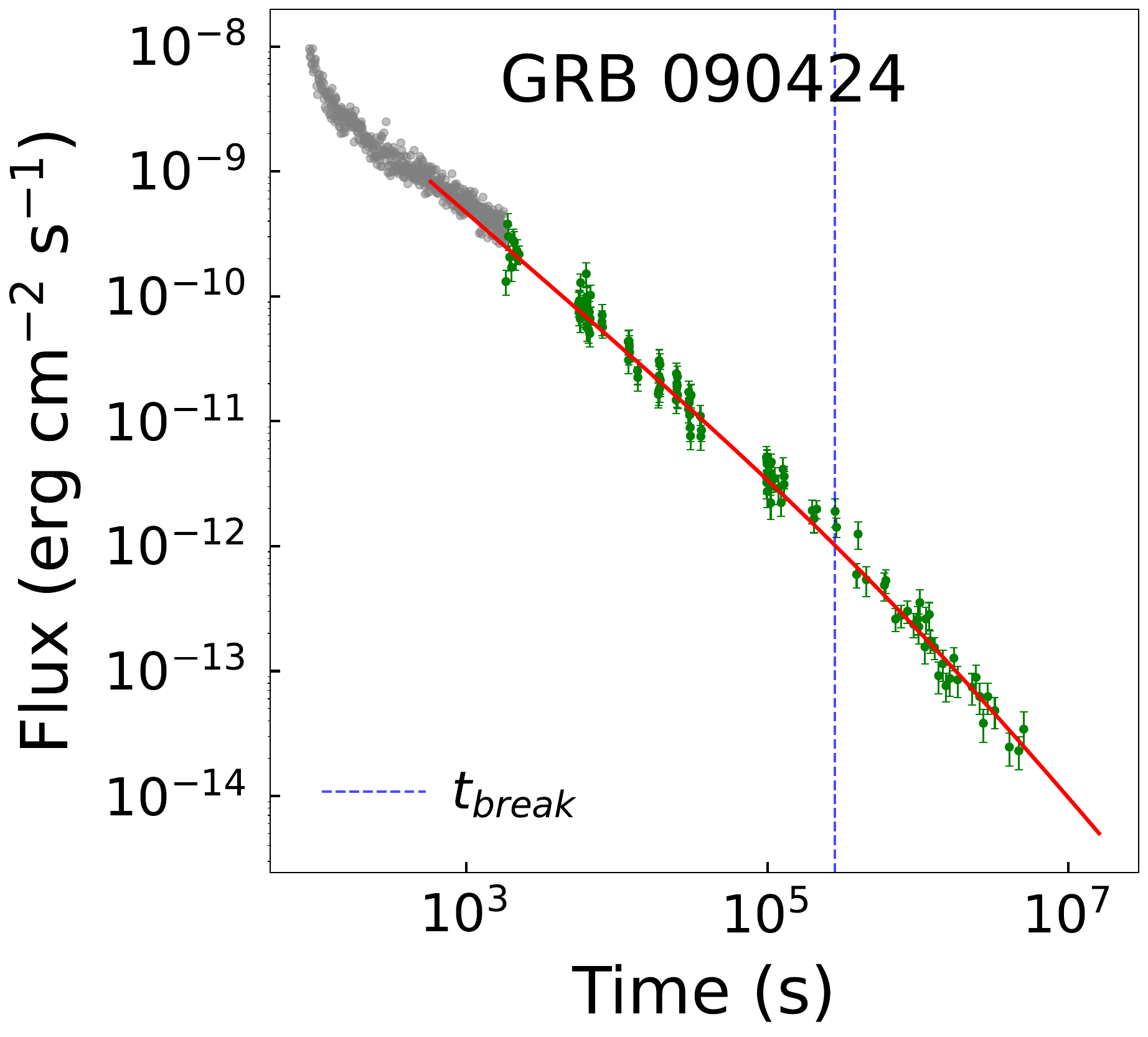}
  \end{subfigure}%
  \hspace{0.01\textwidth}
  \begin{subfigure}[b]{0.23\textwidth}
    \includegraphics[width=\linewidth]{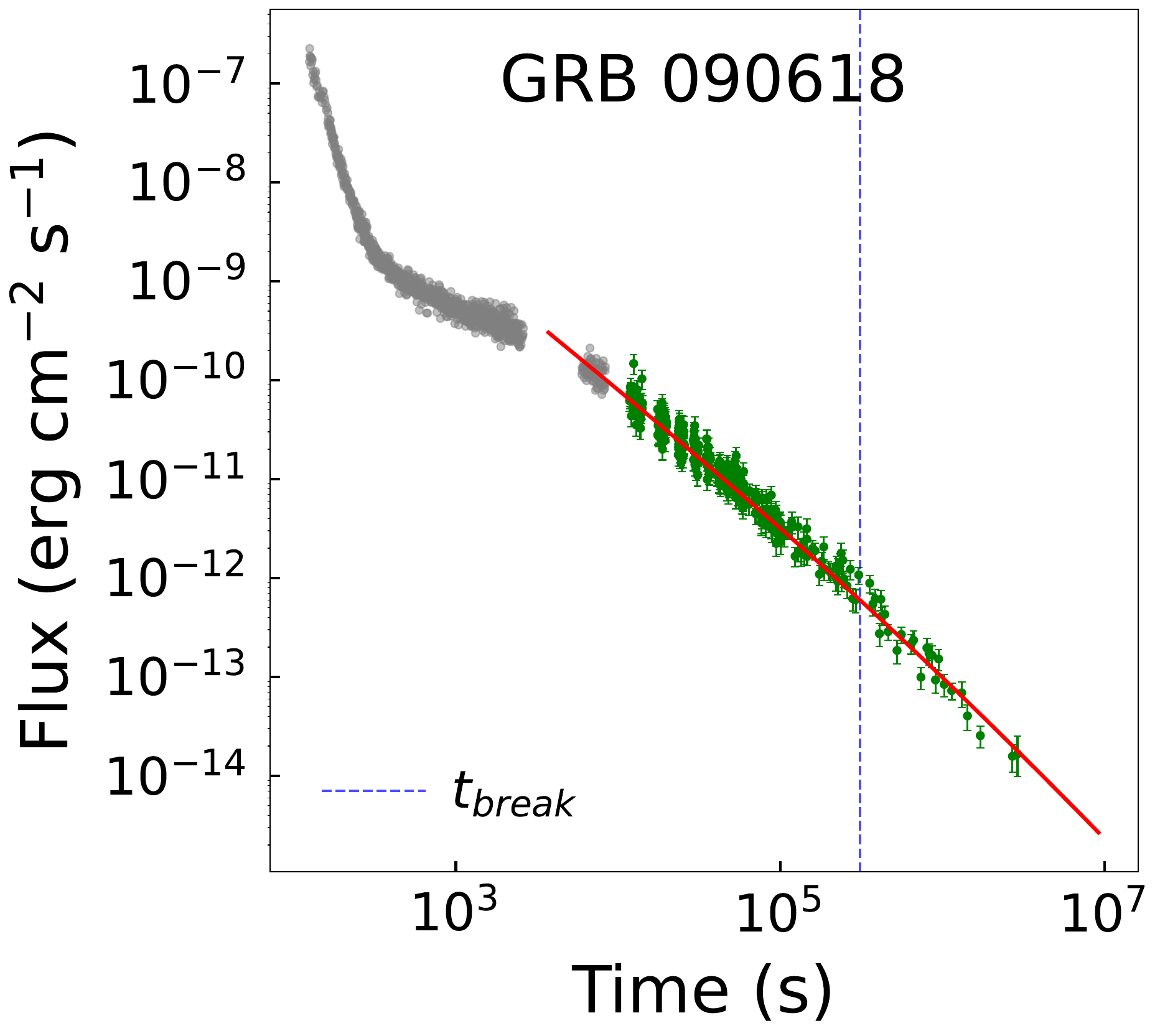}
  \end{subfigure}%
  \hspace{0.01\textwidth}
  \begin{subfigure}[b]{0.23\textwidth}
    \includegraphics[width=\linewidth]{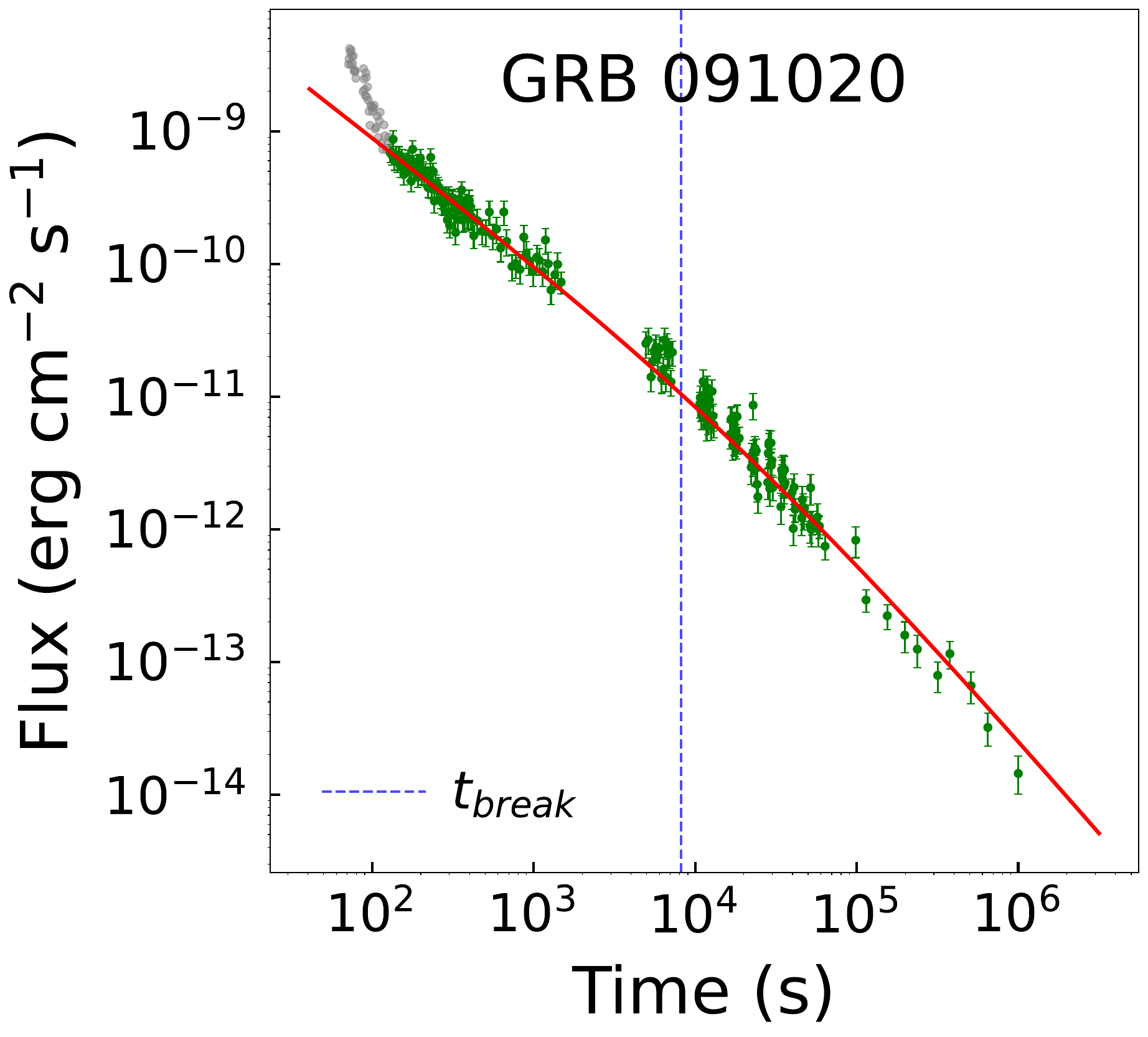}
  \end{subfigure}%
  \hspace{0.01\textwidth}
  \begin{subfigure}[b]{0.23\textwidth}
    \includegraphics[width=\linewidth]{result_091029_20_flux_vs_time.pdf}
  \end{subfigure}%
  \\%
  \begin{subfigure}[b]{0.23\textwidth}
    \includegraphics[width=\linewidth]{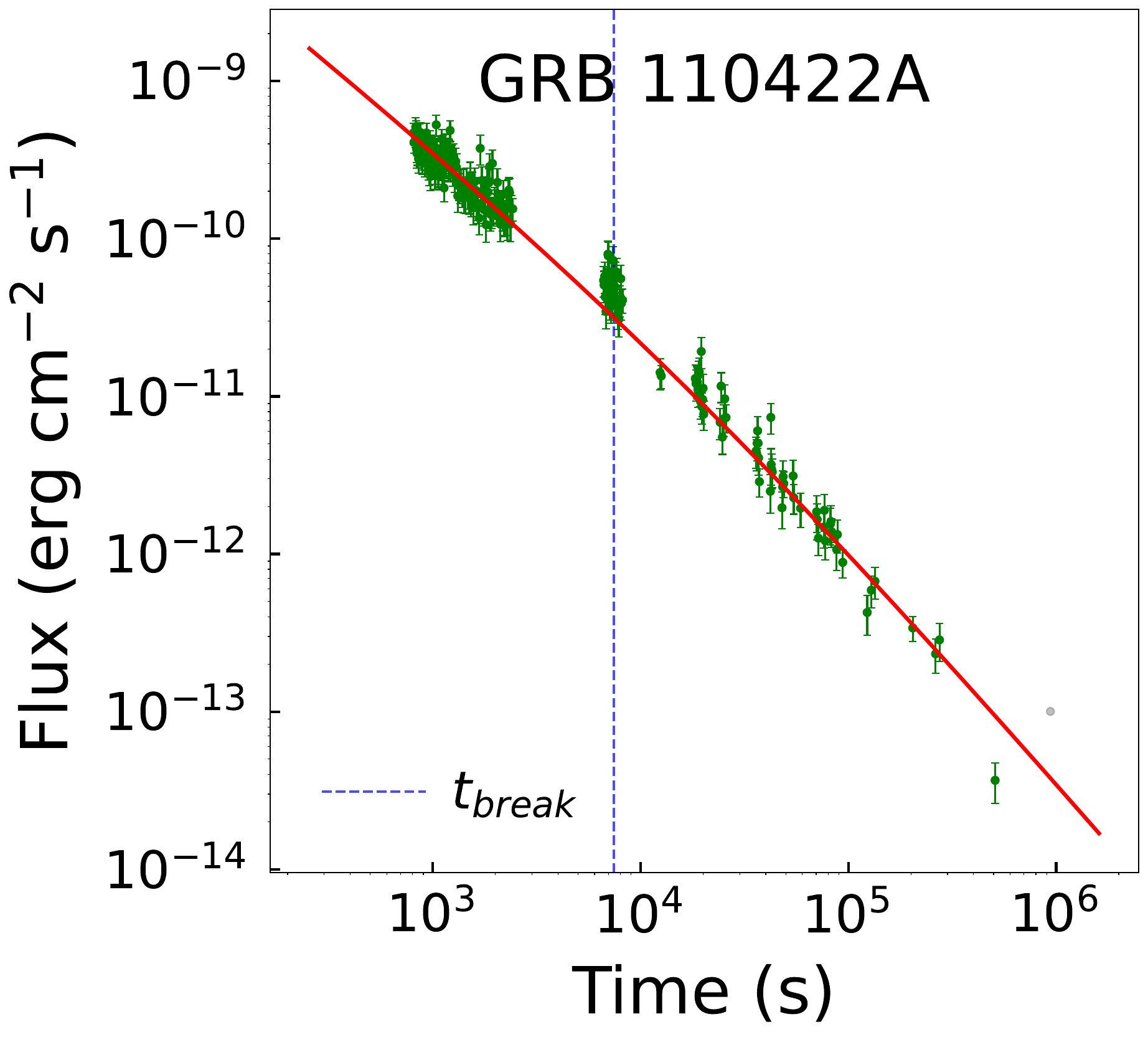}
  \end{subfigure}%
  \hspace{0.01\textwidth}
  \begin{subfigure}[b]{0.23\textwidth}
    \includegraphics[width=\linewidth]{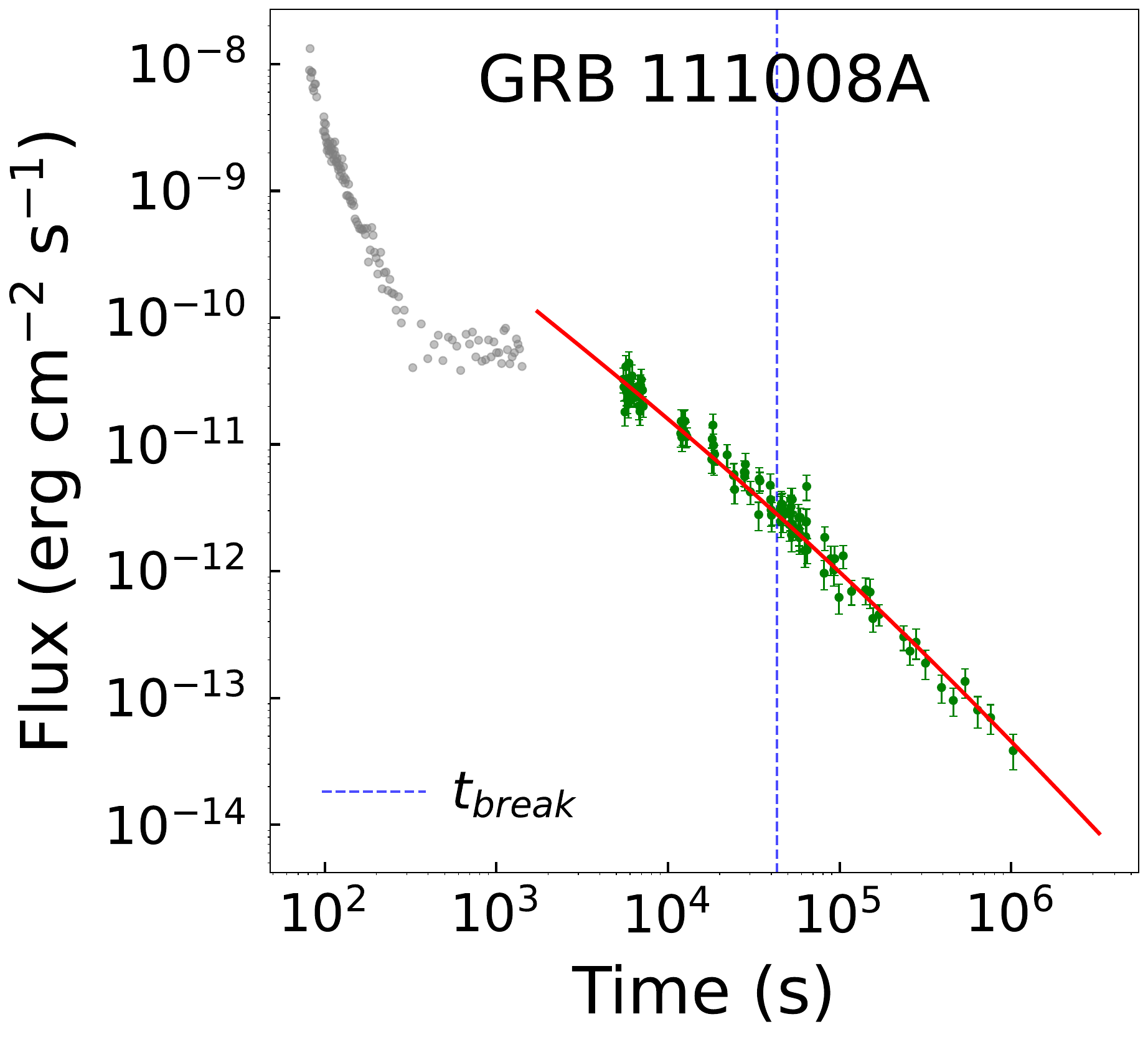}
  \end{subfigure}%
  \hspace{0.01\textwidth}
  \begin{subfigure}[b]{0.23\textwidth}
    \includegraphics[width=\linewidth]{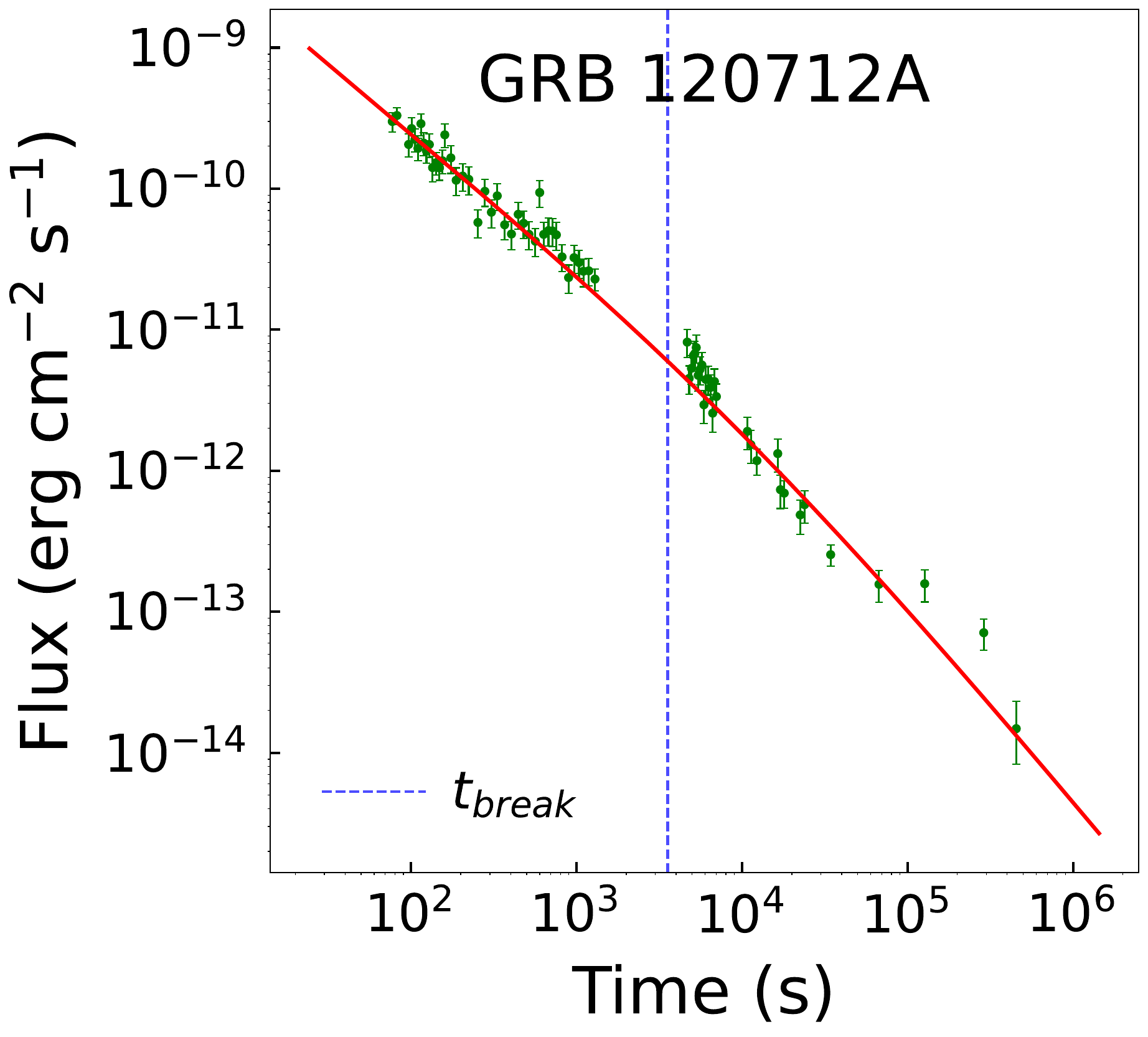}
  \end{subfigure}%
  \hspace{0.01\textwidth}
  \begin{subfigure}[b]{0.23\textwidth}
    \includegraphics[width=\linewidth]{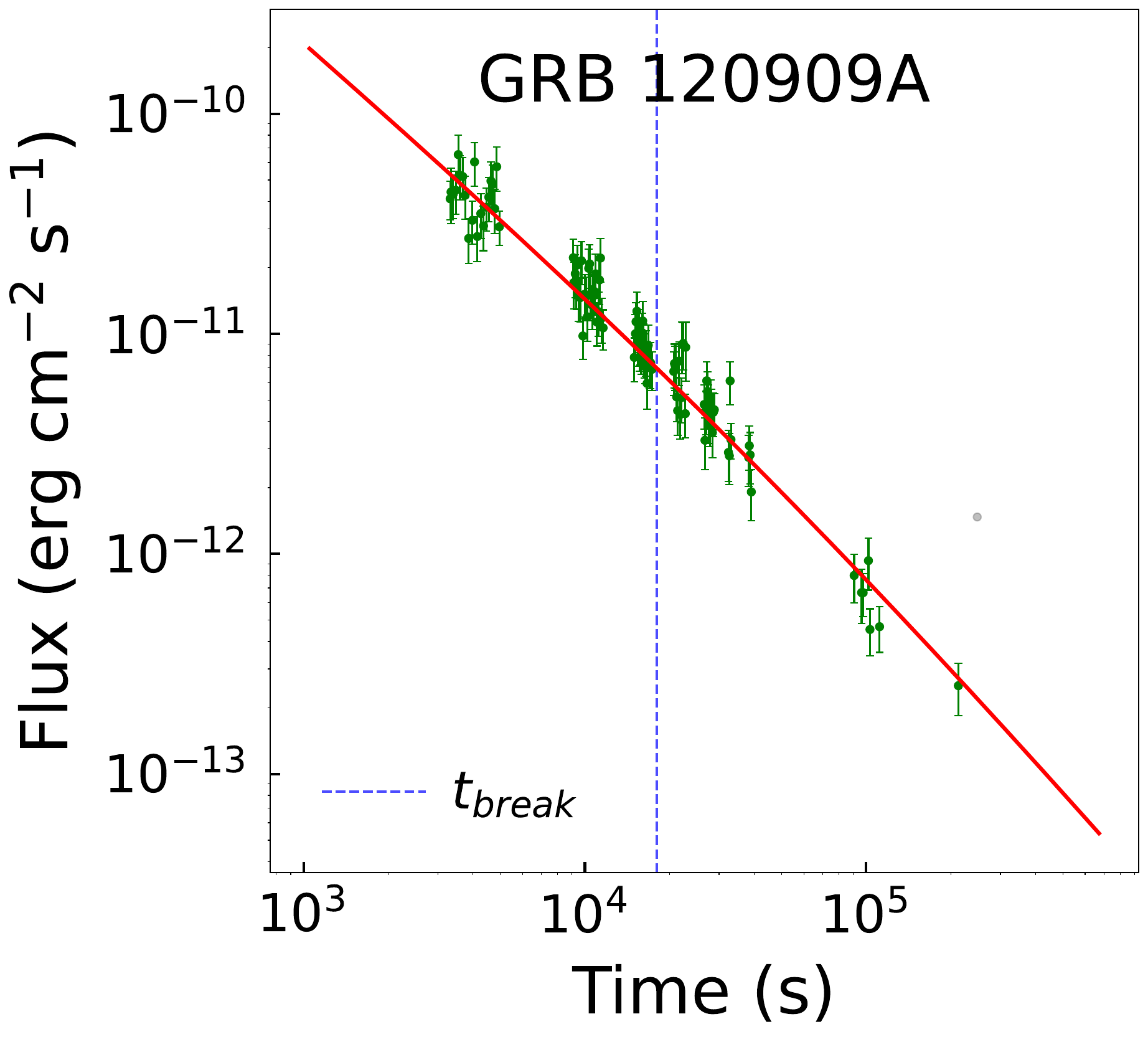}
  \end{subfigure}%
  \\%
  \begin{subfigure}[b]{0.23\textwidth}
    \includegraphics[width=\linewidth]{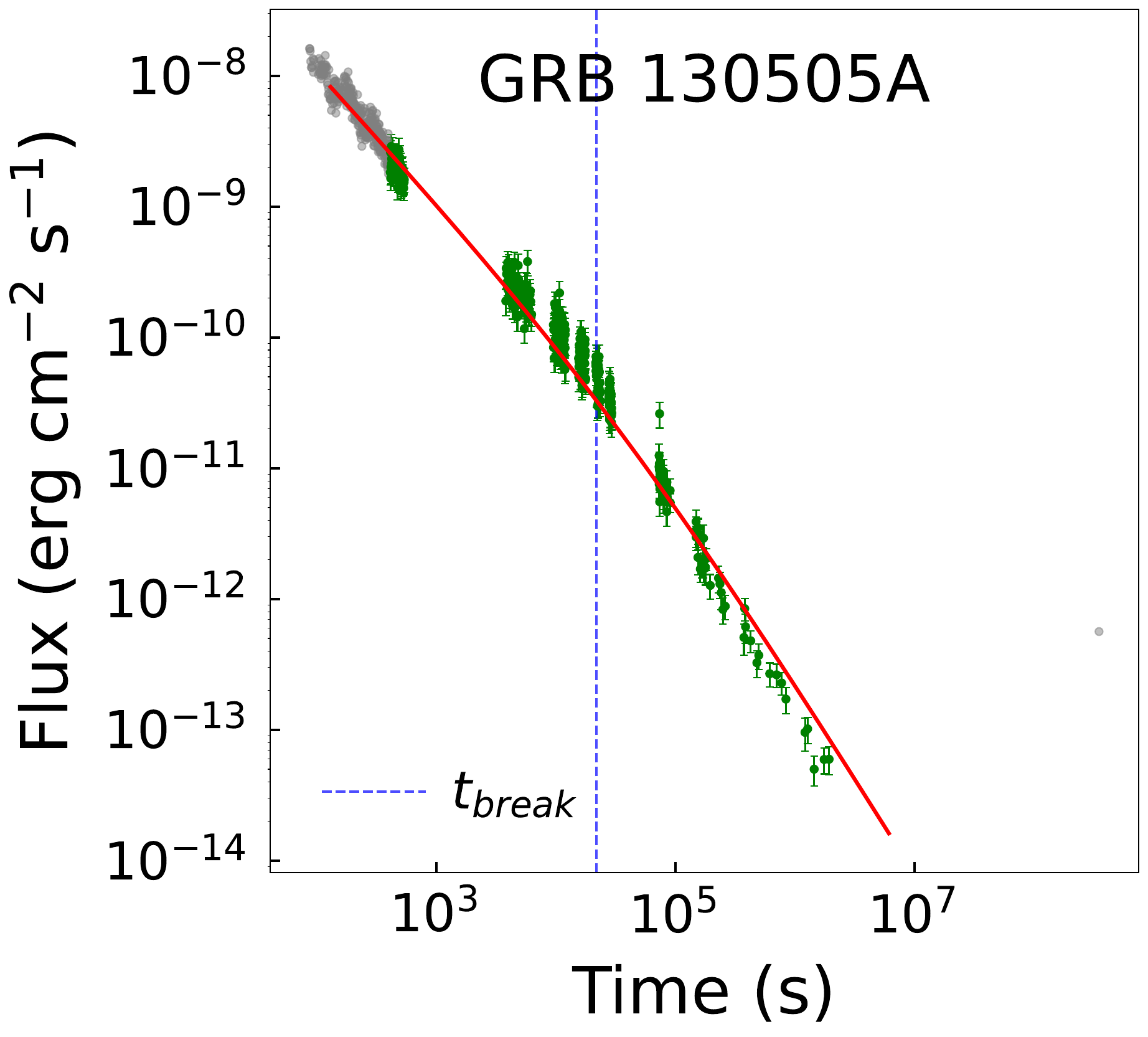}
  \end{subfigure}%
  \hspace{0.01\textwidth}
  \begin{subfigure}[b]{0.23\textwidth}
    \includegraphics[width=\linewidth]{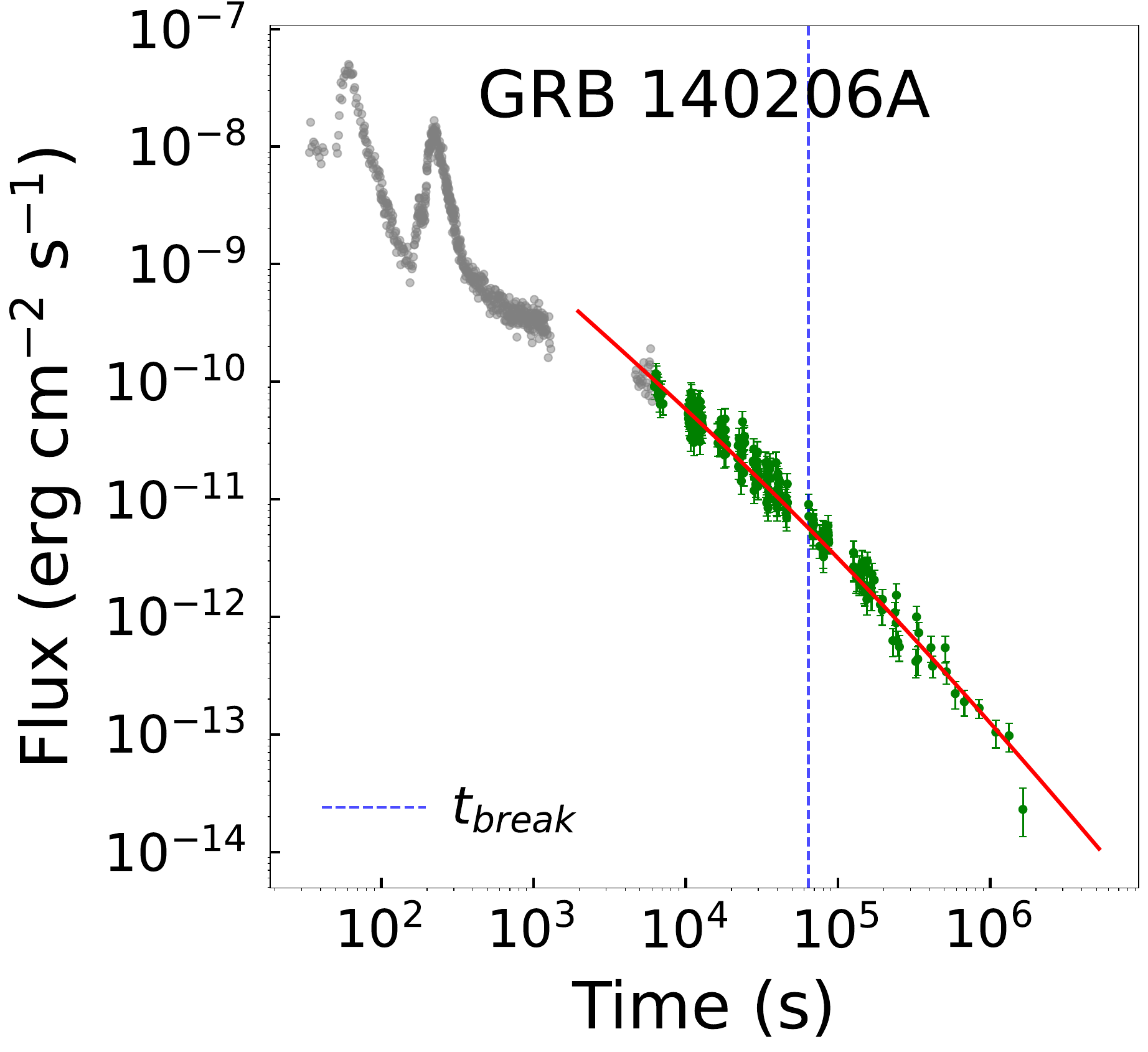}
  \end{subfigure}%
  \hspace{0.01\textwidth}
  \begin{subfigure}[b]{0.23\textwidth}
    \includegraphics[width=\linewidth]{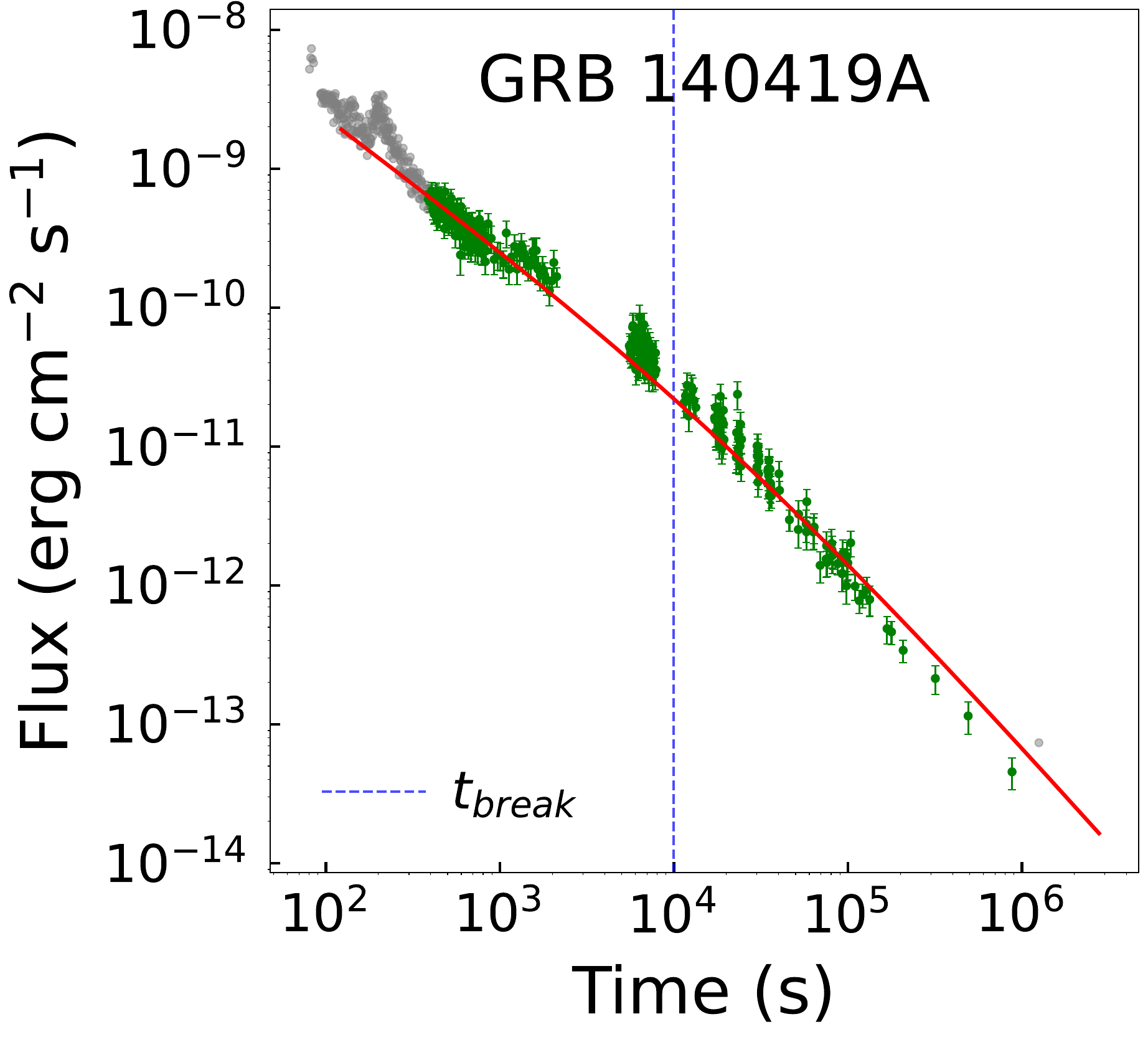}
  \end{subfigure}%
  \hspace{0.01\textwidth}
  \begin{subfigure}[b]{0.23\textwidth}
    \includegraphics[width=\linewidth]{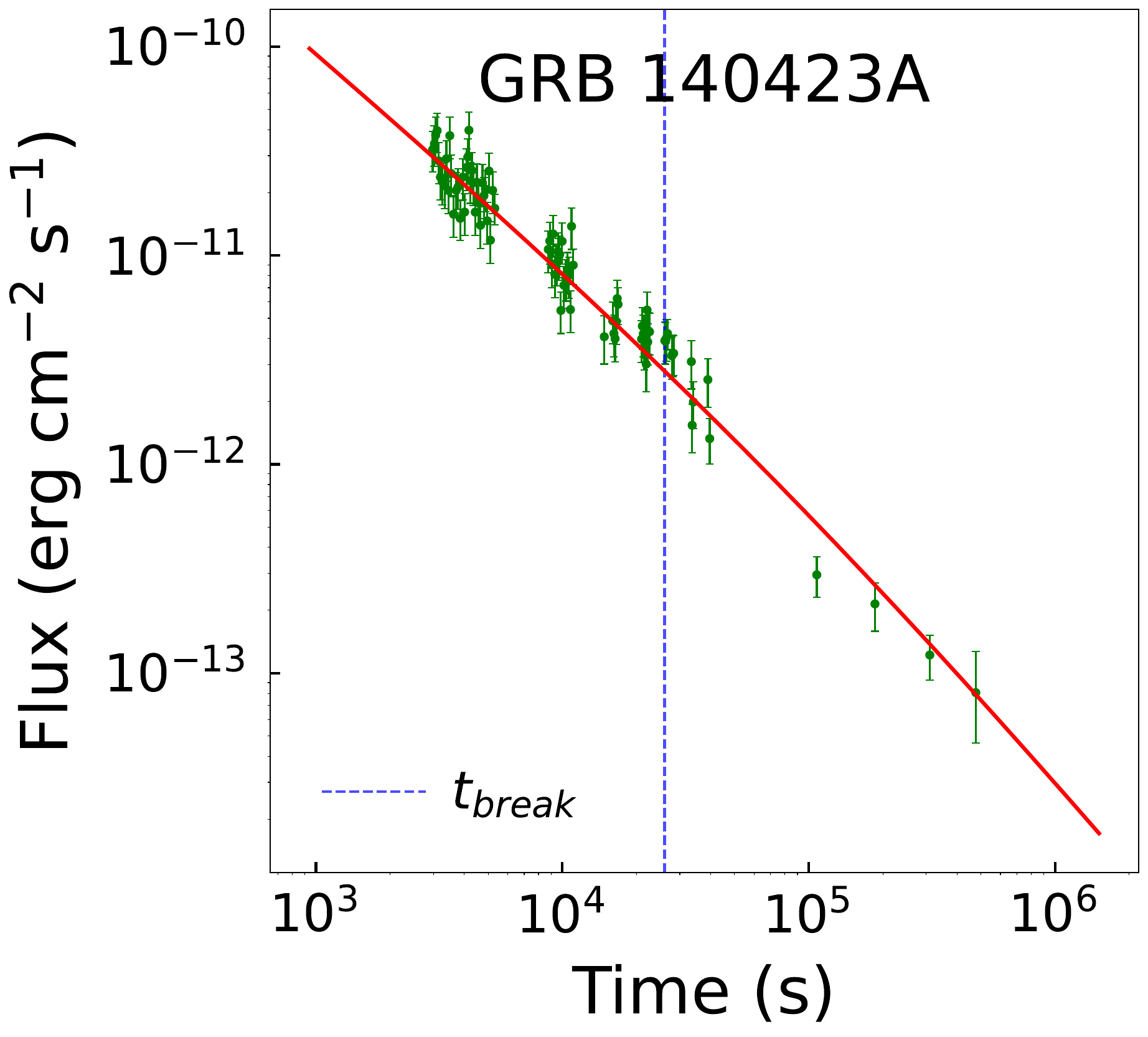}
  \end{subfigure}%
  \\%
  \begin{subfigure}[b]{0.23\textwidth}
    \includegraphics[width=\linewidth]{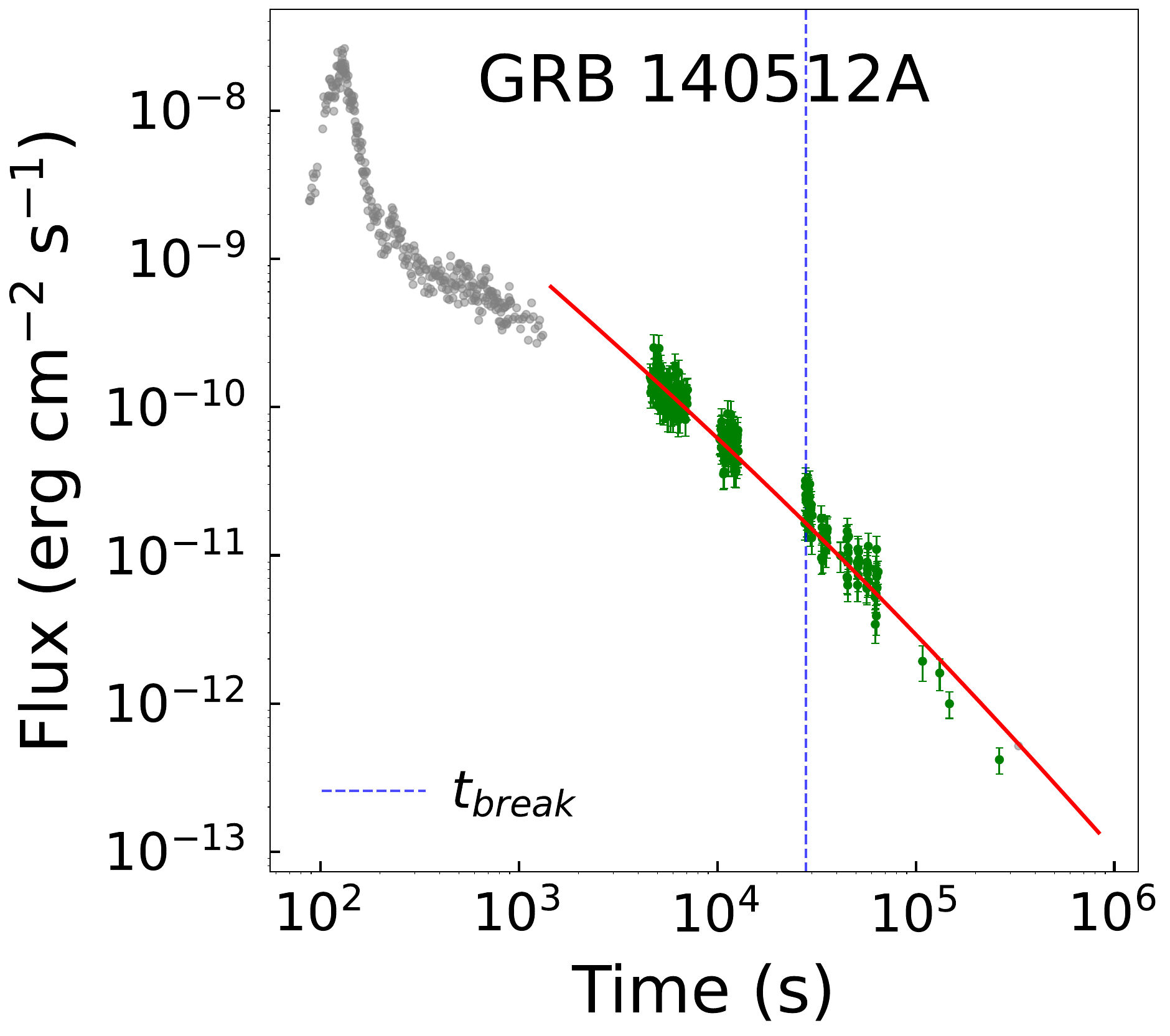}
  \end{subfigure}%
  \hspace{0.01\textwidth}
  \begin{subfigure}[b]{0.23\textwidth}
    \includegraphics[width=\linewidth]{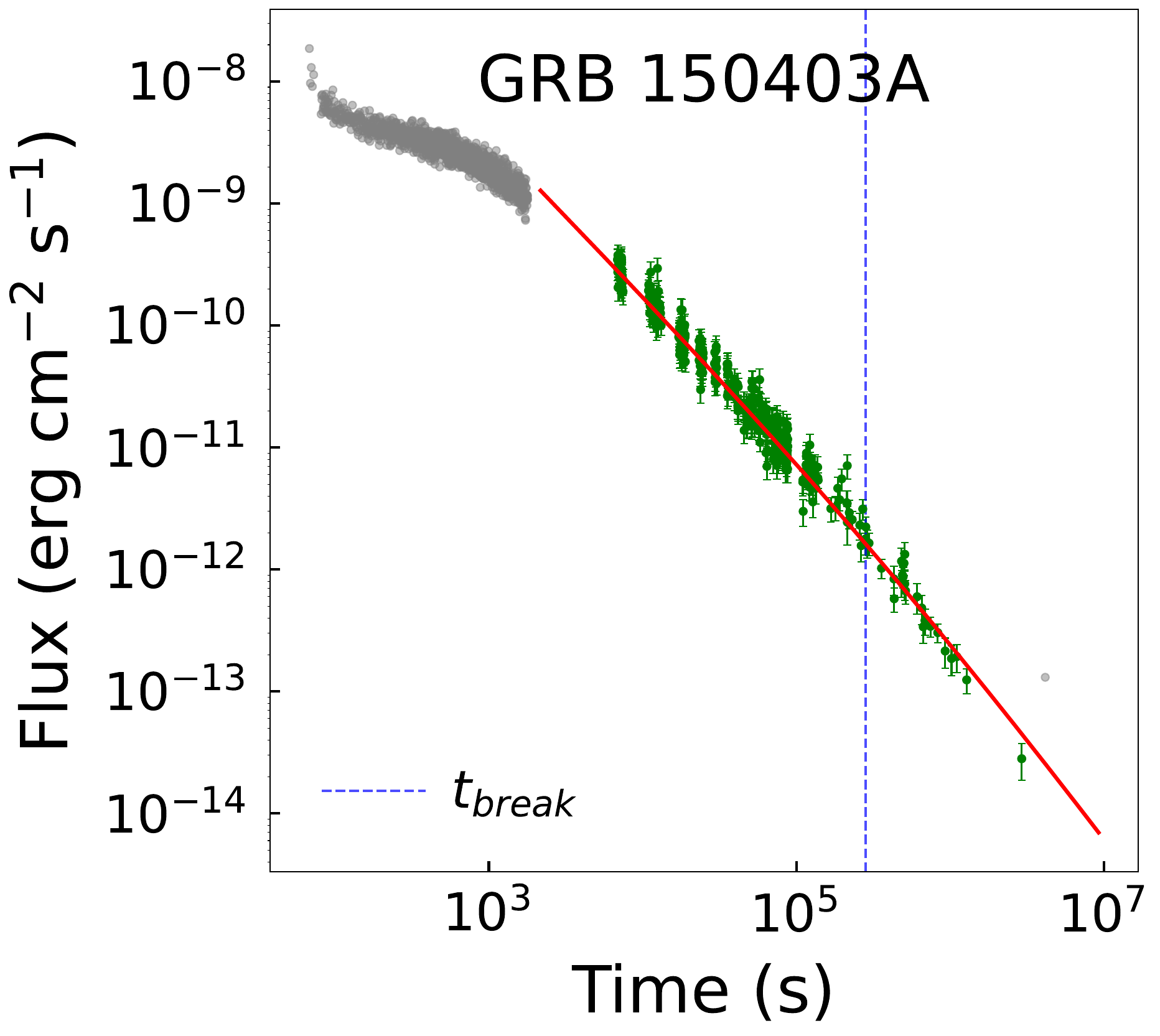}
  \end{subfigure}%
  \hspace{0.01\textwidth}
  \begin{subfigure}[b]{0.23\textwidth}
    \includegraphics[width=\linewidth]{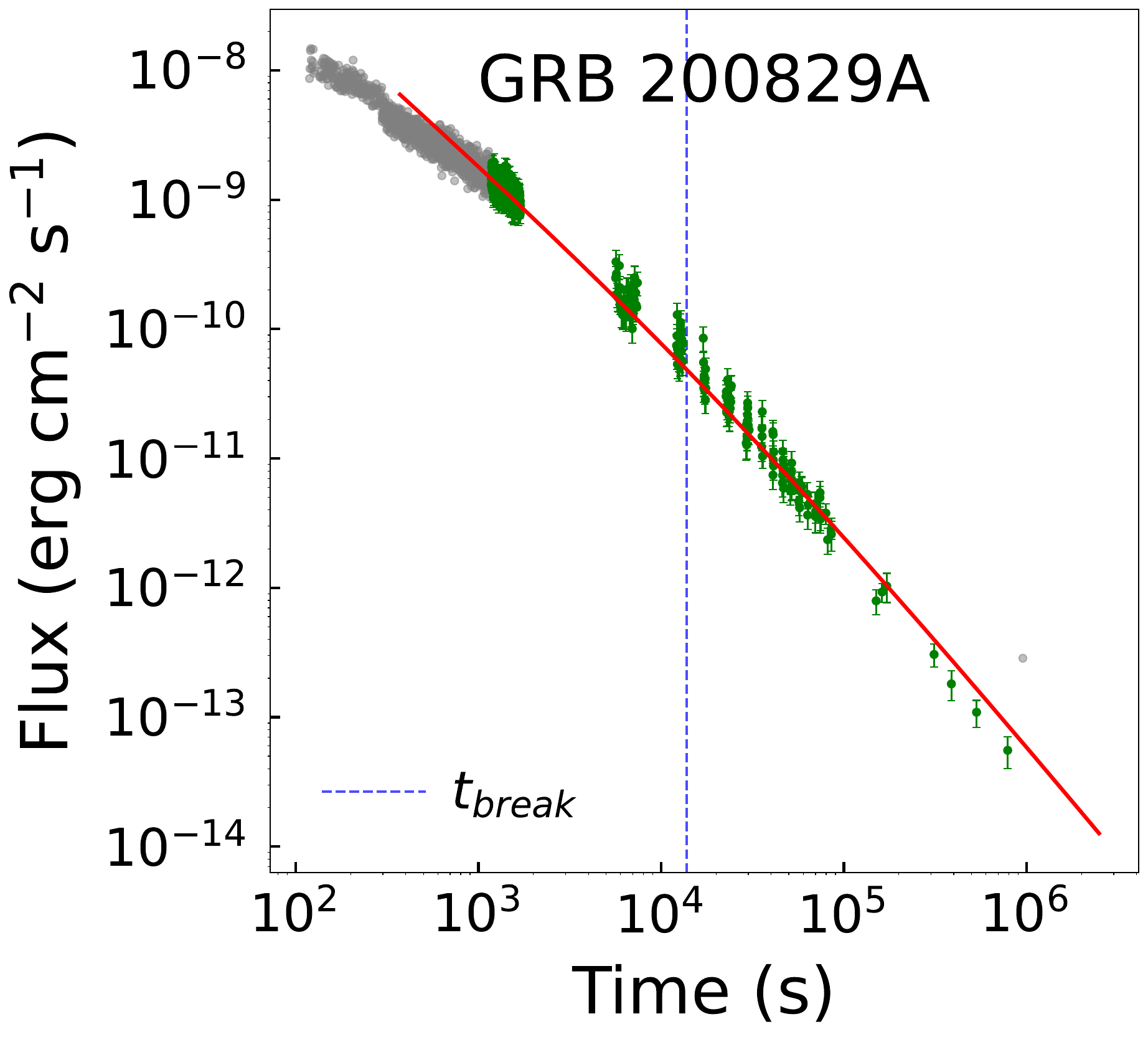}
  \end{subfigure}%
  \hspace{0.01\textwidth}
  \begin{subfigure}[b]{0.23\textwidth}
    \includegraphics[width=\linewidth]{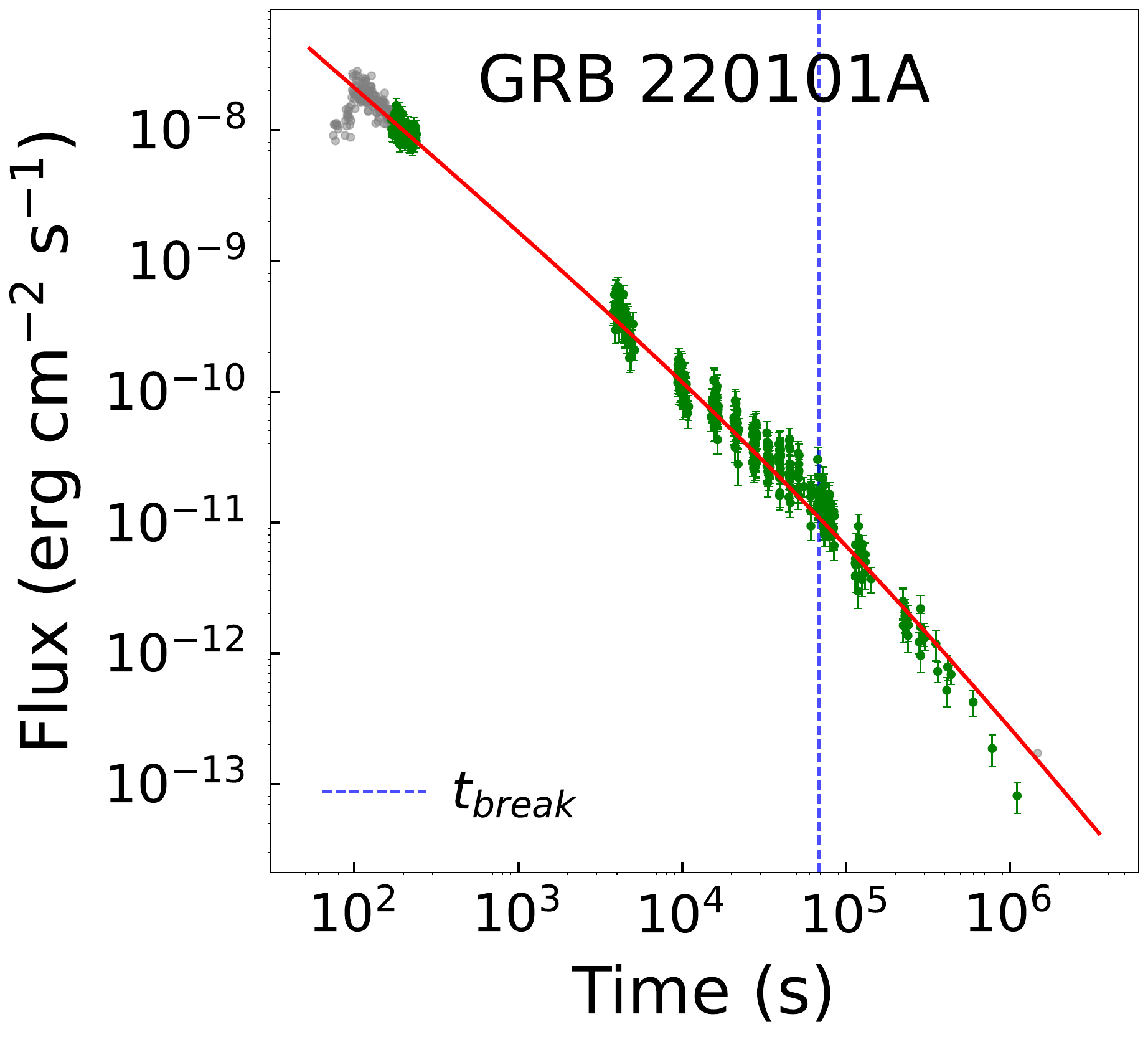}
  \end{subfigure}%
  \\%
  \caption{X-ray afterglow light curve fitting results for model~2 with high-latitude emission. Each panel shows the observed flux (data points) and the best-fit model (solid line) for individual gamma-ray bursts in the stellar wind environment, characterized by density profile $n \propto R^{-2}$.}
  \label{fig:wind_model2_light_curves}
\end{figure}

\begin{figure}[htbp]
\centering
  \begin{subfigure}[b]{0.23\textwidth}
    \includegraphics[width=\linewidth]{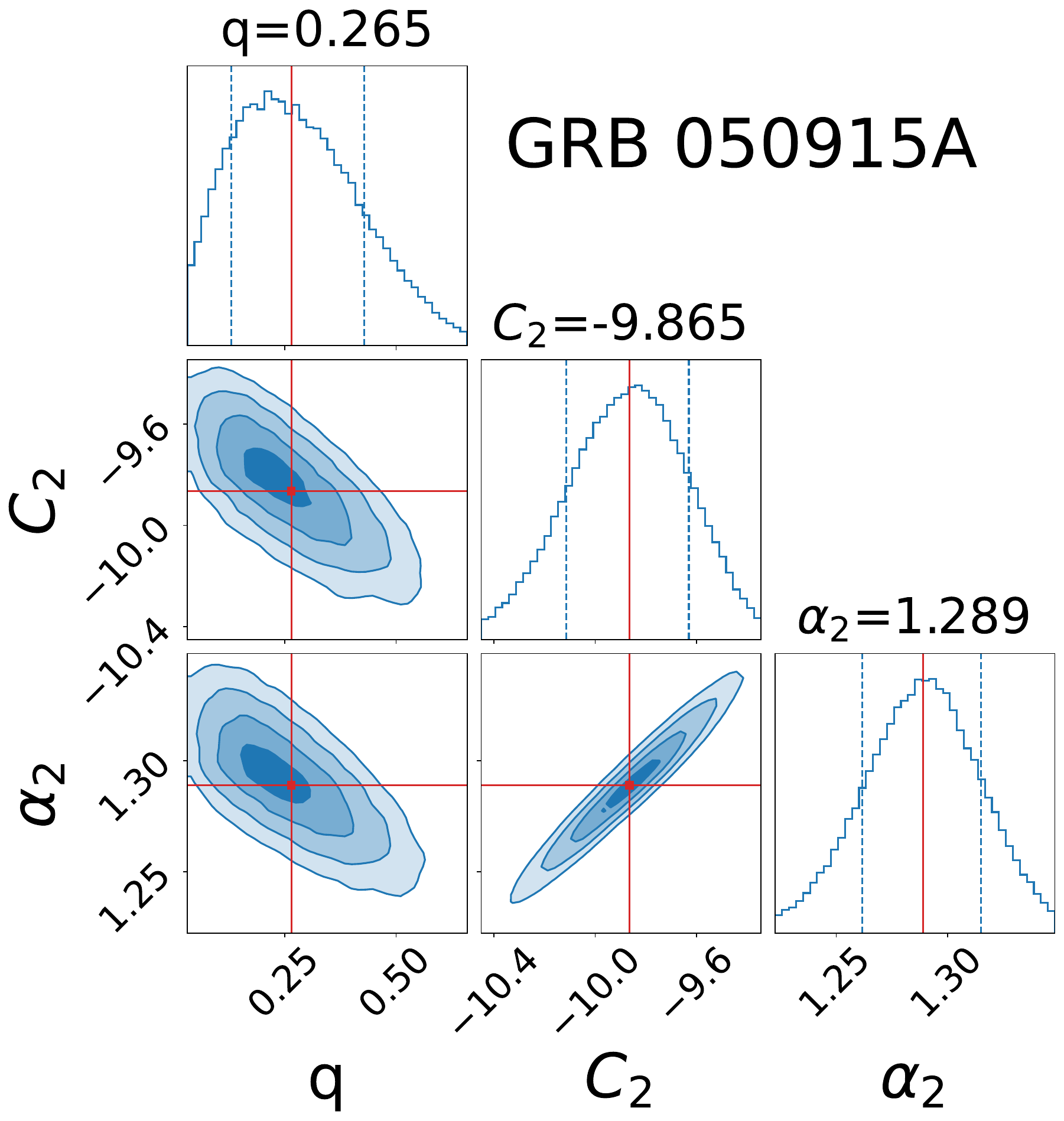}
  \end{subfigure}%
  \hspace{0.01\textwidth}
  \begin{subfigure}[b]{0.23\textwidth}
    \includegraphics[width=\linewidth]{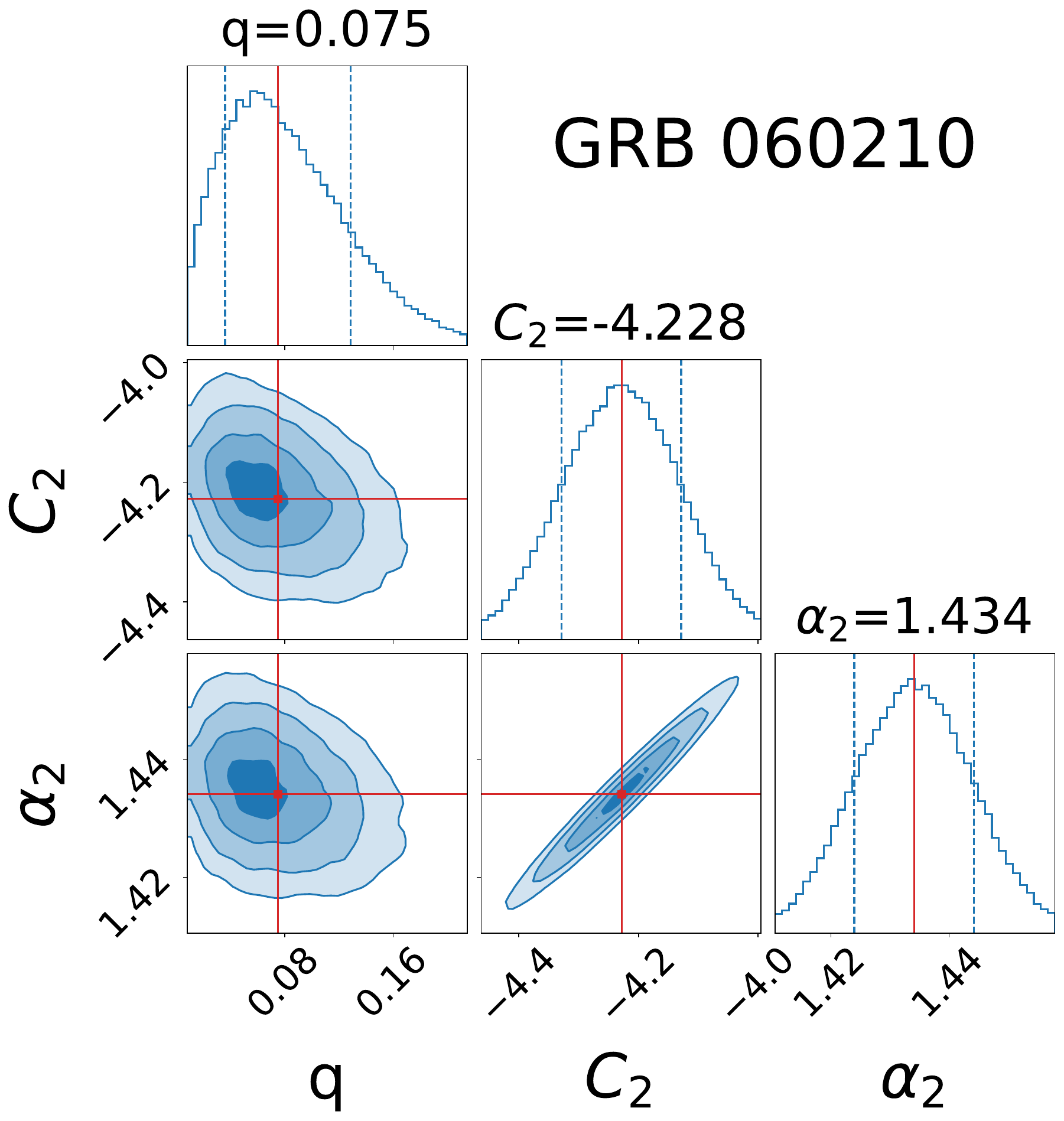}
  \end{subfigure}%
  \hspace{0.01\textwidth}
  \begin{subfigure}[b]{0.23\textwidth}
    \includegraphics[width=\linewidth]{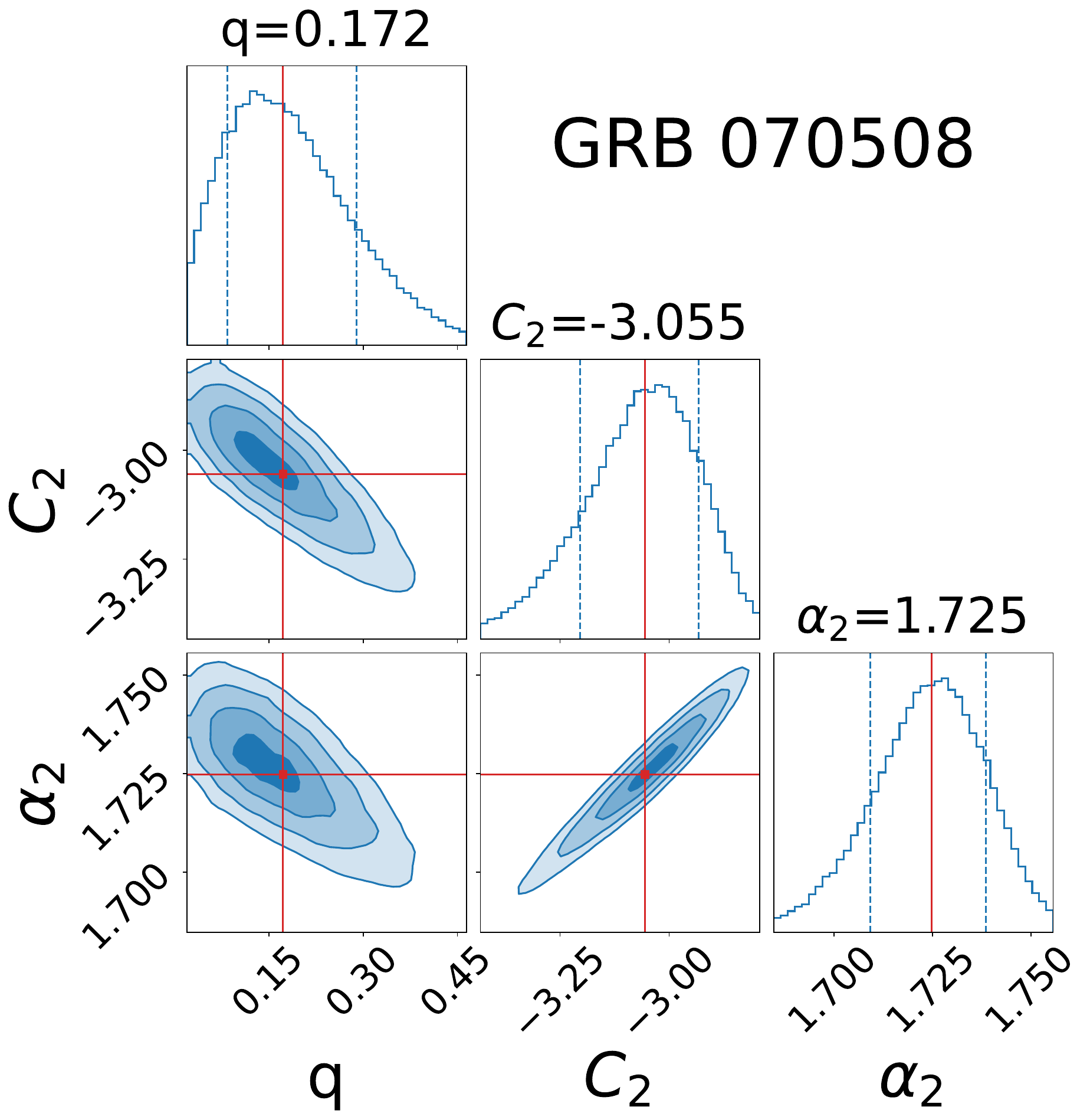}
  \end{subfigure}%
  \hspace{0.01\textwidth}
  \begin{subfigure}[b]{0.23\textwidth}
    \includegraphics[width=\linewidth]{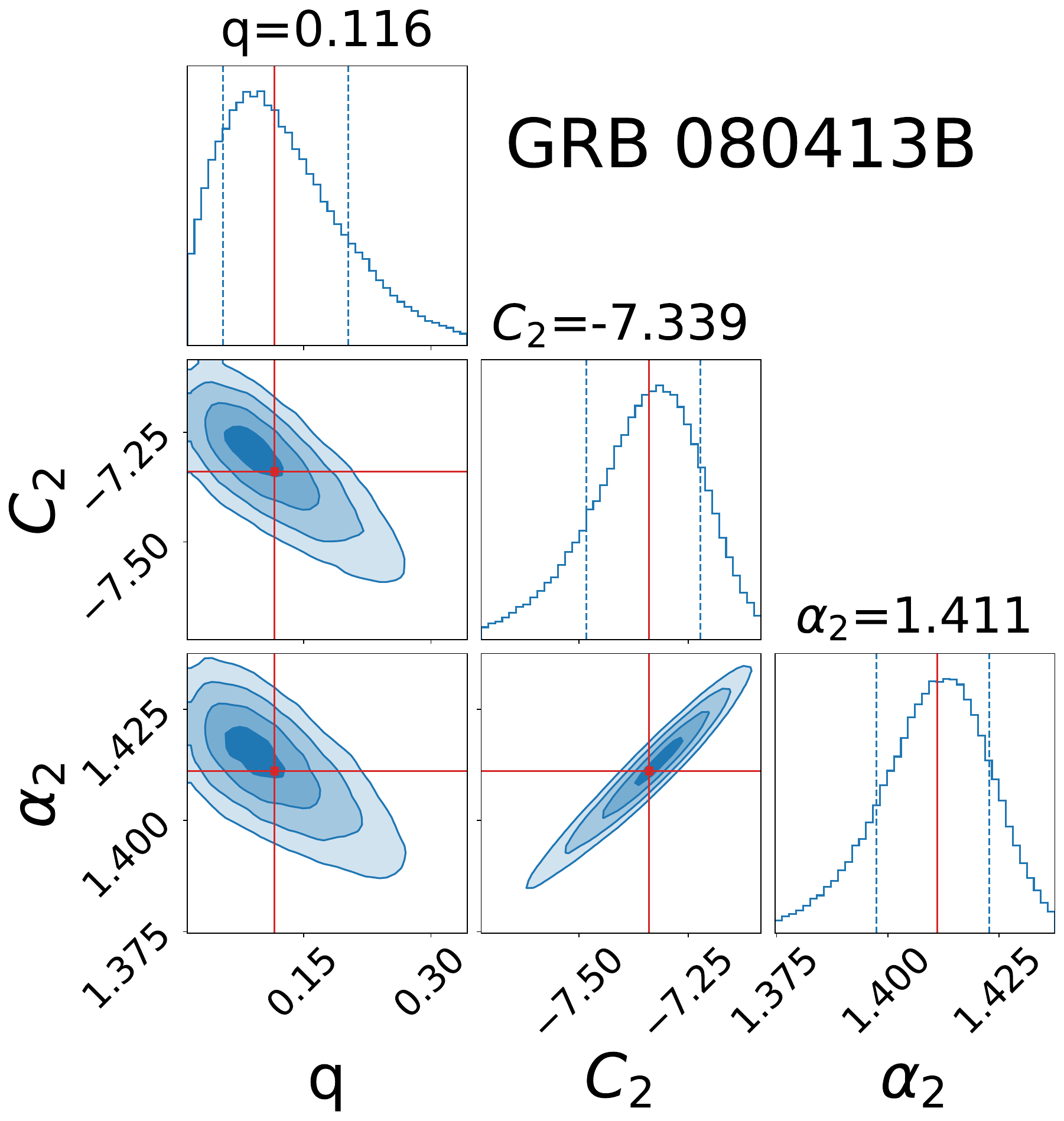}
  \end{subfigure}%
  \\%
  \begin{subfigure}[b]{0.23\textwidth}
    \includegraphics[width=\linewidth]{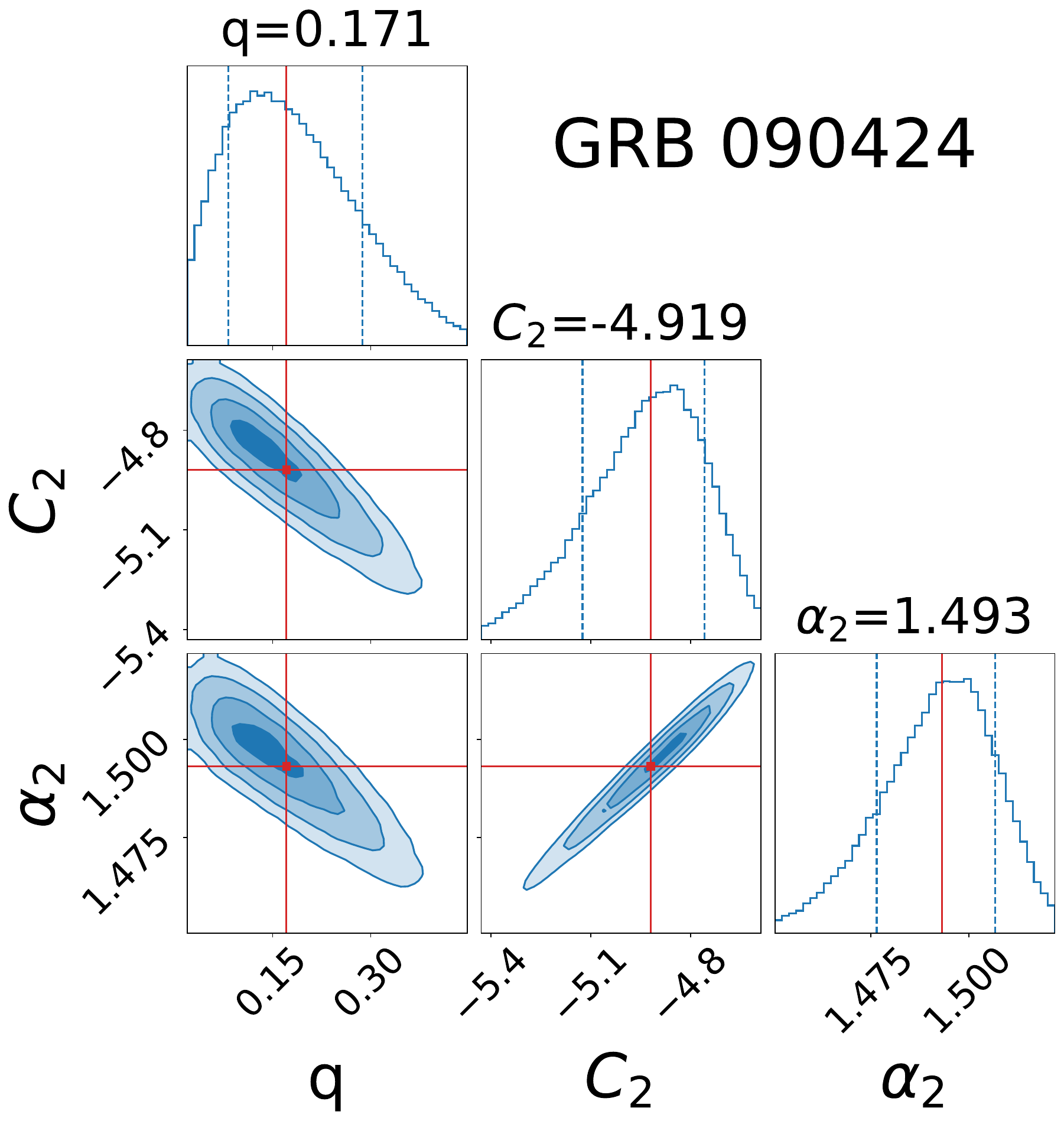}
  \end{subfigure}%
  \hspace{0.01\textwidth}
  \begin{subfigure}[b]{0.23\textwidth}
    \includegraphics[width=\linewidth]{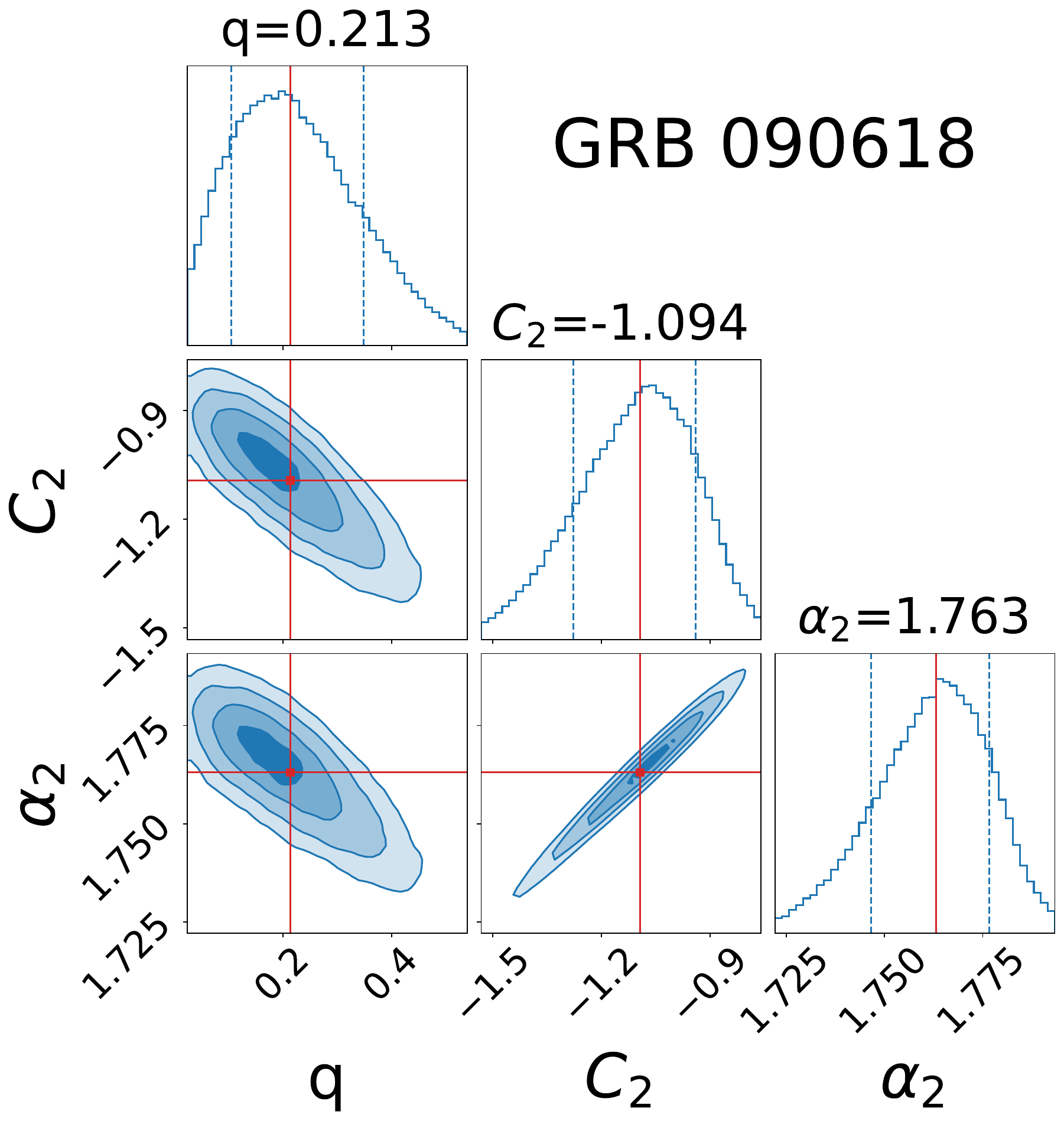}
  \end{subfigure}%
  \hspace{0.01\textwidth}
  \begin{subfigure}[b]{0.23\textwidth}
    \includegraphics[width=\linewidth]{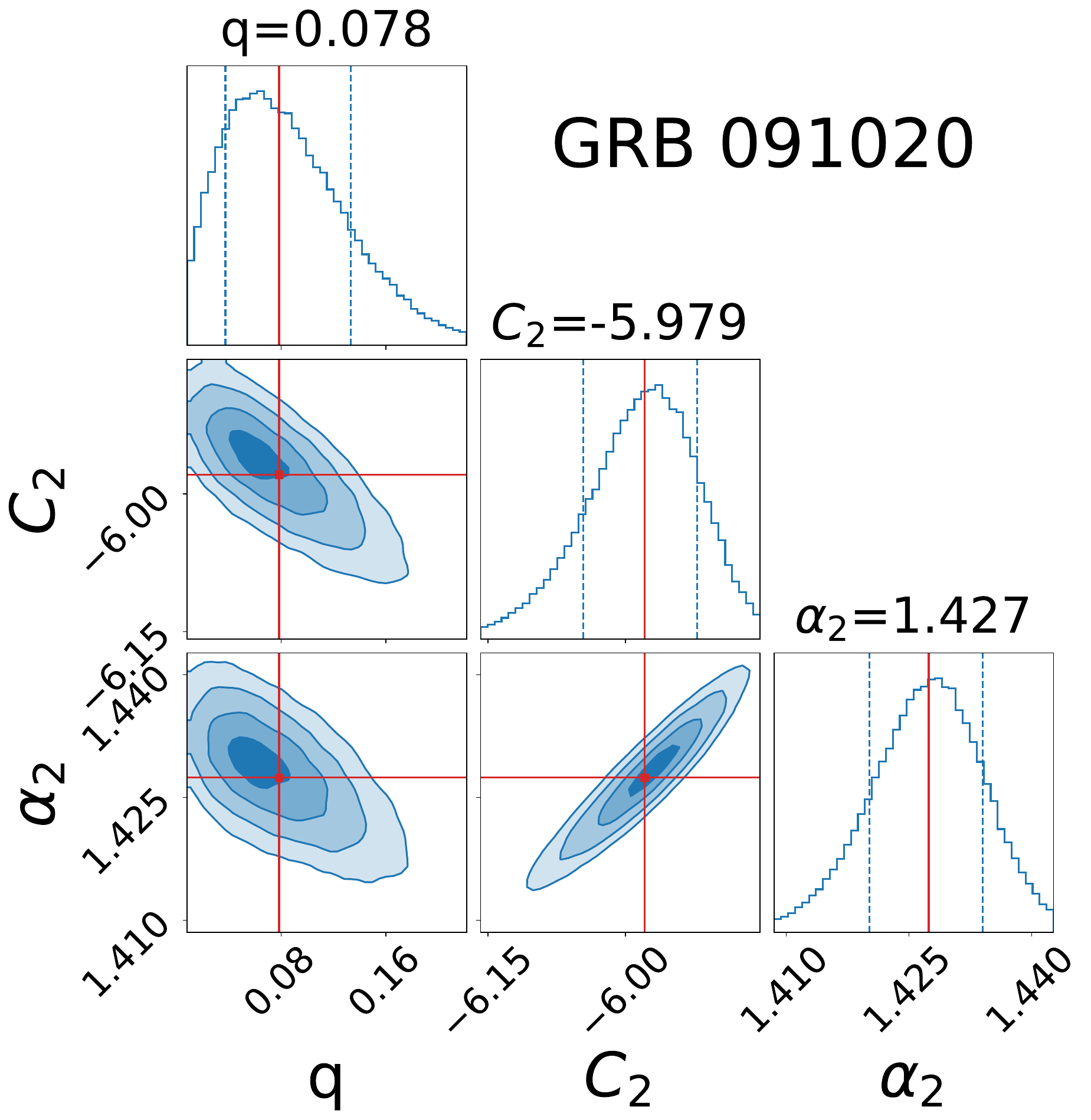}
  \end{subfigure}%
  \hspace{0.01\textwidth}
  \begin{subfigure}[b]{0.23\textwidth}
    \includegraphics[width=\linewidth]{result_091029_20_parameter_distributions.pdf}
  \end{subfigure}%
  \\%
  \begin{subfigure}[b]{0.23\textwidth}
    \includegraphics[width=\linewidth]{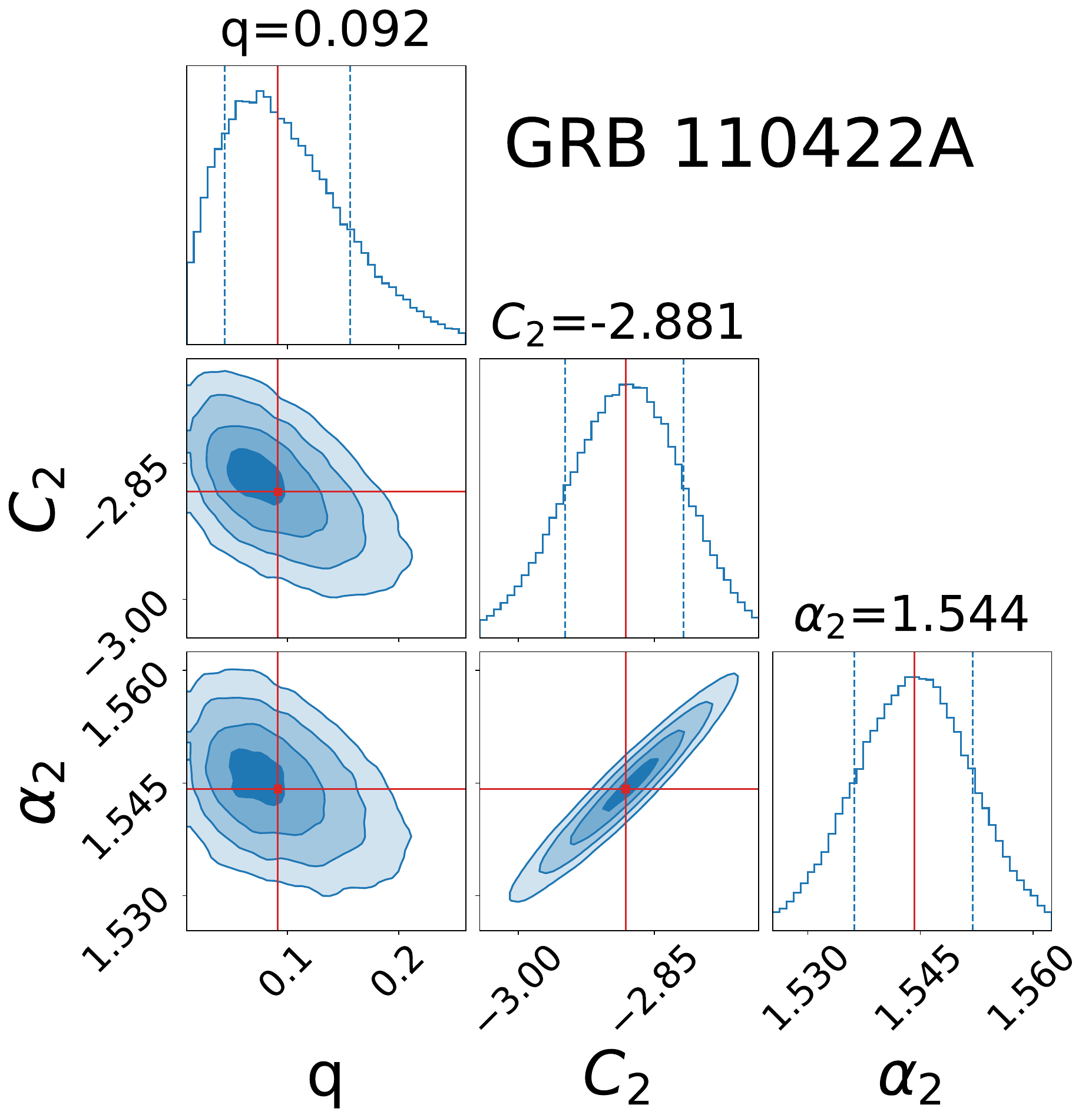}
  \end{subfigure}%
  \hspace{0.01\textwidth}
  \begin{subfigure}[b]{0.23\textwidth}
    \includegraphics[width=\linewidth]{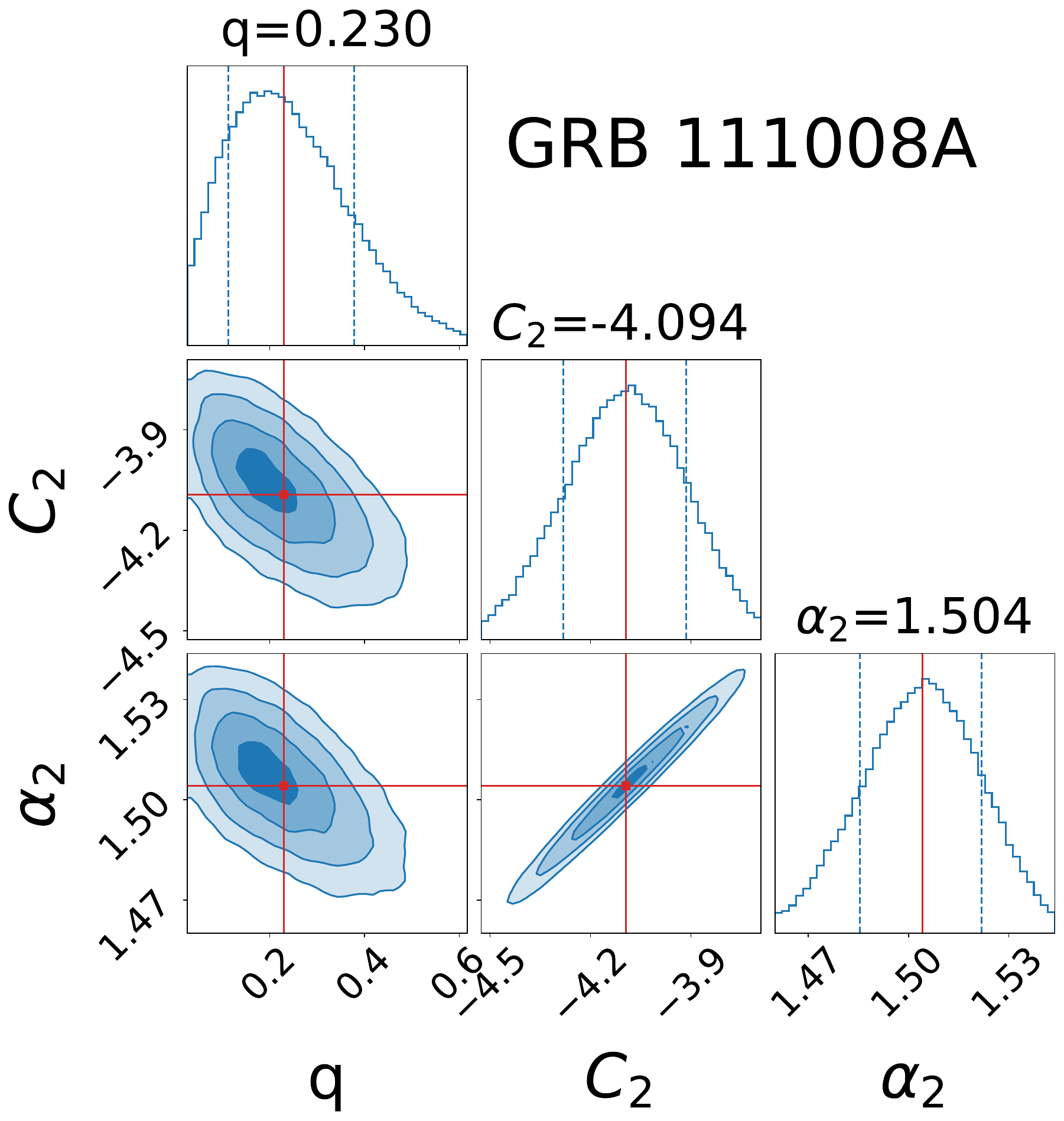}
  \end{subfigure}%
  \hspace{0.01\textwidth}
  \begin{subfigure}[b]{0.23\textwidth}
    \includegraphics[width=\linewidth]{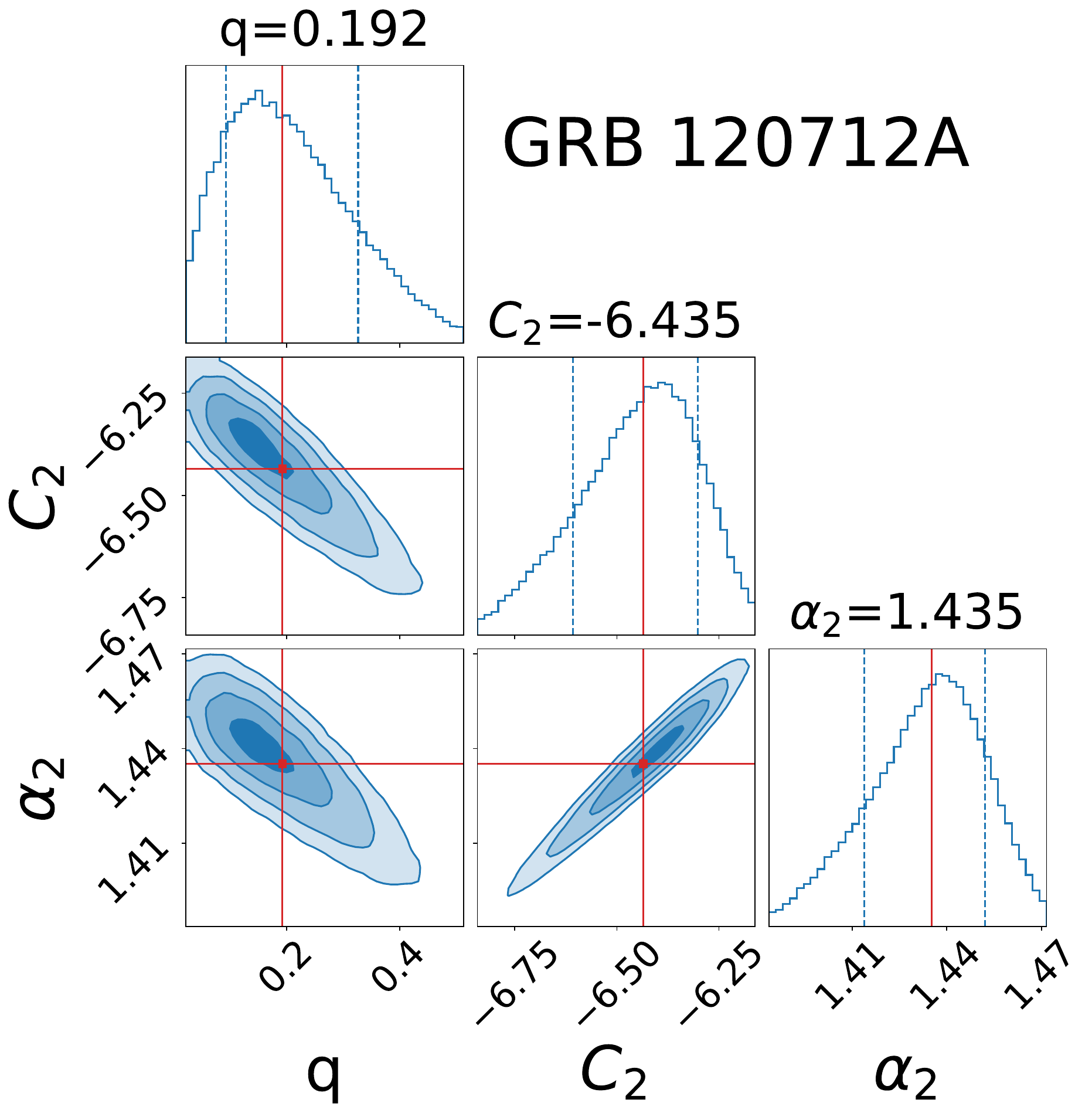}
  \end{subfigure}%
  \hspace{0.01\textwidth}
  \begin{subfigure}[b]{0.23\textwidth}
    \includegraphics[width=\linewidth]{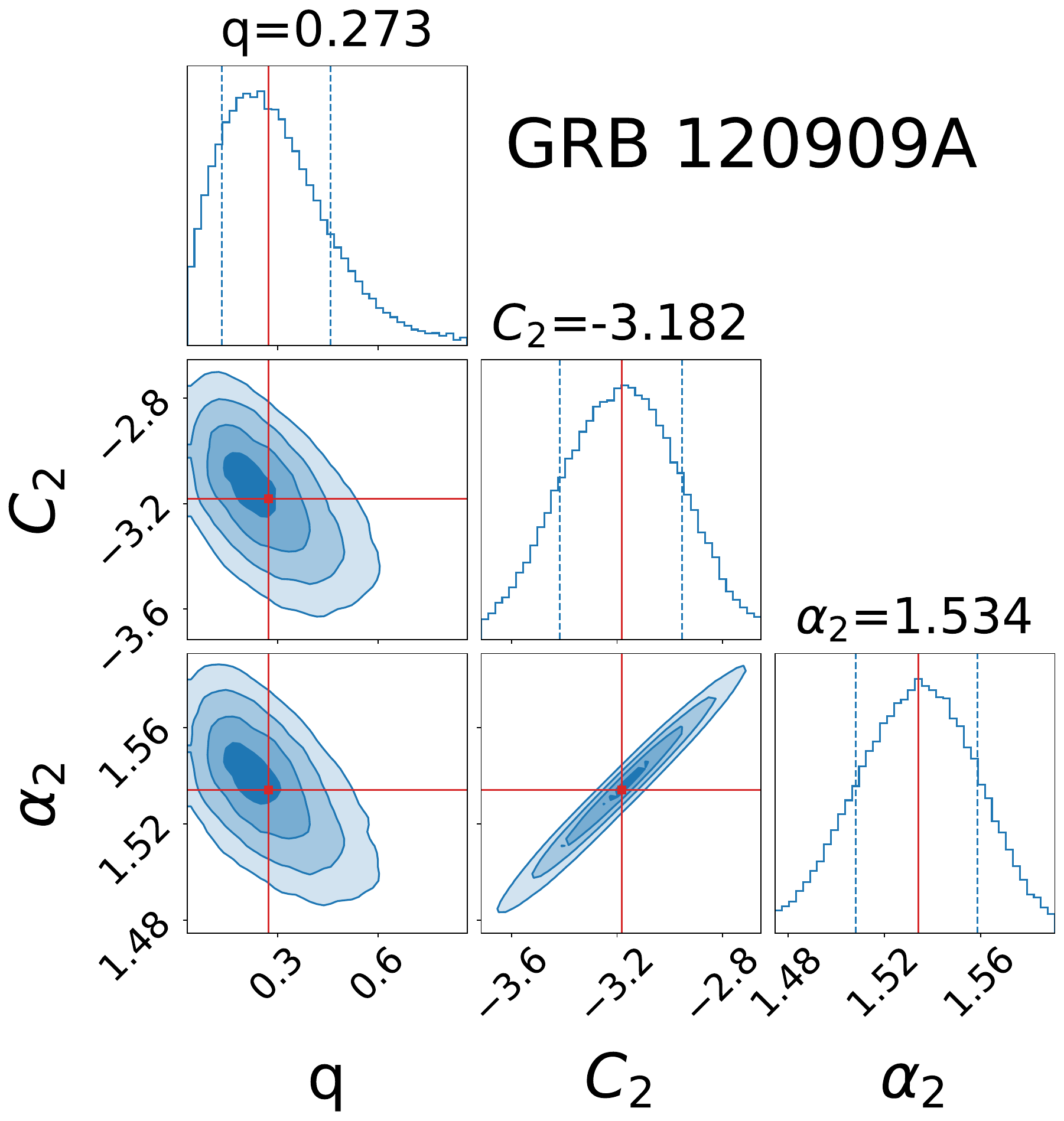}
  \end{subfigure}%
  \\%
  \begin{subfigure}[b]{0.23\textwidth}
    \includegraphics[width=\linewidth]{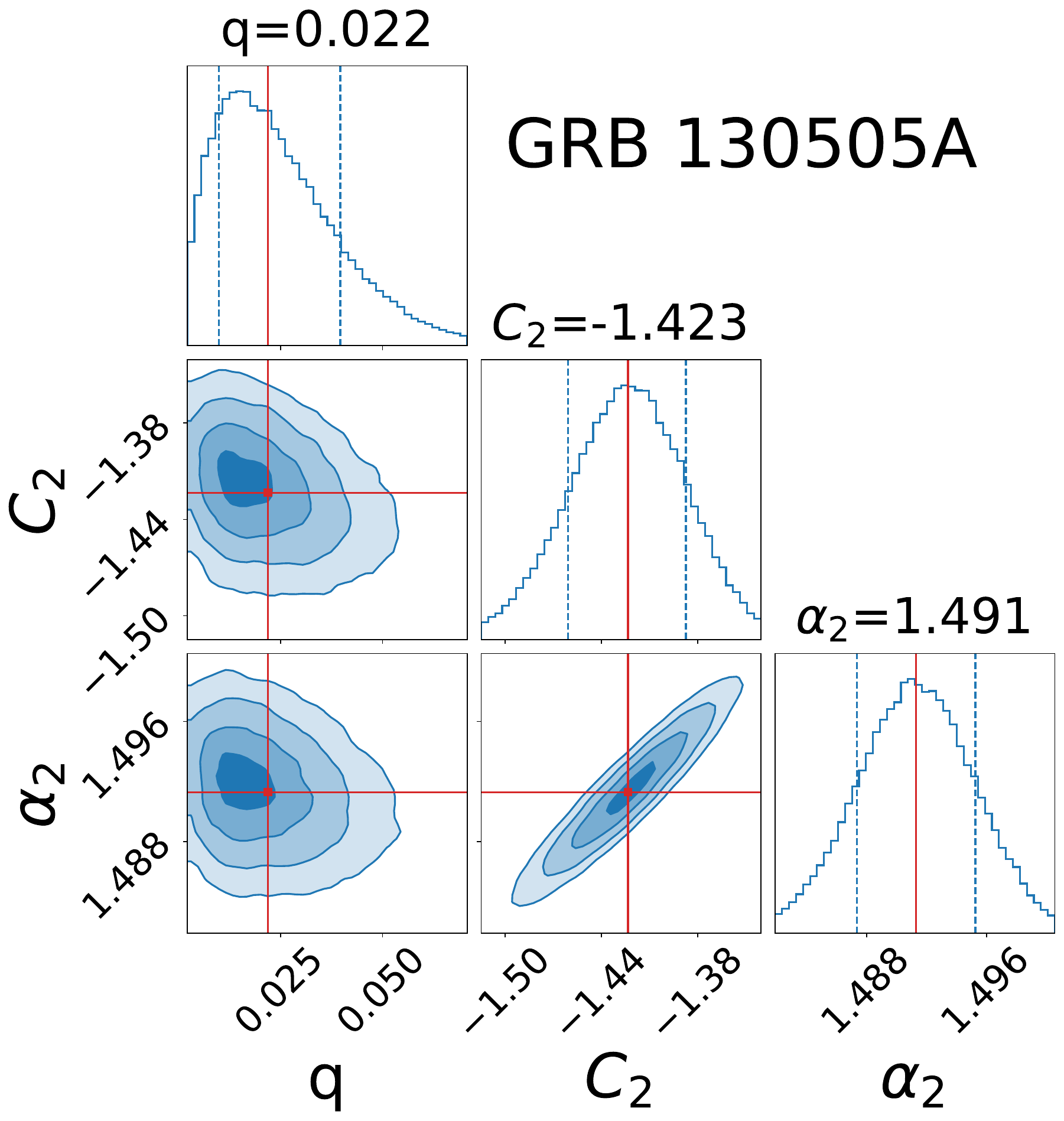}
  \end{subfigure}%
  \hspace{0.01\textwidth}
  \begin{subfigure}[b]{0.23\textwidth}
    \includegraphics[width=\linewidth]{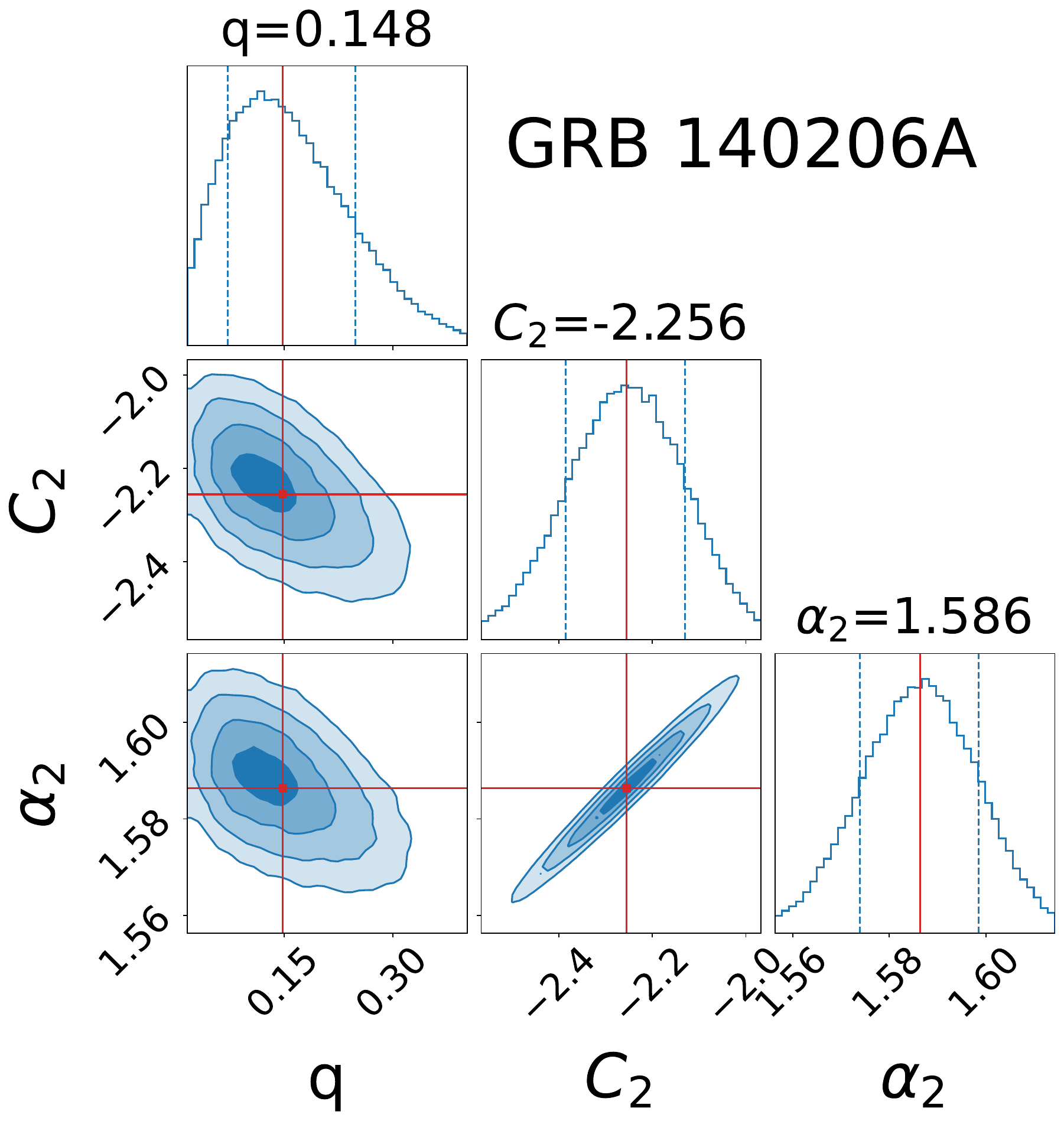}
  \end{subfigure}%
  \hspace{0.01\textwidth}
  \begin{subfigure}[b]{0.23\textwidth}
    \includegraphics[width=\linewidth]{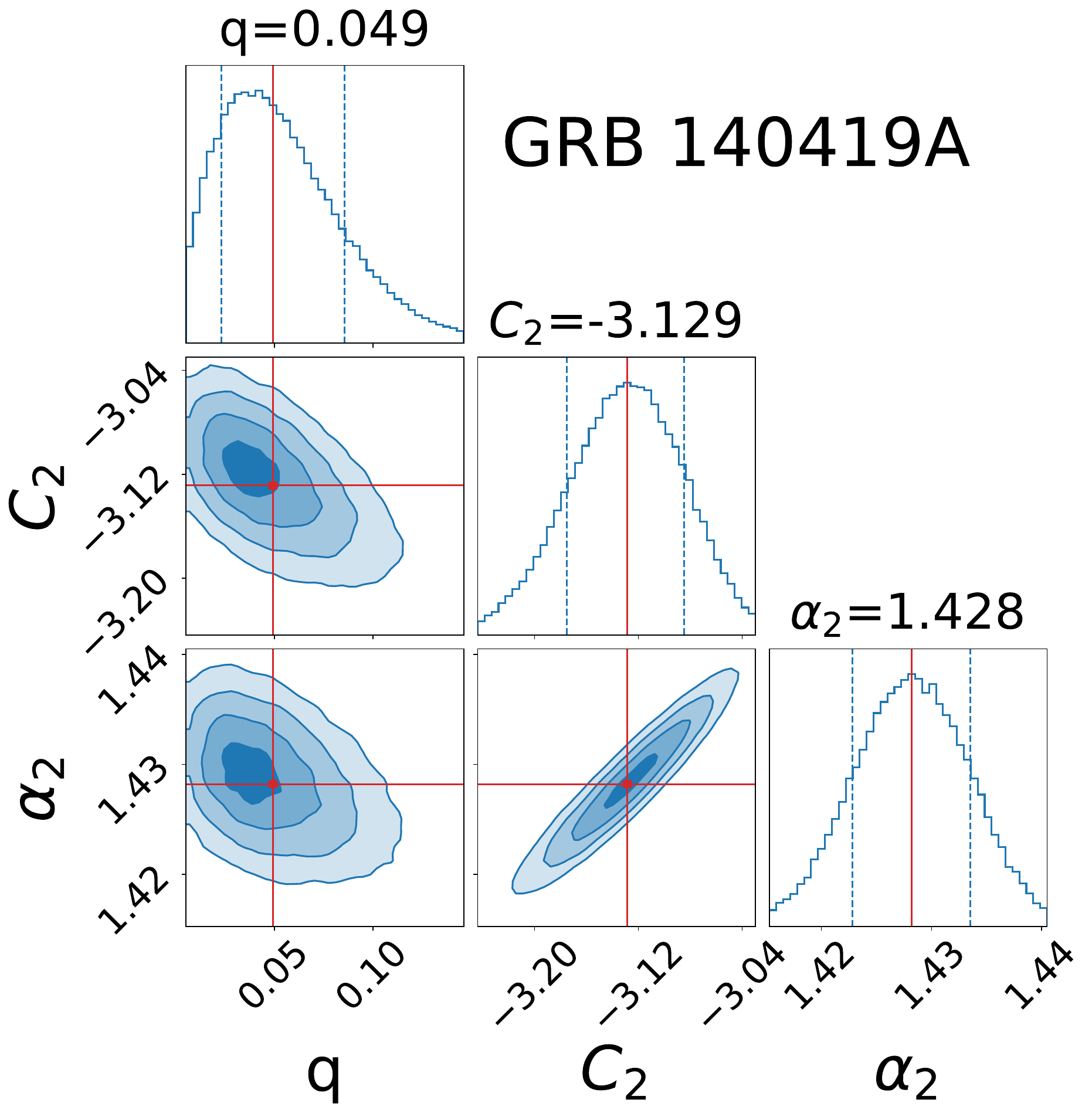}
  \end{subfigure}%
  \hspace{0.01\textwidth}
  \begin{subfigure}[b]{0.23\textwidth}
    \includegraphics[width=\linewidth]{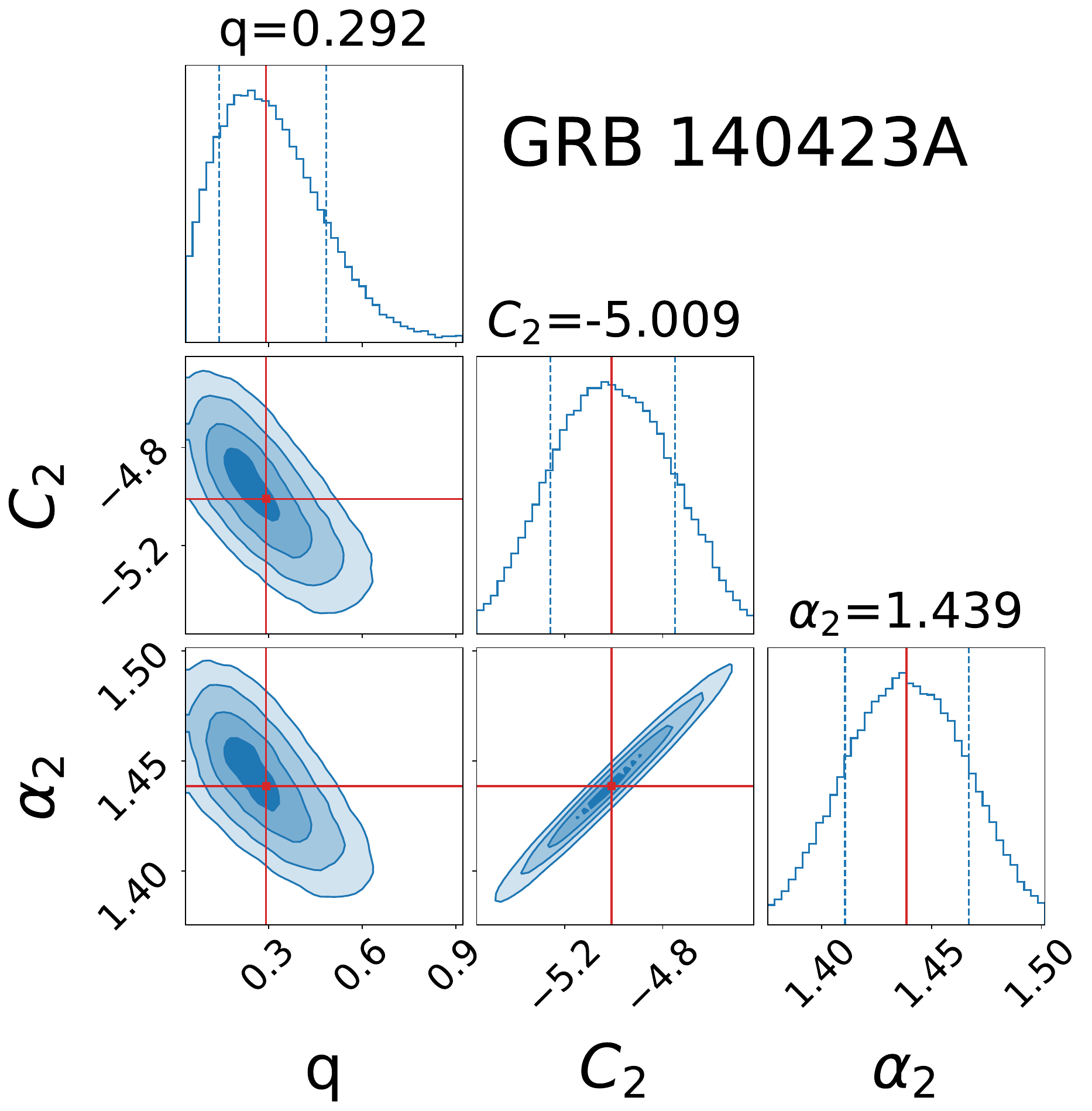}
  \end{subfigure}%
  \\%
  \begin{subfigure}[b]{0.23\textwidth}
    \includegraphics[width=\linewidth]{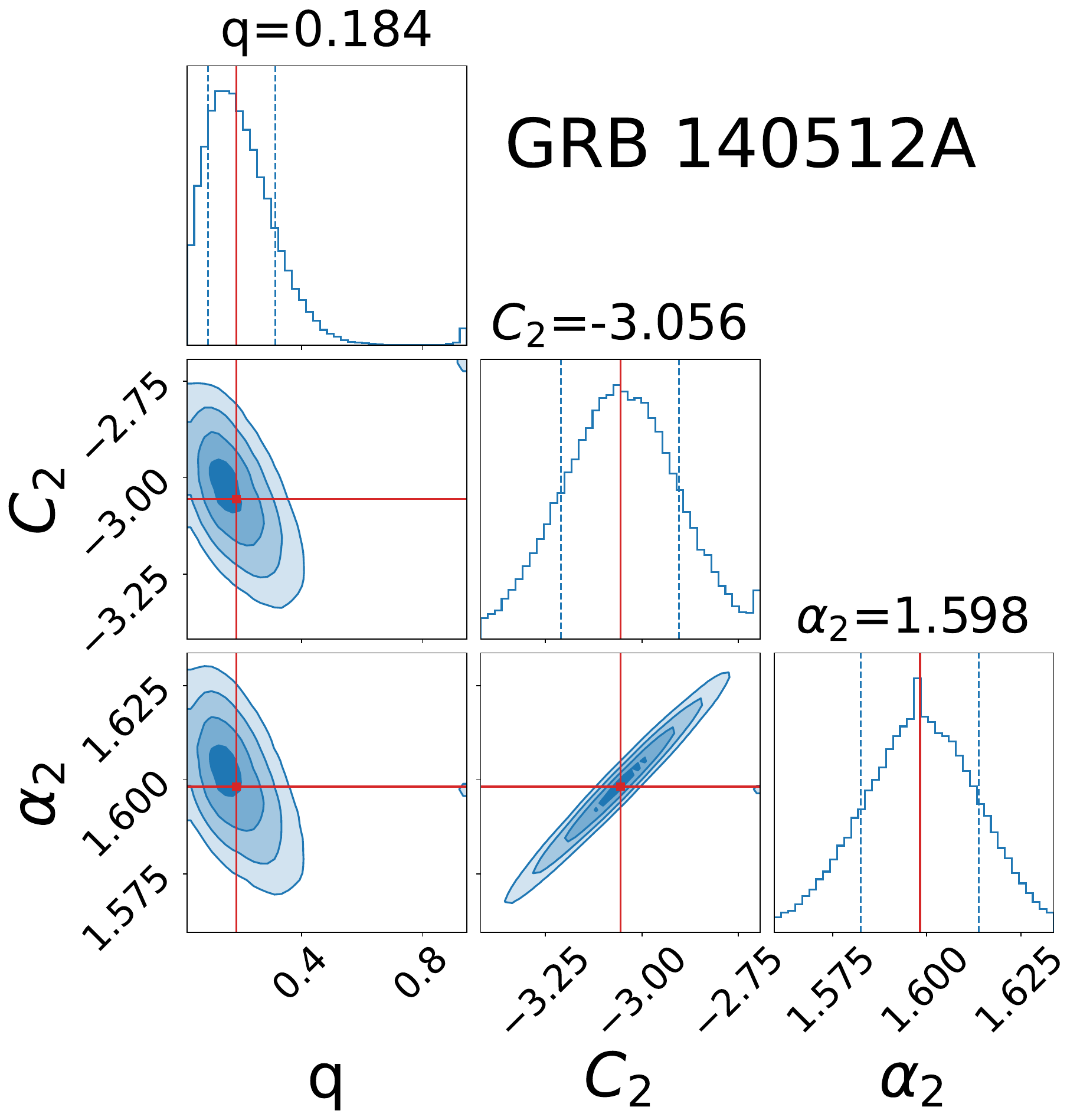}
  \end{subfigure}%
  \hspace{0.01\textwidth}
  \begin{subfigure}[b]{0.23\textwidth}
    \includegraphics[width=\linewidth]{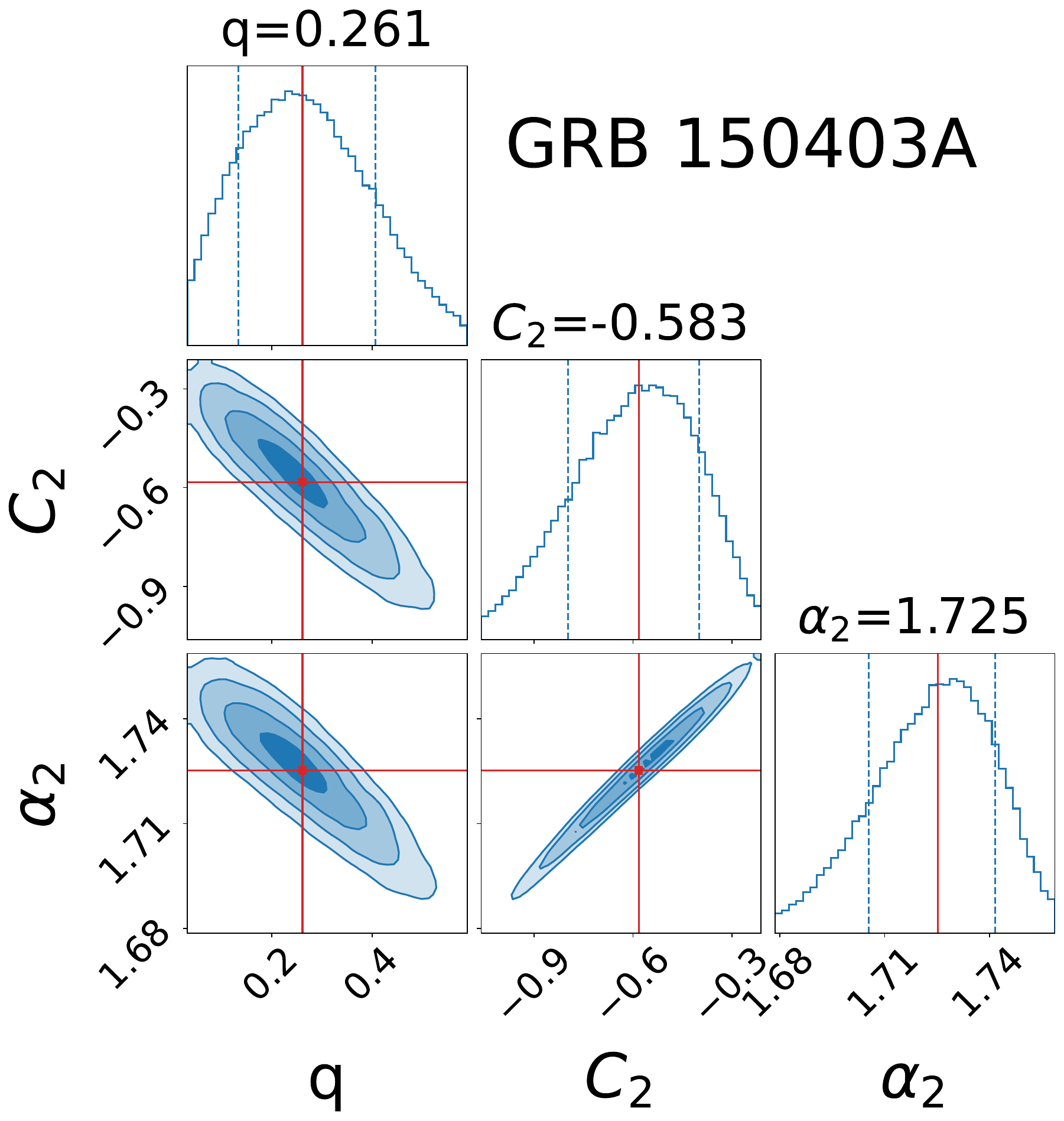}
  \end{subfigure}%
  \hspace{0.01\textwidth}
  \begin{subfigure}[b]{0.23\textwidth}
    \includegraphics[width=\linewidth]{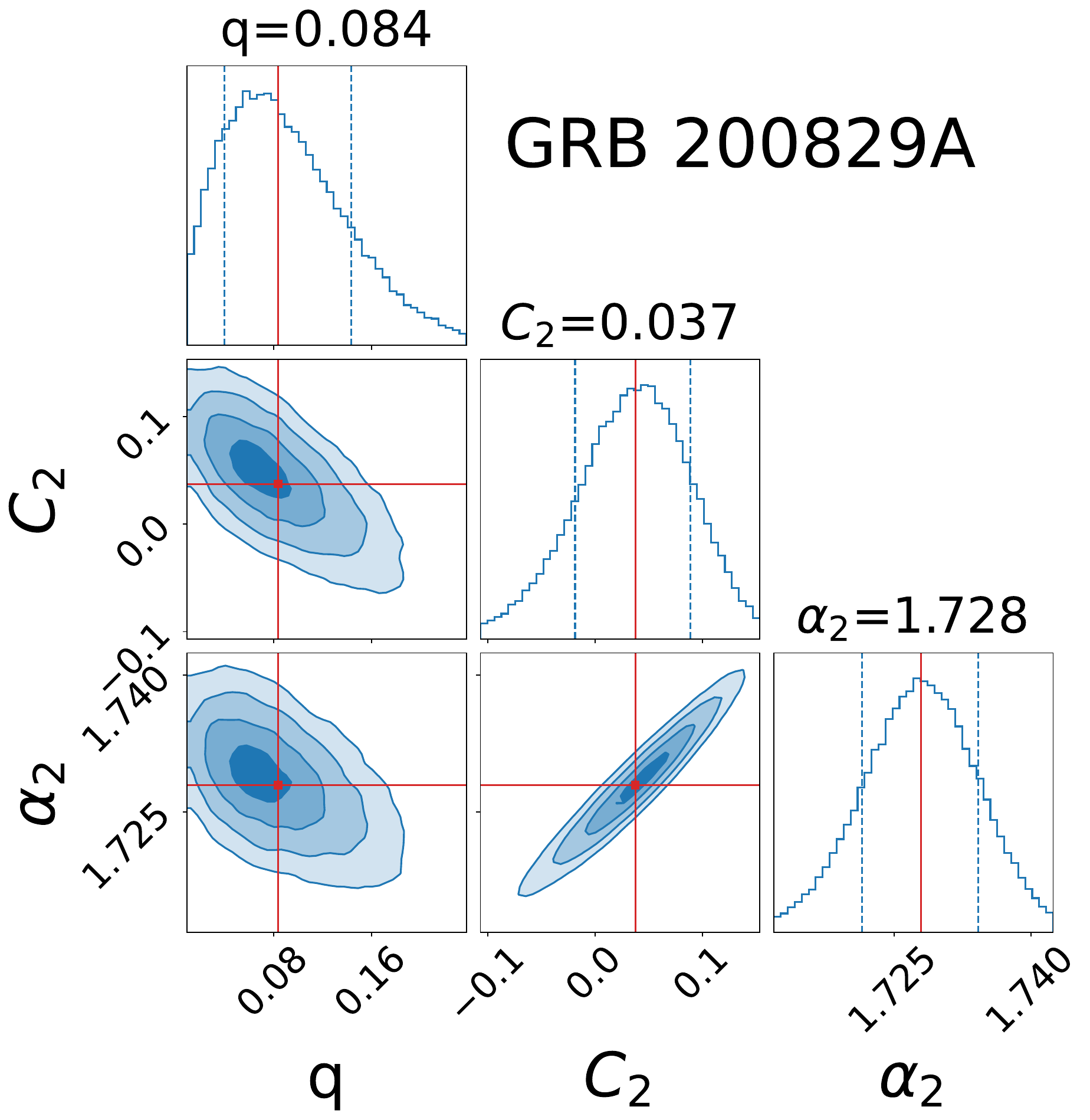}
  \end{subfigure}%
  \hspace{0.01\textwidth}
  \begin{subfigure}[b]{0.23\textwidth}
    \includegraphics[width=\linewidth]{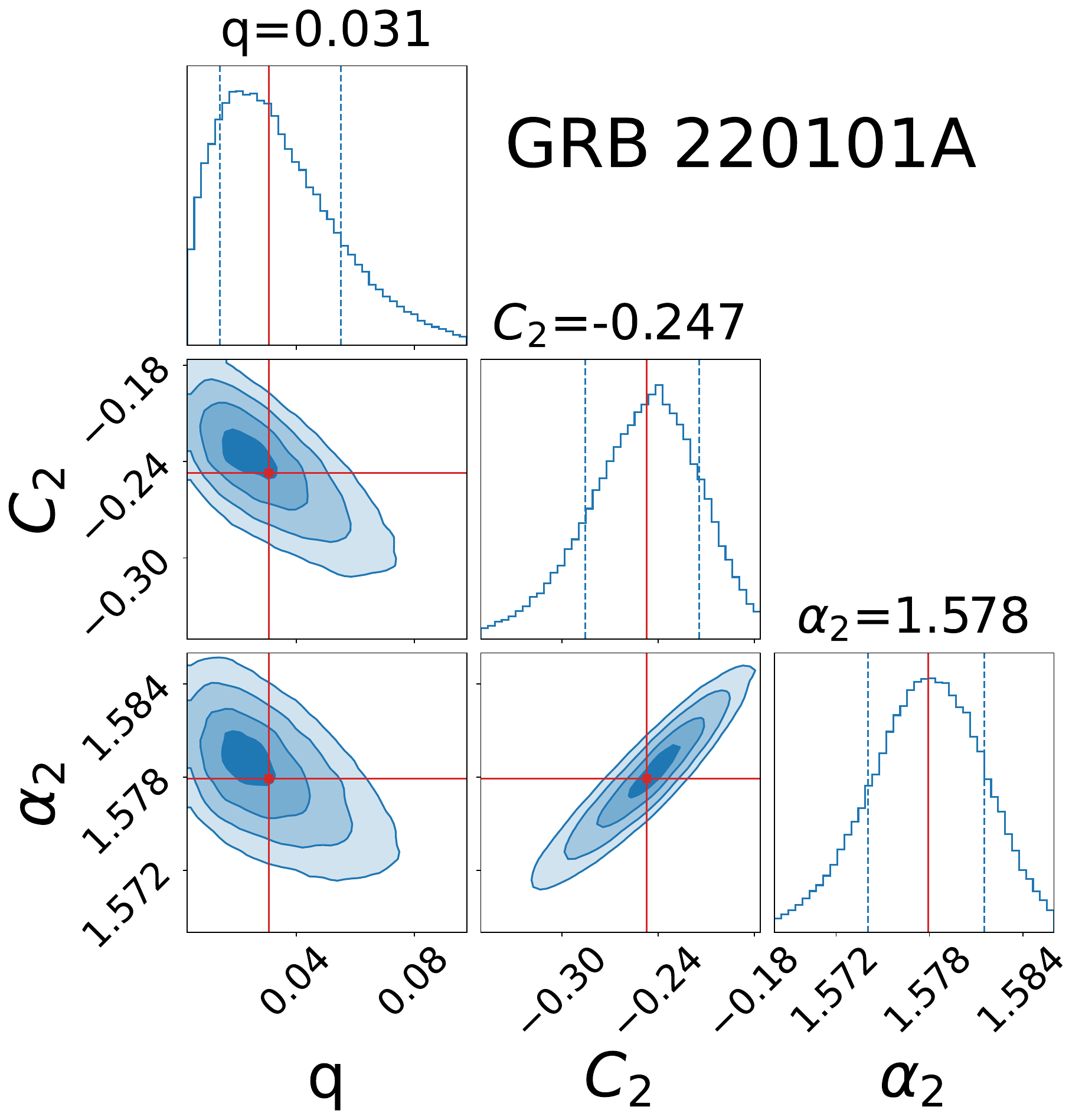}
  \end{subfigure}%
  \\%
  \caption{Posterior parameter distributions for model~2 with full consideration of high-latitude emission. Each panel displays the MCMC sampling results for key parameters including the off-axis ratio $q = \theta_{\text{obs}}/\theta_{\text{jet}}$, decay index $\alpha_2$, and normalization factor in the stellar wind environment.}
  \label{fig:wind_model2_parameter_distributions}
\end{figure}

\clearpage

\bibliographystyle{aasjournalv7}
\bibliography{sample701}

\end{document}